# Origins, Spectral Interpretation, Resource Identification, and Security–Regolith Explorer (OSIRIS-REx) Project

Sample Analysis Plan
UA-PLN-4.5.4-001
Revision 3
June 2023

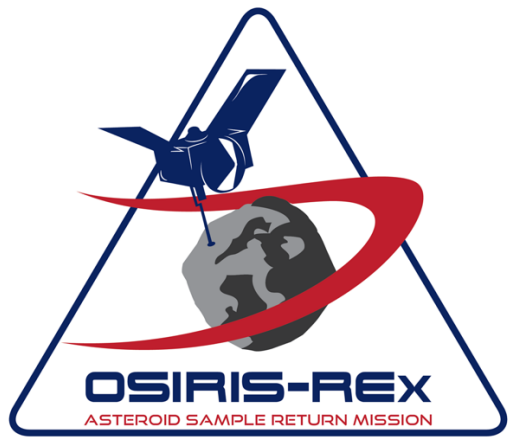

# CM FOREWORD

This document is an OSIRIS-REx PI-Office controlled document. Changes to this document require prior approval of the OSIRIS-REx Principal Investigator.

Questions or comments concerning this document should be addressed to:

OSIRIS-REx PI Office
University of Arizona
1629 E. University Blvd.
P.O. Box 210092
Tucson, AZ 85721-0092

# SIGNATURE PAGE

Original signed by:

Anjani Polit, OSIRIS-REx Mission Implementation Systems Engineer
Jul 3, 2023

Harold Connolly, OSIRIS-REx Mission Sample Scientist
Jul 9, 2023

Dante Lauretta, OSIRIS-REx Principal Investigator
Jul 9, 2023

Jeff Grossman, OSIRIS-REx Program Scientist
Jul 9, 2023

# OSIRIS-REx Sample Analysis Plan

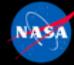 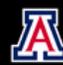 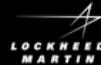

OSIRIS-REx Document

UA-PLN 4.5.4-001

June, 2023

Prepared by The OSIRIS-REx Sample Analysis Team

Sections

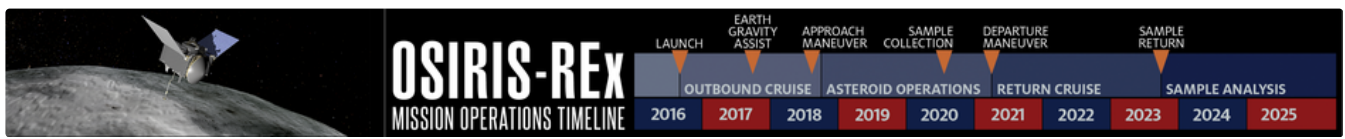

# CHANGE LOG

| Revision/Version | Description of change | Date Approved |
|---|---|---|
| Revision 1 | Initial Release | January 2022 |
| Revision 2 | Post-PS revisions and mid-SART updates<br><br>- Analytical Techniques<br>    - Deleted gamma counting technique<br>    - Added AMS (SEIWG)<br>    - Added Ion Chromatography (SEIWG)<br>    - Added PSFD (SPTAWG)<br>    - Added Lock-In Thermography (SPTAWG)<br>    - Deleted 5.48 Mini Cryogen-Free Measurement System (mCFMS) for Thermal Conductivity (SPTAWG)<br>    - Added Nanoscale Infrared Mapping (NanoIR) (MAPWG)<br>    - Added Thermogravimetric Analysis (TGA) (MAPWG)<br>    - Added Synchrotron-based X-ray Fluorescence Spectroscopy (S-XRF) (MAPWG)<br>    - Added Desorption Electrospray Ionization-Orbitrap Mass spectrometry (DESI-Orbitrap MS) (SOAWG)<br>    - Added 5.62 Synchrotron-based Infrared Spectroscopy (MAPWG)<br>    - Updated 5.5 angle of repose page (SPTAWG)<br>    - Updated 5.1 QRIS (Spectroscopy)<br>    - Added pointer from 5.2 µFTIR to VNMIR<br>    - Added Nano-EA/IRMS to 5.26 EA-IRMS<br>    - Updated 5.36 Gas Chromatography-Mass Spectrometry (GC-MS)<br>    - Updated 5.37 Liquid Chromatography-Mass Spectrometry (LC-MS)<br>    - Updated 5.38 FTICR-MS<br>    - Updated 5.41 Gas Chromatography–Isotopic Ratio Mass Spectrometry (GC-IRMS)<br>    - Updated 5.44 SLS<br>    - Updated 5.46 Differential Scanning Calorimetry<br>- Lab/facility changes<br>    - 5.4 SEM added UC Berkeley<br>    - 5.9 TEM added Molecular Foundry at Lawrence Berkeley National Laboratory<br>    - 5.9 TEM added Dr. Rolf M. Schwiete Cosmochemistry Laboratory at Goethe University Frankfurt – Frankfurt, Germany<br>    - 5.5 FIB-SEM added Molecular Foundry at Lawrence Berkeley National Laboratory<br>    - 5.7 EMPA added Stanford University<br>    - 5.13 XCT added Natural History Museum London<br>    - 5.21 HR-ICP-MS added Natural History Museum, London, UK<br>    - 5.26 Elemental Analysis - Isotope Ratio Mass Spectrometry (EA-IRMS) - The Deines Laboratory Kate Freeman<br>    - 5.36 Gas Chromatography-Mass Spectrometry (GC-MS). Brown Lab for Organic Geochemistry (BLOG), Brown University, Providence, RI, USA - Yongsong Huang (I need to harass to provide information). Furukawa Lab Tohoku University, Sendai, Japan - Yoshihiro Furukawa (I need to harass to verify information, it might be the same as in 5.41 instead)<br>    - 5.37 Liquid Chromatography-Mass Spectrometry (LC-MS). Organic Geochem. & Cosmochem. Lab, Kyushu University, Fukuoka, Japan - Hiroshi Naraoka<br>    - 5.41 Gas Chromatography–Isotopic Ratio Mass Spectrometry (GC-IRMS). Brown Lab for Organic Geochemistry (BLOG), Brown University, Providence, RI, USA - Yongsong Huang. The Deines Laboratory, The Pennsylvania State University, University Park, PA, USA - Kate Freeman. Laboratories for Stable Isotope Geochemistry, California Institute of Technology, Pasadena, CA, USA - Amy Hofmann. Chikaraishi-lab, Hokkaido University, Sapporo, Japan - Yoshihiro Furukawa<br>    - 5.56 Desorption Electrospray Ionization-Orbitrap Mass spectrometry (DESI-Orbitrap MS). Organic Geochem. & Cosmochem. Lab, Kyushu University, Fukuoka, Japan - Hiroshi Naraoka<br>    - 5.17 SIMS added Dept of Earth and Space Sciences, UCLA<br>    - 5.18 NanoSIMS added Stanford University, Washington University St Louis<br>    - 5.19 SHRIMP added Stanford<br>    - 5.2 TIMS added Lawrence Livermore National Laboratory<br>    - 5.21 HR-ICP-MS added Lawrence Livermore National Laboratory, Natural History Museum UK<br>    - 5.22 ICP-MS added Lawrence Livermore, Washington University St Louis<br>    - 5.26 EA-IRMS added Deines Laboratory at Penn State Univ<br>    - 5.36 GC-MS added Brown Lab for Organic Geochemistry, Furukawa Lab Tohoku University Sendai Japan | February 2023 |

- 5.37 LC-MS Added Organic Geochem & Cosmochem Lab Kyushu University Fukuoka Japan
- 5.41 GC-IRMS added Astrobiology Analytical Lab Goddard, Brown Lab for Organic Geochemistry, Deines Laboratory at Penn State Univ, Lab for Stagble Isotope Geochemistry Caltech, Chikaraishi-lab Hokkaido University Sapporo Japan
- 5.44 SLS added UA removed University of Calgary
- 5.46 Differential Scanning Calorimetry removed UA added Nagoya
- Other Text:
    - Updates to 9.1 Coordinated analysis of Targeted Soluble Organics (amino acids, amines, carboxylic acids, aldehydes, ketones, and insoluble cyanide)
- Sample flow diagrams updated
- Update Hypotheses
    - Hypothesis 6, 7, 8, 9, 11 – updated references to reflect papers published since first baselined
    - SPTAWG hypothesis updates
        - Hypothesis 9.3: Bennu's surface experienced mass movement that transported material from the mid latitudes to the equatorial region over the past **0.5 Ma** (Daly et al. 2020; Jawin et al. 2020, 2022).
        - Hypothesis 10.5: Bennu's two distinct boulder lithologies (Hypothesis 1.1) have been compacted and comminuted at different rates due to their different physical and mechanical properties (Cambioni et al., 2021, Rozitis et al. 2020).
        - Hypothesis 11.5: Regolith in Hokioi crater has an angle of friction of 40°. The upper 10-20 cm regolith layer of Hokioi crater has a bulk density of 440–600 kg/m3 and a cohesive strength 0.2–20 Pa (Walsh et al., 2022). The upper 60 cm layer of Hokioi crater has a bulk density of 500-700 kg/m3 and a strength against cratering of < 0.001 Pa (Lauretta et al., 2022).
- Hypothesis traceability to analytical techniques
    - Added synchrotron nano-XRF, TGA, XPS
    - Changed gamma column to AMS, added DESI-Orbitrap MS column,
    - Added hypothesis 10.5, 11.5
    - SPTAWG updates to fill in for new 10.5, 11.5 hyp. Updated a few technique titles that were outdated, added four techniques that were missing (lock in thermography, capacitance dilatometry, shear strength, PSFD).
    - Filled in missing boxes for DESI Orbitrap MS
    - Deleted half circle for uFTIR under hypothesis 3.2
    - SEIWG updates to SIMS (add hyp 2.4), TIMS (10.3, 10.4), M-ICP-MS (10.3, 10.4), AMS (11.3)
    - Changed Vi to Spectroscopy, added new acronyms, moved VNMIR to Spectroscopy
    - Added synchrotron IR spectroscopy, changed VI to VI/QRIS
- WG traceability to hypotheses
    - Added hypothesis 10.5, 11.5
    - Updated hypothesis testing to match other traceability table
- Sample Security Plan added to appendices
- Mass allocation table updates
    - Removed overguide table
    - Added new thin section for SwRI (in Spectroscopy section)
    - Moved XCT from Physical to Pet
    - Added thermal diffusivity and conductivity, shear strenght to physical section
    - Updated team members in physical section
    - SOAWG updates
    - Updated row numbers and row references
- Overguide plan
    - Removed mass table
    - Added overguide process

| | |
|---|---|
| Rev 3 | Rev 3 changes |
| | - 5.64 new SOAWG analytical technique similar to LCMS |
| | - 5.62 Synchrotron-based Infrared Spectroscopy - additional information to fully populate the new analytical technique page. |
| | - Update facility information on 5.60 TGA with more information on the Natural History Museum instruments. |
| | - Updates to µGIS/spatial sections: |
| |     - Section 10.0: |
| |         - GIS transformations unique to thjs system: Removed obsolete nomenclature about "look-up" tables and replaced with details about World Files. |
| |         - Visualization Capabilities: Made clarifications about viz-tool software, specifically that the primary tool will be through ArcgGIS Enterprise, not ArcMap. |
| |         - Required Coordinate System Transformation Metadata: Removed out-of-date nomenclature and streamlined things a bit. |
| |     - Section 11.0: |
| |         - Coordinate Systems and Data Registration: |
| |             - Removed reference to 3D coordinate system |
| |             - Updated/added graphics |
| |             - Small tweaks to descriptions |
| |         - Spatial Data Registration: |

- - - Updated/added graphics to reflect current fiducial design and implementation.
    - Updated description to correctly describe current fiducial design and implementation.
  - Data Visualization:
    - Updated graphics to reflect ArcGIS Enterprise interface, which was not set up during the last draft of this document.
    - Small tweaks to descriptions.
- Updates to SPTAWG Technique Descriptions
  - Structured light scanning (SLS) - small updates to all sections, especially sample impact and data quality. Updated stated acceptable sample size to "approximately 3 mm to 4 cm."
  - Angle of Repose (ARM) - updates to all sections to reflect focus on passive angle of repose measurement.
  - Direct Shear Strength Measurement - small updates to all sections to clarify that measurement will not record displacement data.
- Updated 5.56 to add the potential for Hofmann instrument augmentation using the same or related technique.
- Updated 5.61 S-XRF description to include example figures for 1- and 2-D data sets
- (SEIWG) added 5.65 Tof-SIMS to the Analytical Techniques section
- (SEIWG) made small edits to baseline mass table (no masses changed, only details), technique traceability table, and flow diagram (edits sent to the owners of the respective documents).
- SSAWG edits on epoxy and polished samples
- Updated instrument in 5.4 SEM,. 5.15, Raman, 5.26 EA-IRMS, 5.36 GCMS, and 5.41 GC-IRMS
- Added a citation [Hamilton et al., 2022] to Hypotheses 1.2 and 1.4 and to the references.
- Added link to SOAWG detailed sample flow diagram in §9
- SAMIS updates in Section 10 to reflect the latest developments and plans.
- Removed defunct section on SAMIS (content was merged into Section 10) and renumbered the following sections.
- Minor edits to SART page to remove out of date information.
- Minor edits to overguide page to add reference to sample allocation form and clarify that WGL, PI, and Curation need to approve the requests.
- Hypothesis to analytical technique traceability updates
  - Added Ar/Ar, CE, TOF-SIMS
- Changed duplicate 5.59 technique to 5.66
- Added PSP information to appendix
- Added glovebox exhaust volatile capture proposal to appendix

# 1.0 INTRODUCTION AND SCOPE

The Origins, Spectral Interpretation, Resource Identification, and Security–Regolith Explorer (OSIRIS-REx) spacecraft arrived at its target, near-Earth asteroid (101955) Bennu, in December 2018. After one year of operating in proximity, the team selected a primary site for sample collection. In October 2020, the spacecraft descended to the surface of Bennu and collected a sample. The spacecraft departed Bennu in May 2021 and will return the sample to Earth in September 2023 (Lauretta et al. 2017, 2021). The analysis of the returned sample will produce key data to determine this B-type asteroid's history and that of its components and precursor objects. The main goal of the OSIRIS-REx Sample Analysis Plan (SAP) is to provide a framework for the Sample Analysis Team (SAT) to meet the following Level-1 mission requirement:

- **L1.5b** *Analyze the returned sample to determine:*
  - *presolar history*
  - *formation age*
  - *nebular and parent-body alteration history*
  - *relation to known meteorites*
  - *organic history*
  - *space weathering*
  - *resurfacing history, and*
  - *energy balance in the regolith of Bennu.*

To achieve this goal, this plan: (i) establishes a hypothesis-driven framework for coordinated sample analyses; (ii) defines the analytical instrumentation and techniques to be applied to the returned sample; (iii) provides guidance on the analysis strategy for baseline, overguide, and threshold amounts of returned sample, including a rare or unique lithology; (iv) describes the data storage, management, retrieval, and archiving system; (v) establishes a protocol for the implementation of a micro-geographical information system to facilitate co-registration and coordinated analysis of sample science data; (vi) outlines the plans for Sample Analysis Readiness Testing (SART); and (vii) provides guidance for the transfer of samples from curation to the SAT.

The SAP also addresses the supporting Level-1 mission requirements:

- **L1.2**: *Document the contamination of the sample acquired from collection, transport, curation, and distribution*
- **L4.6:** *Science Data Management – The OSIRIS-REx PI shall be responsible for initial analysis of the scientific data, its subsequent delivery to an appropriate data repository, the publication of scientific findings, and communication of results to the public. Additionally, the OSIRIS-REx PI shall be responsible for collecting engineering and ancillary information necessary to validate and calibrate the scientific data prior to depositing it in a NASA approved data repository. The OSIRIS-REx science data base shall be made available to the science community without restrictions or proprietary data rights of any kind.*

The OSIRIS-REx mission will return the following types of samples:

- At least 60 g of bulk regolith, per mission requirements
- Surface particles adhering to 24 contact pads (2.4 cm$^2$ each)
- Witness coupons, mounted in strategic locations to document any contamination collected in flight during different phases of the mission
- The sampler head of the spacecraft's Touch-and-Go Sample Acquisition Mechanism (TAGSAM; Bierhaus et al. 2018), which includes the annular collection chamber that holds the bulk sample, encircled by the contact pads on the baseplate
- The Sample Return Capsule (SRC), in which the TAGSAM head and samples are stowed (Bierhaus et al. 2018)

Of the asteroidal material returned from Bennu, 75% will be archived by NASA, and the remaining 25% will be available to the OSIRIS-REx Sample Analysis Team (SAT), including any Participating Scientists, for immediate coordinated analysis.

This plan focuses on analysis of the bulk regolith sample, the surface contact-pad samples, the witness coupons, and the TAGSAM head. The SRC analysis plan is outlined in a separate document: SAMPLE RETURN CAPSULE COMPONENTS SCIENCE AND ENGINEERING

ANALYSIS PLAN - PART 2. This plan will only focus on SRC analysis related to measurements of collected sample that is adhering to SRC components in the event of a threshold scenario and to assess the effects of return cruise and Earth entry on the samples.

The "Baseline" Program is designed to meet the Level-1 requirement based on analysis 15 g of bulk sample and three contact pads (7.2 cm$^2$), equivalent to 25% of the mission requirement to return 60 g of bulk sample and 26 cm$^2$ contact surface area (~12 pads).

The "Overguide" Program becomes feasible if substantially more bulk sample and contacted surface area are returned. For planning purposes, the initial Overguide Plan was based on analysis of 62.5 g of bulk sample and six contact pads; that is, it assumes that a total of 250 g bulk sample was collected and all 24 pads made surface contact. These assumptions are based on spacecraft observations during and after the sampling event (Ma et al. 2021).

The "Threshold" Program is driven by analysis of 3.75 g and one contact pad. The Threshold Program covers contingencies in which sample is lost due to an anomaly during Earth return. This program is also applicable to characterization of a minor lithology in a nominal return scenario.

These three analyses programs are designed to provide the mission with flexibility based on the nature of the returned sample. It is conceivable that elements of the Baseline, Overguide, and Threshold plans will all be implemented.

The SAT will follow two guiding principles throughout the analysis process. First, in order to maximize the scientific return of the samples, the science team will coordinate its efforts and analyses. This philosophy means that as analyses are performed, the resulting data products will be delivered to the project for ingest into the sample science database. This centralized data storage ensures that all information from sample analysis is available to all members of the team. Second, the team will prepare for the unexpected. Regardless of how well we plan, we expect to make new discoveries and to continue to be surprised by Bennu, as we were during the spacecraft encounter (e.g., Lauretta & DellaGiustina et al. 2019). To better deal with the unexpected, we define a means of how we will make decisions about such situations and detail a protocol that will be followed.

# 2.0 OSIRIS-REx APPROACH TO SAMPLE ANALYSIS

Laboratory studies of astromaterials have yielded detailed insights into the properties, origins, and evolutionary histories of a wide range of Solar System bodies, including the Moon, Mars, comets, and asteroids. Yet, the parent bodies of meteorites and cosmic dust are generally unknown, genetic and evolutionary relationships among asteroids and comets are unsettled, and links between laboratory and remote observations lack ground truth. The OSIRIS-REx mission will shed light on such relationships by enabling detailed and coordinated laboratory analyses of material from asteroid (101955) Bennu for which the geological context is well-characterized thanks to OSIRIS-REx observations.

To achieve the mission's driving Level-1 requirement to return and analyze a sample of Bennu, we describe herein an hypotheses-driven Sample Analysis Plan (SAP) to constrain the origin and history of asteroid Bennu and its source materials. Each hypothesis is mapped to either the fundamental nature of Bennu as determined from spacecraft remote sensing and astronomical characterization or to a specific period in Bennu's history.

The hypotheses are organized into two levels:

- Level-1 hypotheses, which are fundamental to mission science and are based on:
    - observational data from astronomical characterization and spacecraft-based instruments
    - our understanding of carbonaceous chondrites, specifically those that are close spectral analogs of Bennu
    - models of Solar System formation and evolution
- Level-2 hypotheses:
    - Derive from Level-1 hypotheses
    - Map to the Working Groups (WGs) of the Sample Analysis Team (SAT)
    - Are traceable to distinct testable measurement objectives

These hypotheses, detailed in Section 3.0, relate to the origin and history of Bennu, its parent asteroid, and the source materials present in the protoplanetary disk. It is our intent to test these hypotheses through the analyses of returned samples including both the bulk sample and the material from the contact pads. In some cases, evaluation of the hypotheses will require sample analysis data combined with spacecraft-based information and/or models of asteroid formation and evolution.

# 3.0 DRIVING HYPOTHESES

This plan is driven by hypotheses about Bennu's composition, origin, formation, and evolution, as well as the effects of collection, Earth return, and recovery on the samples. The level-1 and level-2 hypotheses is based upon our best understanding of Bennu and its most likely counterparts among meteorites and cosmic dust samples. They consider the full historical arc of asteroid Bennu, beginning with stellar nucleosynthesis and dust formation in ancient stars, through chemical processes in the Galaxy and nascent solar nebula, planetary formation, and geological history.

The major thematic areas are: the Fundamental Nature of Bennu, the Pre-accretionary Epoch, the Parent-Body Epoch, the Bennu Epoch, and the OSIRIS-REx Epoch. Sample analyses related to the Fundamental Nature of Bennu test the predictions of the returned sample as determined by telescopic and spacecraft-based remote sensing. Pre-accretionary Epoch studies quantify and characterize materials that predate the formation of the Solar System, such as circumstellar dust grains and interstellar organic materials. In addition, these analyses include determining where and how mineral and organic components formed in the protoplanetary disk. The Parent-Body Epoch considers ancient planetary processes within Bennu's parent asteroid, such as the timing and mode of aqueous and thermal alteration and whether telltale clues in the samples tell us where Bennu's components were housed in the parent body prior to disruption. The Bennu Epoch addresses how Bennu's surface materials formed after Bennu was liberated from its parent body and were subsequently altered by exposure to space. It includes study of how Bennu came to be in its current orbital configuration from processes such as YORP and Yarkovsky effects. It also focuses on surface alteration resulting from exposure to the space environment, connecting the detailed microstructures at grain surfaces to space-weathering processes. Finally, OSIRIS-REx Epoch studies characterize sample modification resulting from interaction with the spacecraft components and the curation facility. Tables mapping each hypothesis to the relevant analytical techniques are presented at the end of this section.

## The Fundamental Nature of Bennu

**Hypothesis 1: Remote sensing of Bennu's surface has accurately characterized its mineral, chemical, and physical properties.**

**Hypothesis 2: Bennu contains prebiotic organic compounds.**

## The Pre-accretionary Epoch: The Pre-accretionary Environment and Formation of Bennu's Parent Asteroid

**Hypothesis 3: Bennu contains presolar material derived from diverse sources.**

**Hypothesis 4: Bennu's parent asteroid formed beyond the snow line by accretion of material in the protoplanetary disk.**

## The Parent-Body Epoch: Scattering into the Inner Solar System, Geological Activity, Surface Modification, and Catastrophic Disruption

**Hypothesis 5: Geological activity occurred in the interior of Bennu's parent asteroid early in solar system history.**

**Hypothesis 6: Bennu's parent body experienced >3 billion years of solar system history before being destroyed in a catastrophic disruption.**

## The Bennu Epoch: Bennu's Formation, Dynamical, and Geological Evolution

**Hypothesis 7: Bennu is a rubble pile that formed by re-accumulation of material from the catastrophic disruption of a precursor asteroid.**

**Hypothesis 8: The Yarkovsky effect pushed Bennu far enough inward to reach a dynamical resonance, which flung it out of the main belt and onto a terrestrial planet–crossing orbit.**

**Hypothesis 9: Bennu has experienced surface processing throughout its history.**

**Hypothesis 10: The physical, chemical, and spectral properties of Bennu's surface materials have been modified by exposure to the space environment.**

**Hypothesis 11: The Hokioi Crater in which the Nightingale sample site is located was recently formed and contains relatively unweathered material.**

## The OSIRIS-REx Epoch: Sample Collection and Earth Return

**Hypothesis 12: OSIRIS-REx asteroid proximity operations, the TAG event, and Earth return modified the collected samples, TAGSAM, and the SRC.**

## Traceability Matrices

**Hypotheses mapped to analytical techniques**

**Hypotheses mapped to the sample science working groups**

## 3.1. Hypothesis 1: Remote sensing of Bennu's surface has accurately characterized its mineral, chemical, and physical properties.

### Hypothesis 1.1: Bennu's surface primarily consists of two distinct lithologies defined by their normal albedo, thermomechanical properties, and spectral slope.

- Normal albedo (DellaGiustina et al. 2020)
  - Low reflectance "dark" boulders (3.5 to 4.9% normal reflectance)
  - High reflectance "bright" boulders (4.9 to 7.4% normal reflectance)
- Thermal properties (Rozitis et al. 2020b)
  - Low thermal inertia (~180–250 J $m^{-2}$ $K^{-1}$ $s^{-1/2}$)
  - Higher thermal inertia (~400–700 J $m^{-2}$ $K^{-1}$ $s^{-1/2}$)
- Spectral slope from 1.0 to 2.5 µm (Barucci et al. 2020)
  - Blue mean spectrum (98% of the surface)
  - Less blue with a flatter slope (2% of the surface)
  - Note: spectra split into two groups at 2.7σ confidence level

### Hypothesis 1.2: Bennu's bulk mineralogy and chemistry consist of:

- Mg-rich phyllosilicates (Hamilton et al. 2019, 2021, 2022)
- Magnetite (Lauretta and DellaGiustina et al. 2019; Hamilton et al. 2019, 2021, 2022)
- Mixtures of organics and carbonates (Simon et al. 2020)
  - Total organic carbon >1 wt.% (Kaplan et al. 2021)
  - Three distinct carbonates: calcite, dolomite-breunnerite, and magnesite (Kaplan et al. 2020)
- Less than about 10 vol% anhydrous silicates (Hamilton et al. 2021, 2022)

### Hypothesis 1.3: Bennu's surface contains minor amounts of exogenous material from diverse sources including:

- Material with the same distinctive pyroxene composition as the howardite–eucrite–diogenite (HED) meteorites that originate from asteroid (4) Vesta (DellaGiustina and Kaplan et al. 2021)
- Material with spectral properties similar to those of ordinary chondrite meteorites that are linked to S-type asteroids (Le Corre et al. 2021)

### Hypothesis 1.4: Bennu's dominant lithologies are comparable in bulk mineralogy, petrology, and composition to the most aqueously altered carbonaceous chondrites [petrologic types ≤2.4 in the Rubin et al. (2007) classification, or types 1 and 2 in the Van Schmus and Wood (1967) classification] (Hamilton et al. 2019, 2021; Goodrich et al. 2019; Praet et al. 2021; Kaplan et al. 2020; Hamilton et al. 2022).

### Hypothesis 1.5: Bennu's surface contains material with physical properties that are distinct from characterized meteorites.

- The low-reflectance, low-thermal-inertia boulders have estimated porosities of ~49–55% and tensile strengths of ~0.10–0.15 MPa (Rozitis et al. 2020b).

- The high-reflectance, higher-thermal-inertia boulders have estimated porosities of ~24–38% and tensile strengths of ~0.31–0.78 MPa (Rozitis et al. 2020b).
- Centimeter-scale particles have compressive shear strengths of 5±3 MPa. Particle strength scales as a function of size, with a power law coefficient of –0.25 (Ballouz et al. 2020).
- Carbonates occur in large veins >1 cm thick and tens of centimeters in length (Kaplan et al. 2020).

## 3.2. Hypothesis 2: Bennu contains prebiotic organic compounds.

Hypothesis 2.1: Bennu contains organic molecules that are used in terrestrial biology including protein amino acids, peptides, polyols, purines and pyrimidines, amphiphiles, carboxylic acids, their precursors, and side products from chemical reaction networks (Deamer and Pashley 1989; Shimoyama and Ogasawara 2002; Callahan et al. 2011; Furukawa et al. 2019; Glavin et al. 2021).

- Bennu contains frozen chemical reaction networks, catalytic cycles, autocatalytic cycles, or pathways. Such pathways include organic/inorganic chemistry, photochemistry, and radical chemistry and reactive species such as nitriles, amides, and nitrogen heterocycles. (Aponte et al. 2017).

Hypothesis 2.2: Enantiomeric excesses are present for some chiral molecules and are of the same handedness as found in life (i.e., excesses of L-amino acids and D-sugars) (Cooper and Rios 2016; Glavin et al. 2021).

Hypothesis 2.3: Bennu contains soluble organic matter (SOM) with abundances, distributions, isotopic, and enantiomeric compositions that are similar to the most aqueously altered CI and CM chondrites (Cronin and Chang 1993; Sephton 2002; Schmitt-Kopplin et al. 2010; Glavin et al. 2018; Simon et al. 2020; Kaplan et al. 2021), specifically:

- Amino acid abundances (both prebiotic and abiotic) of 100 to 10,000 ppb (Glavin et al. 2021)
- Aliphatic organics with H/C > 0.3 (Kaplan et al. 2021)
- Polycyclic aromatic hydrocarbons (± peripheral functional groups)
- Highly sulfurized SOM (Schmitt-Kopplin et al. 2010)
- Nitrogen-rich molecules (Schmitt-Kopplin et al. 2010)
- Organometallic compounds (Ruf et al. 2017).

Hypothesis 2.4: Bennu contains insoluble organic matter (IOM), a chemically complex, macromolecular material, with similar elemental and isotopic compositions, structures, and morphologies as that in the most aqueously altered CI and CM chondrites (Cody and Alexander 2005).

## 3.3. Hypothesis 3: Bennu contains presolar material derived from diverse sources.

Hypothesis 3.1: The abundance, isotopic, and chemical compositions of presolar materials in the returned sample are similar to those in the CI and CM chondrites (e.g., Davidson et al. 2014; Zinner 2014)

Hypothesis 3.2: Bennu contains dust grains that condensed in the gaseous envelopes around ancient stars or the ejecta of novae or supernovae (Lauretta et al. 2015).

Hypothesis 3.3: Bennu contains dust grains that formed in the interstellar medium or were modified by processing in the interstellar medium (Lauretta et al. 2015).

Hypothesis 3.4: Bennu contains organic matter that formed in the interstellar medium (Lauretta et al. 2015; Busemann et al. 2006). Such organics are carriers of excess D, $^{15}$N, and/or $^{13}$C resulting from photochemical processing of organics and their precursors at low temperatures (Kerridge 1983; Alexander et al. 2007).

## 3.4. Hypothesis 4: Bennu's parent asteroid formed beyond the snow line by accretion of material in the protoplanetary disk.

Hypothesis 4.1: The initial constituents of Bennu's parent asteroid formed over several million years of solar system history, starting at a $t_0$ of 4.567 Gyr (Kita et al. 2013; Pape et al. 2019; Connelly et al. 2012)

Hypothesis 4.2: Bennu's bulk elemental composition reflects that of its main parent asteroid and is similar to the composition of the Sun, with depletions in moderately to highly volatile elements (e.g., Braukmüller et al. 2018; Wood et al. 2019).

Hypothesis 4.3: The initial constituents of Bennu's parent asteroid were materials that were inherited from the protosolar molecular cloud or were formed and altered in the protoplanetary disk, largely composed of:
- presolar grains
- chondrules
- refractory inclusions
- crystalline and amorphous silicates
- metals
- sulfides
- oxides
- phosphates
- organic compounds (SOM and IOM)
- ices (e.g., $H_2O$, $CO_2$, $NH_3$, etc.)

Hypothesis 4.4: Bennu's parent asteroid accreted in the outer protoplanetary disk, beyond Jupiter, as recorded by distinct isotopic anomalies in a variety of elements (e.g., O, Ca, Ti, Cr, Ni, Zr, Mo, Ru, Pd, Ba, Nd, etc.) (Rowe et al. 1994; McKeegan et al. 1998; Trinquier et al. 2007; Leya et al. 2008; Warren 2011; Connelly et al. 2012; Kruijer et al. 2020; Mezger et al. 2020; Schrader et al. 2020).

## 3.5. Hypothesis 5: Geological activity occurred in the interior of Bennu's parent asteroid early in solar system history.

Hypothesis 5.1: Bennu's parent asteroid was heated by the decay of short-lived radionuclides for ~10 Myr after accretion (e.g., Miyamoto 1991; Steele et al. 2017).

Hypothesis 5.2: Bennu's parent asteroid experienced impact processing over its geological history, resulting in brecciation, impact shock, localized heating, and possibly asteroid-scale heating (Rubin et al. 1995; Walsh et al. 2019; DellaGiustina and Emery et al. 2019; DellaGiustina et al. 2020).

Hypothesis 5.3: Heating of Bennu's parent asteroid resulted in hydrothermal alteration in which melted ice reacted with the initial constituents (Hypothesis 4.3) at 25-375°C, establishing the current bulk mineralogy (Hypothesis 1.2) and organic chemistry (Hypothesis 2.3 & 2.4; Ghosh et al. 2006; Krot et al. 2015; King et al. 2017; Hamilton et al. 2019; Praet et al. 2021).

- Some hydrothermal alteration occurred in a closed system with elemental transport limited to small distances (millimeter to centimeter scales) (Braukmüller et al. 2018; Hamilton et al. 2021; Krietsch 2020).
- Other alteration occurred in a large, open system with elemental transport spanning meters to kilometers (Young et al. 1999; Kaplan et al. 2020).
- Vein deposition in open-system hydrothermal conditions decreased porosity and increased thermal conductivity by filling pore spaces and binding clasts together (Kaplan et al. 2020; Rozitis et al. 2020b).
- SOM was modified during hydrothermal alteration, including mineral/organic reactions, synthesis of new soluble organic compounds, polymers, and amplification of enantiomeric excesses (Hayatsu et al. 1980; Yabuta et al. 2007; Glavin et al. 2018; Pizzarello and Williams 2012).
- IOM formed through the polymerization of organic molecules, predominantly formaldehyde, during hydrothermal alteration (Cody et al. 2011).

Hypothesis 5.4: Bennu's parent asteroid was thermally stratified with distinct hydrological flow systems, resulting in zones of alteration that differed in mineralogy and oxygen isotope ratios (Young 2001; Palguta et al. 2010).

Hypothesis 5.5: Hydrothermal alteration was affected by collisions, such that hydrological systems within the interior of the parent body were suddenly altered (Vernazza et al. 2014).

## 3.6. Hypothesis 6: Bennu's parent body experienced >3 billion years of solar system history before being destroyed in a catastrophic disruption.

Hypothesis 6.1: Bennu's parent asteroid was scattered into the inner main asteroid belt, either through Jupiter's growth on a fixed orbit and/or by inward migration of Jupiter (Walsh et al. 2012; Raymond and Izidoro 2017; Kruijer et al. 2020).

Hypothesis 6.2: During its residence time in the main asteroid belt, Bennu's parent asteroid accreted meter-scale or larger exogenic material, resulting in inter-asteroid mixing at macroscopic scales (DellaGiustina and Kaplan et al. 2021; Le Corre et al. 2021; Tatsumi et al. 2021).

Hypothesis 6.3: Surface regolith was gradually overturned and mixed by 'gardening' processes on Bennu's parent asteroid (Melosh 2011; Walsh et al. 2019).

Hypothesis 6.4: Bennu's parent asteroid experienced a catastrophic disruption in the main asteroid belt that produced either the Eulalia family 830 +370/–100 Myr ago or the New Polana family 1400 ± 150 Myr ago (Campins et al. 2010; Walsh et al. 2013; Bottke et al. 2015; Michel 2001, 2003; Barnouin et al. 2019, Walsh et al. 2019, Scheeres et al. 2019).

## 3.7. Hypothesis 7: Bennu is a rubble pile that formed by re-accumulation of material from the catastrophic disruption of a precursor asteroid.

Hypothesis 7.1. Bennu is a rubble pile composed of fragments dominated by a single parent asteroid. After formation, Bennu may have experienced multiple disruption and re-accretion events (Bottke et al. 2015, 2020a; Barnouin et al. 2019; Scheeres et al. 2019).

Hypothesis 7.2: Bennu's spinning top–like shape and rubble-pile structure were established during accretion of fragments from the catastrophic disruption of a precursor asteroid (Keil et al. 1994; Michel and Ballouz et al. 2020).

Hypothesis 7.3: Bennu's surface contains material that formed at different depths within its parent asteroid (Michel and Ballouz et al. 2020; DellaGiustina et al. 2020; Rozitis et al. 2020b).

Hypothesis 7.4: Bennu retained its hydration state because it experienced minimal impact-induced heating during catastrophic disruption (Hamilton et al. 2019; Kitazato et al. 2019; Michel and Ballouz et al. 2020; Praet et al. 2021).

Hypothesis 7.5: The crater-retention age of Bennu's surface is 0.05 - 1 Ga, reflecting its history of impact processing in the main asteroid belt (Walsh et al. 2019; Bierhaus et al. 2022).

## 3.8. Hypothesis 8: The Yarkovsky effect pushed Bennu far enough inward to reach a dynamical resonance, which flung it out of the main belt and onto a terrestrial planet–crossing orbit.

Hypothesis 8.1: The Yarkovsky effect is controlled by Bennu's shape, rotation state, and thermal and physical regolith properties (i.e., thermal conductivity, specific heat capacity, bulk density, grain density, porosity, strength, and the thermal emission phase function) (Chesley et al. 2014; Rozitis et al. 2020b, 2022).

Hypothesis 8.2: Bennu's shape and internal structure were modified by YORP spin-up and mass-shedding events long after the final accretion of fragments from the catastrophic disruption of its precursor asteroid. Cohesive forces are comparable to gravitational and friction forces and thus determined the regions on Bennu that plastically deformed at a given spin rate (Sánchez and Scheeres 2014; Bottke et al. 2015; Scheeres et al. 2019, 2020).

Hypothesis 8.3: Bennu has been a near-Earth object for 1.75 ± 0.75 Myr (Ballouz et al. 2020).

Hypothesis 8.4. Bennu had a semi-major axis that resulted in a closer perihelion than at present, causing extreme surface heating, mobilization of labile elements, partial devolatilization, and mass fractionation of volatile species (Delbo and Michel 2011; Springmann et al. 2019).

## 3.9. Hypothesis 9: Bennu has experienced surface processing throughout its history.

Hypothesis 9.1: Bennu's surface materials experienced impacts in near-Earth space, producing craters ranging in size from millimeters to tens of meters (Bottke et al. 2020b; Ballouz et al. 2020; DellaGiustina et al. 2020; Bierhaus et al. 2022).

Hypothesis 9.2: Bennu's surface materials experienced thermal stress, volatile release, and fracturing in near-Earth space (Delbo et al., 2014; Molaro et al. 2015; Molaro et al. 2020; Delbo et al., 2022).

Hypothesis 9.3: Bennu's surface experienced mass movement that transported material from the mid latitudes to the equatorial region over the past 0.5 Ma (Daly et al. 2020; Jawin et al. 2020, 2022).

Hypothesis 9.4: Bennu's surface contains centimeter-scale platy particles that were ejected in discrete events and re-impacted. Ejected particles have distinct phase functions with forward scattering peaks at large phase angles (>160°) and a shallow phase function slope (~0.015 mag/deg) at medium phase angles (70–120°) (Lauretta and Hergenrother et al. 2019; Hergenrother et al. 2020a,b).

Hypothesis 9.5: Volatile compounds migrated from high-temperature low-latitude regions of Bennu's surface and condensed in cold traps at high latitudes (Rozitis et al. 2020a).

Hypothesis 9.6: Electrostatic lofting, seismic shaking, and other mechanisms preferentially removed particles smaller than 1 mm from Bennu (Hartzell 2019; Hartzell et al. 2022).

## 3.10. Hypothesis 10: The physical, chemical, and spectral properties of Bennu's surface materials have been modified by exposure to the space environment.

Hypothesis 10.1: Exposure to the space environment modified the spectral properties of Bennu's surface materials (DellaGiustina et al. 2020; Deshapriya et al. 2021; Lantz et al. 2018; Laczniak et al. 2020; Thompson et al. 2020).

- Unweathered materials are spectrally redder in the VIS to NIR, brighter in the NIR, and darker and bluer in VIS wavelengths than Bennu's average spectral slope. The 2.7-µm absorption feature band minimum occurs at shorter wavelengths and has a sharper shape compared to weathered material.
- Mildly weathered materials are brighter in the near-UV wavelengths relative to the VIS NIR than Bennu's average spectra.
- Highly weathered materials are brighter overall and more neutrally sloped than Bennu's average. The 2.7-µm absorption band minimum occurs at longer wavelengths.
- Space weathering results in mid-IR spectral changes including reduced restrahlen band strength and shifts in band positions (Brunetto et al. 2020).
- Bennu's surface materials have a very thin particulate coating that modifies the mid-infrared emissivity but does not significantly affect apparent thermal inertia (Hamilton et al. 2021; Rozitis et al. 2020b).

Hypothesis 10.2: Space weathering changed the chemistry and mineralogy of optically active surfaces.

- Phyllosilicates transformed to nanophase and microphase sulfides (troilite and pentlandite) and nanophase magnetite, with limited abundances of nanophase and microphase iron. (Trang et al. 2021; Gillis-Davis et al. 2017; Thompson et al. 2019, 2020).
- Exposure to different ionizing radiations (x-rays, gamma rays, solar wind, GCR, etc.) released SOM from IOM, destroyed amino acids while maintaining the original enantiomeric and carbon-isotopic ratios, and drove reaction pathways (Chan et al. 2019; Adam et al. 2021; Glavin et al. 2021).
- Aliphatic organics were converted to aromatic molecules. Samples with an aliphatic spectral signature have space exposure ages less than a few million years (Kaplan et al. 2021; Thompson et al. 2020).
- IOM lost H and O due to ion bombardment. It may also have reacted with weathering products from phyllosilicates such as O-radicals and oxygenated functional groups. (De Gregorio et al. 2010; Le Guillou et al. 2013; Laurent et al. GCA 2014; Laurent et al. 2015).

Hypothesis 10.3: Bennu's surface exposure history is dominated by its time in near-Earth space (Hypothesis 8.3) (Ballouz et al. 2020; Krietsch 2020).

Hypothesis 10.4: Individual particles may exhibit both space-weathered and unweathered surfaces and contain various amounts of solar wind, reflecting exposure to the space environment at the uppermost layers of the regolith (Matsumoto et al. 2016).

Hypothesis 10.5: Bennu's two distinct boulder lithologies (Hypothesis 1.1) have been compacted and comminuted at different rates due to their different physical and mechanical properties (Cambioni et al., 2021; Rozitis et al. 2020).

## 3.11. Hypothesis 11: The Hokioi Crater in which the Nightingale sample site is located was recently formed and contains relatively unweathered material.

Hypothesis 11.1: Hokioi crater is part of a population of small (≤25 m) red craters on Bennu that are less than $10^5$ years old and represent the youngest component of the global crater population (DellaGiustina et al. 2020).

Hypothesis 11.2: Hokioi crater contains abundant centimeter-sized and smaller particles with a particle size frequency distribution described by a power law with a slope of –2.2 ± 0.1 (Burke et al. 2021).

Hypothesis 11.3: Material from regions upslope and north migrated into Hokioi Crater and material originally within the crater migrated downslope and to the south (Jawin et al. 2020).

Hypothesis 11.4: The regolith in Hokioi crater is composed of particles with higher density and higher thermal conductivity compared to the bulk properties of the boulders from which they derived (Rozitis et al. 2020, 2022; see also Nakamura et al., 2022).

Hypothesis 11.5: Regolith in Hokioi crater has an angle of friction of 40°. The upper 10-20 cm regolith layer of Hokioi crater has a bulk density of 440–600 $kg/m^3$ and a cohesive strength 0.2–20 Pa (Walsh et al., 2022). The upper 60 cm layer of Hokioi crater has a bulk density of 500-700 $kg/m^3$ and a strength against cratering of < 0.001 Pa (Lauretta et al., 2022).

## 3.12. Hypothesis 12: OSIRIS-REx asteroid proximity operations, the TAG event, and Earth return modified the collected samples, TAGSAM, and the SRC.

Hypothesis 12.1: Samples will contain contaminants from the spacecraft and ground activities (Dworkin et al. 2018), including:

- remnants of gas and metal from the thrusters
- fragments of the pyro valves on the gas bottles
- remnants of the nitrogen/helium gas mixture used during collection
- fragments of the TAGSAM head
- trace amounts of Braycote lubricant used in assembly of the SRC
- traces of SRC heatshield ablation products
- evidence of ATLO and the launch environment
- traces of air, soil, water, or other materials from the recovery area at the Utah Test and Training Range
- chemical evidence of curatorial processing
- minimal contamination by terrestrial biology, which can provide useful data to distinguish abiotic vs. biosignatures (Glavin et al., 2019; Smith et al., 2021).

Hypothesis 12.2: The sample was modified by mechanical agitation during collection and Earth entry. Agitation affected the relative size frequency distributions of different sample lithologies with different strength properties. Some samples lost their space exposed surfaces owing to abrasion.

Hypothesis 12.3: The collected sample experienced temperatures up to 75 °C after collection and through Earth return.

Hypothesis 12.4: The SRC may contain samples that are outside the TAGSAM. The sample canister air filter may contain volatiles that outgassed from the samples in the sample canister.

Hypothesis 12.5: The TAGSAM head contains meteoroid impact pits resulting from exposure to the space environment between launch cover deployment (August 2018) and sample stow (November 2020).

# Traceability to Working Groups

**Table 3.2:** Science Working Groups Mapped to Level-2 Hypotheses.

| | Driving Hypothesis | SSAWG | MAPWG | SEIWG | SOAWG | SPTAWG | PADWG |
|---|---|---|---|---|---|---|---|
| | **Hypothesis 1: Remote sensing of Bennu's surface has accurately characterized it mineral, chemical, and physical properties** | | | | | | |
| 1.1 | Bennu's surface primarily consists of two distinct lithologies defined by their normal albedo, thermomechanical properties, and spectral slope | ◉ | | ◉ | ◉ | ◉ | ◉ |
| 1.2 | Bennu's bulk mineralogy and composition consists of: Mg-rich phyllosilicates, magnetite, mixtures of organics and carbonates, <10% anhydrous silicates. | ◉ | ◉ | ◉ | ◉ | | ◉ |
| 1.3 | Bennu's surface contains minor amounts of exogenous material from diverse sources including HED and S-type asteroids | ◉ | ◉ | ◉ | | | ◉ |
| 1.4 | Bennu's dominant lithologies are comparable in bulk mineralogy, petrology, and composition to the most aqueously altered carbonaceous chondrites | ◉ | ◉ | ◉ | ◉ | | ◉ |
| 1.5 | Bennu's surface contains material with physical properties that are distinct from characterized meteorites. | | | | | ◉ | ◉ |
| | **Hypothesis 2: Bennu contains prebiotic organic compounds** | | | | | | |
| 2.1 | Bennu contains organic molecules that are used in terrestrial biology including protein amino acids, peptides, polyols, purines and pyrimidines, amphiphiles, carboxylic acids, their precursors, and side products from chemical reaction networks | | | | ◉ | | |
| 2.2 | Enantiomeric excesses are present for some chiral molecules and are of the same handedness as found in life | | | | ◉ | | |
| 2.3 | Bennu contains soluble organic matter (SOM) with abundances, distributions, isotopic, and enantiomeric compositions that are similar to the most aqueously altered CI and CM chondrites | | | ◉ | ◉ | | |
| 2.4 | Bennu contains insoluble organic matter (IOM), a chemically complex, macromolecular material, with similar elemental and isotopic compositions, structures, and morphologies as that in the most aqueously altered CI and CM | | ◉ | ◉ | ◉ | | |
| | **Hypothesis 3: Bennu contains presolar material derived from diverse sources** | | | | | | |
| 3.1 | The abundance, isotopic, and chemical compositions of presolar materials are similar to those in the CI and CM chondrites | | ◉ | ◉ | | | |
| 3.2 | Bennu contains dust grains that condensed in the gaseous envelopes around ancient stars or the ejecta of novae or supernovae | | ◉ | ◉ | ◉ | | |
| 3.3 | Bennu contains dust grains that formed in the interstellar medium or were modified by processing in the interstellar medium | | ◉ | ◉ | | | |
| 3.4 | Bennu contains organic matter that formed in the interstellar medium. Such organics are carriers of excess D, 15N, and/or 13C resulting from photochemical processing of organics and their precursors at low temperatures | | ◉ | ◉ | ◉ | | |
| | **Hypothesis 4: Bennu's parent asteroid formed beyond the snow line by accretion of material in the protoplanetary disk** | | | | | | |
| 4.1 | The initial constituents of Bennu's parent asteroid formed over several million years of solar system history, starting at a $t_0$ of 4.567 Gyr | | ◉ | ◉ | | | |
| 4.2 | Bennu's bulk elemental composition reflects that of its parent asteroid and is similar to the composition of the Sun, with depletions in moderately to highly volatile elements | | | ◉ | ◉ | | |
| 4.3 | The initial constituents of Bennu's parent asteroid were materials that were inherited from the protosolar molecular cloud or were formed and altered in the protoplanetary disk | | ◉ | ◉ | ◉ | | |
| 4.4 | Bennu's parent asteroid accreted in the outer protoplanetary disk, beyond Jupiter, as recorded by distinct isotopic anomalies in a variety of elements | | | ◉ | ◉ | | |
| | **Hypothesis 5: Geological activity occurred in the interior of Bennu's parent asteroid early in solar system history** | | | | | | |
| 5.1 | Bennu's parent asteroid was heated by the decay of short-lived radionuclides for ~10 Myr after accretion | ◉ | ◉ | ◉ | | | |
| 5.2 | Bennu's parent asteroid experienced impact processing over its geological history, resulting in brecciation, impact shock, localized heating, and possibly asteroid-scale heating | | ◉ | ◉ | ◉ | | |
| 5.3 | Heating of Bennu's parent asteroid resulted in hydrothermal alteration in which melted ice reacted with the initial constituents at 25-375°C, establishing the current bulk mineralogy | ◉ | ◉ | ◉ | ◉ | ◉ | |
| 5.4 | Bennu's parent asteroid was thermally stratified with distinct hydrological flow systems, resulting in zones of alteration that differed in mineralogy and oxygen isotope ratios | ◉ | ◉ | ◉ | ◉ | | |
| 5.5 | Hydrothermal alteration was affected by collisions, such that hydrological systems within the interior of the parent body were suddenly altered | ◉ | ◉ | ◉ | | | |
| | **Hypothesis 6: Bennu's parent body experienced >3 billion years of solar system history before being destroyed in a catastrophic disruption** | | | | | | |
| 6.1 | Bennu's parent asteroid was scattered into the inner main asteroid belt, either through Jupiter's growth on a fixed orbit and/or by inward migration of Jupiter | ◉ | ◉ | ◉ | | | |
| 6.2 | During its residence time in the main asteroid belt, Bennu's parent asteroid accreted meter-scale or larger exogenic material, resulting in inter-asteroid mixing at macroscopic scales | ◉ | ◉ | ◉ | | | |
| 6.3 | Surface regolith was gradually overturned and mixed by 'gardening' processes on Bennu's parent asteroid | ◉ | ◉ | ◉ | | | |
| 6.4 | Bennu's parent asteroid experienced a catastrophic disruption in the main asteroid belt that produced either the Eulalia family 830 +370/−100 Myr ago or the New Polana family 1400 ± 150 Myr ago | ◉ | ◉ | ◉ | | | |
| | **Hypothesis 7: Bennu is a rubble pile that formed by re-accumulation of material from the catastrophic disruption of a precursor asteroid** | | | | | | |
| 7.1 | Bennu is a rubble pile composed of fragments dominated by a single parent asteroid. After formation, Bennu may have experienced multiple disruption and re-accretion events | ◉ | ◉ | ◉ | | | |
| 7.2 | Bennu's spinning top–like shape and rubble-pile structure were established during accretion of fragments from the catastrophic disruption of a precursor asteroid | | ◉ | ◉ | | ◉ | |
| 7.3 | Bennu's surface contains material that formed at different depths within its parent asteroid | ◉ | ◉ | ◉ | ◉ | ◉ | |
| 7.4 | Bennu retained its hydration state because it experienced minimal impact-induced heating during catastrophic disruption | ◉ | ◉ | ◉ | ◉ | | |
| 7.5 | The crater-retention age of Bennu's surface is 0.1 - 1 Ga, reflecting its history of impact processing in the main asteroid belt | | ◉ | ◉ | | | |
| | **Hypothesis 8: The Yarkovsky effect pushed Bennu far enough inward to reach a dynamical resonance, which flung it out of the main belt and onto a terrestrial planet-crossing orbit** | | | | | | |
| 8.1 | The Yarkovsky effect is controlled by Bennu's shape, rotation state, and thermal and physical regolith properties | | | | | ◉ | |
| 8.2 | Bennu's shape and internal structure were modified by YORP spin-up and mass-shedding events long after the final accretion of fragments from the catastrophic disruption of its precursor asteroid. | | | | | ◉ | |
| 8.3 | Bennu has been a near-Earth object for 1.75 ± 0.75 Myr | | | ◉ | | ◉ | |
| 8.4 | Bennu had a semi-major axis that resulted in a closer perihelion than at present, causing extreme surface heating, mobilization of labile elements, partial devolatilization, and mass fractionation of volatile species | | | ◉ | ◉ | | |
| | **Hypothesis 9: Bennu has experienced surface processing throughout its history** | | | | | | |
| 9.1 | Bennu's surface materials experienced impacts in near-Earth space, producing craters ranging in size from millimeters to tens of meters | ◉ | ◉ | ◉ | | ◉ | |
| 9.2 | Bennu's surface materials experienced thermal stress, volatile release, and fracturing in near-Earth space | | ◉ | ◉ | | ◉ | |
| 9.3 | Bennu's surface experienced mass movement that transported material from the midlatitudes to the equatorial region over the past 0.2 Ma | ◉ | ◉ | ◉ | | | |
| 9.4 | Bennu's surface contains centimeter-scale platy particles that were ejected in discrete events and re-impacted. | ◉ | ◉ | ◉ | | ◉ | |
| 9.5 | Volatile compounds migrated from high-temperature low-latitude regions of Bennu's surface and condensed in cold traps at high latitudes | | | | ◉ | | |
| 9.6 | Electrostatic lofting, seismic shaking, and other mechanisms preferentially removed particles smaller than 1 mm from Bennu | ◉ | ◉ | ◉ | | ◉ | |
| | **Hypothesis 10: The physical, chemical, and spectral properties of Bennu's surface materials have been modified by exposure to the space environment** | | | | | | |
| 10.1 | Exposure to the space environment modified the spectral properties of Bennu's surface materials | ◉ | ◉ | ◉ | ◉ | ◉ | ◉ |
| 10.2 | Space weathering changed the chemistry and mineralogy of optically active surfaces | ◉ | ◉ | ◉ | ◉ | | ◉ |
| 10.3 | Bennu's surface exposure history is dominated by its time in near-Earth space | ◉ | ◉ | ◉ | | | ◉ |
| 10.4 | Individual particles may exhibit both space-weathered and unweathered surfaces and contain various amounts of solar wind, reflecting exposure to the space environment at the uppermost layers of the regolith | ◉ | ◉ | ◉ | ◉ | | ◉ |
| 10.5 | Bennu's two distinct boulder lithologies have been compacted and comminuted at different rates due to their different physical and mechanical properties. | | ◉ | | | ◉ | |
| | **Hypothesis 11: Hokioi crater was recently formed and contains relatively unweathered material** | | | | | | |
| 11.1 | Hokioi crater is part of a population of small (≤25 m) red craters on Bennu that are less than 105 years old and represent the youngest component of the global crater population | ◉ | ◉ | ◉ | | | |
| 11.2 | Hokioi crater contains abundant centimeter-sized and smaller particles with a particle size frequency distribution described by a power law with a slope of –2.2±0.1 | ◉ | ◉ | | | ◉ | |
| 11.3 | Material from regions upslope and north migrated into Hokioi Crater and material originally within the crater migrated downslope and to the south | ◉ | ◉ | ◉ | | | |
| 11.4 | The regolith in Hokioi crater is composed of particles with higher density and higher thermal conductivity compared to the bulk properties of the boulders from which they derived | | ◉ | ◉ | | ◉ | |
| 11.5 | Regolith in Hokioi crater has an angle of friction of 40°. The upper 10-20 cm regolith layer of Hokioi crater has a bulk density of 440–600 kg/m3 and a cohesive strength 0.2–20 Pa. The upper 60 cm layer of Hokioi crater has a bulk density of 500-700 kg/m3 and a strength against cratering of < 0.001 Pa. | | | | | ◉ | |
| | **Hypothesis 12: OSIRIS-REx asteroid proximity operations, the TAG event, and Earth return modified the collected samples, TAGSAM, and the SRC** | | | | | | |
| 12.1 | Samples will contain contaminants from the spacecraft and ground activities | ◉ | ◉ | ◉ | ◉ | | |
| 12.2 | The sample was modified by mechanical agitation during collection and Earth entry. Agitation affected the relative size frequency distributions of different sample lithologies with different strength properties. Some samples lost their space exposed surfaces owing to abrasion | ◉ | ◉ | | | ◉ | |
| 12.3 | The collected sample experienced temperatures up to 75 °C after collection and through Earth return | ◉ | | | | | |
| 12.4 | The SRC may contain samples that are outside the TAGSAM. The sample canister air filter may contain volatiles that outgassed from the samples in the sample cannister. | ◉ | ◉ | ◉ | ◉ | | |
| 12.5 | The TAGSAM head contains micrometeorite impact pits resulting from exposure to the space environment between launch cover deployment and sample stow | ◉ | | ◉ | | | |

# Traceability to Analytical Techniques

**Table 3.1:** Analytical Techniques Mapped to Level-2 Hypotheses.

[Table: Hypothesis traceability matrix linking testing according to analytical technique. Columns grouped into Spectroscopy, Mineralogy and Petrology, Elements and Isotopes, Organic Chemistry, and Physical and Thermal Testing. Rows list Driving Hypotheses 1–12 with sub-hypotheses (1.1–12.5). Cells contain filled circles (tests fully) or open circles (tests partially).]

*Technique:*
**Spectroscopy:** VI = visual inspection; VNMIR = visible, near-infrared, and mid-infrared spectroscopy

**Mineralogy and Petrology:** uXRD = X-ray diffraction; SEM = scanning electron microscopy; FIB-SEM = focused ion-beam scanning electron microscopy; EBSD = electron backscatter diffractometry; EMPA = electron microprobe analysis; CL = cathodoluminescence spectroscopy; TEM = transmission electron microscopy; EDS = energy dispersive X-ray spectroscopy; TGS = thermogravimetric analysis; XPS = X-ray photoelectron spectroscopy

**Elements and Isotopes:** SIMS = secondary ion mass spectrometry; NanoSIMS = nanoscale secondary ion mass spectrometry; SHRIMP = sensitive high-resolution ion microprobe; TIMS = thermal ionization mass spectrometry; HR-ICPMS = high-resolution inductively coupled plasma mass spectrometry; MC-ICP-MS = Multi-Collector Inductively Coupled Plasma Mass Spectrometry; LA-ICP-MS = Laser ablation Inductively Coupled Plasma Mass Spectrometry; XRF = X-ray Fluorescence Spectroscopy; EA-IRMS = Elemental Analysis – Isotope Ratio Mass Spectrometry; NGMS = noble gas mass spectrometry; SNMS = Secondary Neutral Mass Spectrometry; APT = Atom Probe Tomography; IC = Ion Chromatography; Ni-NGMS = Neutron Irradiation noble gas mass spectrometry; RI-TOF-NGMS = Resonance Ionization time-of-flight noble gas mass spectrometry; Ar/Ar = $^{40}$Ar/$^{39}$Ar geochronology and thermochronology; TOF-GMS = Time-of-Flight secondary ion mass spectrometry

**Organic Chemistry:** GC-MS = gas chromatography mass spectrometry; uLIMS = micro two-step laser desorption mass spectrometry; LCMS = liquid chromatography mass spectrometry; CE = Capillary Electrophoresis; FT-ICR-MS = Fourier Transform Ion Cyclotron Resonance Mass Spectrometry; XANES = X-ray Absorption Near-Edge Structure Spectroscopy; NMR = Nuclear Magnetic Resonance Spectroscopy; Raman = Raman Spectroscopy; GC-IRMS = Gas Chromatography Isotope Ratio Mass Spectrometry; EA-IRMS = Elemental Analysis – Isotope Ratio Mass Spectrometry; µFTIR = micro Fourier Transform Infrared Spectroscopy; DESI-Orbitrap MS = Desorption Electrospray Ionization-Orbitrap Mass spectrometry

**Physical Testing:** Mass = mass measurement; 3D Scan = 3D Laser Scanning Pycnometry = Gas Pycnometry; AFM = atomic force microscopy; DSC = differential scanning calorimetry; PSFD = particle size frequency distribution

● tests fully     ○ tests partially

# 4.0 EXPECTED NATURE OF THE RETURNED SAMPLE

The OSIRIS-REx spacecraft sampled asteroid Bennu on 20 October 2020 using a Touch and Go (TAG) maneuver, in which TAGSAM contacted the surface for several seconds before the spacecraft backed away (Lauretta et al. 2022). TAGSAM consists of a sampler head at the end of a 3-m-long articulated arm. The sampler head was designed to collect two types of sample, which visual inspection confirmed. The first is a bulk sample (2 cm or smaller in shortest dimension) through the fluidization of the regolith achieved by the release of high-purity nitrogen gas that helps to funnel grains into the main collecting area, held in place by a mylar flap. The second is the capture of millimeter-sized particles on 24 'contact pad' made of stainless-steel Velcro® that are embedded into the surface of the TAGSAM baseplate. Based on what we have learned from the OSIRIS-REx science campaign during asteroid operations, we can make several confident predictions about the returned sample that will assist in the preparations needed for the analysis phase.

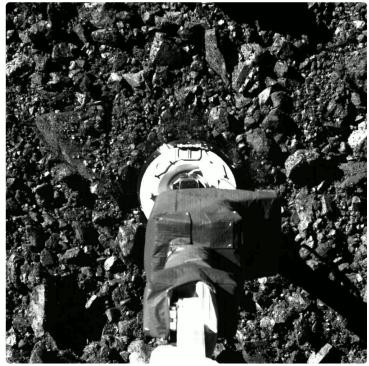

Figure 4.1. This series of 16 SamCam images shows TAGSAM touching down and sinking into the surface of Bennu, then backing away.

The mission team performed two separate types of analyses to constrain the mass of the sample collected. The first set of analyses focused on characterization of visible regolith in and on the TAGSAM head after sampling. This effort resulted in a total estimated mass of ~400 g. The team used angular momentum change during TAGSAM arm deployment before and after TAG to estimate the sample mass. The alternative SMM results showed the samples collected by the spacecraft were well above the requirement (60 g), with a best estimate of 250 ± 101 g (Ma et al. 2020).

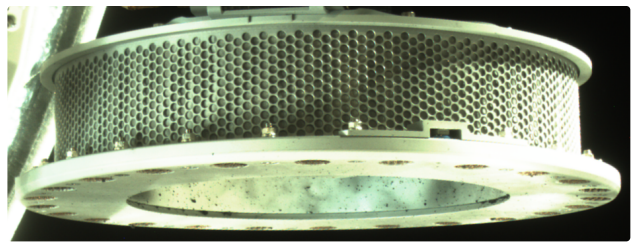

Figure 4.2. StowCam image of the TAGSAM head taken after sampling. Regolith particles up to 4 mm in diameter are present adhering to the contact pads. Additional material is visible on other surfaces such as the baseplate.

Post-sampling images showed revealed that the TAGSAM was leaking material. The estimated mass of sample stowed reflects this mass loss (Ma et al. 2020).

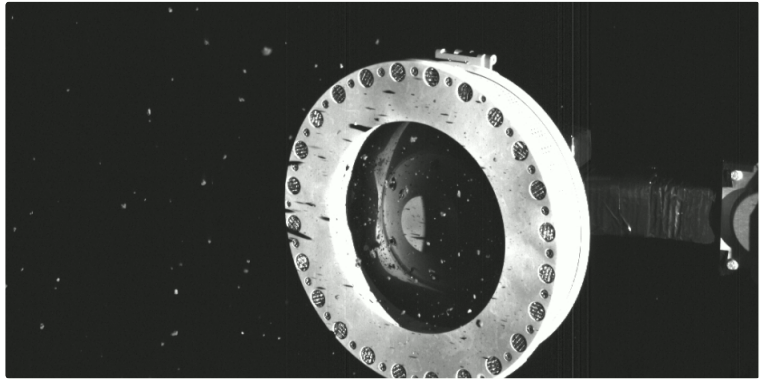

**Figure 4.3**. This series of three SamCam images shows the TAGSAM head full of regolith collected from Bennu. Some of these particles can be seen escaping into space, owing to the mylar flap being wedged open by larger particles that did not fully pass through it.

The particles sizes of the returned material is expected to range from sub-micron to 3 cm, the largest particle observed in post-TAG images. The returned material will be binned into the following particle size range categories: (1) fines, 1–100 µm; (2) intermediate, 100–500 µm; and (3) coarse, ≥500 µm. Particles could consist of fine-grained friable clusters, or alternatively, they could be breccias or clasts. We predict that the sample will contain different types of lithic components analogous to chondrites. Based on reflectance and infrared spectroscopy, we expect that such components will be similar to what we find in carbonaceous chondrites, i.e., chondrules, calcium-aluminum–rich inclusions (CAIs), and matrix. Such components could be contained within clusters or occur as individual particles. We expect the sample to contain: (1) hydrated silicates, (2) carbonates, (3) magnetite, and (4) organic components. Some particles or aggregates also could be non-chondritic and igneous in nature, similar to howardite–eucrite–diogenite (HED) meteorites, at least for pyroxene-bearing material. Whether the constituent materials occur together as assemblages or in monomineralic particles is to be determined. Nonetheless, we are confident to a high level that the material collected in both the bulk and contact pads will be heterogenous in nature.

Both the TAGSAM and the SRC have witness plates mounted in strategic locations to document any contamination collected in flight during different phases of the mission. The SRC also contains an air filter which was used to permit the TAGSAM canister to equalize in pressure during launch and landing. The filter was designed to exclude water vapor, particles, and organics from entering the sample canister and contaminating the sample. It also has the possibility of capturing volatiles from the sample. The SRC is also equipped with sets of non-reversible temperature indicators to record the maximum temperature the sample and hardware experienced. Three sets of strips were deployed. One for sample science: 100, 110, 120, and 130°F (38, 43, 49, 54°C) as well as two for hardware: 260, 280, 300, 320, 340, 360, 380, and 400°F (127-204°C); and 350, 360, 370, and 380°F (177-193°C). Note that the strips on the SRC backshell were heated to 125°C before launch.

In addition to the flight witness plates, contamination knowledge samples were collected and archived prior to launch. These consist of a set of silicon wafers and aluminum foils which were exposed during spacecraft integration and testing. 25% of each of these was analyzed, in addition to the liquid monopropellant analyses and gas analyses. The remaining 75% was archived at NASA's Johnson Space Center (JSC). Also at JSC, 419 items from the spacecraft and the integration and test environment are archived. These range from gloves, to paints, to monopropellant dried residue and blanks, to a thruster. These are described in Dworkin et al. (2018). These materials are available to the sample analysis team for parallel analyses with samples; allocation will follow similar procedures as samples, since these are also limited. OSIRIS-REx will strive to preserve 75% of these materials for the permanent NASA collection.

# 5.0 ANALYTICAL TECHNIQUES

The SAT is divided into working groups organized by analytical objectives and techniques. Each working group is responsible for specific data products, which are linked to one or more hypotheses. The working group leads are responsible for assigning data products to individual team members. They are also responsible for monitoring the sub-allocation of sample material within their group and for tracking and reporting the status of sample analysis and data product completion to the PI.

The following sample science working groups are responsible for defining the analytical techniques used in this plan:

- Sample Spectral Analysis (SSAWG)
- Sample Mineralogy and Petrology (MAPWG)
- Sample Elemental and Isotopic Analysis (SEIWG)
- Sample Organic Analysis (SOAWG)
- Sample Physical and Thermal Analysis (SPTAWG)

Other working groups focus on supporting analyses of the returned hardware and for data management. These groups use the analytical techniques under the other WGs.

- Contact Pad Analysis (PADWG)
- Sample Data Management and Archiving (SADAWG)
- Contamination Control and Knowledge (CCWG)
- Touch-and-Go Sample Acquisition Mechanism (TAGSAMWG)
- Sample Return Capsule Science (SRCWG)

Measurements of the returned sample will be coordinated.  This section describes the techniques that will be used to analyze the returned sample, as well as the data products these techniques will generate. Each technique supports testing of one or more hypotheses. Details for each technique are included as an appendix.

## Spectral Analysis

- 5.1 Quantitative Reflectance Imaging
- 5.2 Visible, near-infrared, and mid-infrared (VNMIR) spectroscopy

## Mineralogy and Petrology

- 5.3 Visible Light Microscopy
- 5.4 Scanning Electron Microscopy (SEM)
- 5.5 Focused-Ion-Beam Scanning-Electron Microscopy (FIB-SEM)
- 5.6 Electron backscatter diffraction (EBSD) and transmission Kikuchi diffraction (TKD)
- 5.7 Electron microprobe analysis (EMPA)
- 5.8 High-resolution cathodoluminescence (HR-CL)
- 5.9 Transmission Electron Microscopy (TEM)
- 5.10 Energy-dispersive X-ray Spectroscopy (EDS)
- 5.11 Electron Energy-Loss Spectroscopy (EELS)
- 5.12 X-ray Absorption Near-Edge Structure (XANES)
- 5.13 micro-X-ray computed tomography (XCT)
- 5.14 X-ray Diffraction (XRD)
- 5.15 Raman Vibrational Spectroscopy

<: 

# 6.0 BASELINE SAMPLE MASS ALLOCATION

The Baseline Sample Analysis Plan is based on the successful return of 60 g of bulk sample and 12 (out of 24) contact pads with surface sample. The Baseline Plan defines the program for nominal analysis of two major lithologies and up to six minor lithologies. Mass allocation for analysis of minor lithologies (up to 3.75 g) is addressed in the Threshold Sample Analysis Section of this plan.

**Table 6.1:** Mass distribution for the 15 g of sample allocated to the OSIRIS-REx SAT.

# 7.0 OVERGUIDE SAMPLE ANALYSIS PROCEDURE

Overguide analyses becomes feasible if substantially more bulk sample and contacted surface area are returned. If 250 g of bulk sample was collected and all 24 pads made surface contact, as indicated by spacecraft observations, the overguide plan would include analysis of up to 62.5 g of bulk sample and six contact pads. All mass requests and overguide analysis plans are considered notional and depend on the nature and total mass of the sample. Decisions on the overguide allocations will be made after sample return.

Procedure for allocating overguide mass:

- The Astromaterials Curator will write a letter to both the OSIRIS-REx Program Scientist and the Chief Scientist for Astromaterials Curation within SMD stating the official sample mass. The Program Scientist and Chief Scientist for Astromaterials Curation will write an additional memo for the record stating this is the official bulk sample mass.
- Determination, post-curation certification of mass, that we are in overguide situation (>60g of bulk sample and >12 contact pads with surface sample) and PI announcement to the SAT that we are in an overguide situation.
- Initial focus will be on the baseline allocations.
- MSS announces that applications are now being accepted for overguide analysis. Until that time no overguide requests will be considered.
  - Applications shall come in the sample allocation form, with supporting documentation on the rationale for the request, including an explanation for why this allocation is needed to address the SAP hypotheses.
  - Proposals must first be approved by the WGLs before before forwarding to MSS for discussion with the PI, both of whom may consult the SSC as needed on the decision process.
  - Any conflicts will be resolved following the Rules of the Road.
- PI and Curation must approve the allocation request, with Program Scientist concurrence.
- Announcement to the team about the overguide allocation decisions will include publication of the approved overguide mass table.
- The policy will be to reserve approximately 50% of overguide SAT allocation for future overguide requests during the two-year analysis campaign. This process will be repeated as many times as the MSS and PI determine is necessary during the two years of OSIRIS-REx sample analysis.

# 8.0 THRESHOLD SAMPLE MASS ALLOCATION

The "Threshold" Program is driven by analysis of 3.75 g and a single contact pad. The Threshold Program covers contingency scenarios in which sample is lost due to an anomaly during Earth return. This program is also applicable to characterization of a minor lithology in a nominal return scenario.

**Table 8.1:** Mass distribution for the 3.75 g of threshold sample allocated to the OSIRIS-REx SAT.

| | | | | | | | | | Bulk Sample Allocation | | | | | | | |
|---|---|---|---|---|---|---|---|---|---|---|---|---|---|---|---|---|
| | | | | | | Mass per Test | | | Number of Tests | | | Mass per Investigation | | | Non-Destructive | |
| | | | | | CBE | Test Contingency | SubTotal | | CBE | Spare | Total | CBE | Contingency | Total | Yes = 0 No = 1 | Consumed Quantity |
| Row # | Test Objective | Team Member | Instrument | Institution | Follow-on Analysis (Y/N/P) - if Y - which primary sample and notes; If Partial (P) describe | g | % | g | g | | | | g | % | g | g | |
| | | | | **Bulk Elemental Abundances** | | | | | | | | | | | | |
| 7 | Major element bulk abundances | Lauretta | Nu AttoM / Thermo Scientific Element 2 | UBC | N - Primary mass allocation | 0.010 | 0% | 0.000 | 0.010 | 10 | 2 | 12 | 0.100 | 0% | 0.000 | 0.100 | 1 | 0.100 |
| 8 | Trace element bulk abundances | Lauretta | Nu AttoM / Thermo Scientific Element 2 | UBC | N - Primary mass allocation | 0.100 | 0% | 0.000 | 0.100 | 2 | 0 | 2 | 0.200 | 0% | 0.000 | 0.200 | 1 | 0.200 |
| | | | | | **Total Bulk Elemental Abundances:** | | | | | | | | 0.300 | 0% | 0.000 | 0.300 | | 0.300 |
| | | | | **Physical & Thermochemical Properties** | | | | | | | | | | | | |
| 11 | Bulk density, grain density, porosity | Macke/Righter/Hildebrand | Gas Pycnometry | JSC | N - Primary mass allocation | 1.617 | 0% | 0.000 | 1.617 | 1 | 0 | 1 | 1.617 | 0% | 0.000 | 1.617 | 0 | 0.000 |
| 12 | Bulk thermal conductivity | Ryan | Spherical cell bulk thermal conductivity analyzer | UA | N - Primary mass allocation | 0.400 | 0% | 0.000 | 0.400 | 1 | 0 | 1 | 0.400 | 0% | 0.000 | 0.400 | 0 | 0.000 |
| 13 | Bulk heat capacity | Biele/Opeil/Takeda | Differential scanning calorimetry | UA/Boston College | N - Primary mass allocation | 0.030 | 0% | 0.000 | 0.030 | 2 | 1 | 3 | 0.060 | 0% | 0.000 | 0.060 | 0 | 0.000 |
| 14 | Coefficient of thermal expansion | Opeil | Cantilever dilatometer | Boston College | N - Primary mass allocation | 0.070 | 0% | 0.000 | 0.070 | 1 | 0 | 1 | 0.070 | 0% | 0.000 | 0.070 | 0 | 0.000 |
| 15 | Cohesion between particles | Hoover/Sanchez | Particle cohesion with AFM | ASU | N - Primary mass allocation | 0.001 | 0% | 0.000 | 0.001 | 10 | 1 | 11 | 0.010 | 0% | 0.000 | 0.010 | 0 | 0.000 |
| 16 | Elastic modulus, crushing strength, critical flaw length at failure, and fracture behavior | Hoover | Compression test | ASU | N - Primary mass allocation | 0.001 | 0% | 0.000 | 0.001 | 20 | 2 | 22 | 0.020 | 0% | 0.000 | 0.020 | 0 | 0.000 |
| 17 | Young's modulus, shear modulus, Poisson's ratio, Lamé's constant, and bulk modulus | Hildebrand | Seismic velocity and rock ultrasonic constant measurement | U-Calgary | N - Primary mass allocation | 0.001 | 0% | 0.000 | 0.001 | 10 | 1 | 11 | 0.010 | 0% | 0.000 | 0.010 | 0 | 0.000 |
| 18 | Angle of repose and angle of internal friction | Ryan/Ballouz/Sanchez | Angle of repose measurement | JSC | Y - Observations during curation processing | 0.000 | 0% | 0.000 | 0.000 | 0 | 0 | 0 | 0.000 | 0% | 0.000 | 0.000 | 1 | 0.000 |
| 19 | Bulk density, grain density, porosity | Ryan/Macke/Righter | Structured Light Scanner | JSC | Y - Observations during curation processing | 0.000 | 0% | 0.000 | 0.000 | 0 | 0 | 0 | 0.000 | 0% | 0.000 | 0.000 | 1 | 0.000 |
| 20 | Grain and matrix thermal conductivity | Daly | Scanning thermal microscopy (SThM) with AFM | York U | Y - UA thin sections - Row 25 & 26 | 0.000 | 0% | 0.000 | 0.000 | 0 | 0 | 0 | 0.000 | 0% | 0.000 | 0.000 | 1 | 0.000 |
| 21 | Indentation modulus; indentation hardness; ductility; elastic work, inelastic work, and total work | Hoover | Nano and micro-indentation | ASU | Y - UA thin sections - Row 25 & 26 | 0.000 | 0% | 0.000 | 0.000 | 0 | 0 | 0 | 0.000 | 0% | 0.000 | 0.000 | 1 | 0.000 |
| | | | | | **Total Physical and Thermal Properties Analysis:** | | | | | | | | 2.187 | 0% | 0.000 | 2.187 | | 0.000 |
| | | | | **General Petrography and Petrology** | | | | | | | | | | | | |
| 24 | Internal density distribution | Righter | X-ray computed tomography | JSC | Y - Pycnometry sample - row 11 | 0.000 | 0% | 0.000 | 0.000 | 0 | 0 | 0 | 0.000 | 0% | 0.000 | 0.000 | 0 | 0.000 |
| 25 | Sample texture/petrography/mineralogy | Connolly | Cameca SX-100 Electron Microprobe | UA | Y - Pycnometry sample - row 11 | 0.000 | 0% | 0.000 | 0.000 | 1 | 0 | 1 | 0.000 | 0% | 0.000 | 0.000 | 0 | 0.000 |
| 26 | Sample texture/petrography/mineralogy | Lauretta | Cameca SX-100 Electron Microprobe | UA | Y - Pycnometry sample - row 11 | 0.000 | 0% | 0.000 | 0.000 | 1 | 0 | 1 | 0.000 | 0% | 0.000 | 0.000 | 0 | 0.000 |
| 27 | Sample texture/petrography/mineralogy | Libourel | JEOL JSM 700F FEG-SEM | Cote d'Azur | Y - Pycnometry sample - row 11 | 0.000 | 0% | 0.000 | 0.000 | 1 | 0 | 1 | 0.000 | 0% | 0.000 | 0.000 | 0 | 0.000 |
| 28 | Sample texture/petrography/mineralogy | Bland | TESCAN CLARA FE-SEM etc. | Curtin U | Y - Pycnometry sample - row 11 | 0.000 | 0% | 0.000 | 0.000 | 1 | 0 | 1 | 0.000 | 0% | 0.000 | 0.000 | 0 | 0.000 |
| 29 | Sample texture/petrography/mineralogy | Russell | TESCAN CLARA FE-SEM etc. | NHM | Y - Pycnometry sample - row 11 | 0.000 | 0% | 0.000 | 0.000 | 1 | 0 | 1 | 0.000 | 0% | 0.000 | 0.000 | 0 | 0.000 |
| 30 | Sample texture/petrography/mineralogy | McCoy | JEOL JXA-8530F+ | NMNH | Y - Pycnometry sample - row 11 | 0.000 | 0% | 0.000 | 0.000 | 1 | 0 | 1 | 0.000 | 0% | 0.000 | 0.000 | 0 | 0.000 |
| 31 | Sample texture/petrography/mineralogy | Righter | JEOL JXA-8530F & Cameca SX-100 | JSC | Y - Pycnometry sample - row 11 | 0.000 | 0% | 0.000 | 0.000 | 1 | 0 | 1 | 0.000 | 0% | 0.000 | 0.000 | 0 | 0.000 |
| 32 | Sample texture/petrography/mineralogy | Yurimoto | FE-SEM of JEOL JSM-7000F | Hokkaido | Y - Pycnometry sample - row 11 | 0.000 | 0% | 0.000 | 0.000 | 1 | 0 | 1 | 0.000 | 0% | 0.000 | 0.000 | 0 | 0.000 |
| | | | | | **Total Petrography and Petrology:** | | | | | | | | 0.000 | 0% | 0.000 | 0.000 | | 0.000 |
| | | | | **Nanoscale Mineralogy** | | | | | | | | | | | | |
| 35 | In situ mineralogy and surface chemistry | Zega | Hitachi TEM | UA | Y - UA thin sections - Row 25 & 26 | 0.000 | 0% | 0.000 | 0.000 | 1 | 0 | 1 | 0.000 | 0% | 0.000 | 0.000 | 0 | 0.000 |
| 36 | In situ mineralogy and surface chemistry | Keller | JEOL 2500SE TEM | JSC | Y - JSC thin section - Row 31 | 0.000 | 0% | 0.000 | 0.000 | 1 | 0 | 1 | 0.000 | 0% | 0.000 | 0.000 | 0 | 0.000 |
| | | | | | **Total Mineralogy:** | | | | | | | | 0.000 | 0% | 0.000 | 0.000 | | 0.000 |
| | | | | **Isotopic Abundances** | | | | | | | | | | | | |
| 39 | Noble gas abundances and isotopic ratios | Busemann | Noble Gas Mass Spectrometer | ETH-Zurich | N - Primary mass allocation | 0.020 | 0% | 0.000 | 0.020 | 1 | 1 | 2 | 0.020 | 0% | 0.000 | 0.020 | 0 | 0.000 |
| 40 | Noble gas abundances and isotopic ratios | Marty | Noble Gas Mass Spectrometer | Nancy | N - Primary mass allocation | 0.025 | 0% | 0.000 | 0.025 | 1 | 1 | 2 | 0.025 | 0% | 0.000 | 0.025 | 1 | 0.025 |
| 41 | Noble gas abundances and isotopic ratios | Marty | LA-NGMS | Nancy | N - Primary mass allocation | 0.010 | 0% | 0.000 | 0.010 | 3 | 1 | 4 | 0.030 | 0% | 0.000 | 0.030 | 1 | 0.030 |
| 42 | Halogen | Clay | NI-NGMS | Manchester | N - Primary mass allocation | 0.002 | 0% | 0.000 | 0.002 | 5 | 1 | 6 | 0.010 | 0% | 0.000 | 0.010 | 1 | 0.010 |
| 43 | Xe | Gilmour | RI-TOF-NGMS | Manchester | N - Primary mass allocation | 0.001 | 0% | 0.000 | 0.001 | 3 | 1 | 4 | 0.003 | 0% | 0.000 | 0.003 | 1 | 0.003 |
| 44 | Ar-Ar | Jourdan | Step heating | Curtin | N - Primary mass allocation | 0.010 | 0% | 0.000 | 0.010 | 1 | 1 | 2 | 0.010 | 0% | 0.000 | 0.010 | 1 | 0.010 |
| 45 | Noble gas abundances and isotopic ratios | Yurimoto | LIMAS | Hokkaido | Y - Japan thin section - Row 32 | 0.000 | 0% | 0.000 | 0.000 | 1 | 1 | 2 | 0.000 | 0% | 0.000 | 0.000 | 0 | 0.000 |
| | | | | | **Total Noble Gas Isotopic Abundances:** | | | | | | | | 0.098 | 0% | 0.000 | 0.098 | | 0.078 |
| 47 | Volatile/organic in-situ abn+isotopes | PHUB/AN/LP | Nanosims/SIMS | UA/JSC/Nancy | N - Primary mass allocation | 0.100 | 0% | 0.003 | 0.103 | 1 | 1 | 2 | 0.100 | 0% | 0.000 | 0.100 | 0 | 0.000 |
| 48 | H, C, N, O bulk & isotopic abundances | Franchi | Stable isotope-MS | Open U | N - Primary mass allocation | 0.100 | 0% | 0.000 | 0.100 | 1 | 1 | 2 | 0.100 | 0% | 0.000 | 0.100 | 1 | 0.100 |
| 49 | Cr & Ti isotopes | Busemann/Schönbächler | TIMS & HR-ICP-MS | ETH Zurich | N - Primary mass allocation | 0.020 | 0% | 0.000 | 0.020 | 1 | 1 | 2 | 0.020 | 0% | 0.000 | 0.020 | 1 | 0.020 |
| 50 | In situ stable isotopes | Connolly | Cameca 1280 SIMS | U-Hawaii | Y - UA thin sections - Row 25 & 26 | 0.000 | 0% | 0.000 | 0.000 | 1 | 1 | 2 | 0.000 | 0% | 0.000 | 0.000 | 0 | 0.000 |
| 51 | In situ stable isotopes volatiles (H,C,O,N,S etc) | Barnes | Cameca nanoSIMS II | UA | Y - UA thin sections - Row 25 & 26 | 0.000 | 0% | 0.000 | 0.000 | 1 | 1 | 2 | 0.000 | 0% | 0.000 | 0.000 | 0 | 0.000 |
| 52 | Presolar grain minerology/isotopic composition | Haenecour | Cameca nanoSIMS II | UA | Y - UA thin sections - Row 25 & 26 | 0.000 | 0% | 0.000 | 0.000 | 1 | 1 | 2 | 0.000 | 0% | 0.000 | 0.000 | 0 | 0.000 |
| 53 | In situ stable isotopes volatiles (H,C,O,N,S etc) | Piani | Cameca 1280 HR SIMS | Nancy | Y - France thin section - Row 27 | 0.000 | 0% | 0.000 | 0.000 | 1 | 1 | 2 | 0.000 | 0% | 0.000 | 0.000 | 0 | 0.000 |
| 54 | In situ stable isotopes volatiles (H,C,O,N,S etc) | Franchi | Cameca NanoSIMS 50L | OU | Y - UK thin section - Row 29 | 0.000 | 0% | 0.000 | 0.000 | 1 | 1 | 2 | 0.000 | 0% | 0.000 | 0.000 | 0 | 0.000 |
| 55 | Presolar grain minerology/isotopic composition | Franchi | Cameca NanoSIMS 50L | OU | Y - UK thin section - Row 29 | 0.000 | 0% | 0.000 | 0.000 | 1 | 1 | 2 | 0.000 | 0% | 0.000 | 0.000 | 0 | 0.000 |
| 56 | Presolar grain minerology/isotopic composition | Nguyen | Cameca NanoSIMS 50L | JSC | Y - JSC thin section - Row 31 | 0.000 | 0% | 0.000 | 0.000 | 1 | 1 | 2 | 0.000 | 0% | 0.000 | 0.000 | 0 | 0.000 |
| 57 | In situ stable isotopes | Yurimoto | Cameca 1280 SIMS/L | Hokkaido | Y - Japan thin section - Row 32 | 0.000 | 0% | 0.000 | 0.000 | 1 | 1 | 2 | 0.000 | 0% | 0.000 | 0.000 | 0 | 0.000 |
| | | | | | **Total Stable Isotopic Abundances:** | | | | | | | | 0.220 | 30% | 0.000 | 0.220 | | 0.120 |
| 59 | Radiogenic age date (Al-Mg) | Connolly | Cameca 1280 SIMS | Hawaii | Y - UA thin sections - Row 25 & 26 | 0.000 | 0% | 0.000 | 0.000 | 0 | 0 | 0 | 0.000 | 0% | 0.000 | 0.000 | 0 | 0.000 |
| 60 | Radiogenic age date (Al-Mg) | Barnes | Cameca nanoSIMS II | UA | Y - UA thin sections - Row 25 & 26 | 0.000 | 0% | 0.000 | 0.000 | 0 | 0 | 0 | 0.000 | 0% | 0.000 | 0.000 | 0 | 0.000 |
| 61 | Radiogenic age date (Al-Mg) | Marty + Piani | Cameca 1280 HR SIMS | Nancy | Y - France thin section - Row 27 | 0.000 | 0% | 0.000 | 0.000 | 1 | 1 | 2 | 0.000 | 0% | 0.000 | 0.000 | 0 | 0.000 |
| 62 | Radiogenic age date (U-Pb) | Ireland | SHRIMP | ANU | Y - Australia thin section - Row 28 | 0.000 | 0% | 0.000 | 0.000 | 1 | 1 | 2 | 0.000 | 0% | 0.000 | 0.000 | 0 | 0.000 |
| 63 | Radiogenic age date (Al-Mg) | Yurimoto | Cameca 1280 SIMS | Hokkaido | Y - Japan thin section - Row 32 | 0.000 | 0% | 0.000 | 0.000 | 0 | 0 | 0 | 0.000 | 0% | 0.000 | 0.000 | 0 | 0.000 |
| | | | | | **Total Radiogenic Isotopic Abundances:** | | | | | | | | 0.000 | 30% | 0.000 | 0.000 | | 0.000 |
| | | | | **Bulk Organic Abundances** | | | | | | | | | | | | |
| 68 | In-situ organic distributions | Haenecour | Cameca nanoSIMS II | UA | Y - UA thin sections - Row 25 & 26 | 0.000 | 0% | 0.000 | 0.000 | 1 | 0 | 1 | 0.000 | 0% | 0.000 | 0.000 | 1 | 0.000 |
| 69 | In-situ organic distributions | Nguyen | Cameca NanoSIMS 50L | JSC | Y - JSC thin section - Row 31 | 0.000 | 0% | 0.000 | 0.000 | 1 | 0 | 1 | 0.000 | 0% | 0.000 | 0.000 | 1 | 0.000 |
| 70 | Bulk C, N, and H abundance and isotopes | Foustoukos/Alexander | EA-IRMS | Carnegie | N - Primary mass allocation | 0.020 | 30% | 0.006 | 0.026 | 10 | 2 | 12 | 0.200 | 0.0% | 0.000 | 0.200 | 1 | 0.200 |
| | | | | | **Total Bulk Organic Abundances:** | | | | | | | | 0.200 | 0.0% | 0.000 | 0.200 | | 0.000 |
| | | | | **Soluble Organic Analysis** | | | | | | | | | | | | |
| 74 | HCN abundance | Aponte | LC-MS | GSFC | N - Primary mass allocation | 0.010 | 0% | 0.000 | 0.010 | 10 | 2 | 12 | 0.100 | 0% | 0.000 | 0.100 | 1 | 0.100 |
| 75 | SOM abundance, chemical structure, elemental composition | Schmitt-Kopplin | Cryo 800MHz Bruker 1H-NMR | HelmholtzZentrum Muenchen | N - Primary mass allocation | 0.020 | 0% | 0.000 | 0.020 | 1 | 0 | 1 | 0.020 | 0% | 0.000 | 0.020 | 1 | 0.020 |
| 76 | Desorbed organic fingerprinting | Schmitt-Kopplin | FTICR-MS | HelmholtzZentrum Muenchen | N - Primary mass allocation | 0.005 | 0% | 0.000 | 0.005 | 1 | 0 | 1 | 0.005 | 0% | 0.000 | 0.005 | 1 | 0.005 |
| 77 | Hypervolatiles abundance and C isotopes | Aponte | GC-MS/IRMS | GSFC | N - Primary mass allocation | 0.100 | 0% | 0.000 | 0.100 | 1 | 0 | 1 | 0.100 | 0% | 0.000 | 0.100 | 1 | 0.000 |
| 78 | Soluble organic separation and fingerprinting | Schmitt-Kopplin | (HILIC-RP)LC-MS*2 | HelmholtzZentrum Muenchen | P residue can be used for bulk IOM elemental analyses or solvent-tolerant mineral analyses | 0.010 | 0% | 0.000 | 0.010 | 2 | 0 | 2 | 0.020 | 0% | 0.000 | 0.020 | 1 | 0.020 |
| 79 | Amino acid/amines distribution, abundance and chirality | Glavin/Dworkin/Parker | LC-MS | GSFC | P residue can be used for bulk IOM elemental residue can be used for bulk IOM elemental analyses or solvent-tolerant mineral analyses + PSG | 0.010 | 0% | 0.000 | 0.010 | 10 | 2 | 12 | 0.100 | 0% | 0.000 | 0.100 | 1 | 0.100 |
| 81 | Carboxylic/hydroxy acids distribution, abundance and chirality | Simkus | GC-MS | GSFC | P residue can be used for bulk IOM elemental analyses or solvent-tolerant mineral analyses + PSG | 0.000 | 0% | 0.000 | 0.000 | 0 | 0 | 0 | 0.000 | 0% | 0.000 | 0.000 | 1 | 0.000 |
| 82 | Aliphatic/aromatic hydrocarbons distribution and abundance | Aponte | GCxGC-MS | GSFC | P residue can be used for bulk IOM elemental analyses or solvent-tolerant mineral analyses + PSG | 0.020 | 0% | 0.000 | 0.020 | 10 | 2 | 12 | 0.200 | 0% | 0.000 | 0.200 | 1 | 0.200 |
| 83 | Peptides and fragile amino acid distribution and abundance | Parker | LC-MS | GSFC | P residue can be used for bulk IOM elemental analyses or solvent-tolerant mineral analyses + PSG | 0.020 | 0% | 0.000 | 0.020 | 10 | 2 | 12 | 0.200 | 0% | 0.000 | 0.200 | 1 | 0.200 |
| | | | | **Insoluble Organic Analysis** | | | | | | | | | | | | |
| 86 | Presolar grains from residues | Haenecour/Nguyen | Nano-SIMS | UA/JSC | P follow-on to analysis of organics from residues | 0.000 | 0% | 0.000 | 0.000 | 1 | 0 | 1 | 0.000 | 0% | 0.000 | 0.000 | 1 | 0.000 |
| | | | | | **Total Compound Specific Organic Abundances:** | | | | | | | | 0.745 | 0% | 0.000 | 0.745 | | 0.645 |
| | | | | **Spectroscopy** | | | | | | | | | | | | |
| 92 | Identification of Spectral Features | Hamilton | microFTIR | SwRI | Y - Pycnometry sample - row 11 | 0.000 | 0% | 0.000 | 0.000 | 1 | 0 | 1 | 0.000 | 0% | 0.000 | 0.000 | 0 | 0.000 |
| | | | | | **Total Spectral Properties Analysis:** | | | | | | | | 0.000 | 0% | 0.000 | 0.000 | | 0.000 |
| | | | | | **Total Allocated Mass:** | | | | | | | | 3.750 | 0% | 0.000 | 3.750 | | 1.143 |

# 9.0 COORDINATED ANALYSIS FLOW DIAGRAMS

A critical aspect of the sample analysis program is coordination of laboratory results to enable a comprehensive understanding of Bennu's formation and history. This coordinated analytical approach will minimize the amount of sample that is consumed from destructive analysis and also provide multiple types of data sets to aid in testing hypotheses and determining the origins and history of Bennu. Coordinated analysis requires pre-defined, hand-offs of sample from one lab to the next. This diagrams below illustrate the flow of sample material from the curation facility to various labs to additional labs around the world.

# OSIRIS-REx Sample Analysis Sequence Flowchart

Last updated: 05/15/2023

[Full-page flowchart depicting the OSIRIS-REx sample analysis sequence, beginning with Sample Recovery at UTTR and branching through Environmental Samples from UTTR and SRC Components, Opening of the Sample Canister, Processing of TAGSAM Samples, First Sample Analysis for quick-look sample(s) (Tier 1, Tier 2, Tier 3), Sample allocation from Curation, Sample Analysis for the Contact Pads, Baseline Mass Allocation, Bulk Elements and Isotopes, Bulk Physical & Thermal Properties, Bulk Organics, In-situ Analyses (Sample texture/petrography/mineralogy, Physical and Thermal Properties, Bulk & In-situ spectral features), Nanoscale Mineralogy, Elements and Isotopes (Presolar Grains, In-situ IOM, Stable isotopes, Radiogenic isotopes, Noble gases), with a legend of abbreviations at bottom right.]

# OSIRIS-REx MAPWG

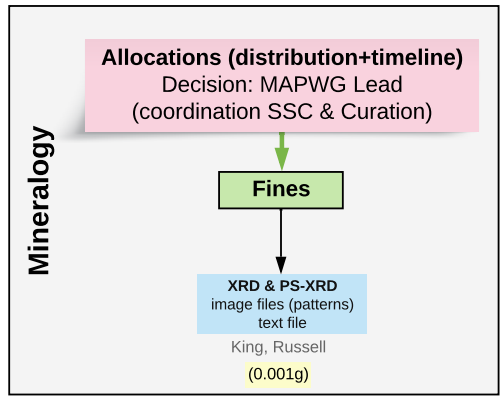
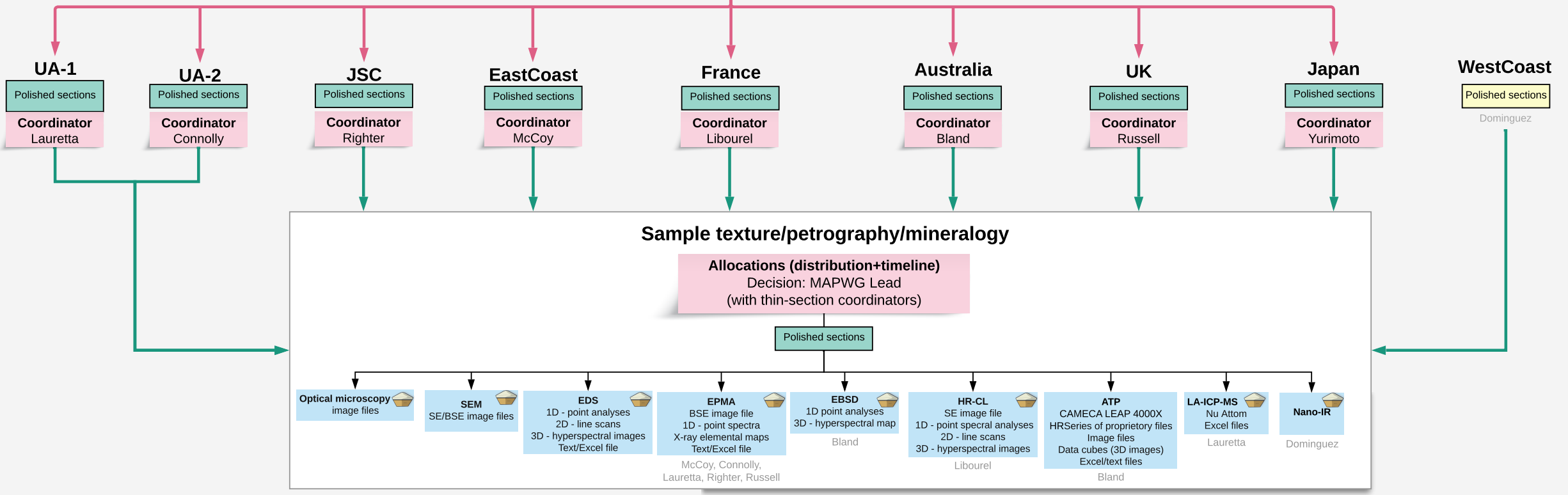
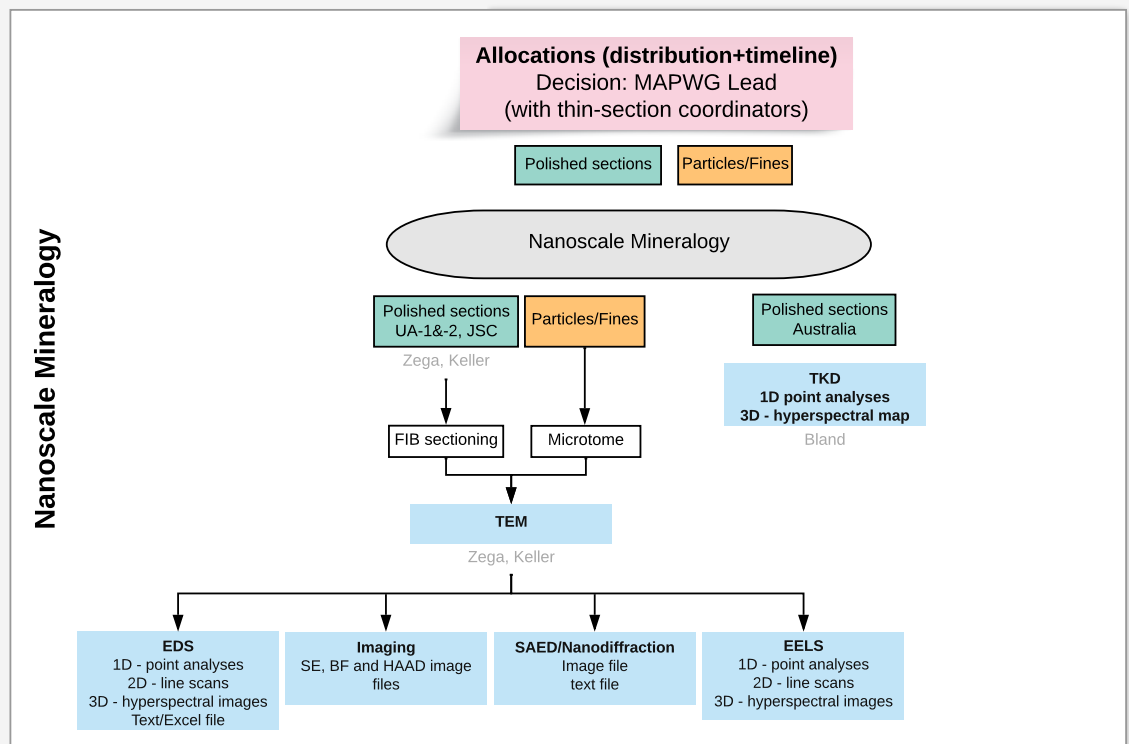

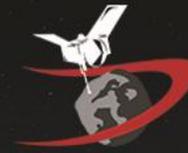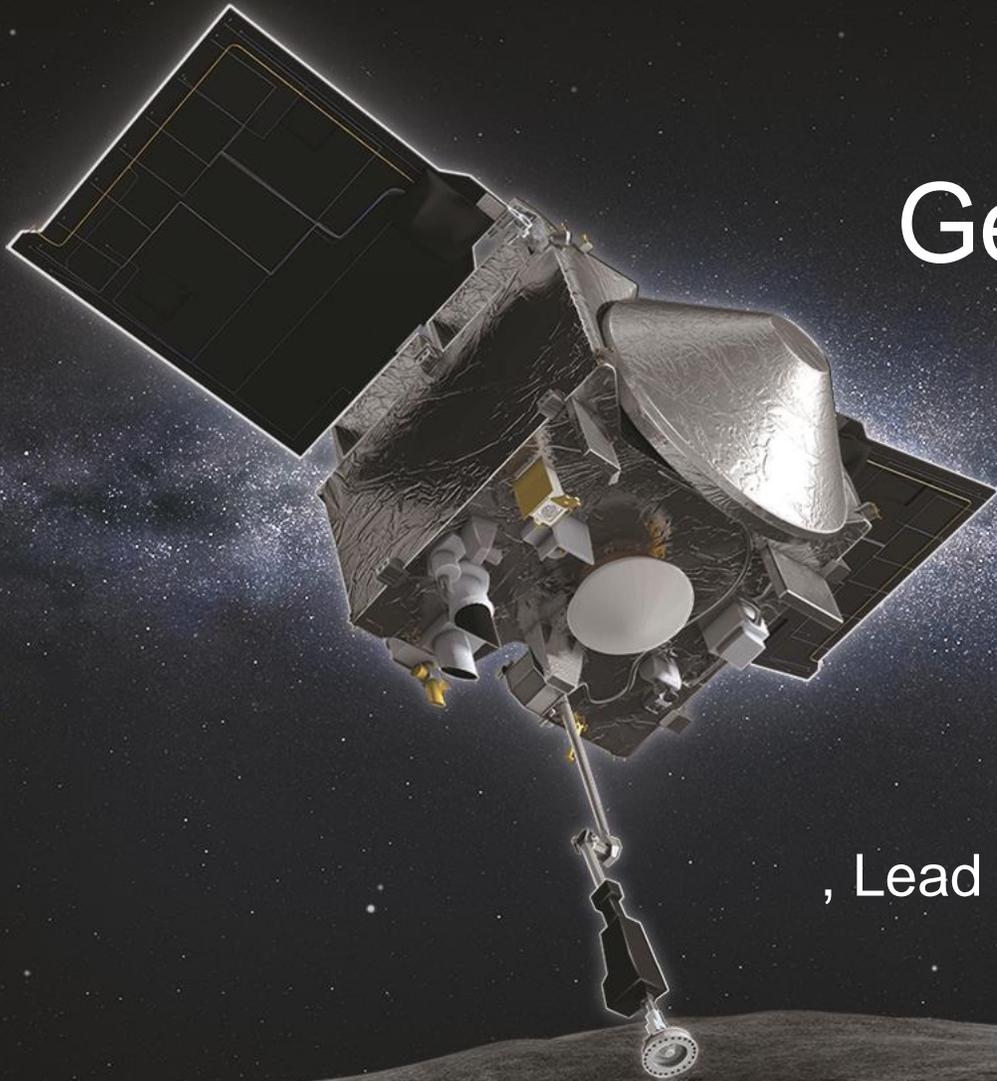

# Geographic Coordinator Analysis Flow Diagrams for Individual Grains and Polished Sections

Prepared by Deputy Lead Scientist Sara S. Russell
Mission Sample Scientist Harold C. Connolly Jr.,
, Lead Scientist Tom Zega, Deputy Lead Scientist Tim McCoy
and the Geographic Coordinators

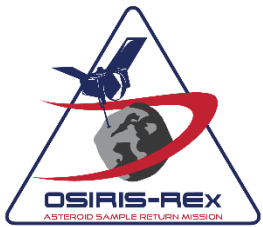
# Geographic coordinators

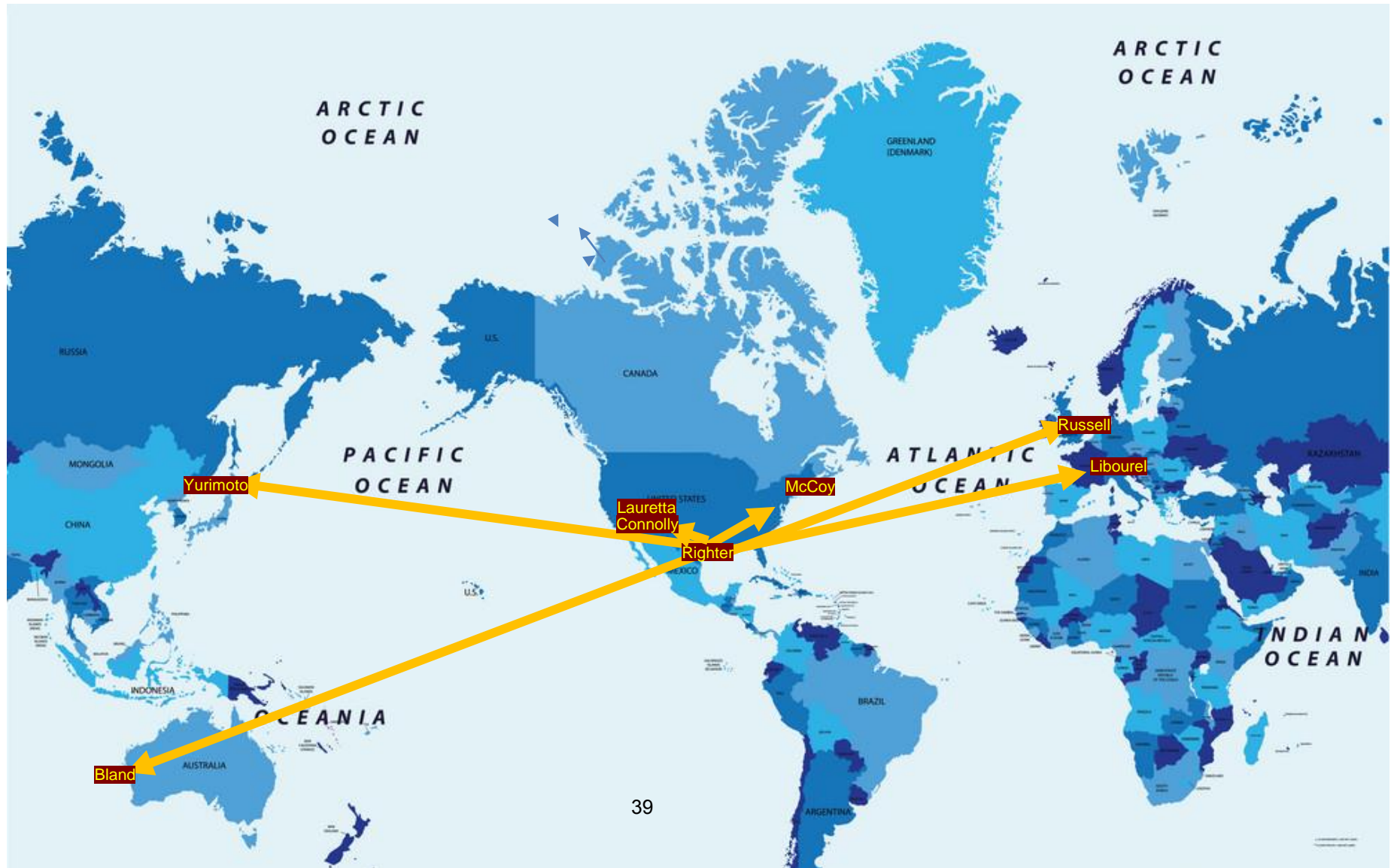

# Pre- and initial characterization of polished sections for decision caucus

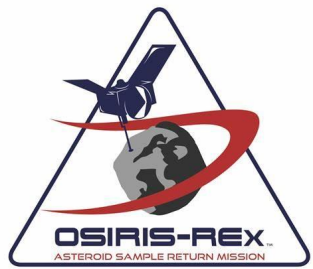
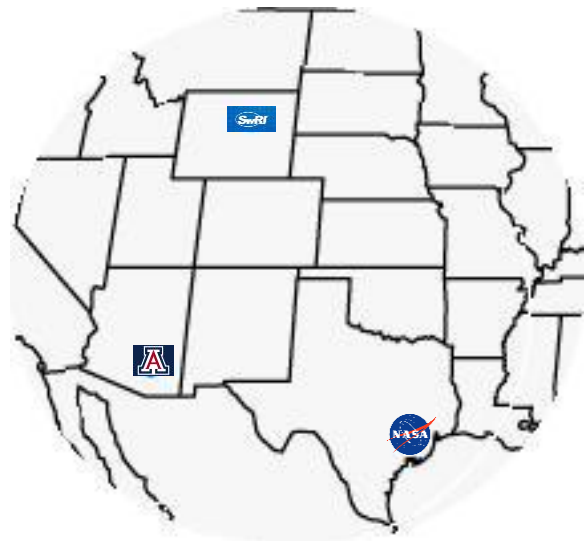

**Curation**
- microXCT → SOI preparation
- Polished section production
- Optical microscopy base maps → For pre-characterization and implementation of μGIS

**Hamilton**
- Optical microscopy
- μFTIR → Initial-characterization of lithologies

**UA + 8 leads**
- C-coat, optical images, μGIS
- E-beam BEI mosaic → Additional minimal initial-characterization of lithologies
- Decision point of additional pre-characterization → Caucus and allocation decision point

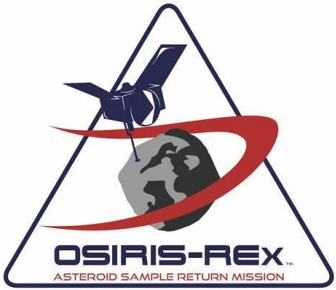

# Initial characterization for fine and intermediate particles

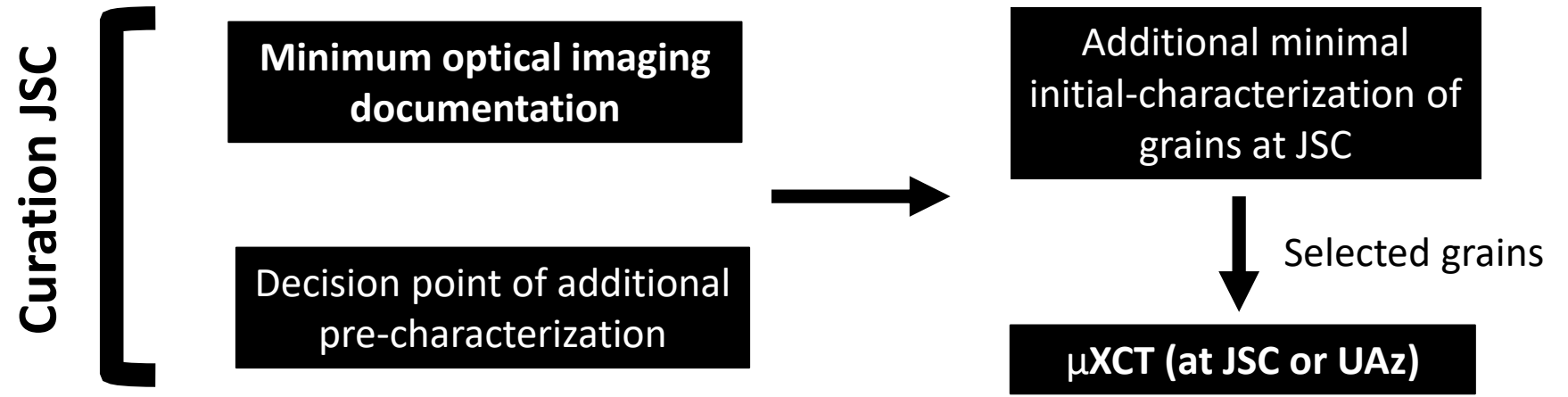

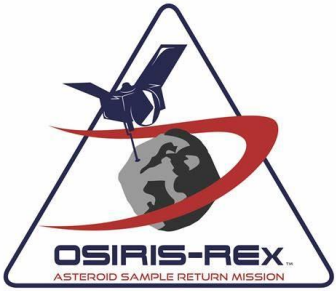

# Timeline

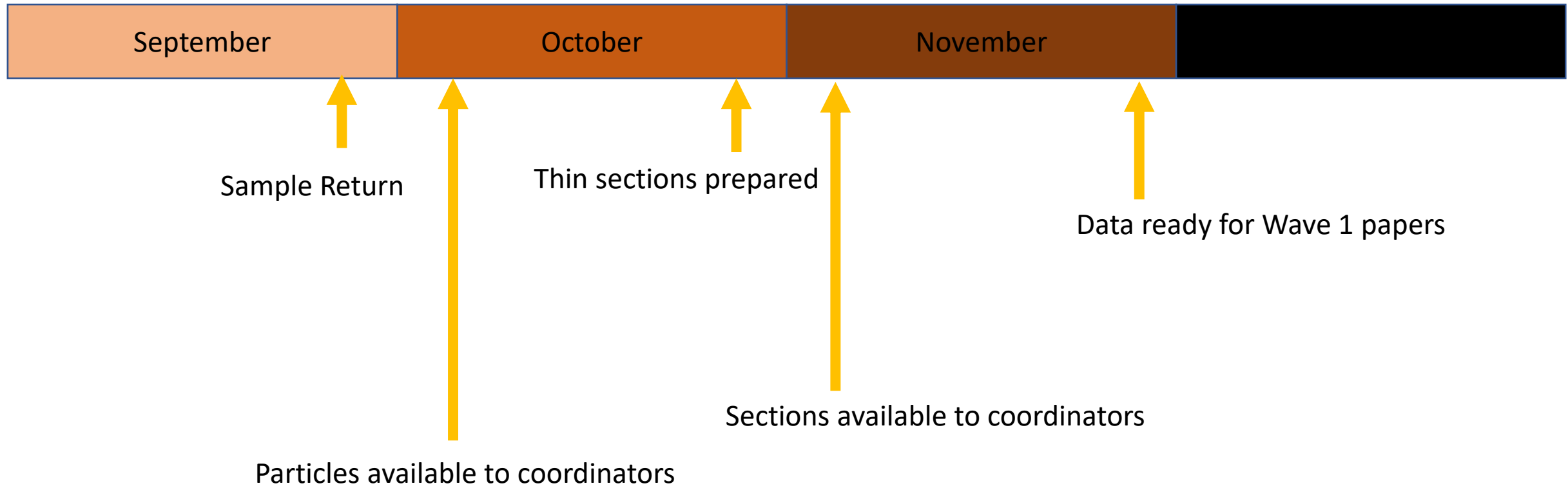

- Sample Return (September)
- Particles available to coordinators (early October)
- Thin sections prepared (late October)
- Sections available to coordinators (early November)
- Data ready for Wave 1 papers (late November)

# Arizona Plan- Lauretta and Connolly coordinators

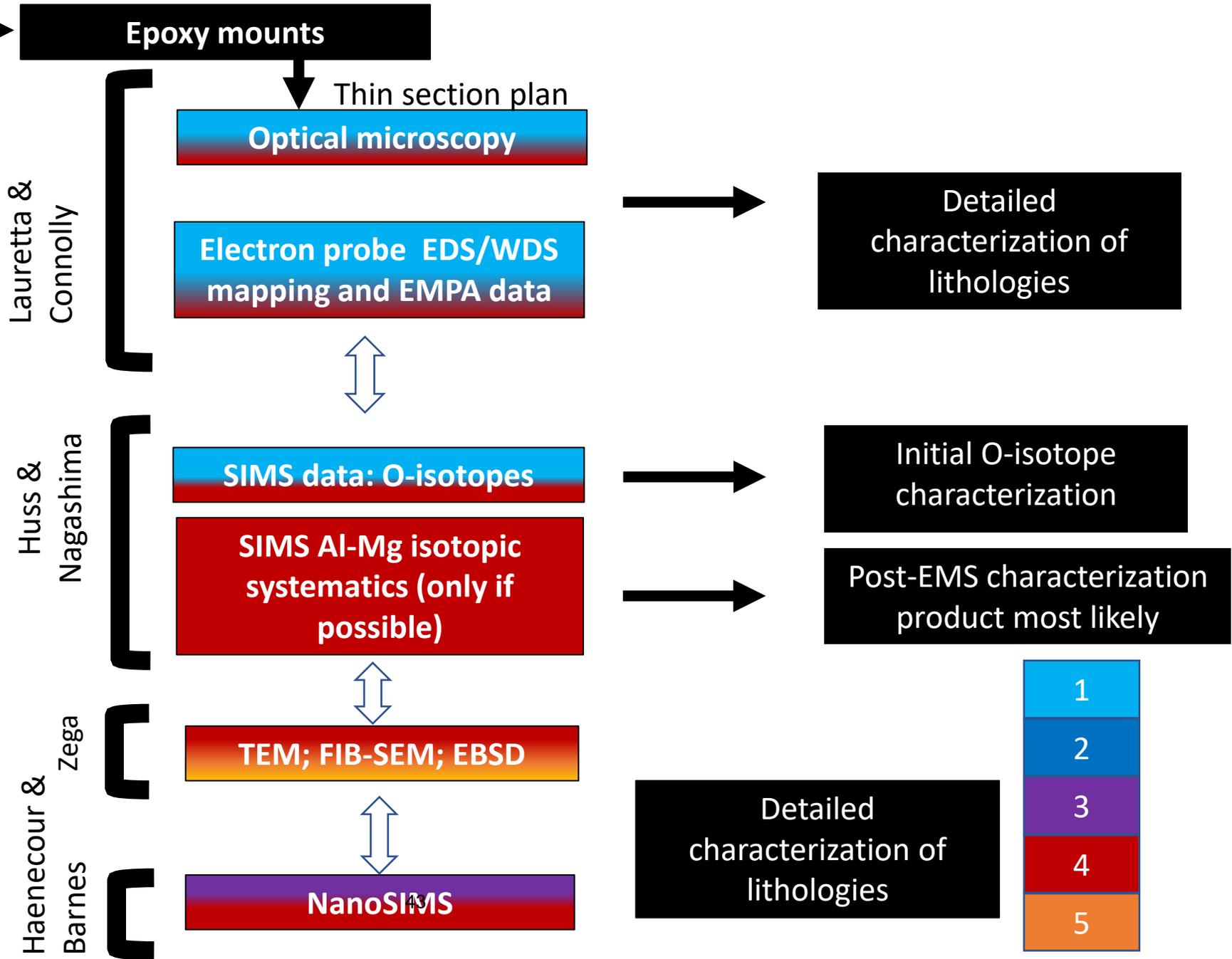

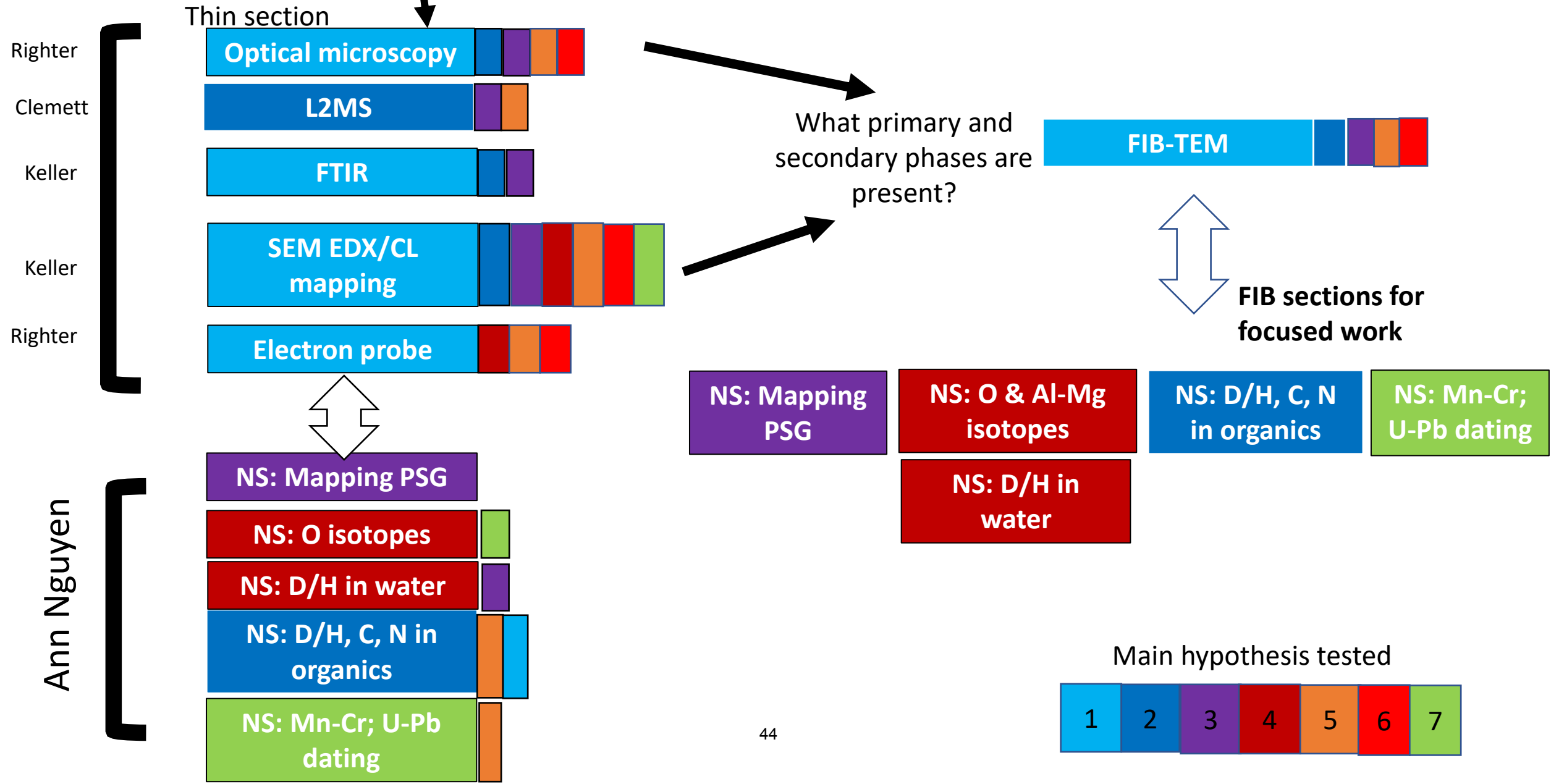

# Smithsonian Plan - McCoy coordinator

| Individual grains | Thin Section | Main hypothesis tested |
|---|---|---|

**McCoy (Corrigan, Postdoc collabs)**

- Optional Micro-CT
- VLM (as received, surface mounted, epoxy impregnated) | VLM
- In house epoxy impregnation
- SEM (surface mounted, epoxy impregnated) | SEM
- EPMA (epoxy impregnated) | EPMA
- Indium mounting of selected grains

**Univ. of Hawaii**

- SIMS: D/H in Phyllosilicates | SIMS: D/H in Phyllosilicates
- SIMS: Mn-Cr in carbonates | SIMS: Mn-Cr in carbonates
- Optional: Al-Mg CAIs | Optional: Al-Mg CAIs

Main hypothesis tested: 1, 4, 5, 6, 7, 9

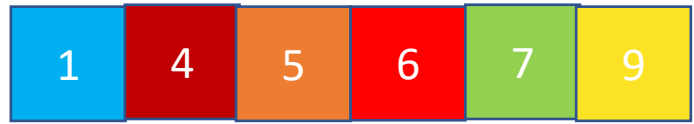
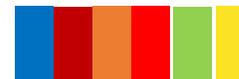
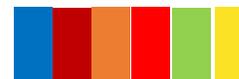
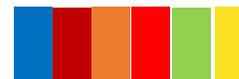
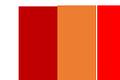
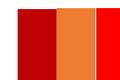
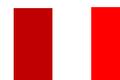

# Japan plan- Yurimoto coordinator

**Individual grain**

**Optical microscopy** → **Divided several grains**

- Mounted → Thin section plan
- Embedded into metal

**Yurimoto:**
- Optical microscopy
- SEM EDX/EBSD/CL mapping
- Electron probe

**Kawasaki:**
- O & Mg isotopes

**Sakamoto:**
- Mapping of C, N & O isotopes
- Mapping of trace elements

**Bajo:**
- Mapping of solar wind noble gas

**Main hypothesis tested:**
1
2
3
4
5

# France/Germany plan - Libourel coordinator
### ≈ 100 mg of grains from Bennu

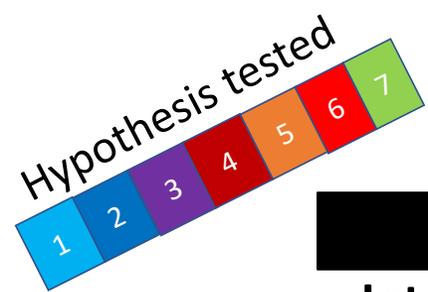

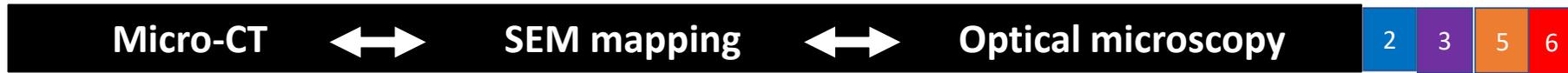

Micro-CT ↔ SEM mapping ↔ Optical microscopy | Thin section

**Intermediates > 100 μm** | **Fines < 100 μm**

Libourel:
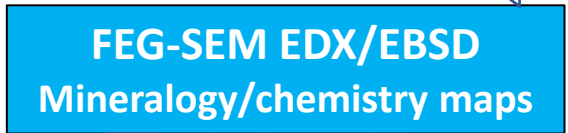
**FEG-SEM EDX/EBSD** — Mineralogy/chemistry maps
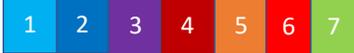

Portail:
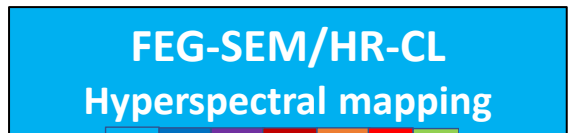
**FEG-SEM/HR-CL** — Hyperspectral mapping
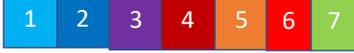

Brenker:
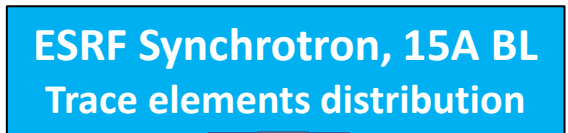
**ESRF Synchrotron, 15A BL** — Trace elements distribution
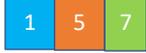

Piani:
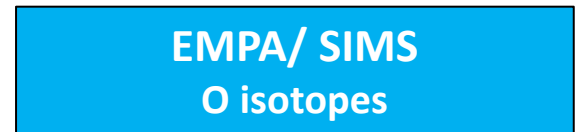
**EMPA/ SIMS** — O isotopes
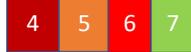

Libourel:
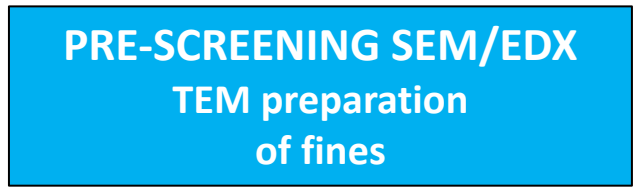
**PRE-SCREENING SEM/EDX** — TEM preparation of fines

Brenker:
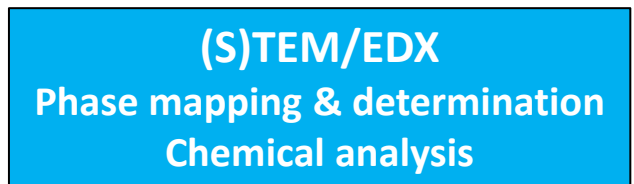
**(S)TEM/EDX** — Phase mapping & determination, Chemical analysis
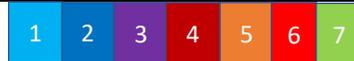

Thin section — Libourel/Portail:
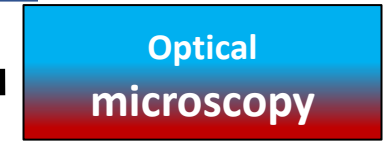
**Optical microscopy**

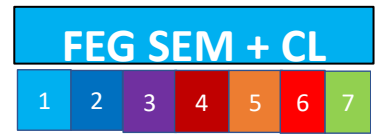
**FEG SEM + CL**
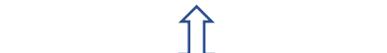

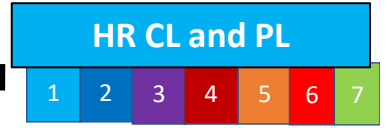
**HR CL and PL**
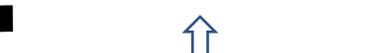

Brenker:
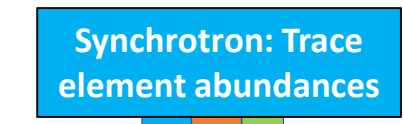
**Synchrotron: Trace element abundances**
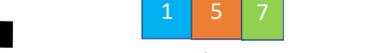

Piani:
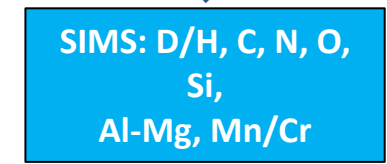
**SIMS: D/H, C, N, O, Si, Al-Mg, Mn/Cr**
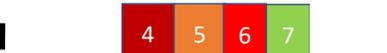

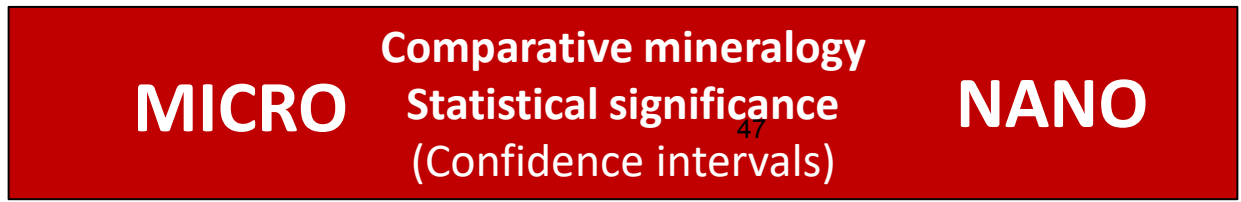
**MICRO** — Comparative mineralogy, Statistical significance (Confidence intervals) — **NANO**

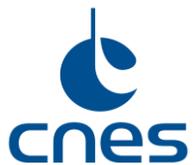

# Ozzie thin section plan - Bland coordinator

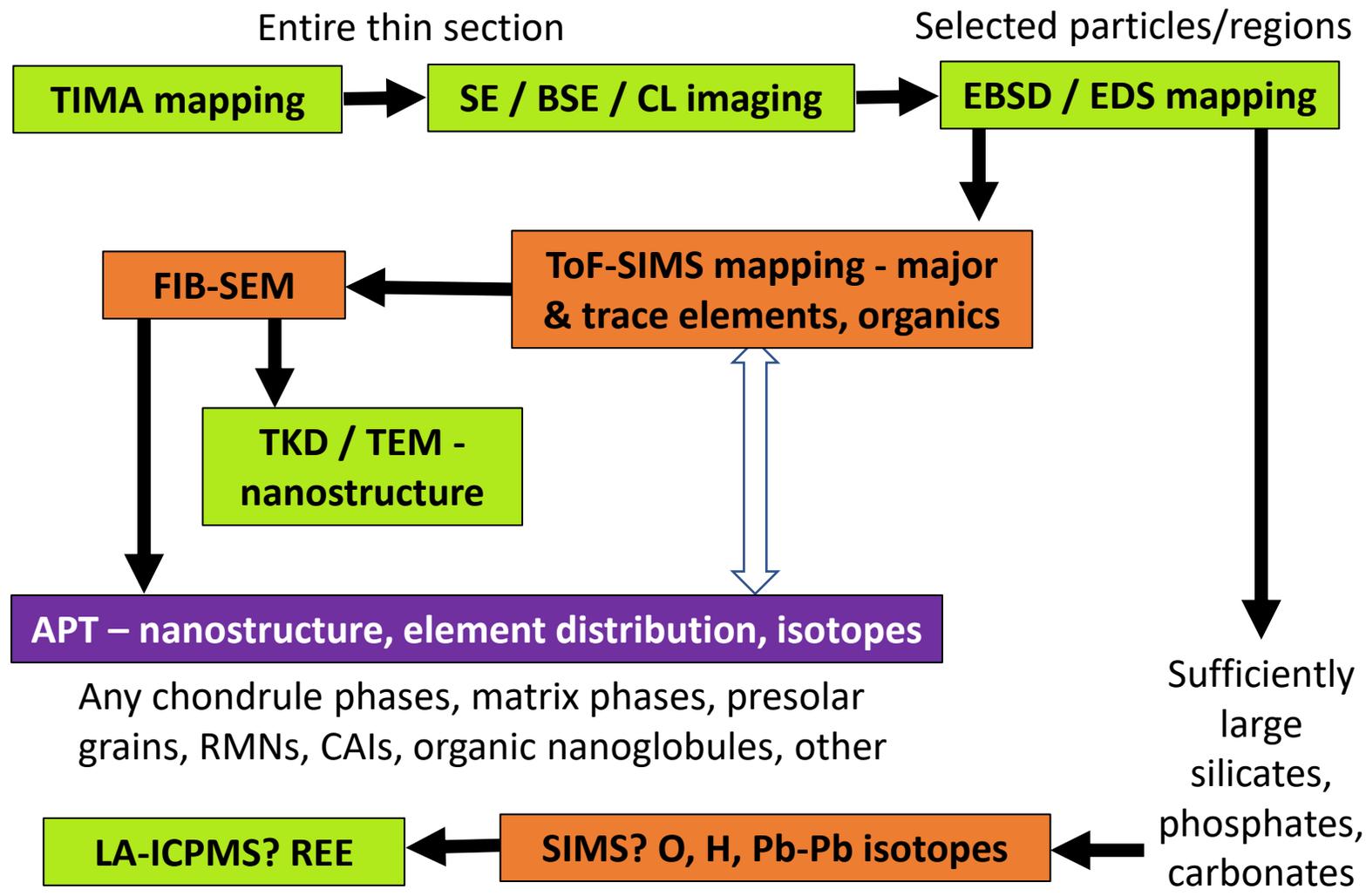

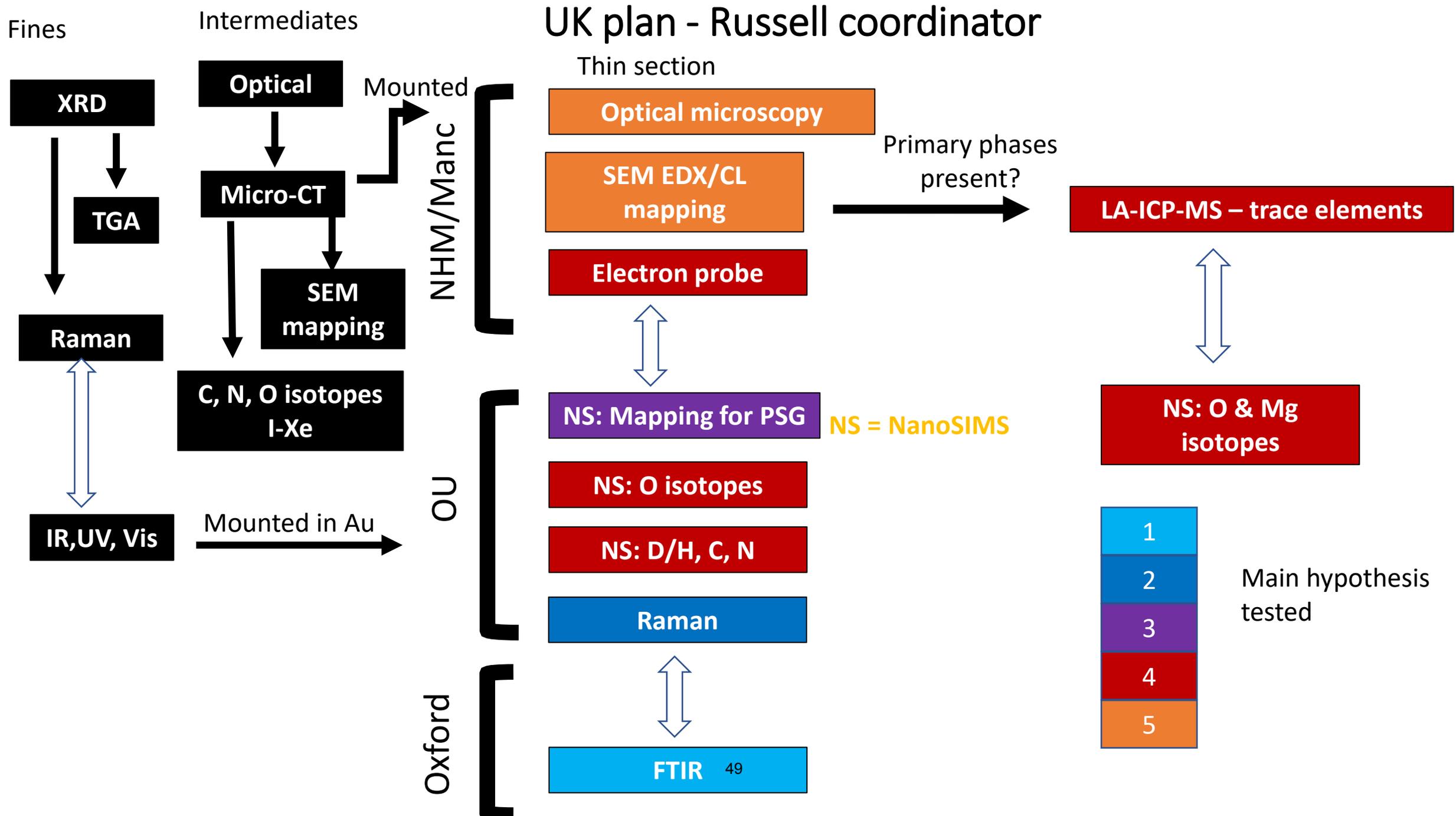

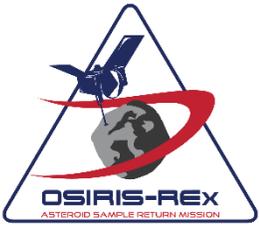

# Data available for Phase 1 papers (Lauretta et al. and Connolly et al.) by end November, 2023

- From all thin sections pre distribution –
  - μCT, micro-FTIR, BSE mosiac
- From individual grains and thin sections-
  - Optical microscopy, SEM, electron probe data
    ⟹ Mineralogy, petrology and mineral chemistry on several clasts

- Additionally:
  - In Situ Oxygen isotopes (UAz, Japan, UK, Australia)
  - L2MS (JSC)
  - HR-CL (France), CL (Japan)
  - EBSD (Uaz, Australia and Japan)

  - μCT on individual grains

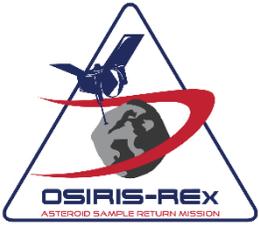

# Data available for MAPWG paper (Zega-McCoy et al.) by end February, 2024

- TEM from sections at UAz, JSC, Australia and France sections
- FIB-SEM, EBSD from UAz sections
- FTIR from JSC/UK sections (also for SSAWG)

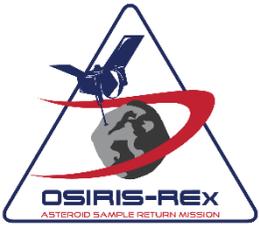

# Data available for SEIWG paper (Barnes-Nguyen et al.) by end Feb

All data/techniques from Phase 1, plus:
- Trace element data from UAz section (SIMS), Japan section (SCAPS), UK section (LA-ICP-MS), France section (ESRF), Australia section (APT, TOF-SIMS, LA-ICP-MS)
- Map PSG (UAZ, JSC, UK, Australia)
- Al-Mg (JSC, France, UK)
- Mn-Cr (JSC, Smithsonian, France)
- D/H, C, N (JSC, Smithsonian, Japan, UAZ, UK, France, Australia)
- Si isotopes (France)
- Solar Wind noble gases (Japan)
- U-Pb (JSC)
- Pb-Pb (Australia)

If impact glasses are found:
- Ar-Ar dating from Smithsonian section

# OSIRIS-REx SSAWG

## Bulk & In-situ spectral features

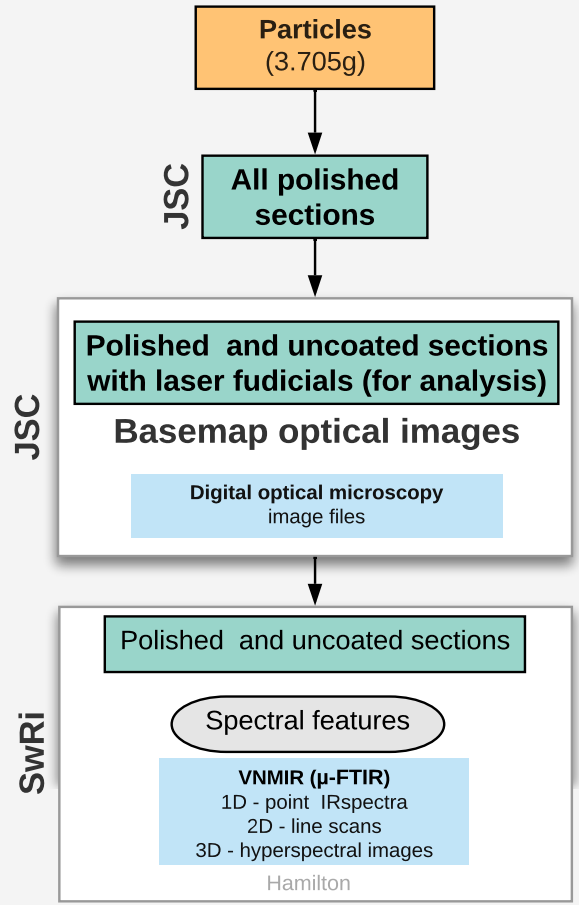
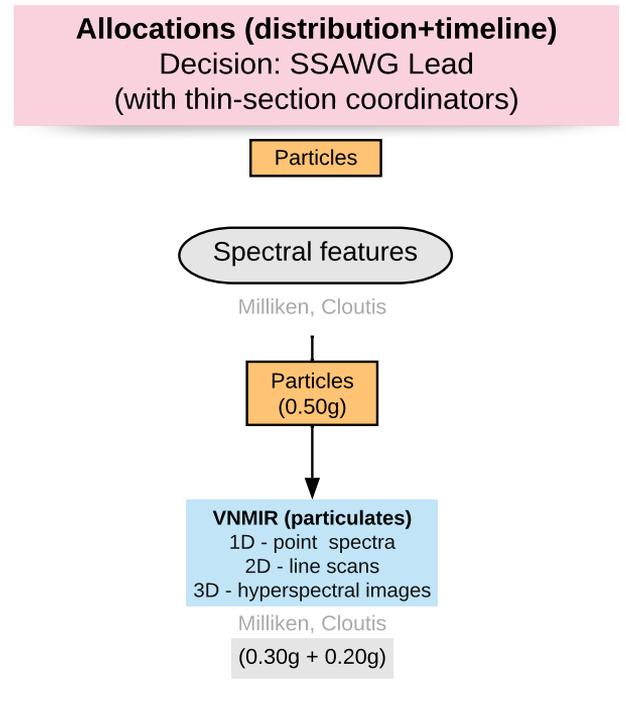

# OSIRIS-REx SEIWG

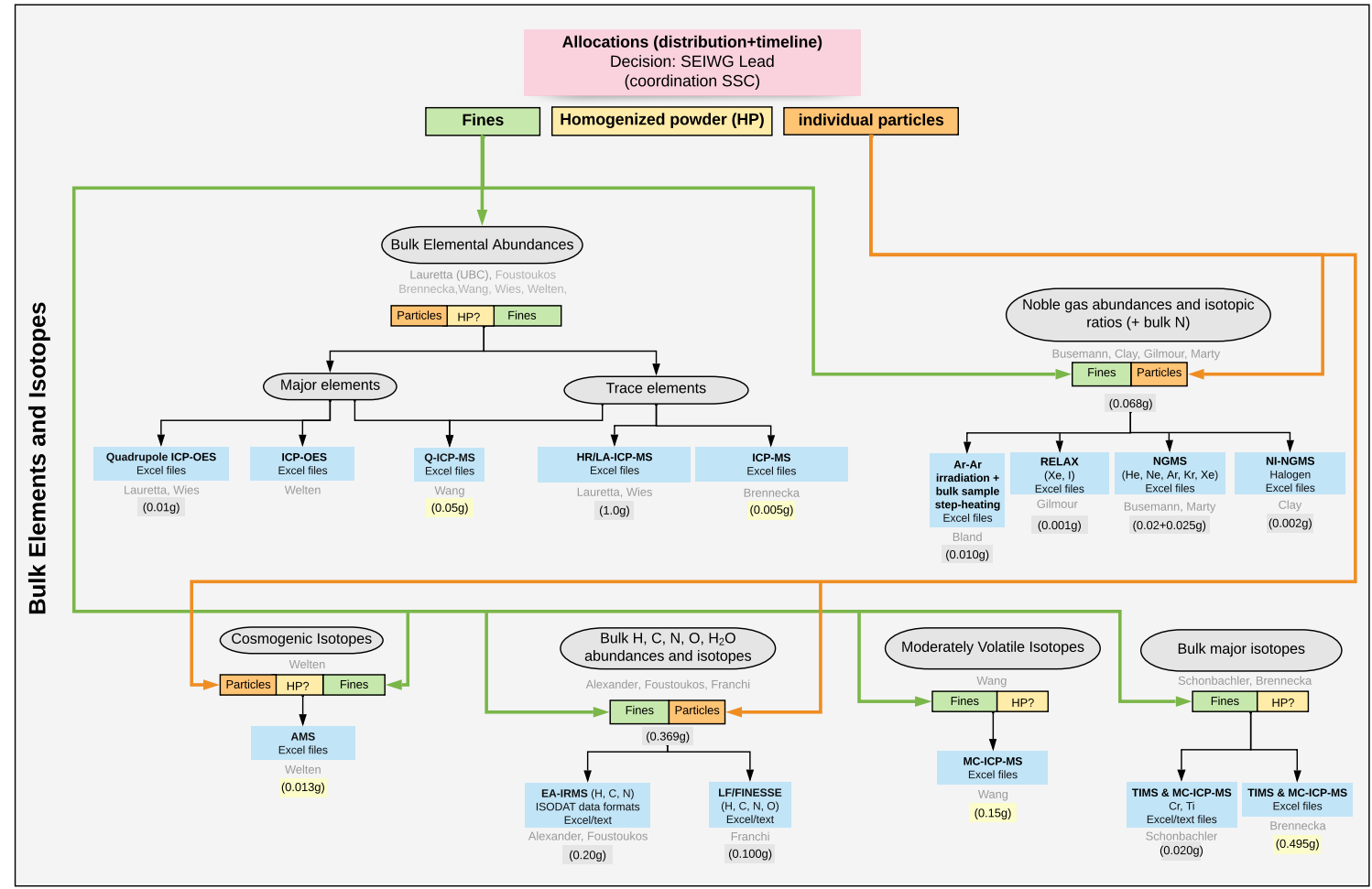

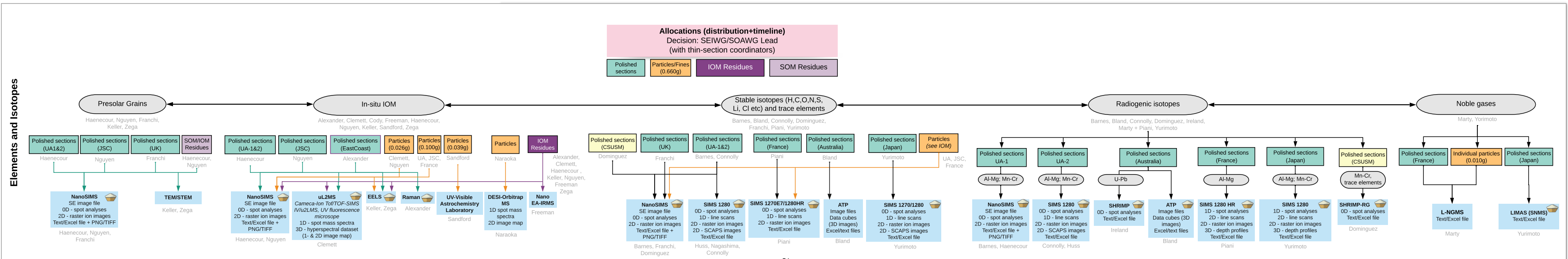

# OSIRIS-REx SOAWG

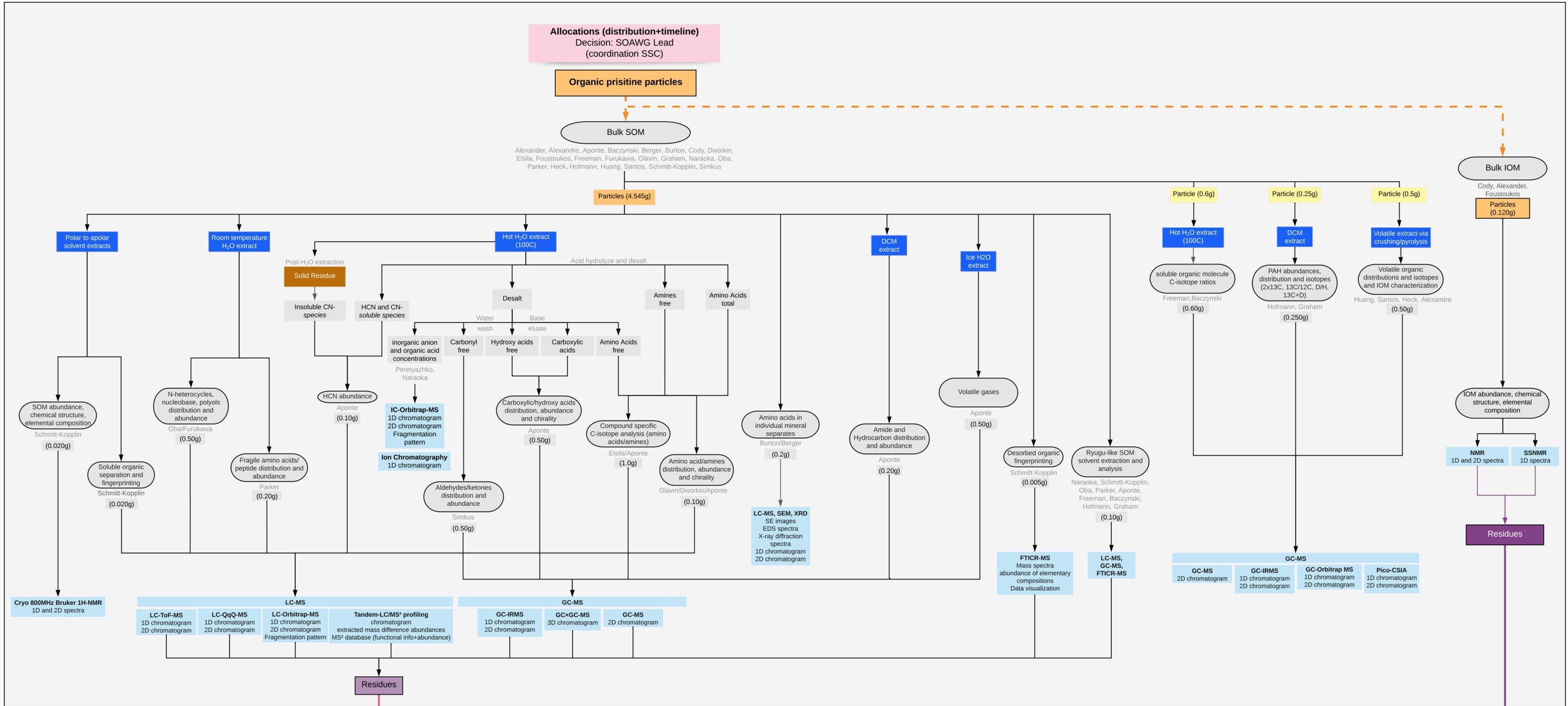

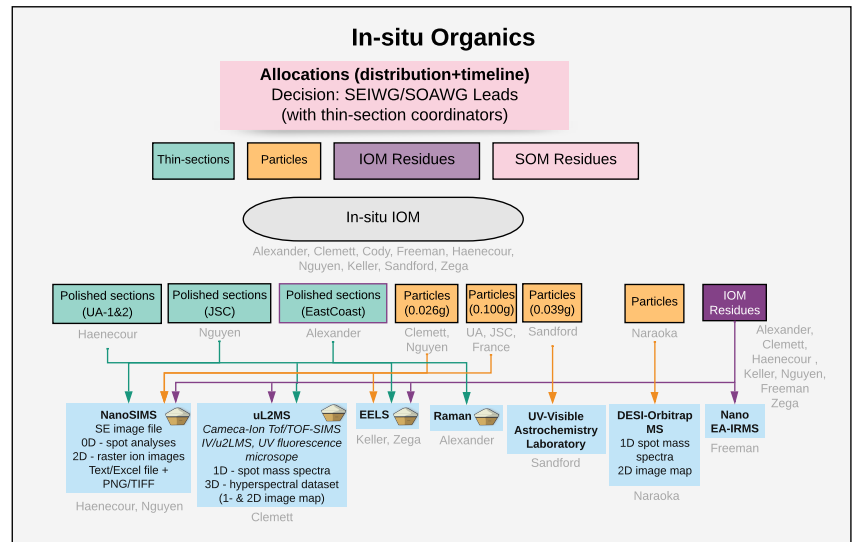

# OSIRIS-REx SPTAWG

## Bulk Physical & Thermal Properties

**Allocations (distribution+timeline)**
Decision: SPTWG Lead
(coordination SSC)

(0.692g)

- Individual Particles
- Fines
- Homogenized Powder (HP)

### Mechanical
Hildebrand, Hoover, Opeil

**Particles**

- Coefficient of thermal expansion
  - **Cantilever dilatometer**
  - Opeil
- Young's modulus, shear modulus, Poisson's ratio, Lamé's constant and bulk modulus
  - **Seismic Velocities and Rock Ultrasonic Elastic Constants**
  - Hildebrand
  - (0.110g)
- Elastic modulus, crushing strength, critical flaw length at failure, and fracture behavior
  - **Compression test**
  - Hoover
  - (0.001g)
- Direct Shear Strength Measurement
  - Hildebrand

### Thermal conductivity
Daly, Ryan

- Particles → Thermal conductivity
  - **Lock-in thermography**
  - Tanaka, Nagano, Ryan
  - (0.150g)
  - **Spherical Cell Thermal Conductivity Analyzer**
  - Siegler, Ryan
  - (0.400g)
- HP? → Heat capacity
  - **Differential scanning calorimeter**
  - Excel/text files
  - Biele and Opeil, Ryan
  - (0.030g)

### Granular
Hoover, Sanchez

- Fines, Particles
  - Cohesion between particles
    - **AFM Particle Cohesion**
    - Hoover, Sanchez
    - (0.001g)
  - Angle of repose, angle of friction
    - **Angle of repose measurements**
    - Sanchez, Ryan

---

## All samples at JSC

- Mass measurement
  - text files
- Sample shape
  - 3D laser scan
  - image file
- Bulk density, grain density, porosity
  - Gas pycnometry
  - Excel/text file
  - Macke, Righter
  - (3.705g)

---

## Physical and Thermal Properties

**Allocations (distribution+timeline)**
Decision: SPTWG Lead
(coordination Connolly/Lauretta)

**Particles**

### Thermal conductivity
- Grain and matrix thermal conductivity
  - **SThM with AFM**
  - Daly

### Mechanical
- Indentation modulus, indentation hardness, ductility, elastic work, inelastic work and total work, creep curves, storage modulus and loss modulus
  - **Nano- and micro-indentation**
  - Hoover

## 9.1 Coordinated Analyses of Targeted Soluble Organics

## Coordinated analysis of amino acids, amines, carboxylic acids, aldehydes, ketones, and insoluble cyanide

Coordination of sample analysis can lead to increased sample efficiency and a better ability to compare analytical results. Briefly, the coordinated method involves a hot-water extraction, as described in the primary amino acid section. It is also possible to collect volatiles prior to hot water extraction, either by direct gas sampling and/or by first disaggregating the sample in ice/water. The extract is then split into several aliquots as shown in the figure below. A portion is used for analysis of amines and free amino acids using the AccQ•Tag method. A second portion is hydrolyzed and analyzed for total amino acids. The bulk (~90%) of the extract is desalted, as described in the primary amino acid section. The $NH_4OH$ eluate, containing amino acids, is analyzed using amino acid protocols. The water eluate, however, is split into three portions for analysis of hydroxy acids, carboxylic acids, and aldehydes/ketones. The solid residue can be analyzed for insoluble cyanide-containing species. This split increases the amount of material allocated for each individual compound class but does not impact the ability to compare with the targeted analyses used in meteorite studies and will be verified during SART.

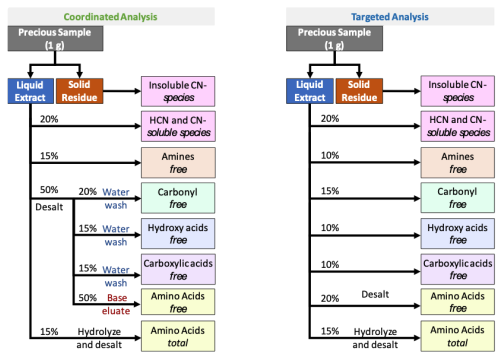

Schematic showing sample splits for coordinated and targeted (traditional) analyses of multiple soluble organic compound classes in a single extract.

## Hayabusa2 coordinated analysis of soluble organic material

The Hayabusa2 soluble organic matter analyses were performed from a single organic and a single water extract to maximize the number of species that could be detected from a small amount. This trades the improved detection for the ability to cross-compare with targeted extractions used on meteorites. To better compare with Ryugu soluble organic matter analyses, a small sample of Bennu material will be extracted the same way as Ryugu material and compared with the analytical techniques available.

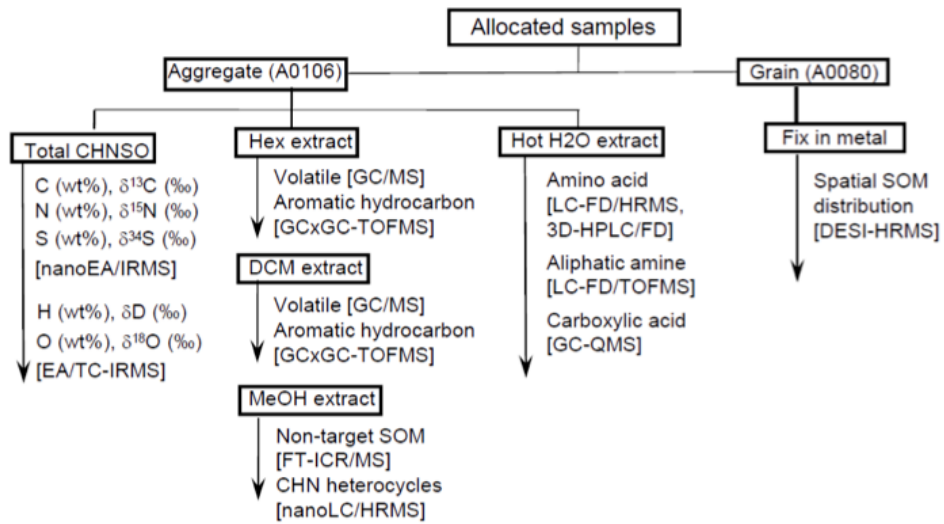

Hyabusa2 extraction method. OSIRIS-REx will focus on the Hexane and Hot Water extraction trees. Others as sample size permits. (Naraoka et al. 2023)

# 10.0 DATA STORAGE AND MANAGEMENT

**Scope and Description**

This section addresses the roles and responsibilities of various agents in the production and management of OSIRIS-REx sample analysis data (referred to as the Sample Analysis Data System). It describes the strategy to collect, store, manage, and share data generated during the sample analysis phase of the mission, maintain system integrity and accessibility, and describes the plan to deliver applicable data to the data preservation archive. It also describes and defines the automatic spatial registration workflow and data transformations.

**Sample Analysis Data System**

The OSIRIS-REx Sample Analysis Data System comprises four entities:

- **Sample Analysis Micro-Information System (SAMIS):** The main project-level hub for all OSIRIS-REx sample analysis data and the interface between all other entities listed below.
- **Sample Analysis Team (SAT) – individual labs and data product producers:** The SAT is divided into working groups organized by analytical objectives. Each working group is responsible for specific sample analysis data products. The SAT will send data products to SAMIS for additional processing and distribution. Additional information about the role of the SAT in data management can be found in the Sample Analysis Appendix to the Science Data Management Plan (UA-PLN-9.4.4-004).
- **Astromaterials Acquisition and Curation Office, part of the Astromaterials Research and Exploration Science (ARES) Division at Johnson Space Center (henceforth referred to as JSC)**: Responsible for the initial sample, sub-sample, split creations and characterization, and assigning sample identification numbers. JSC will produce and manage Preliminary Examination (PE) data for each sample. A subset of this information will be delivered to SAMIS for each sample allocated to OSIRIS-REx Sample Analysis. A detailed description of the data management interface between SAMIS and JSC can be found in the Curation to SAMIS Interface Control Document (UA-ICD-4.5.3-001).
- **Astromaterials Data System (Astromat):** Astromat will register data products received from SAMIS with with digital object identifiers (DOIs) and return the DOIs to SAMIS for access by the SAT members. A detailed description of the interface between SAMIS and Astromat can be found in the Astromat to SAMIS Interface Control Document (UA-ICD-4.5.4-001).

During the sample analysis phase, data will move among these four entities, as illustrated in Figure 1.

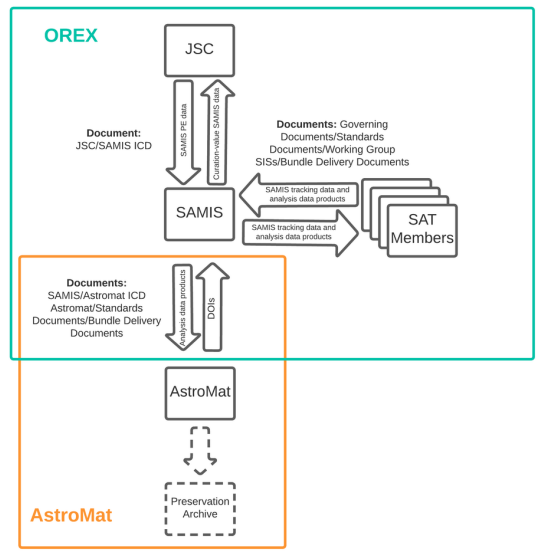

Figure 1: The flow of data between SAMIS, JSC, the SAT, and Astromat. Everything within the green (upper) box is covered by the OSIRIS-REx Sample Analysis Plan (SAP). Everything in the orange (lower) box is the responsibility of Astromat in coordination with NASA headquarters (HQ).

**Sample Analysis Micro-Information System (SAMIS)**

SAMIS is a custom-designed tool for collaborative and coordinated analysis of the sample returned from Bennu. It collects, stores, processes, and distributes all data pertaining to the sample analysis phase of the mission. SAMIS consists of four main components:

- *Sample Analysis Tracking Application (SATA)* - Tracks the physical location and state of the sample. Used to keep track of sample location and schedule and maintain the quality of a sample as it moves between labs. SATA also helps ensure that the analytical schedule for each sample is maintained.
- *Sample Analysis Desktop Application (SADA)* - A web-based app used to upload, download, search, share, and view the sample analysis data generated by all labs and instruments on the mission.
- *SAMIS Server* - Controls SATA and SADA access to the SAMIS Database.
- *SAMIS Database* - A relational database with a GIS (spatial) extension. The system stores all data from the sample analysis phase of the mission, including coordinate system applied to samples for clear documentation of where sample analysis occurred. It performs automatic spatial registration between datasets.

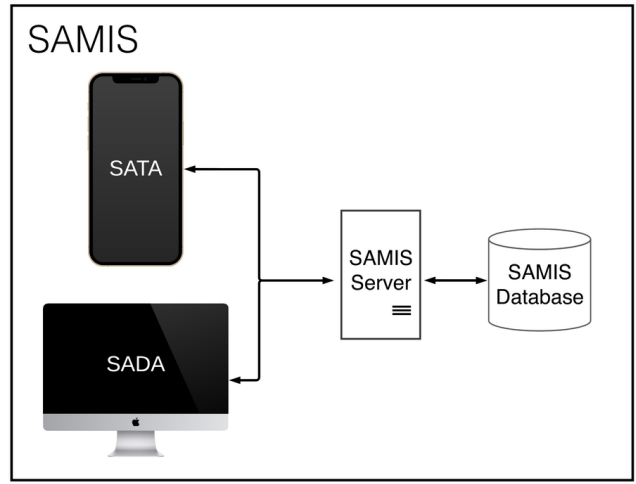

Figure 2: Diagram showing the flow of information between the four main parts of SAMIS.

## Mapping to L1 Requirements

**Requirement L1.2: Document the contamination of the sample acquired from collection, transport, curation, and distribution**

- The Sample Analysis Tracking Application (SATA) contributes to requirement **L1.2** because the phone app is designed to track any physical changes (expected and unexpected) or processes performed on the samples as they move between the labs during the sample analysis phase of the mission. This information is stored in the SAMIS Database and accessible via the SADA or SATA.
- The collection of all of the analytical data and corresponding metadata serves as a **record of the contamination and destructive processes** an individual sample or sub-sample has undergone due to these types of processes. By spatially co-registering analytical data to a common coordinate system, we are able to precisely document the extent of a destructive process on a thin section surface. Scientists will be able to spatially query an area of interest on a thin section basemap and quickly visualize the nearby locations of previous analysis. This will **inform the sample scientists about the integrity of the sample, as it pertains to their unique analysis needs, down to an area-of-interest level of precision.**

**Requirement L4.6: Science Data Management** – The OSIRIS-REx PI shall be responsible for initial analysis of the scientific data, its subsequent delivery to an appropriate data repository, the publication of scientific findings, and communication of results to the public. Additionally, the OSIRIS-REx PI shall be responsible for collecting engineering and ancillary information necessary to validate and calibrate the scientific data prior to depositing it in a NASA approved data repository. The OSIRIS-REx science data base shall be made available to the science community without restrictions or proprietary data rights of any kind.

- The SAMIS contributes to the **initial analysis of the scientific data** by being the central hub for all data pertaining to the analysis of OSIRIS-REx samples during the analysis phase of the mission. The spatial capabilities of the SAMIS contribute to this requirement by transforming disparate datasets into a common data type (vectors or rasters with a corresponding attribute table). This allows researchers access each other's data products without needing special programs or domaine knowledge to locate or decipher the data.
- The SAMIS contributes to the data's **subsequent delivery to an appropriate data repository** by using PDS inspired metadata requirements, non-proprietary, widely accessible data formats when possible, and open, documented standards. It also permits curation, both at the time of archive and for future references, all information gathered on samples to better inform on future allocations as needed to the scientific community.
- The SAMIS contributes to the **publication of scientific findings, and communication of results to the public** by making the data accessible to all sample scientists on the mission. It spatially registers datasets from different instruments and labs to one another for ease of scientific comparison by means of searching for data by location and by spatial queries such as mapping where things are, finding what's nearby or inside an area of interest, mapping densities, or mapping change over time. It can visualize the spatially registered data in a visualization tool which will allow for the simple and intuitive production of figures and graphics for use in scientific publications and for communications and public engagement materials.
- The SAMIS contributes to the **collection of engineering and ancillary information necessary to validate and calibrate the scientific data** by collecting, maintaining, and allowing for the search and distribution of any metadata generated during analysis (this includes sample identification metadata and analysis metadata). It collects additional analysis information such as reference standards used and any information about the physical condition of the sample at the time of analysis via the SATA. The spatial capabilities of the SAMIS contribute to this requirement by capturing location ancillary data of an analysis process. By keeping track of the locations of the different analyses within a single coordinate system, researchers are able to maintain the integrity of their results by assessing the destructive processes that have been performed in that area before and compare their results quickly and easily to other analyses performed in the same area of interest.

## Data

"Data" in this context is defined as information that is passed between and among the elements of the OSIRIS-REx Sample Analysis System. Different types of data have different uses within the Sample Analysis System and are treated differently by the SAMIS.

### Types of Data

Data generated during OSIRIS-REx sample analysis can be categorized as follows:

- **SAMIS sample identification data** - Images and information collected by the Preliminary Examination Team (PET) or the SAT for the purpose of identification when a new sample, subsample, or split is created (e.g., sample image, description of sample condition, sample mass). SAMIS sample identification data created by the SAT will be reported to JSC at the end of the analysis phase of the mission to be included in the sample catalog.
- **SAMIS tracking data** - Data pertaining to the physical state and location of the sample acquired using the SATA. A subset of these data will be reported to JSC at the end of the mission to be included in the sample catalog.
- **Analysis data** - Data produced from the analysis of samples allocated to the SAT, as well as any supporting calibrations, blanks, standards, and proxy analyses. These data are formatted by the SAT to comply with OSIRIS-REx Sample Analysis Data Standards Documents, Software Interface Standards Documents, and Bundle Delivery Documents. Analysis data will be processed by SAMIS into Astromat compatible data products, assigned DOIs, and archived with Astromat. The Astromat to SAMIS Interface Control Document (UA-ICD-4.5.4-001) has a detailed description of this transformation and the ultimate data product format standard. Analysis data includes the following subtypes:
  - **Raw data** - Data uploaded from an instrument after the analysis in a format that is accessible without the use of proprietary software and with as minimal processing as possible.
  - **Processed data** - Data that have had subsequent scientific processing, but do not combine different data products or reduce a product down to the point where it could be considered a publication-ready scientific result.

- **Derived data:** Data that is created by combining and processing existing data into a higher-level, publication-ready product. Derived data are being defined on an analytical technique-by-analytical technique basis as products that could be considered a scientific result that will be published in a paper with a methods section and include a list of input product DOIs. These data are not being ingested by the SAMIS.

**Standards Used**

The SAMIS team has worked closely with the SAT and the Astromat team to develop a set of internal protocols, referred to as Data Standards. These protocols will allow any Data Product produced by an OSIRIS-REx SAT member as part of the Sample Analysis effort to be delivered to the ADA in a consistent format and with consistent documentation with the explicit goal of ensuring compliance with FAIR principles as far as possible within the timeline given.

SAT members are required to format their products as specified in a set of data product documents: Software Interface Specification (SIS) documents (UA-SIS-4.5.4-010 through UA-SIS-4.5.4-18), Bundle Delivery Documents (BDDs) (UA-SIS-4.5.4-020 through UA-SIS-4.5.4-80), and Data Standards Documents (DSDs) (UA-SIS-4.5.4-002 through UA-SIS-4.5.4-007). These documents are designed to establish a standard, archivable format to which all data product types (e.g., images, tables, documents) will conform, while also ensuring that any additional metadata fields and ancillary information are captured for each analytical technique (e.g., *backscatter electron* image, quantitative chemical analysis tabular data collected during electron microprobe analysis (EMPA)). The standards defined in these documents are being produced in concert with the development of the interface between SAMIS and Astromat to minimize the amount of additional processing needed to transform OSIRIS-REx data into an archive-ready form.

**Additional SAMIS Data Processing**

Most science data processing during the sample analysis phase of the mission happens at the individual labs by members of the SAT, and these data will be delivered to the SAMIS without the need for further processing. There are two main exceptions. The first is the transformation of analysis data with spatial components from the instrument coordinate system to the sample basemap coordinate system and the transformation of those data into spatial data types. The second is the transformation of the SAT formatted data products into data product packages for SAMIS download and submission to Astromat. A detailed explanation of this transformation and resulting format can be found in the Astromat to SAMIS Interface Control Document (UA-ICD-4.5.4-001).

**Data Quality and Validation**

The SAMIS will validate all analysis data products submitted by SAT members. This validation process includes the following automated data integrity checks:

1. Confirmation that the required data product components identified in the corresponding BDD are present.
2. Confirmation that each data product component is in one of the acceptable formats identified in the corresponding DSD.
3. Confirmation that the .yaml metadata files have the required metadata fields as defined in the corresponding DSD and they are in the correct format.

The SAMIS will also verify that a subset of the required keyword values are matched to controlled vocabularies maintained in the SAMIS Database either by checking user input or by adding the keywords automatically to ensure correctness and consistency. These keywords are listed in the Sample Analysis Appendix to the Science Data Management Plan (UA-PLN-9.4.4-004).

**Data Release**

It is the responsibility of the SAT to ensure the scientific corrections and quality of any submitted data. Data delivered to the SAMIS will be given a three-week grace period before being automatically transferred to Astromat via a nightly batch delivery process. During this three-week period, data submissions can be deprecated and re-submitted to the SAMIS should any data quality issues be found. After the three-week period, the Data Products will be delivered to the ADA and will remain in a calibration and validation state, meaning the DOI will be findable and metadata will be viewable, but the data itself will not be viewable or downloadable until the SAT member releases them via the process described in section 4.2.4 of the Astromat to SAMIS Interface Control Document (UA-ICD-4.5.4-001).

Data will undergo an Astromat-led peer review. It is the responsibility of the submitting SAT member to respond to ADA peer review as needed. The workflow for data submission, review, and release is detailed in section 4.2 of the Astromat to SAMIS Interface Control Document (UA-ICD-4.5.4-001).

Any data products that have not been published at the end of the Close Out Phase of the OSIRIS-REx mission will automatically be released and made accessible to the public by Astromat.

## Coordinate Systems and Data Registration

A unique feature of the SAMIS is the integration of Geographic Information System (GIS) components which are intended to make data viewing and sharing more accessible and intuitive. The integration of these capabilities and their use by the SAT are described in the following sections.

Prior to sample analysis, the sample and sub-samples will be imaged for the purpose of generating basemaps with defined baseline coordinate systems to which all applicable data (rasters and geometry) will be registered. Instead of using the planetary (largely Earth-based) coordinate reference systems automatically built into the PostGIS extension of Postgres, SAMIS will automatically spatially register data through the use of known fiducial locations and a First Order Polynomial (Affine) Transformation that ultimately transforms geometry coordinates and produces parameters for World Files which transform rasters on-the-fly based on pixel location, scale, and orientation. Once an initial registration process is complete for a given dataset, shapefiles with updated geometry and World Files will automatically be generated and included in data downloads.

Spatial data registration across the project is largely based on the generation of thin-section basemaps and the careful design and implementation of fiducials (or sample mount markers) on thin-sections themselves to be used as tie-points. This process is described in more detail in the following sections.

**Coordinate Systems**

The SAMIS includes 3 main types of coordinate systems through which data can be accessed and visualized depending on user need and applicability:

- **2D X,Y Instrument Coordinate Systems**– These coordinate systems are defined per instrument and thin-section and are meant to take advantage of whatever spatial information and/or reference frames a given instrument and software are already using. The primary purpose of these coordinate systems is to spatially register data output from a single analysis in preparation for registration to thin-section basemaps. These coordinate systems will also preserve sample orientation information within instruments and provide useful context when revisiting completed analyses.

- **2D X,Y Thin Section Basemap Coordinate Systems** – These coordinates systems are defined per thin-section and are meant to be the basis for registering and visualization data output from any instrument and any analysis across the project. Coordinate systems are defined using basemaps generated at JSC ahead of thin-section distribution and analysis.

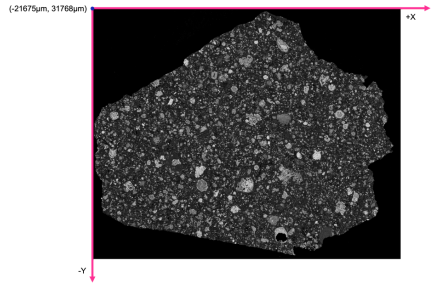

Example UArizona EMPA (CAMECA) data registered in the instrument coordinate frame. The x,y coordinates of the upper-left corner are defined based on instrument output.

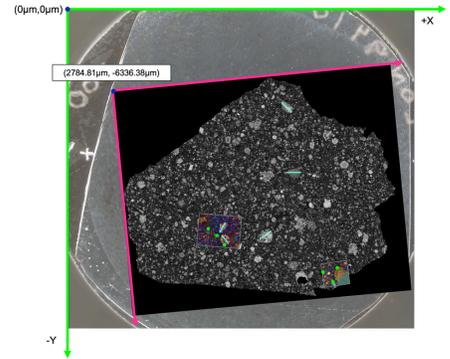

Example UArizona EMPA (CAMECA) data along with measurements from other data sessions registered to the basemap. Note that the x,y coordinates of the upper-left corner have been transformed from the instrument coordinate system into the basemap coordinate frame.

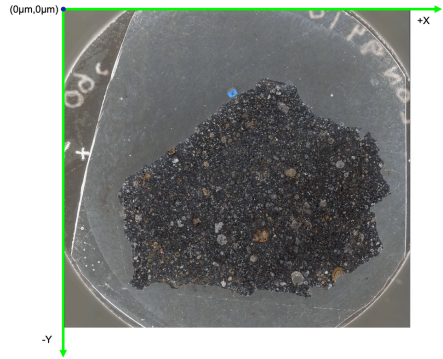

Example reflected light digital microscope basemap image with defined basemap coordinate system. The upper-left corner of a basemap is the origin (0,0).

**Spatial Data Registration**

Analysis data across the project are registered to common basemaps through the use of strategically placed fiducials on sample mounts and first order polynomial (affine) transformation algorithms. Three fiducials with a simple crosshair and label design are etched around the edge of the sample, defining nominal sample orientation with North (N) being up or the top of the sample and East (E) and West (W) fiducials being etched approximately orthogonally on either side of the sample. The x,y coordinates of the fiducial crosshairs are recorded in the basemap coordinate frame ad stored in SAMIS. With each subsequent analysis, sample scientists will locate the fiducials and record the x,y coordinates of the crosshair in the instrument coordinate frame and deliver that information to SAMIS along with the data they have collected. The fiducial coordinates in the basemap coordinate and in the instrument coordinate frame become inputs into a first order polynomial (affine) transformation that is automatically kicked off when spatial data is delivered to SAMIS, ultimately spatially registering data from a given analysis session to the basemap. Fiducials are designed and implemented to be sufficiently small to minimize crosshair location error in the instruments as much as possible, but large enough that they can be reasonably identified in basemap imagery. With this design fiducials are not visible to the naked eye. Fiducials are etched sufficiently close to the sample edge to make fiducal location as easy as possible, but not close enough to interfere with, damage, or alter the sample in any way.

**Example of fiducial design and implementation:**

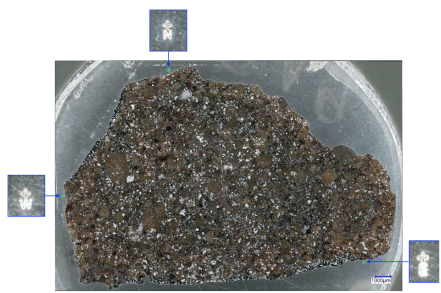

Example thin-section fiducial locations displayed on digital microscope basemap image.

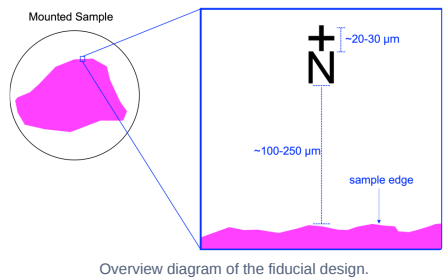

Overview diagram of the fiducial design.

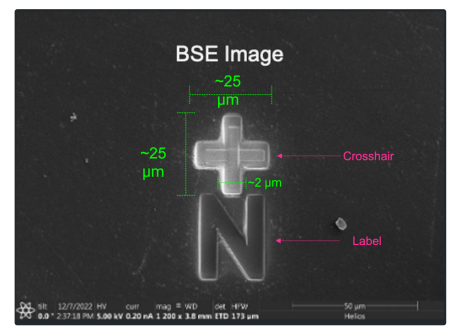

Example fiducial etched at UArizona with FIB with annotated fiducial dimensions.

## Data Visualization

The 2D visualization tool will allow team members to visualize, analyze, and share spatially registered sample analysis data project-wide and fully online. Rather than build a visualization tool from scratch, SAMIS will take advantage of the ArcGIS Enterprise software system, the backbone for running the Esri suite of applications in addition to custom applications, through UArizona software licensing. This software system provides an infrastructure for visualization, analytical tools, and data sharing that is well-established in the GIS industry and will come at no cost to the project.

ArcGIS Enterprise will be available through the Sample Analysis Desktop Application (SADA) and includes a "portal" that will be customized to the needs of our team. The portal will include a gallery for browsing data, maps, and tools and a Map tool which will enable 2d visualization of multiple spatially registered layers, the generation of thin-section maps, and further data analysis to be performed including (but not limited to) taking measurements and viewing graphs and attribute tables. ArcGIS allows a seamless connection between our PostgreSQL database and gives us the option to access more powerful Esri GIS applications, like ArcMap or ArcGIS Pro, which can be used for georegistration and other advanced spatial processing and analysis.

**Example of data visualization in ArGIS Enterprise:**

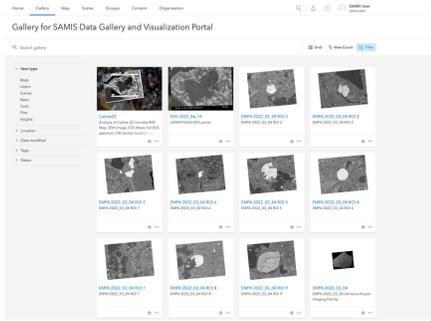
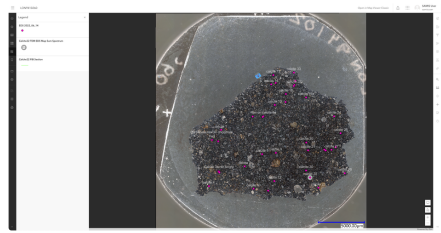

ArcGIS Enterprise Map Tool.

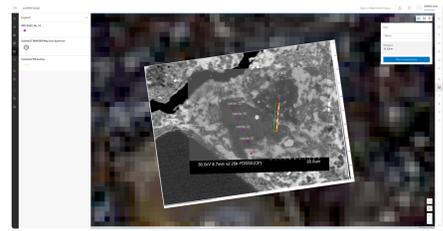

Zoom-in of sample map in Map Tool displaying FIB section location and other measurements

ArcGIS Enterprise Data Gallery page.

## SAMIS Backup and Security

*System Availability and Backup:*

- The SAMIS database will be backed up initially by PostgreSQL backup to disk and then by NetBackup of the database filesystem.  The database can be restored from tape in case of disaster recovery.
- Normal backup and recovery of all system parts (DB, app, website) will be done at the file system level, but will also be written to tape via NetBackup.
- Version control and backups of the SATA, SADA, and SAMIS Server codebases will be maintained by the Bitbucket repository.
- The SAMIS Server will automatically restart after a fatal error. All fatal errors will be logged and a notification will be sent.

*System Integrity:*

- User creation will only be available via the SADA using an administrator account. Any new user must make a request through a Working Group lead. Roles and group assignment within the database will be used to define individual user privileges for all applications.
- Each sample analysis team member will get one user account that will be valid for SAMIS/SATA/SADA.
- There will be a single database user account assigned to the SAMIS server.
    - Individual users will not have database user accounts. All database interaction will be facilitated through the SAMIS server.
- Users of SATA and SADA will have a single username/password for application access.
- Users of SATA and SADA will use Two Factor Authentication for access.
- For Database security, we are following the Principle of Least Privilege. Users, programs, and processes will have only the minimum privileges necessary to perform their functions.
    - Individual users of application = no database access
    - Production application database user = read, write, create, remove objects and data, execute permissible procedures and functions.
    - Developer user = Production DB read-only; Development DB read, write, create, remove objects and data, execute permissible procedures and functions.
    - Database Admin user = privileged/ superuser for administration of the database.
- No VPN access will be required to use this system through the SATA or SADA.
- The database is ACID (atomicity, consistency, isolation, durability) and MVCC (multi-version concurrency control) compliant ensuring transactional data consistency and integrity during system errors. The database uses WAL files (write-ahead logging) which log records describing the changes to data into files prior to permanently committing the change.

## Development of the SAMIS

The SAMIS development team uses an agile software development methodology for the development of the system which is an iterative and adaptive way of approaching a project. The SAMIS development cycles are organized into roughly 3 month blocks with end dates that correspond to wider project milestones, such as the UA Sample Analysis Readiness Test (SART) and the wider SART, and real-life events like long holiday breaks and University closures. Development cycles consist of:

- **Defining requirements:** Requirements are kept on the Sample Analysis Software Development Confluence pages and are revisited after each development cycle.
- **UI Design and Prototyping:** Low fidelity prototyping with UI design considerations are done using LucidChart, which is a collaborative diagramming software. The SAMIS team developed low-fidelity prototypes of the app and the visualization tool based on the requirements and collaboratively gathered feedback from the users based on those prototypes. That feedback was used as additional input into the requirements.
- **Development:** During the development phase of the cycle, the SAMIS team holds weekly and as-needed planning and design meetings, weekly and as-needed code review meetings, and twice weekly checkins with the full development team (including the Sys Admin, DB admin, GIS Engineer, and Sample Archiving and Data Specialist. Frequent meetings with MSS Connolly and members of analytical team are also conducted to ensure successful communication of the vision is implemented under the PI's overall guidance.
- **Internal Testing:** After each development phase of the cycle, the SAMIS team performs two weeks of internal testing. Both developers and non-software members of the SAMIS development team participate in the testing.
- **External Testing:** After internal testing is complete, the SAMIS team performs "external" testing on the system with members of the SAT who are not involved in SAMIS development. The testing includes use cases to guide the selected SAT members through the features needing testing and unique-to-each-cycle forms for gathering user feedback.
- **Repeat:** Integrate user feedback and add next set of features.

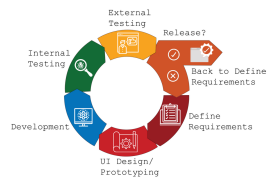

In additional to these steps in the cycle the SAMIS development team holds weekly bi-weekly meetings with the SADAWG. The SAMIS team also holds bi-weekly meetings with Astromat and quarterly in-person TIMs. The SAMIS team holds monthly and as-needed meetings with JSC.

# 11.0 SAMPLE ANALYSIS READINESS TEST

The Sample Analysis Readiness Test (SART) is a pre-sample return activity with the major goal of certifying the team to perform integrated, coordinated analysis of a limited mass of returned sample. The SART is designed to test the abilities and readiness of the OSIRIS-REx Sample Analysis Team (SAT) to complete the following mission L1 requirement: "Return and analyze a sample of pristine carbonaceous asteroid regolith in an amount sufficient to study the nature, history, and distribution of its constituent minerals and organic material." It is implied in the L1 requirement that analysis will be done in a fashion to preserve as much of the returned material as possible for future generations; to do that, we must be prepared to the level we can be and that requires the implementation of the SART.

All members of the SAT and their laboratories are required to participate and become fully certified before analysis of the returned sample.

The SART will begin no later than June 2022 and will be completed by June 2023. Continuously during the SART, lessons learned will be captured in the SAP and Curation Plan, and implemented in real time if possible. A summary of lessons learned will be presented at the Sample Analysis Readiness Review.

The SART is not an end-to-end, real-time test of all the analysis process that will be applied to the unknown returned sample, but a series of focused endeavors to test the following:

1. Effectiveness of the design and implementation of the integrated, coordinated analysis of the returned samples through investigation of analog materials with a goal of minimizing damage to the material while optimizing the science return;
2. Ability of the SAT to effectively archive data collected with instrumentation on analog materials, either newly collected or through a series of previously collected test data;
3. Ability of the SAT to successfully utilize the Sample Analysis Tracking Application and the Sample Analysis Desktop Application to track and maintain accurate accounting of samples and analysis progress from curation through laboratory analysis and back to curation;
4. Ability of the SAT to prepare and ship samples on a timely basis between curation and distributed SAT laboratories;
5. Ability of the SAT to successfully analyze data using the MicroGIS for applicable samples;
6. Ability of the overall management structure to perform effectively in conducting and expediting the analysis process;
7. Overall communications of the SAT.

## Required certifications to be completed through the Sample Analysis Readiness Test

All members of the Sample Analysis Team (SAT) are required to be fully certified through the Sample Analysis Readiness Test (SART) before they can participate in the analysis of the returned sample. Certification is defined as having the following:

1. A signed copy of the formal agreement to abide by the revised OSIRIS-REx Rules of the Road (ROTR).
2. User registration and familiarity with the Sample Analysis Tracking Application (SATA) and Sample Analysis Desktop Application (SADA).
3. Demonstrated proficiency in the handling, care, and storage of extraterrestrial material with minimal contamination and loss of allocated sample.
4. Demonstrated proficiency in any lab-specific sample preparation techniques (e.g., application of conductive coatings, sub-sampling, acid digestion, surface preparations, etc.).
5. Demonstrated proficiency in analyzing low-petrologic-type carbonaceous chondrites (type-1 or 2) with minimal contamination and loss of sample. Heritage studies are acceptable if all personnel, facilities, and instrumentation are still in place.
6. Documentation of all calibration standards and analytical techniques in the mission data repository.
7. Documentation of all data products that will be delivered to the mission, including raw, calibrated, higher-level, and foundational data products (as defined below), along with appropriate metadata in the mission data repository.

8. Demonstrated application and documentation of error analysis used for calibration and production of quantitative higher-level data.

9. Certification of proficiency in use of the SAMIS system to deliver required data including

  · Raw data (measurement signal as the instrument records it)

  · Calibrated data (application of all corrections and calibrations to raw data to produce a science-ready data product)

  · Documentation of all algorithms and/or software used to produce calibrated and higher-level data products.

  · Higher-level products (those that combine multiple data calibrated data products)

  ·  Foundational data products (coordinate systems, basemaps, etc.)

  · Appropriate metadata

  · Physical sample tracking data acquired using the sample tracking system

10. Completed loan agreement with NASA Johnson Space Center – Astromaterials Research and Exploration Science (ARES) division.

## Samples used for the Sample Analysis Readiness Test

### Materials analogous to asteroid Bennu

The Sample Analysis Readiness Test (SART) will utilize, but will not be limited to, the following samples for preparation and analysis: (1) Sutter's Mill (C), 15 g, curated by the University of Arizona, and (2) Murchison (CM2), 12 g, curated by the National History Museum of the Smithsonian.

NASA JSC curation will participate in the preparation, allocation and shipping of the SART samples to the Sample Analysis Team (SAT). Participation of JSC curation is critical to having a fully certified SAT that can participate in the analysis of the returned sample by OSIRIS-REx.

In addition to the focused, active SART through use of the above stated meteorites, our solid relationship with the Hayabusa2 SAT and JAXA curation will permit a full exchange of many aspects of their curation and sample analysis activities to be shared with the OSIRIS-REx team with the goal of helping to improve preparedness for the return of samples from asteroid Bennu.

### Materials analogous to SRC components

In addition to the meteorite samples discussed above, the SAT needs to practice on the analysis of the SRC filter and all flight witness coupons through use of analogous and/or archived materials.

## SAMIS Testing

SADAWG is planning for SAMIS training and testing during SART in FY22-23, as well as UArizona internal testing (referred to as 'mini-SART') in FY21. Below, we describe the general testing requirements for different testing subsets. These tasks are only baseline needs for SAMIS testing during SART based on the SADAWG vision and requirement documents. The SAMIS testing requirements for the SART will be adjusted as needed during the upcoming development cycles and beta testing before SART.

### Testing Requirements for All Labs and All SAT Members

These testing requirements are for any lab that will be receiving Bennu sample during the sample analysis phase of the mission.

SATA/SADA

- Completing the SAMIS user training (login credentials will be granted on completion)
- Accessing the SAMIS via SADA and SATA in lab environment
- Performing required steps for sample receipt using either the SATA or SADA

- Performing required steps for updating sample condition based on the analysis technique. This can be realistic, but fabricated information (*no sample analysis required*)
- Performing steps to begin and end sample analysis. This can be realistic, but fabricated information *(no sample analysis required)*
- Performing steps to return sample

SADA only

- Registering instruments
- Uploading each data format used by an instrument (*either new SART sample analysis data or previously acquired data but need to have all original raw/process data and metadata*)
- Upload corresponding and required metadata
- Locating data on the SADA
- Downloading data from the SADA
- Reviewing data and changing data quality flags on uploaded data
- **SADA Visualization Tool**
    - Searching and locating data using the visualization tool
    - Displaying instrument data and verifying the automatic registration performed by the SAMIS Database is correct
    - Manual registration or manual adjustment to the automatic registration
    - Annotating data within the visualization tool
    - Exporting visualized data to other formats (publishable maps, PNGs, TIFF, excel tables)

**Testing Requirements for JSC/UA/Goddard**

These testing requirements are for any lab that will be making samples or subsamples.

- Creation of a new sample with sample ID/QR code and ingestion into SAMIS for sample tracking
    - bulk sample
    - powder
    - thin section
- Creation of a new thin section coordinate system
-
    - Sample mounting technique and visual coordinate system anchor on physical sample
    - Sample basemap imaging and ingestion of basemap imaging into SAMIS for creation and verification of basemap coordinate system
    - Creation of new subsample with new sample ID/QR code and ingestion into SAMIS for sample tracking

Ingestion of subsample imaging and/or analysis along with required registration metadata into SAMIS. Verification of SAMIS automatic registration of subsample data to parent sample basemap

**Testing requirements for Working Group Leads**

SADA only

- Checking sample analysis upload/timeline and data submission progress
- Verifying submitted data and changing data quality flags on uploaded data
- Notifying the system of an issue with submitted data

**Testing requirements for the UArizona internal SART**

These testing requirements are for UArizona's internal SART ('mini-SART') scheduled for late 2021. This list is not comprehensive as there will be substantial overlap with the testing requirements in the previous sections.

SATA/SADA

- Notifying the system of the planned destruction of a sample or portion of a sample
- Notifying the system of the unplanned destruction of a sample
- Notifying the system when a sample has been held at the same lab too long
- Notifying the system and following the procedure for when a sample is lost in the mail
- Notifying the system and following the procedure for when a sample has sustained damage in the mail (+upload photo documentation)
- Notifying the system and following the procedure for when a sample has been damaged in the lab
- Following the procedure for when a QR code or ID cannot be read, or needs to be reprinted
- Following the procedure for when the wrong sample has been received
- Following the procedure for when a change to a sample analysis flow/schedule has to be changed/requested

# 12.0 PRELIMINARY EXAMINATION (PE)

The detailed implementation of the PE is described in the OSIRIS-REx Curation Plan.

The PE phase of the mission will begin once the sample canister is opened at JSC. The purposes and goals of PE are:

1. Disassembly of the sample canister and removal of samples, witnesses, and space-exposed hardware to be curated.
2. Open the sample canister assembly by releasing the latches and driving the motor
3. Remove flight witness plates and other SRC components.
4. Perform the initial documentation of the sample canister interior.
5. Video-document sample canister disassembly (with appropriate International Traffic in Arms Regulations/Export Administration Regulations (ITAR/EAR) status determined in advance).
6. Interface with the respective WGs to define the needs of the SAT.
7. Initial characterization using non-destructive methods and isolation of materials for various purposes.
8. Selection of returned samples for the SAT to address mission science goals (up to 25 wt. percent, including selection of samples for rapid analysis (including destructive analysis) by the SAT for releasing initial results), and samples archived by NASA (75 wt. percent) to include samples for international agencies (CSA 4 wt. percent and JAXA wt. 0.5percent), samples for remote storage at White Sands Complex (WSC) and samples for hermetically sealed or cold storage.
9. Generation and publication of a catalog with a level of detail about each sample that is sufficient for members of the scientific community to make informed requests of materials to conduct their scientific investigations.

The Preliminary Examination Team (PET) is comprised of:

- OSIRIS-REx mission Principal Investigator (PI) (Team Leader)
- All members of the OSIRIS-REx Sample Science Council (SSC)
- OSIRIS-REx SRCSTIWG Lead (flight hardware)
- OSIRIS-REx Contact Pad Lead
- Lockheed Martin hardware specialists (two)
- At least one JAXA representative on OSIRIS-REx
- One CSA representative on OSIRIS-REx

# 13.0 DECISION MAKING AUTHORITY AND PROCESSES DURING SAMPLE ANALYSIS

## Sample Science Decision Authority

This section provides the guiding principles for key decisions related to analysis of the OSIRIS-REx samples. The University of Arizona (UA) is responsible for mission leadership with Dr. Dante Lauretta, UA Professor of Planetary Sciences, serving as principal investigator (PI). Dr. Lauretta is accountable to NASA for the success of the investigation, with full responsibility for its scientific integrity and its execution within committed cost and schedule. The PI reports to the Planetary Missions Program Office (PMPO) at Marshall Space Flight Center for programmatic issues and to the Science Mission Directorate (SMD) at NASA Headquarters for scientific issues. He is responsible for the safety and security of the samples allocated to the science team, sample analysis planning and implementation, data analysis, data product generation, data archiving, and publication of the results in the peer-reviewed literature.

JSC provides facilities and personnel for the curation of reference materials and blanks, archived during earlier phases of the mission, and for the receipt and curation of the returned samples and associated hardware. After Earth return, the SRC will be transported to JSC where samples will undergo an assessment by the Preliminary Examination Team. The JSC Astromaterials Curator is responsible for ensuring that decisions are made to maximize the long-term value of the samples. JSC reports to the PI for status on sample allocation for preliminary examination and scientific analysis in support of OSIRIS-REx objectives.

## Decisions Regarding Sample Allocation

As defined in the OSIRIS-REx Curation Plan, during Preliminary Examination, the OSIRIS-REx PI is consulted on all decisions about splitting the sample and identifying material to deliver to international partners. The PI is responsible for the entire loan agreement for team samples. When deciding on the samples to be allocated to the science team, the PI and the OSIRIS-REx Curator will work together to reach concurrence.

The PI makes all decisions related to the sub-allocation of samples to the individual sample science working groups (WGs). Once allocated, the distribution of the sample among the different laboratories supporting each WG is at the discretion of the WG Lead. In the event that there is disagreement within the WG as to sample allocation, an individual team member may protest a WG Lead decision with the PI. In such cases, the PI will allow all team members to present their side of the issue, then make a unilateral decision as to the path forward.

PI decisions are informed by the recommendations of the Sample Science Council (SSC). This council supports the OSIRIS-REx PI in the development and implementation of this plan and in management of the science team. Members of the SSC are appointed by the PI.

## Decisions Regarding Resource Allocations

The OSIRIS-REx PI has the primary responsibility for financial management for mission science activities during sample analysis. The PI Office will establish an estimated resource profile with each US-based organization participating in the sample analysis activities for the Baseline, Overguide, and Threshold programs. For universities and private research labs, resources will flow from the University of Arizona to that organization via subcontract. For US government laboratories, task orders will be issued for science tasks. The PI reserves the right to re-allocate fiscal resources based on our evolving knowledge of the nature of the returned sample. It is possible that some analyses identified in this plan will be descoped or that new analyses will be added to the plan over the course of the sample analysis program. In addition, the PI Office will perform a financial assessment annually. In cases where organizations show a large underrun in spending, the PI may decide to recover and re-allocate those funds to other science activities.

## Decisions Regarding Publications

Per the OSIRIS-REx Rules of the Road (AD-1): "Any team member intending to produce a **written publication**, such as a journal article, conference abstract, or conference proceedings paper, that is:

- funded by the mission,
- based on mission data that have not yet been released to the scientific community,

OR

- based on mission data that have been released for less than six months

must follow the workflow and approval process outlined in the OSIRIS-REx Publication Guide."

> For this purpose, Participating Scientists are considered to be "funded by the mission."

# 14.0 APPENDICES

## 14.1 Acronym List

AFM | atomic force microscop(e/y)

APCI | atmospheric pressure chemical ionization

APPI | atmospheric pressure photoionization

APT | atom probe tomography

BSE | back-scattered electron

CAI | calcium-aluminum–rich inclusion

CCD | charge coupled device

CCWG | Contamination Control/knowledge Working Group

CDD | compact discrete dynode

CI | chemical ionization

CL | cathodoluminescence

CPO | crystallographic preferred orientation

CSIA | compound-specific isotopic analysis

CT | computed tomography

CWG | Curation Working Group

DCM | dichloromethane

DSC | differential scanning calorimeter

EA-IRMS | elemental analysis–isotope ratio mass spectrometry

EBSD | electron backscatter diffraction

EDS [not EDX or other forms, for consistency] | energy-dispersive X-ray spectro(meter/scopy)

EELS | electron energy-loss spectroscopy

EI | electron impact

EM | electromagnetic radiation

EMPA | electron microprobe analysis

ESI | electrospray ionization

FEG | field emission gun

FEM | finite element (model/method)

FIB | focused ion beam

FTICR-MS | Fourier transform ion cyclotron resonance mass spectrometry

FTIR | Fourier transform infrared (spectroscopy)

FWHH | full width at half height

GC | gas chromatography

GC-C-IRMS | gas chromatography–combustion–isotope ratio mass spectrometry

GIS | geographic information system

HR | high-resolution

ICP-MS | inductively coupled plasma mass spectrometry

ICP-OES | inductively coupled plasma optical emission spectrometry

IOM | insoluble organic material

IR | infrared

JSC | Johnson Space Center

LA-ICP-MS | laser-ablation inductively coupled plasma mass spectrometry

LC | liquid chromatography

LIMAS | laser ionization mass nanoscope

MAPWG | Mineralogy and Petrology Working Group

MC-ICP-MS | multi-collector inductively coupled plasma mass spectrometry

mCFMS | mini cryogen-free measurement system

MIR | mid-infrared

MO | molecular orbital

MS | mass spectromet(er/ry)

MSS | Mission Sample Scien(ce/tist)

m/z | mass/charge ratio

nanoIR | nanoscale infrared (mapping)

nanoSIMS | nanoscale secondary ion mass spectrometry

NGMS | noble gas mass spectrometry

NI-NGMS | neutron irradiation noble gas mass spectrometry

NIR | near-infrared

NMR | nuclear magnetic resonance (spectroscopy)

PDS | Planetary Data System

OCAMS | OSIRIS-REx Camera Suite [includes PolyCam, MapCam, SamCam]

OLA | OSIRIS-REx Laser Altimeter

OSIRIS-REx | Origins, Spectral Interpretation, Resource Identification, and Security–Regolith Explorer

OTES | OSIRIS-REx Thermal Emission Spectrometer

OVIRS | OSIRIS-REx Visible and InfraRed Spectrometer

PADWG | Contact Pad Working Group

PAH | polycyclic aromatic hydrocarbon

PDA | photodiode array detector

PE | preliminary examination (of the sample)

PMT | photomultiplier detector

Q, q [prefix] | quadrupole

QqQ | triple-quadrupole

QSA | quasi-simultaneous arrivals

REXIS | Regolith X-ray Imaging Spectrometer

RI | resonance ionization

RMS | root mean square

SADA | Sample Analysis Desktop Application

SADAWG | Sample Analysis Data Archiving Working Group

SAMIS | Sample Analysis Micro-Information System

SARA | Sample Acquisition and Return Assembly

SART | Sample Analysis Readiness Test

SAP | Sample Analysis Plan

SAT | Sample Analysis Team

SATA | Sample Analysis Tracking Application

SATM | Sample Analysis Team Meeting

SDD | silicon drift detector

SE | secondary electron

SEIWG | Sample Elements and Isotopes Working Group

SEM | scanning electron microscop(e/y)

SHRIMP | sensitive high mass resolution ion microprobe

SIMS | secondary ion mass spectrometry

SNMS | secondary neutral mass spectrometry

SNR | signal-to-noise ratio

SOAWG | Sample Organic Analysis Working Group

SOM | soluble organic matter

SOPIE | Science Operational Proficiency Integrated Exercise

SPC | stereophotoclinometry

SPME | solid phase micro-extraction

SPTAWG | Sample Physical and Thermal Analysis Working Group

SRC | Sample Return Capsule

SRCSTIWG | Sample Return Capsule Science Team Interface Working Group

SS [prefix] | solid state

SSAWG | Sample Spectral Analysis Working Group

SSC | Sample Science Council

s-SNOM | scattering scanning near-field optical microscopy (s-SNOM)

STEM | scanning transmission electron microscop(e/y)

SThM | scanning thermal microscop(e/y)

STXM | scanning transmission X-ray microscop(e/y)

S-XRF | synchrotron-bas X-ray fluorescence

TAGCAMS | Touch and Go Camera System [includes NavCam 1 and 2, StowCam]

TAGSAM | Touch and Go Sample Acquisition Mechanism

TEM | transmission electron microscop(e/y)

TGA | thermogravimetric analysis

TIM | technical interchange meeting

TIMS | thermal ionization mass spectrometry

TKD | transmission Kikuchi diffraction

TOF, ToF | time-of-flight

UHP | ultrahigh (purity/performance)

UTTR | Utah Test and Training Range

UV | ultraviolet

VI | visual inspection

VLM | visible light microscopy

VNMIR | visible, near-infrared, and mid-infrared (spectroscopy)

WDS | wavelength-dispersive spectro(meter/metry/scopy)

WG(L) | working group (lead)

XANES | X-ray absorption near edge structure (spectroscopy)

XAS | X-ray absorption spectroscopy

XCT | X-ray computed (micro)tomography

XRD | X-ray diffraction

XRF | X-ray fluorescence

Z | atomic number

µ [prefix] | micro

µ-L$^2$MS | microprobe two-step laser mass spectrometry

## 14.2 Roles and Responsibilities of the Members of the Sample Analysis Data Management and Software Development Team

- **Sample Analysis Data Archiving Working Group (SADAWG) Lead** - Oversees the timely development, testing and implementation of the Sample Analysis Micro-Information System (SAMIS) systems, and coordinates efforts with other Working Groups. During the SART and Sample Analysis Phase, the SADAWG Lead will oversee the timeline of sample data submission to SAMIS, monitor the progress of sample analysis and data submission to SAMIS, facilitate communication between the Sample Analysis Development Team and other working groups, and coordinate potential changes to the sample analysis data submission schedule with the Sample Science Council (see Section 13).
- **Sample Analysis Development Team** is in charge of the design, development, testing, implementation, maintenance, and day-to-day operation of the SAMIS systems.
  - **SAMIS Project Manager** - Monitors the schedule, resources, and project scope. Plans what work needs to be done, when, and who's going to do it. Coordinates efforts between members of the SADAWG technical team and scientific team. Anticipates and deals with changes to the project. Keeps the team organized, motivated, and happy.
  - **Software Engineers** - Design, develop, and test software specific to the needs of the Sample Analysis Team. Collaborate cross-functionally with data scientists, database administrators, GIS specialists, and other technical staff to optimize and iterate on software solutions. Determines operational feasibility by evaluating analysis, problem definition, requirements, solution development, and proposed solutions. Performs code reviews and maintains a version-controlled repository for that code. Documents and demonstrates solutions by developing documentation, flowcharts, layouts, diagrams, charts, code comments, and clear code. Provide ongoing maintenance, support, and enhancements in existing systems and platforms.
  - **System Administrator** – Manages IT systems and networks used for the project. Manages network and server security and server system backups.
  - **Database Administrators** - Plan and oversee the setup of database server including installation and update of database software and management tools. Create and maintain databases for development, testing, and production systems. Monitor and manage capacity and performance of databases. Perform backup and recovery of databases including all necessary system files. Support developers with code reviews, query tuning, data modeling, and schema migration.
  - **GIS Application Development Engineers** - Designs and develops GIS programming applications and tools specific to the microscopy analysis needs of the Sample Analysis Team. Maintains and iterates on the development of these GIS software programs including supporting the writing and execution of software and data processing pipeline tests. Architects a system for transforming microscopy data into a spatial dataset that can utilize the unique capabilities of the system. Design and develop a GIS-based visualization tool. Create demos and guides to explain GIS features and data visualization capabilities to sample scientists. Writes and implements spatial queries and functions.

  - **Data Archive Scientist** – Plans and develops planetary science standards–compliant data dictionaries for use by the OSIRIS-REx Sample Analysis Team. Coordinates with scientists and software team members to ensure data products are properly documented and described for long-term archival. Coordinates between different Planetary Science standards organizations to ensure data products meet all standards. Assists in data product format development to ensure products meet interoperability and long-term storage standards.

## 14.3 The OSIRIS-REx Sample Security Plan

In addition to those aspects of sample security specifically described within the OSIRIS-REx Curation Plan, the mission stipulates that these following procedures must be followed by all members of the SAP who are performing analyses:

- All samples must be shipped under a $N_2$ purge. Approved types of shipping containers can be found [on Confluence], with recommendations on producing $N_2$ purge for sample shipping.
- All Bennu samples that are not being prepared for, or being analyzed, must always remain in an $N_2$ atmosphere.
    - Such an environment can be created within a sample cabinet, like many team members already have, but to comply with the Curation Plan only Bennu samples can be stored in these. Such cabinets are 'off the shelf' and bought from any laboratory supply company in the US such a VWG or Wards.
    - Samples may also be returned to their shipping contains and a $N_2$ purge applied within team members laboratories. For example, 'chip clip' bags can be reused for sample storage in a $N_2$ atmosphere. These bags can be placed into a glovebox, either a standard one or a portable one that has a $N_2$ atmosphere and sealed appropriately.

The major reason to implement the storage of sample in an $N_2$ environment when not being analyzed is to minimize reactions between the Bennu sample (e.g., rocks and minerals, etc.) and ambient Earth atmosphere. It is clear from studies of Hayabusa2 returned sample from asteroid Ryugu, and investigations on recent meteorite falls, that the these samples are altering with exposure to Earth's atmosphere. We want to try to minimize such alteration. Sample will be processed in air at JSC for the production of polished sections, etc., but stored in an $N_2$ environment when not being processed or investigated upon in any way.

## 14.4 Participating Scientist Program Information

In June of 2022, NASA implemented the OSIRIS-REx Participating Scientist Program (SPS) through the selection of 8 proposals with a total of 32 new team members to contribute to the analysis of the returned sample. The complementary research proposed by each PS has been integrated into the SAP and all PS members have been integrated into the sample analysis team. The Announcement of Opportunity indicated that a total of 3 g, above and beyond the OSIRIS-REx required 15 g of Bennu regolith needed to meet its L1 objectives, would be made available to the PS teams for research. If the mission was confirmed to be in an overguide mass allocation scenario, potentially as much as 10 g could be made available to PS teams for analysis.

The mass needed for the successful implementation of the PS analysis program could either come from the OSIRIS-REx overguide allocation or as a separate allocation from sample not allocated to the OSIRIS-REx sample analysis team. The OSIRIS-REx PI and NASA Program Scientist will decide on which reservoir of Bennu regolith will be tapped for the PS analysis program post-return. Below is the baseline mass table for the entire PS program as defined by the PS PI's and reviewed by relevant WG leads and MSS.

Mass Requirement Table for Participating Scientists:

| Test Objective | Team Member | Instrument | Institution | Follow-on Analysis (Y/N/P) - if Y - which primary sample and notes; If Partial (P) describe | Mass per Test | | | | Number of Tests | | | Mass per Investigation | | | | Non-Destructive | |
|---|---|---|---|---|---|---|---|---|---|---|---|---|---|---|---|---|---|
| | | | | | CBE | Test Contingency | SubTotal | | CBE | Spare | Total | CBE | Contingency | | Total | Yes = 0 | Consumed |
| | | | | | g | % | g | g | | | | g | % | g | g | | |
| **General Petrography and Petrology** | | | | | | | | | | | | | | | | | |
| Sample texture/petrography/mineralogy | King | PS-XRD | NHM | N - Primary mass allocation | 0.001 | 0% | 0.000 | 0.001 | 1 | 0 | 1 | 0.001 | 0% | 0.000 | 0.001 | 1 | 0.001 |
| Sample texture/petrography/mineralogy | Dominguez | NanoFTIR etc. | SD | N - Primary mass allocation | 0.300 | 0%% | 0.000 | 0.300 | 1 | | 1 | 0.300 | 0% | 0.000 | 0.300 | | |
| | | | | Total Petrography and Petrology: | | | | | | | | 0.301 | 0% | 0.000 | 0.301 | | 0.001 |
| **Nanoscale Mineralogy** | | | | | | | | | | | | | | | | | |
| In situ mineralogy and surface chemistry | Dominguez | NanoFTIR etc. | SD | | 0.304 | 30% | 0.091 | 0.395 | 1 | 0 | 1 | 0.304 | 0% | 0.000 | 0.304 | 0 | 0.000 |
| | | | | Total Mineralogy: | | | | | | | | 0.304 | 0% | 0.000 | 0.304 | | 0.000 |
| **Isotopic Abundances** | | | | | | | | | | | | | | | | | |
| Cr & Ti etc. isotopes | Brennecka | ICP-MS, MC-ICP-MS & TIMS | LLNL | N - Primary mass allocation | 0.500 | 0% | 0.000 | 0.000 | 1 | 0 | 1 | 0.500 | 0% | 0.000 | 0.500 | 1 | 0.500 |
| $^{13}C/^{12}C$ & $^{15}N/^{14}N$ | Freeman/Baczynski | CAMECA IMS 1290 | UCLA | N - Primary mass allocation | 0.250 | 0% | 0.000 | 0.000 | 1 | 0 | 1 | 0.250 | 0% | 0.000 | 0.250 | 0 | |
| K, Ci, Zn, & Rb isotopes | Wang | ICP-MS & MC-ICP-MS | WashU | N - Primary mass allocation | 0.050 | 0% | 0.000 | 0.050 | 1 | 0 | 1 | 0.050 | 0% | 0.000 | 0.050 | 1 | 0.050 |
| $^{10}Be, ^{26}Al, ^{36}Cl, \& ^{41}Ca$ | Welten | AMS & ICP-OES | Berkeley | N - Primary mass allocation | 0.037 | 0% | 0.000 | 0.037 | 1 | 0 | 1 | 0.037 | 0% | 0.000 | 0.037 | 1 | 0.037 |
| $^{21}Ne$ | Welten | AMS | Berkeley | Y - Row 39 & 40 from main SAP Baseline SAP table | 0.000 | 0% | 0.000 | 0.000 | 0 | 0 | 0 | 0.000 | 0% | 0.000 | 0.000 | 0 | 0.000 |
| | | | | Total Radiogenic Isotopic Abundances: | | | | | | | | 0.837 | 0% | 0.000 | 0.837 | | 0.587 |
| **Bulk Organic Abundances** | | | | | | | | | | | | | | | | | |
| Volatile organics distribution and isotopes | Huang/Santos/Heck/Alexandre | GCMS and GC-IRMS | Brown University | N - Primary mass allocation | 0.045 | 0% | 0.000 | 0.045 | 10 | 2 | 12 | 0.450 | 0% | 0.000 | 0.450 | 1 | 0.450 |
| | | | | Total Bulk Organic Abundances: | | | | | | | | 0.450 | 0% | 0.000 | 0.450 | | 0.450 |
| **Soluble Organic Analysis** | | | | | | | | | | | | | | | | | |
| Water soluble organic molecule distribution and C-isotopes | Freeman/Baczynski | Pico CSIA | PSU | N - Primary mass allocation | 0.030 | 0% | 0.000 | 0.030 | 10 | 2 | 12 | 0.300 | 0.0% | 0.000 | 0.300 | 1 | 0.300 |
| Distribution and abundance of aliphatic/aromatic hydrocarbons and H-isotopes | Hofmann | GC-Orbitrap MS | Caltech | N - Primary mass allocation | 0.025 | 0% | 0.000 | 0.025 | 10 | 2 | 12 | 0.250 | 0.0% | 0.000 | 0.250 | 1 | 0.250 |
| **Insoluble Organic Analysis** | | | | | | | | | | | | | | | | | |
| Coordinated C, N, and S-isotopes of IOM | Freeman/Baczynski/House | Nano EA-IRMS | PSU | N-primary mass allocation | 0.030 | 0% | 0.000 | 0.000 | 10 | 2 | 12 | 0.300 | 0% | 0.000 | 0.300 | 1 | 0.300 |
| Distribution and abundance of organic fragments derived from IOM | Huang/Santos/Heck/Alexandre | pyrolysis GC-MS | Brown University | N-primary mass allocation | 0.005 | 0% | 0.000 | 0.005 | 10 | 2 | 12 | 0.050 | 0% | 0.000 | 0.050 | 1 | 0.050 |
| | | | | Total Compound Specific Organic Abundances: | | | | | | | | 0.900 | 0.0% | 0.000 | 0.900 | | 0.900 |
| | | | | Total Allocated Mass: | | | | | | | | 2.792 | 0.0% | 0.000 | 2.792 | | 1.938 |

# 14.5 Glovebox Exhaust Volatile Capture

Approved plan for glovebox exhaust volatile capture.

## SIMPLE GLOVEBOX EXHAUST VOLATILE CAPTURE PROPOSAL

### 0. Purpose and Background

This proposal is a follow-on to salvage as much of the science proposed by Huang et al. (Appendix 1) provided that there is no impact to the sample and a *de minimis* impact to curation. Furthermore, it is anticipated that there would be no additional funds required from the project. The science remains the same, to inventory the volatile species released by the OSIRIS-REx regolith sample in the curation nitrogen ($GN_2$) glovebox. What is lost is as follows:

- Capture of volatiles from the canister glovebox. Samples from only the TAGSAM glovebox would be collected.
- Quantitation. Since the mechanism is in a region of dead space exposure is determined by complex fluid dynamics so absolute quantitation is impossible. Though the collection uses adsorbents which collect compound at different efficiencies, the relative abundance of some analytes should be possible.
- Universal collection by volatility. Instead of freezing all compounds, analytes would be captured by selective adsorption on a variety of substrates.
- Temporal resolution. One to two samples would be collected between TAGSAM head disassembly to containerization.
- Risk of the introduction of some ISO 5 room air when the airlock is opened if the curation staff is unable to close the collector valve.

What is gained follows:

- Simplicity of operation, low maintenance, and room temperature shipping.
- Multiple parallel and complimentary sample collectors.
- Samples are shelf-stable for days to months.
- No request for additional labor or procurement.
- Full backup collector.

The sections that follow are the answers to five questions in the email from MSS Connolly on April 24, 2023.

(1) CWG leads must formally sign-off on the document indicating their positive support and level of confidence in implementation
(2) A timeline with milestones for development and implementation
(3) A notional timeline for sampling and analysis with all details related to the analysis including the laboratory(s) and Co-Is to make the analyses
(4) That you agree to oversee all aspects of the proposed implementation and coordination of the analyses
(5) Add a paragraph of how results could be compared to those of other gas sampling techniques, and finally
(6) Address in the proposal the means by which we will publish the results (e.g., add to an existing publication, propose a new one, etc.).

## 1. CWG Leads Sign-off

See Appendix 2 for approval. This appendix also shows additional details of the plumbing design and location in the TAGSAM glovebox.

## 2. Development and Implementation Timeline

The design is a sampler containing a library of adsorbent materials inside a dead volume downstream from the TAGSAM glovebox exhaust port and check valve but upstream of any other glove boxes in building 31. The design has an airlock of two valves to permit its removal or replacement without interrupting the glovebox exhaust (**Figure 2.1**). The development timeline is shown in **Table 2.1**. The library of adsorbents and targeted analytes is shown in **Table 2.2**.

*Figure 2.1.* Schematic of the collector. A. the collection configuration, B. the post-collection configuration. The apparatus has a glovebox side which would be a permanent addition to the glovebox and the collector side which would be hot swapped. C. Shows the various designs at actual size for the collector once the glovebox side and the library of fibers were finalized.

*Table 2.1. Development and implementation timeline.*

| Date (2023) | Activity |
|---|---|
| April 20 | Concept discussed with Graham and Curation |
| April 24 | Authorization from MSS to explore options |
| April 25-26 | Huang declines to discuss backup methods of volatile collection |
| April 30 | Begin discussion with curation on implementation |
| May 3-13 | Graham begins research on efficacy of adsorbents in flowing nitrogen |
| May 4 | Design meeting with JSC |
| May 10 | Discuss publication strategy with Sandford |
| May 10-24 | Graham researches potential adsorbent materials in detail |
| May 15 | Review detailed plan with W. Connelly |
| May 26 | Completion of glovebox side of the design with JSC |
| June 2 | Downselect the collector side of the design |
| June 2 | Oral approval from curation |
| June 7 | W. Connelly receives quote for custom tee for glovebox, Dworkin provides a charge code |
| June 9 | Procurement of glovebox side components, ship directly to JSC |
| June 16 | Written approval from curation |
| June 16 | Finalization of adsorbent material attachment to collector |
| June 16 | Procurement of collector side plumbing (triplicate) |
| June 17 | Submission of this proposal to MSS |
| Late June | Procurement of adsorbent fibers (triplicate) |
| July | Confirm method of activation and analysis of adsorbents at GSFC |
| Mid July | Tap holes and add brackets to the collector end caps to secure the adsorbents |
| Late July | Custom tee delivery to JSC, cleaning, installation of glovebox side components |
| Early August | Ship assembled collectors (triplicate) to JSC |
| Early August | JSC cleans collector exterior |
| August 24 | Deadline for glovebox integration of control collector |

*Table 2.2. Adsorbent library that would be deployed. The cross design is necessary to hold the breadth of materials.*

| Substrate | Adsorbent material | Analyte | Adsorbed analyte stability |
|---|---|---|---|
| SPME fiber | 250μm Polydimethylsiloxane (PDMS) | 60-275 Da volatiles | 10 days |
| | Divinylbenzene (DVB)/PDMS/carboxen (activated carbon resin) | 60-275 Da volatile and semi-volatiles | 10 days |
| | DVB/PDMS | 60-300 Da amines, polars, aromatic semi-volatiles | 10 days |
| | Carboxen (activated carbon resin)/120μm PDMS | 30-225 Da highly volatiles | 10 days |
| | Polyacrylate | 80-300 Da polar, semi-volatiles | 10 days |
| Cartridge | RAD168/$H_3PO_4$ microporous polyethylene | $NH_3$ | 12 months |
| | RAD170/Zinc acetate impregnated polyethylene | $H_2S$ | 6 months |
| | RAD165/Stainless steel net with 2,4-dinitrophenylhydrazine coated Florisil | Aldehydes | 2 months |
| | RAD145/Stainless steel net with Carbograph 4 (graphitized charcoal) | Benzene, toluene, ethylbenzene, xylene (VOC/BTEX) | 6 months |
| Steel mesh bag | 20-45 mesh Supelco Carboxen 569 | $C_2$-$C_5$ highly volatiles | Unknown |

## 3. Sampling and Analysis Notional Timeline

A blank (control) collector would be deployed by August 24 for exposure of the empty TAGSAM glovebox exhaust during the 1-month quiescent cleanroom settling period. The blank would be exchanged for the second collection unit before the TAGSAM head is moved to the TAGSAM glovebox as convenient to curation, around September 24. The blank solid phase microextraction (SPME) fibers would need to be shipped to GSFC and analyzed at this time to retain the analytes (**Table 2.2)**, though the cartridges could wait to be analyzed at once. The second collection unit would collect for at least two weeks, possibly until containerization. The third collection unit would be held in reserve if needed or could be swapped for the second unit. It is anticipated that all collections would conclude after sample containerization, approximately one month. However, after analysis the adsorbents can be reused. So additional sampling would be possible if it is desired by the Project.

Analysis would take place at the Astrobiology Analytical Laboratory at GSFC and performed by Heather Graham. Graham has experience with the collection and analysis of adsorbents deployed in cleanrooms at GSFC and JSC. Analysis would take place on one of the GC-MS instruments at GSFC already described in the SAP using the existing data archiving procedures already in place.

## 4. Team Roles, Responsibilities, and Governance

Analysis would be overseen by MPS Jason Dworkin; he takes responsibility for the success of the work and the compliance with JSC requirements and schedule. The modification of the TAGSAM glovebox exhaust would be performed by JSC curation using components specified by JSC curation. The device would be assembled and loaded by Heather Graham. JSC curation would clean the exterior of the device to permit entry in the ISO 5 cleanroom. The device would be installed and removed by JSC curation and shipped to GSFC for analysis.

## 5. Comparison of Results

The results would be compared with the canister air filter results, which would have been exposed to outgassing of the bulk sample. These would also be compared to the volatile experiments of Aponte and Huang, who will each extract individual ~500 mg samples via aqueous sonication and direct introduction to a GC-MS and SPME capture and GC-MS analysis, respectively. These four different views of sample volatiles could generate a more complete picture of the volatile chemistry of Bennu. Furthermore, the air samples analyzed as part of the SRC retrieval campaign will provide useful contamination background and may also collect sample outgassing in the UTTR cleanroom.

If there is a sample return anomaly, then the exhaust gas collection could inform the level of environmental contamination that remains on the sample.

## 6. Publication

If the results of the study are significant, we propose that would be published in a new manuscript, Graham-SA-2, with an expected submission date of late FY24. Note that Graham-SA-1 has a delivery date of FY25, so this plan is compliant with the one first-author publication per person per year policy in Phase F. Upon acceptance of this proposal, the publication plan will be submitted to Chief Editor Wolner for processing and approval.

A potential title would be "Volatile Compound Capture and Characterization During Curation of Returned Sample from Asteroid Bennu". Such a date would permit time to analyze the samples and not conflict with Graham-SA-1 in FY25 (**Table 6.1**). If the results do not warrant a publication, the data could contribute to Sandford-SA-1 "Detection of Captured Volatiles in the SRC Canister Filter" in FY24, Aponte-SA-1 "Abundance and isotopic composition of hypervolatiles and volatile organics in samples returned from Bennu by OSIRIS-REx" in FY24, and/or Huang-SA-1_PS "Comprehensive non- (or micro-) destructive structural and isotopic analyses of volatile organics in Bennu: implications for the origins and synthetic pathways of prebiotic organic compounds" in FY24.

*Table 6.1* Sample Analysis Publication Plan additional information for the proposed FY24 publication of Graham-SA-2 by Graham, et al. "Volatile Compound Capture and Characterization During Curation of Returned Sample from Asteroid Bennu"

| Working Groups | Hypothesis | Level 2 Hypotheses | Data Products (Responsible WG) |
|---|---|---|---|
| SOAWG, SRCWG | Direct: 2<br>Direct: 4<br>Indirect: 3<br>Indirect: 9.5<br>Indirect: 12 | 2.1, 2.3, 3.1, 4.3, 9.5, 12.1 | Compound identification of volatiles species using GC-MS. Individual data products: Raw data files, and CSV files listing compound identifications, abundance calculations, and x-y pairs for chromatograms (SOAWG/SRCWG). |

# 7. Abbreviations and Acronyms

| | |
|---|---|
| BTEX | Benzene, Toluene, Ethylbenzene, Xylene |
| CCWG | Contamination Control Working Group |
| Co-I | Co-Investigator |
| CSV | Comma Separated Values |
| CWG | Curation Working Group |
| Da | Dalton |
| DVB | Divinylbenzene |
| FY | Fiscal Year |
| GC-MS | Gas Chromatography Mass Spectrometry |
| $GN_2$ | Nitrogen gas |
| GSFC | Goddard Space Flight Center |
| ISO | International Organization for Standardization |
| JSC | Johnson Space Center |
| MPS | Mission Project Scientist |
| MSS | Mission Sample Scientist |
| OSIRIS-REx | Origins, Spectral Interpretation, Resource Identification, Security, Regolith Explorer |
| PDMS | Polydimethylsiloxane |
| SAP | Sample Analysis Plan |
| SOAWG | Sample Organic Analysis Working Group |
| SPME | Solid phase Microextraction |
| SRC | Sample Return Capsule |
| SRCWG | SRC Working Group |
| TAGSAM | Touch and Go sample Acquisition Mechanism |
| UTTR | Utah Test and Training Range |
| VOC | Volatile organic compounds |
| μm | Micrometer |

# Appendix 1. Huang parent proposal

# Proposal to sample the volatile compounds from the curation glove box

**Yongsong Huang, Ewerton Santos, Philipp Heck, Jason Dworkin and Daniel Glavin**

**January 31, 2023**

## 1. Introduction

It is well known that carbonaceous chondrites contain abundant volatile compounds, most of which are likely organic molecules (Sokol, 2020). For example, according to Murphy (2020) "the famous Murchison meteorite that fell in Australia in 1969, often described as a mudball, has a particularly powerful pungency, variously compared by meteoriticists as similar to methylated spirits, wet hay and tar. The reason for the not-so-heavenly scent of meteorites lies in the organic and other carbon compounds they contain, especially true for the carbonaceous chondrite group of meteorites (of which Tagish Lake, Aguas Zarcas and Murchison are all members). Although some non-carbonaceous meteorites can pack a punch too." Notably, the pre-rain Aguas Zarcas fragments also gave off a pungent "Murchison-like" ordor (a compost-like scent) (https://www.lpi.usra.edu/meteor/metbull.php?code=69696) and Brussells sprout ordor (Weisberger, 2019).

Unfortunately, no one has properly collected these volatile compounds for analyses. We have no idea what volatile organics are contained in carbonaceous chondrites. Notably, these volatile compounds evaporate quickly, so by the time systematic analyses begin on these samples, most of these compounds have already been lost. Rarely had any curation facilities have adequate measures to systematically prevent these volatiles from vaporization. Falling though the sky and between the time of fall and sample collection, a considerable amount of time would have passed, these volatiles would have gone.

## 2. Hypotheses

Asteroid sample from Bennu collected by the OSIRIS-REx mission offers an unprecedented opportunity to examine these volatiles in materials that experienced minimal exposure to terrestrial contamination. The samples represent the most pristine carbonaceous astromaterials analogous to CM or CI chondrites. It is likely that the samples would contain large amounts of volatiles as well. Analyses of these volatiles will enable the investigation of hypotheses 2.1, 3.4, 4.3, and 9.5 (Table 1). The analysis of the SRC air filter in the baseline sample analysis plan investigates these hypotheses as well as contamination hypotheses 12.1 and 12.4. However, the limited flow of Bennu-infused air out of the SRC air filter and the combination of launch and UTTR influx air that enables the testing of hypothesis 12.1 makes this a challenging measurement. However, the capture and analysis of volatiles released in the first days of curation after the SRC is opened provides a novel opportunity to capture volatiles to test these hypotheses and address the unknowns in meteoritic odors. The results could be published as a part of the publication Sandford-SA-1 "Detection of Captured Volatiles in the SRC Canister Filter" or in Huang-SA-1_PS "Comprehensive non- (or micro-) destructive structural and isotopic analyses of volatile organics in Bennu: implications for the origins and synthetic pathways of prebiotic organic compounds" depending on the nature of the two investigations (OSIRIS-REx Publication Plan).

**Table 1. Hypotheses to be tested by this investigation (Section 3.0).**

| |
|---|
| **2.1: Bennu contains organic molecules that are used in terrestrial biology…their precursors, and side products from chemical reaction networks** |
| **3.4: Bennu contains organic matter that formed in the interstellar medium…** |
| **4.3: The initial constituents of Bennu's parent asteroid were materials that were inherited from the protosolar molecular cloud or were formed and altered in the protoplanetary disk…** |
| **9.5: Volatile compounds migrated from high-temperature low-latitude regions of Bennu's surface and condensed in cold traps at high latitudes…..** |

One of the main characteristics of prebiotic organic syntheses is the formation of numerous low molecular weights, volatile organic compounds such as C1 to C5 carboxylic acids, aldehydes, ketones, amines, alcohols, nitriles, and hydrocarbons. The compounds are building blocks for the larger molecules such as amino acids which may be formed by Strecker synthesis during the later aqueous alteration on the parent body (Hypothesis 2.1). A large number of small volatile organic compounds have been spectroscopically observed in the interstellar clouds. But the exact identities of these volatiles are difficult to assess without mass spectrometric analyses of physical samples on Earth. Many of the important questions related to the three hypotheses above may be answered by systematically analyzing volatiles and hyper volatiles from Bennu (Hypothesis 3.4). The compositions and abundances of these volatile species in the exhaust may bear significant similarity to those in our PSP proposal (Huang et al. 2022), so if Bennu releases volatiles as anticipated the measurements should be achievable with the proposed implementation.

## 3. Proposed Implementation

While it is possible significant amount of such volatiles are contained and adsorbed within the Bennu sample matrices, a fraction of these volatiles may start to rapidly degas in the glove box during the encapsulation of Bennu material for examination (which may take days or weeks) at the Johnson's Space Center curation facility, if the pungent smell from the Murchison or Aguas Zarcas meteorites is an indication of the outgassing that could occur with the returned Bennu samples. Currently, there is no plan to collect the volatiles from Bennu samples during the initial examination. The glove box housing the Bennu samples would be continuously purged with nitrogen gas via an exhaust pipe, and any sample outgassing products swept away with the nitrogen purge and lost if not captured.

We propose to add a simple and non-invasive system to collect and trap most of the volatiles that are vented from the glove box. This design accepts trades to eliminate any risk to the sample and minimize the impact to the curation flow and facility. The proposed system design is shown in Figure 1. Basically, a condenser trap will be used to condense the volatiles from the glove box exhaust pipe. The curation system adds no back pressure which could endanger the sample and requires no other modifications. We will use liquid argon (boiling point -186°C) for cryogenic trapping, which would not condense nitrogen gas (boiling point -196°C). The condensers will be periodically swapped at a pre-determined cadence (TBD) as the curation flow and schedule permit, and sealed. The primary goal here is to capture any changes to the composition of Bennu volatile outgassing products over time. Personnel from Brown University can be available to tend the condensers as the JSC curation team prefers. The sealed samples will be taken to Brown University for analysis.

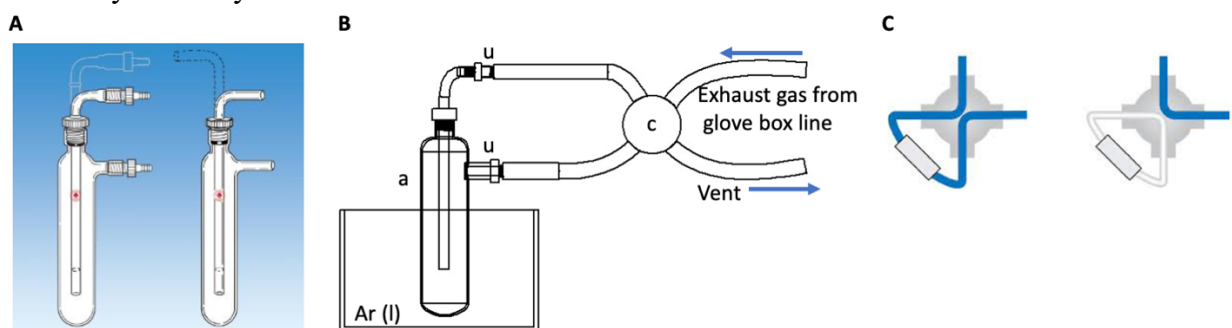

**Figure 1.** A. Avantor trap available at https://us.vwr.com/store/product/11817200/vacuum-trap-hose-connection-glass-adapter-ace-glass-incorporated. B. Condenser chilled with a liquid argon dewar; u are unions for cold trap with threads and Teflon seal (after sampling, a dead cap will be used for sealing off and transport); c is the 4-way valve. C. The 4-way valve has two flow paths for the 4-way valve, the left is the condensing position, and the right permits the condenser swap without interfering with the exhaust flow. Periodically, the traps will be replaced with new ones, so that a time series set of samples can be collected. Such time series samples will establish what and by how much different volatiles are emitted from Bennu samples at different times during the initial curation process.

The Brown team will purge the volatile compounds to GCMS (Technique 5.36) for structure identification and GC-IRMS (Technique 5.41) for compound specific isotopic ratio measurements. Brown has all the facilities for the proposed analyses. We have a Frontier Lab cryogenic trap to retain the volatile organics in the GC column front during purging of the compounds from the collection bottle (Figure 2). The cryotrapped compounds are released into the GC column after completion of purging, hence sharp chromatographic peaks will be achieved. An Agilent SuperWax GC column or equivalent would be used for analyses. This column resolves volatile polar organics very well and requires no derivatization. In addition to purging all compounds into GCMS, we can also use SPME (or SPME arrow with larger sample capacity) and SPME on-fiber derivatization to deliver smaller portions of compounds (so that the same samples can be analyzed by multiple instruments with repetition), and to target individual volatile compound classes such as carboxylic acids, amines, aldehydes and ketones.

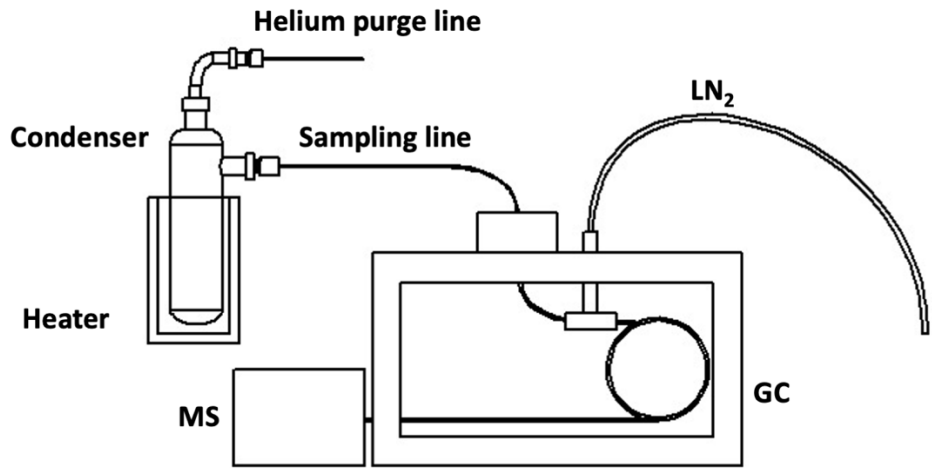

**Figure 2.** The sample testing configuration at Brown University uses components already present.

# Appendix 2. Approval by CWG and CCWG

| 16 June 2023 | **Glovebox Exhaust Characterization** |
|---|---|

Nicole Lunning    Kevin Righter    Christopher Snead    Wayland Connolly    Jason Dworkin

This memo is **approval** and evaluation of the materials and implementation of the exhaust sample collection from the TAGSAM glovebox. A design was presented for the collection of compounds which may be entrained in exhaust nitrogen with no impact to the sample and minimal impact to the curation workload. This design would collect gases on a best-effort basis downstream from the glovebox. The collection would be on adsorbent materials inside an airlock made of 2-inch 316 stainless steel and Viton sanitary fittings with no opportunity for exposure to the sample nor the cleanroom environment (Figure 1) even in the event of a nitrogen gas ($GN_2$) failure, glove leak, or power outage.

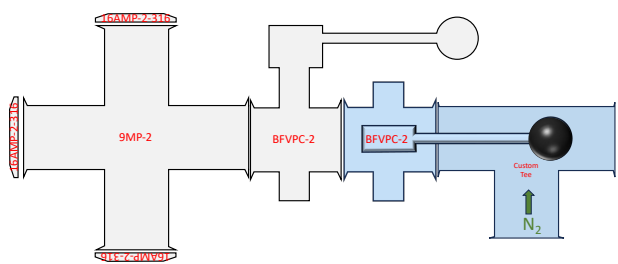

Figure 1. The blue components, a custom tee and valve would be fitted to the TAGSAM glovebox exhaust line after the check valve. The white components would contain the adsorbents and can be attached and removed via the two valves without disrupting the exhaust flow or exposing the interior to the cleanroom. Red text indicates sanitaryfittings.us part numbers, gaskets and clamps are omitted for clarity.

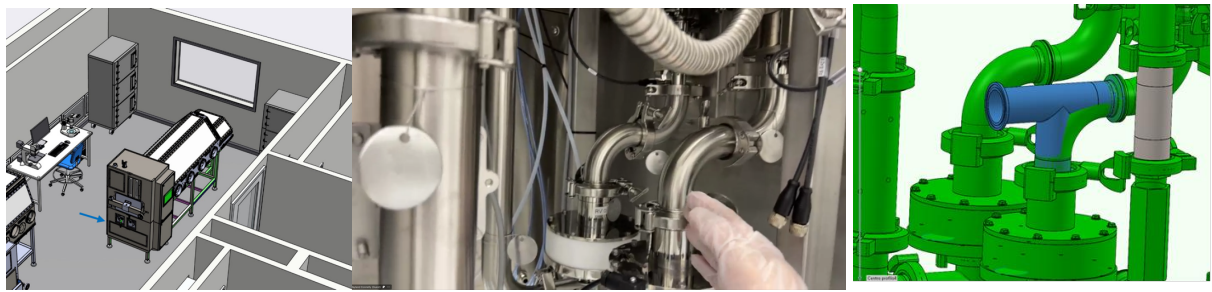

Figure 2. Left, the arrow points to the approximate location of the tee. Center photo of location of the tee to be added after the exhaust check valve at the far end of the glovebox from the exhaust connection to the sample chamber. Right, the design of the custom tee. Some clamps and gaskets are omitted for clarity.

All parts will be cleaned and installed by curation personnel. One custom tee from DF Siena (dfsiena.it), the same vendor at produced the exhaust plumbing on the glovebox would be procured and installed with an off-the shelf approved valve and end cap for when not in use (sanitaryfittings.us). Three collectors would be assembled and supplied by GSFC from the same approved vendor (sanitaryfittings.us). The first would be attached in August prior to the start of the cleanroom quiescent period, to collect a blank. The second would be attached prior to the arrival of the sample and valved off until the TAGSAM head is placed in the glovebox. The third is a backup but can be installed part way through the TAGSAM disassembly. The gas collection operation would conclude after sample containerization. However, the valved off and capped custom tee can remain on the glovebox.

Curation accepts that on a best effort basis that the collector valve be closed when the glovebox airlock is used to minimize the introduction of room air into the collector, since there can be backflow into the collector from the airlock. Curation accepts that the valve should be closed when the airlock is cleaned to avoid the introduction of cleaning solvents (hydrogen peroxide solution and/or isopropanol) from the airlock into the collector.

Sample collectors would be shipped, or hand carried to GSFC for disassembly and analysis. Since the design has significant dead space and it is understood that the material collected will be relative at best and qualitative at worst.

## 5.1 - 5.65 Details of Analytical Techniques

## Spectroscopy

Spectral measurements are the overseen by the Sample Spectral Analysis Working Group (SSAWG). These analyses are aimed at measuring the diagnostic signatures of both organic and inorganic compounds to help test OSIRIS-REx hypotheses.  This section describes the following techniques.

- 5.1 Visual Inspection
- 5.2 Visible, near-infrared, and mid-infrared (VNMIR) spectroscopy

# 5.1 Quantitative Reflectance Imaging

**Overview:**

Quantitative reflectance imaging of the returned sample will occur soon after the SRC is opened at NASA-JSC. This characterization provides quantitative information on the reflectance of the sample for discrete wavelengths comparable to the filter set on MapCam (Rizk et al. 2018). The objective of this investigation is to rapidly assess the lithological diversity using normal reflectance and spectral band ratios, similar to the color imaging of Bennu (DellaGiustina et al. 2020). After evaluating the number of distinct lithologies, we will organize our scientific investigations with reference to our coordinated analysis investigation plan to ensure characterization of all major lithologies and a subset of minor lithologies, depending on the heterogeneity of the material.

**How it works:**

QRIS is an 8-color imager designed to image the returned sample in the sample trays, as well as in TAGSAM. QRIS's nominal field of view (with its "1X" 12 mm focal length lens) is 60 degrees. When positioned ~30 cm above the target, it images a 35 cm square region (large enough for the TAGSAM or the sample tray assembly). QRIS was originally required to image with a pixel scale of 0.1 mm/pixel over a 40 cm FoV. Updated tray design have shrunk the required field of view to 35 cm, so the pixel scale has correspondingly shrunk by ~12% (~.08 mm/pixel), but is often colloquially referred to as 1X or .1 mm/pixel. The 12 mm focal length lens can be replaced with 25 mm or 50 mm lenses, which provide magnification of approximately 2X and 4X respectively. The 25 mm lens shrinks pixel scale by approximately 50%, and the 50 mm lens shrinks pixel scale by approximately 75%. Calibration images are taken with the 12 mm lens, and therefore calibration for images taken with the other two lenses are not calibrated as extensively.

While imaging, the sample is illuminated by two LED lightboxes. The lightboxes have LED strips mounted at the top and white glass diffusers at the bottom. The diffusers spatially mix the LED lights (which otherwise come from discreet physical locations, causing spectral variations across an image). The lightboxes each have 8 narrowband LED strips and 1 broadband 'white' LED strip. The 8 colors are UV, violet, blue, green, amber, red, deep red, and infrared. See the QRIS cal document for detailed spectra.

This QRIS camera has a 12-bit detector, which is not enough dynamic range to cover the dark expected sample (~4% albedo), bright inclusions (as bright as 30% albedo, based on remote sensing of Bennu), and very bright metallic surfaces in the tray assembly. To accommodate this range of reflectances, QRIS images in each color with a range of exposure times designed to take well-exposed images of the brightest and darkest expected targets. In post-processing, these images are combined into a single high dynamic range image.

Calibrated reflectance standards are included in the QRIS field of view for in-situ radiometric calibration. These standards are used to convert the HDR images (acquired in units of DN) to units of reflectance.

**Sample preparation:**

None - images will be acquired while the samples are in the pristine curation cabinets.

**Sample impact:**

None

**Data quality:**

QRIS with its 12 mm lens will image the sample with ~0.1 mm/pixel resolution. 25 mm and 50 mm lenses will be ~2X and ~4X smaller.

**Data products:**

Images comparable to MapCam Level-0 (raw, uncalibrated images), Level-2 (calibrated reflectance), and higher-level products (e.g., color ratios).

**Facility(ies):**

The imaging system is being designed at the University of Arizona and will be delivered to JSC prior to sample return.

# 5.2 Visible, near-infrared, and mid-infrared (VNMIR) spectroscopy

## Overview

Visible (V), near-infrared (NIR), and mid-infrared (MIR) spectroscopic techniques are used for the inorganic (e.g., mineralogical) and organic characterization of samples.  VNMIR spectroscopy is best suited to the identification of compounds present (at the measurement's spatial scale) in percent-level or greater abundances, with detectability varying as a function of characteristics such as sample preparation and phase composition. Derivation of absolute phase abundances is possible in some situations for VNIR spectra (with non-linear modeling) and in most situations for MIR spectra (with linear modeling). Measurements can be acquired with Fourier transform infrared (FTIR) interferometric spectrometers, among other designs, under a range of environmental conditions. Typically measurements are made in ambient conditions, although in some cases, it is desirable to measure samples in a specific environment to simulate remote sensing measurements.

Visible wavelengths range from ~0.38 to 0.70 µm. Near-infrared wavelengths begin at ~0.70 µm; we define the end of the OSIRIS-REx Visible-Infrared Spectrometer (OVIRS) spectral range, ~4.2 µm, as the end of the NIR. In this range, electronic (e.g., charge transfer and crystal-field splitting) and molecular vibrational absorptions are the most common features. The NIR region is suited for identifying the presence of Fe- and $H_2O/OH^-$-bearing minerals, as well as C-bearing compounds such as carbonates and organics. The MIR is generally defined as ranging from ~5 to ~50 µm. Absorptions in this region are the result of photons interacting with fundamental vibrational modes (phonons) in the sample. The NIR and MIR are both suited to measuring phase abundances down to a few vol%. VNMIR wavelengths cannot all be measured simultaneously or with the same instrument. Most commonly, a VNIR spectrum (out to ~2.5 µm) can be acquired with a single source, detector, and beamsplitter combination. Longer wavelengths (2.5 to 4 µm) require different hardware. For the MIR, different sources, detectors, and beamsplitters are used, and lead to different spectral measurement ranges such as ~2.5 to 14 (or 25) µm and ~5 to 50+ µm.

## How it works

VNMIR spectroscopy typically involves: (i) measuring natural or artificial light reflected off of a sample (reflectance); (ii) measuring light emitted by the sample (emission); or (iii) the transmission of artificial light through a sample (transmission). Several fundamental interactions of light with materials are measured as a function of wavelength. The positions, widths, and strengths of absorption features in a spectrum provide diagnostic information about the material's composition, structure, and/or physical state.

*Reflectance spectra* are commonly acquired by using an artificial light source to illuminate a sample at the desired wavelength range (e.g., quartz tungsten halogen lamp for visible wavelengths, Globar source for IR wavelengths). The light is typically arranged to be incident on a sample at a certain angle or range of angles. A detector that is sensitive to a desired wavelength range is used to measure the amount of light reflected off of the sample, commonly at a certain angle or range of angles. If these incidence and emergence angles are relatively uniform then the measurement is "bi-directional", whereas if the angular range is broader (a cone of incident or emergent light) then the measurement is "bi-conical". Factors other than composition can influence reflectance spectra, such as the viewing geometry of the measurement, particle size, roughness, crystal orientation, etc.

*Emission spectra* differ from reflectance measurements in that there is no incident light source; the sample itself is the source. This technique requires the use of custom hardware configurations. Laboratory spectrometers in use by the OSIRIS-REx team for emission measurements are equipped with uncooled (but thermoelectrically temperature-controlled) detectors, and the sample is heated (commonly to ~80°C, but this temperature is selectable) so as to increase its radiant energy relative to that of the detector for optimal signal-to-noise ratio (SNR). These measurements are generally considered to be hemispherical. Emission and reflectance spectra may be qualitatively thought of as inverses of each other; both measure the optical constants *n* (refractive index) and *k* (extinction or absorption coefficient). Quantitative comparisons of reflectance and emission spectra, however, require specific instrumental configurations.

*Transmission spectra* are measured by using an artificial light source to illuminate a sample. The light transmitted through the sample is then measured by the detector. Only the absorption coefficient is measured and transmission spectra are therefore not directly comparable to reflectance and emission measurements. Absorption spectra can be thought of as the inverse of transmission spectra.

## Sample preparation

Details of sample mass and preparation can vary by laboratory and measurement configuration. Ideally, the material should fill the field of view of the detector, but this may not be possible for very small samples in reflectance or emission, in which case it is desirable to use dishes, cups, or substrates that are coated with a material of very low reflectivity (high emissivity), such as certain IR-black paints. When measuring particulate samples it is common to record the particle size range of the sample and to arrange the powdered sample so that the surface is relatively flat. Similarly, when measuring intact chips or non-particulate samples, it is desirable to have the surface of the sample relatively horizontal or at least in a position where the relative incidence and emergence angles can be estimated if the measurement is not hemispherical. Rock surfaces may be irregular and do not require careful geometric alignment, although many sample holders allow for flat surfaces of slabs or cut rocks to be viewed from a perpendicular perspective.

*Unprepared bulk samples* of particulates, rock chips, or small rocks (up to ~6 cm in diameter for certain emission instrument configurations) are typically suitable for reflectance and emission spectroscopy. For reflectance measurements, the sample is placed in a dish, cup, or on a substrate such that the incident light illuminates the material and the detector is commonly arranged to measure the reflected light at the desired angle(s). For emission measurements, the sample is placed in a temperature-controlled, heated cup, and the energy emitted by the sample is directed into the spectrometer via a series of one or more mirrors.

*Thin and thick sections, potted butts, and cut slabs* (up to ~1 cm in thickness) are particularly well suited to reflectance analysis by FTIR microscopes (μ-FTIR). The use of uncovered thin and thick sections >30 μm will not result in measurement of the silica slide underneath the sample in a reflectance measurement. In the case of thin and thick sections and potted butts, μ-FTIR measurements permit some cross-correlation with other methods that use these preparations (e.g., electron microprobe analysis). Round and rectangular sections can be accommodated by the μ-FTIR systems described below.

*Prepared bulk samples* include particulate materials derived from sieving of natural particulates into discrete particle size ranges or from sieving of crushed solid samples. The particle size range can be selected to improve the SNR in reflectance measurements, or to acquire measurements at multiple particle sizes to observe changes in spectral behavior as a function of particle size. Transmission measurements typically require the preparation of fine-particulate (few microns) size fractions that are subsequently mixed with an IR-transparent medium (such as KBr, which separates the particles) and pressed into pellets or plates for analysis. These samples must be optically thin so as to avoid saturation of the absorption bands.

## Sample impact

VNMIR spectroscopic measurements generally are considered non-destructive because they do not typically alter the chemistry or structure of the sample during the measurement. However, some measurement conditions may result in modification of the sample – although such conditions are not planned for the returned sample, it is worth noting what can happen. If a measurement transfers significant energy into the sample, through an intense illumination source or excessive heating of the sample, there is the possibility of damaging sensitive materials or volatiles. The potential for sample damage generally increases with increasing energy (decreasing wavelength) of the incident light. Long exposure times at longer wavelengths also have the potential to alter samples; e.g., illuminating a sample with a strong lamp or synchrotron beam for a long time could result in loss of water or other volatiles. These issues can be minimized by decreasing illumination times (shorter integration times), using more sensitive detectors, or co-adding a series of short-duration measurements to reduce random noise. For emission measurements to remain non-destructive, the sample temperature must be carefully controlled to ensure that heating leading to loss of water and other volatiles does not occur. Measurements made in vacuum could cause some materials to degas. Outside the measurement environment, the act of transferring particulate sample material between containers (shipping, measurement, back to shipping) can result in some small loss of mass, such as for the smallest particle sizes that are inclined to cling to surfaces. Repeated transfer of material may also affect particle morphology and size as the number of grain-to-grain contacts erodes asperities. These changes in the physical nature of the sample can result in different spectral characteristics in subsequent measurements.

## Data quality

Detectability and data quality vary as a function of sample physical characteristics and preparation, spectral and spatial resolution, wavelength stability, radiometric calibration, phase composition, and SNR. There are no universal detection limits by mineral or compound. A general rule of thumb for penetration depth is ~10× the wavelength of investigation, but this can vary with sample surface roughness, coatings, etc.

Spectral resolution in conventional laboratory instruments is user-selectable. An FTIR system measures in units of frequency called wavenumbers ($cm^{-1}$); these are related to wavelength by the relationship $cm^{-1} = 10,000/\mu m$. VNIR measurements are usually discussed in terms of wavelength (nanometers or microns), which are much smaller numbers than wavenumbers in this region). In the VNIR, 1-nm spectral sampling (2-nm resolution) is feasible, although sampling on the order of 2 to 5 nm is more common. In the MIR, spectral sampling in the laboratory is commonly selected to be 2 $cm^{-1}$ (4 $cm^{-1}$ resolution). In FTIR systems, higher resolution typically requires longer integration times to obtain the best SNR.  For the returned sample, laboratory measurements will be acquired at spectral resolutions meeting or exceeding the resolutions of the OVIRS (2 nm sampling below 2.4 µm and 5 nm sampling between 2.4 and 4.3 microns) and OTES (8.66 $cm^{-1}$ sampling) instruments.

Spatial resolution varies considerably with instrumental configuration. The hardware in standard µ-FTIR systems may allow for apertures as small as 10 µm/pixel, but the practical useful limit is defined by wavelength-dependent diffraction, which means a larger spot for longer wavelengths (~14 to 25 µm in the MIR, depending on the combined range of the installed detector and beamsplitter). For benchtop FTIR systems, diffraction limits still apply but are not usually the limiting factor, as spot sizes range from a few millimeters to several centimeters.  For the returned sample, spot sizes at NASA Ames (pressed pellets), RELAB, Oxford, and Winnipeg (all powders) will follow the standard procedure for those labs. At SwRI-Boulder, µ-FTIR spectral maps on thin sections will vary in spatial resolution depending on the nature of the sample; reconnaissance maps have a nominal spot size of 300 µm and targeted areas may merit smaller spot sizes.

Wavelength stability and radiometric calibration also affect data quality. Wavelength stability generally is not a factor in FTIR systems, where a highly stable, visible-wavelength laser is used to count fringes in the interferometer and control the moving mirror. Many labs nevertheless will use a standard (e.g., polystyrene film) to periodically confirm wavelength accuracy.

Radiometric calibration varies with instrumental configuration. VNIR reflectance measurements commonly are ratioed to a standard material such as diffuse or polished gold plates, pressed Halon, or pressed Spectralon. Accurate knowledge of the absolute reflectance values as a function of wavelength for these standards determines in part the accuracy of the absolute reflectance values. MIR reflectance measurements commonly use diffuse or polished gold calibration targets. MIR emission measurements rely on carefully calibrated blackbody targets whose temperatures are well controlled and known throughout the duration of their measurement (other known targets may be employed as well). Accuracy and precision vary with instrument configuration, but published data show that reproducibility is high and variability is not significant compared to the vast majority of spectral features of interest.

## Data products

The primary measurement data products expected from VNMIR measurements include reflectance, emissivity, and/or transmission spectra of returned samples in any of the following forms, depending on laboratory and instrument:

- 1D point analyses (individual spectra from a specific sample or sample location)
- 2D line scans (individual spectra plotted as a function of distance, e.g., from µ-FTIR)
- 3D (hyperspectral) spectrum images (2D maps of a sample in which every pixel contains a spectrum in the third dimension, also known as an image cube)

These products can be used qualitatively and quantitatively (under the right conditions) to identify which minerals and compounds are present and, where x-y mapping capability exists, their distribution. Products are delivered in a number of formats, including XML, ASCII, HDF, binary with header files, and delimited text. The products that will be produced by each lab (with facilities described in the next section) are described below:

NASA Ames (µ-FTIR): There are two primary data products aimed at organics characterization: 1) 1D point analyses measured in transmission using pressed pellets containing dispersed powder; 2) maps of individual particles or clusters of particles. The manufacturer's OMNIC Picta software produces spectral and map data formats that are proprietary so spectra will be exported to .CSV files, and maps will include images of the field of view and "heat maps" of various band intensities.

Brown University (RELAB): The primary data product will be a "standard" RELAB spectral measurement – 1D point analyses measured on particulates in reflectance at VNMIR wavelengths. [TBR: 1D and 2D products can be delivered as text files whereas 3D products can be delivered as binary image cubes with separate header files that contain the necessary metadata.]

Oxford University (Emission): The primary data product from the Oxford lab is a 1D thermal emission (MIR) spectrum of particulates measured in a simulated asteroid environment. The manufacturer's software uses proprietary file formats, but there is an option for an ASCII-based XML format with standard fields and headers. Additional housekeeping data (temperature and possibly pressure) are also recorded as date stamped comma-separated value ASCII files. These ancillary files contain information on the temperature environment of the chamber during the measurements and are used as part of the data reduction to generate calibrated emissivity spectra.

Southwest Research Institute (µ-FTIR): The primary data products from the SwRI lab are for mineralogical characterization and include 1D (spot), 2D (line), and 3D (map) spectra acquired from 2.5 to 25 microns (NMIR) on thin sections, thick sections, potted butts, and possibly chips. The manufacturer's OMNIC Picta software saves data in proprietary formats but data can be exported to .CSV files (1D point spectra and 2D line scans) as well as ENVI .dat/.hdr files (3D maps). These are reformed into HDF-formatted files for this lab's archive, which enables the addition of ancillary information and CCD mosaic images for referencing the positions of spectral measurements.

University of Winnipeg (ASD): The primary data product is a 1D spectrum of each sample at VNIR wavelengths. The spectrometer initially produces files in proprietary formats, but can easily be converted to ASCII or text-delimited files.

## Facilities

Most of the facilities below are equipped with multiple spectrometers; this description focuses on the equipment that is anticipated to produce the primary, unique product of each lab for OSIRIS-REx sample analysis. Other resources in each lab can be brought to bear as sample availability and time allow. Note that the NASA Ames facility will primarily focus on organic analyses, whereas the others will focus on inorganic analyses. Also, although it may appear that the NASA Ames and SwRI-Boulder µ-FTIR systems are the same, they are not. They have different detectors and different measurable wavelength ranges and so are complementary systems with somewhat different capabilities.

NASA Ames Research Center, Mountain View, CA, USA - Scott Sandford

Thermo Scientific Nicolet iN10 MX FTIR microscope: $LN_2$-cooled mercury-cadmium-telluride (MCT) imaging (point) detector with spectral coverage across the 11,700 to 600 cm$^{-1}$ (0.85 to 16.67 µm) range and spectral resolution down to 1 cm$^{-1}$; horizontal spatial sampling down to 5 to 10 µm; transmission or reflectance measurements; computer-controlled aperture and mapping software with motorized stage; mapping over areas up to 10 x 10 mm or better; equipped with a built-in OMNIC Picta spectral library that can be used to assist with assessment of the composition/chemical structure of families of compounds for which no commercial or local laboratory standards are available.

Brown University (RELAB), Providence, RI, USA - Ralph Milliken

Custom bi-directional spectrometer (BDR): reflectance measurements of the wavelength range of 0.3 to 2.6 µm; sampling interval of 1 nm (or more); Xenon and Halogen light sources; detectors include photomultiplier (0.3 to 0.85 µm range) and custom Judson/Teledyne InSb (0.6 to 2.6 µm range); incident, emergent, and normal vector are co-planar (measurements are 'on-axis') but the incident and emergent angles can be independently varied up to ~±70° to study photometric effects. Common reflectance standards are pressed halon or Spectralon™ at VIS-NIR wavelengths and gold (diffuse or mirror) at NIR-MIR wavelengths. Precision <0.25% in reflectance. Features custom data acquisition and processing software.

Oxford University, Oxford, UK - Neil Bowles

Vertex 70V vacuum spectrometer (FTIR) with PASCALE simulated space environment thermal emission spectroscopy chamber. Spectral range: ~6000 to 50 cm$^{-1}$. Spectral resolution: better than 0.4 cm-1. For measurements using the PASCALE chamber, the sample will be measured within a controlled temperature range that simulates the conditions found on Bennu as closely as possible. The PASCALE environment chamber and sample cup temperatures are closely monitored along with the pressure in the chamber. PASCALE uses a PID (proportional, integral, derivative) temperature controller and all users are fully trained in the operation of the chamber to ensure that temperature limits are not exceeded.

Southwest Research Institute, Boulder, CO, USA - Vicky Hamilton

Thermo Scientific Nicolet iN10 FTIR microscope: Point spectrometer (not imaging) equipped for both reflectance (primary) and transmission measurements using one of two built-in detectors and a KBr beamsplitter. Optical geometry yields reflectance spectra that are comparable to emission spectra. Primary measurements will use a liquid $N_2$-cooled, extended-wavelength MCT (MCT-B) detector. Two interchangeable sources permit NIR (~0.85 to 2.5 µm) and MIR measurements (~2.5 to 25 µm) of powdered samples (reflectance of polished and solid samples is measurable only in the MIR as a result of the physics of scattering at shorter wavelengths). Spectral resolution down to 1 cm$^{-1}$;

horizontal spatial sampling down to ~2 µm and selectable aperture sizes from 10 to 300 µm (square or rectangular and rotatable; e.g., a 20 x 90 µm spot at 42° orientation). Spectra of samples are ratioed to comparable measurements of a polished or diffuse gold plate, depending on the sample surface texture. Instrument features computer-controlled, programmable mapping with motorized stage; area mappable in a single session depends on selected detector, spatial resolution, and spectral resolution.

Centre for Terrestrial and Planetary Exploration (C-TAPE), University of Winnipeg, Winnipeg, MB, Canada - Ed Cloutis

ASD HiRes 4 reflectance spectrometer: Point spectrometer that measures reflectance from 350 to 2500 nm with spectral resolution that varies between 2 and 7 nm (depending on wavelength). It consists of three detectors that cover the following wavelength ranges: 350 to 1000, 1000 to 1830, and 1830 to 2500 nm. The three detectors have slightly different field of views, and any offsets at 1000 and 1830 are multiplicatively corrected to the reflectance values of the middle detector (1000 to 1830 nm), which is thermoelectrically cooled. The spectrometer measures reflectance in 1.4-nm steps, which are then internally sampled by the instrument using a cubic spline interpolation to output data at 1-nm resolution. Reflectance is measured at $i$=30° and $e$=0°. Illumination is provided by an in-house 100W quartz-tungsten-halogen bulb which is controlled by a regulated DC power supply and air-cooled to provide stable illumination. The incident light is collimated using an integrating sphere and hollow-core light pipe. The size of the illuminated spot is on the order of 5 mm. Reflectance spectra are acquired by measuring reflectance of a NIST-traceable Spectralon disk (calibration standard: CS), dark current (DC), and sample (S). The measurements are converted to reflectance by:

$$R = CS - DC/(S - DC)$$

This removes the effect of dark current and converts the measurements to absolute reflectance. The same number of calibration standard, dark current, and sample spectra are acquired. Regular measurements are also made of a holmium oxide-doped Spectralon disk; the disk has numerous narrow absorption features and these are used to ensure that the instrument is properly calibrated to wavelength.

# Mineralogy and Petrology

# 5.3 Visible Light Microscopy

**Optical light microscopy**

**Overview:**

Light microscopy is an analytical technique for the visual examination of rough samples using binocular microscopes or polished thin or thick sections in transmitted and/or reflected light using a petrographic microscope. Most light microscopes use a visible wavelength light source, although other wavelengths can be used to identify specific phases activated by those wavelengths (e.g., phosphates by UV light). Transmitted or reflected light can either be viewed by the user and/or captured through a video system for single frame images, mosaics, or videos. Light microscopy is one of the first technique employed in investigation of a sample and is followed by increasingly sophisticated analyses.

**How it works:**

For irregular samples, binocular microscopes provide information on the texture, color, and heterogeneity. A high-intensity light source is directed at the sample, typically as a spot or through a ring source, and the light reflected back is observed. The depth of field can vary considerably based on the specific setup. Beam splitters can be used to allow photography/videography simultaneously with user observations. Although not a technique that allows definitive mineral identification, one can often identify specific minerals or classes of minerals through such observations.

For samples that have either been polished in a thick section or prepared as a ~30-micrometer thin section, visual wavelength light is directed perpendicular to the specimen either from above (reflected light) or below (transmitted light).

Transmitted light is the more common observational mode for most geologists. Such observations are typically performed using petrographic microscopes. Textures of large areas, individual particles, or portions of individual particles and/or grains can be ascertained. In meteoritic materials, for example, the textures of individual chondrules can be quickly determined across a large area in plane polarized transmitted light. In addition, crossed polarization transmitted light produces interference colors and extinction (darkening as the stage is rotated) that can be distinctive of individual minerals. Skilled users can typically distinguish different silicate minerals from one another, as well as from, e.g., carbonates or phosphates, by their shapes, occurrences, interference colors, and extinction. Modal analyses — the determination of the abundance of different phases — can be conducted by observations at points separated by regular spacings. Sizes of individual particles (e.g., chondrules) can be measured using calibrated scales embedded in the optical system of the microscope. In minerals that have been subjected to shock, a variety of features are present that have been calibrated using data from laboratory experiments, allowing the skilled user to determine the peak shock pressure experienced by a sample.

Reflected light optical microscopy is ideal for minerals that are opaque to transmitted light. Within chondritic meteorites, this includes a variety of sulfides and metal phases that are rare to common, depending on the meteorite. Differences in reflectivity and color can be diagnostic of individual phases. In metal samples, etching with a dilute solution of nitric acid in alcohol differentially dissolves the metal phases, revealing mineralogical differences and structures that can be indicative of the cooling history of the sample. Skilled users of reflected light optical microscopy can use it for identification of opaque phases, but many users instead rely on scanning electron microscopy.

**Sample preparation:**

No sample preparation is required for examination in binocular light microscopy, making it an ideal technique for preliminary examination. For transmitted and reflected light optical microscopy, the surface of the sample must be flat, relative to the angle of the interrogating beam. This flatness is typically achieved through grinding, lapping, and mechanical polishing.

**Sample impact:**

Light microscopy is generally a non-destructive technique, although thin or thick section preparation is destructive. Etching to reveal metal structures is destructive in that it requires repolishing to remove.

**Data quality:**

Light microscopy is most applicable to samples in the range from 10 cm to a few hundred microns in size. Resolution varies with sample size and viewing, but features down to a few microns in size can routinely be imaged with optical light microscopy techniques. Accuracy of the technique depends significantly on the experience of the user. Phase identifications and shock classifications/peak shock pressures determined by very skilled users are highly reliable. In all cases, confirmation through other means (e.g., SEM, EMPA, TEM) is desirable if possible.

**Data products:**

The data products expected from optical light microscopy measurements include:

- Photos/videos that can range from single frames to large mosaics or video files
- Phase identifications made in reflected or transmitted light
- Modal analyses of individual minerals/components, such as chondrule abundances
- Size distributions of individual particles/phases
- Shock classifications determined by observation of 25–50 randomized olivine and plagioclase (where present) grains within a meteorite

Most results from optical microscopy are qualitative, although modes, size distributions, and shock effects are typically quantified and can be expressed graphically or compared to well-established calibrations of similar features in a variety of meteorites.

**Facility(ies):**

Most laboratories have one or more binocular or petrographic microscopes. Solely as an example:

Smithsonian Institution National Museum of Natural History, Washington, DC, USA

- Zeiss binocular microscope with video capability suitable for samples up to 10 cm in size, with ring light and multiple lenses.
- Wild M3Z binocular microscope (6.5–40x) with ring light.
- Nikon petrographic microscopes (2) with 5–100x objective lenses equipped with both reflected and transmitted light capabilities.
- Olympus petrographic microscope with 1.5–100x objective lenses with reflected and transmitted light and automated mapping capability.

# 5.4 Scanning Electron Microscopy (SEM)

**Overview:**

Scanning electron microscopy (SEM) is an imaging technique that uses secondary and backscattered electrons to image irregular and polished bulk samples from less than a micron to centimeters in size through beam rastering or stage movement. SEM is generally coupled with one or more of energy-dispersive X-ray spectroscopy (EDS), wavelength-dispersive spectroscopy (WDS), cathodoluminescence imaging, or electron backscatter diffraction (EBSD); these techniques are covered separately.

**How it works:**

A high-energy beam of electrons (generally ≤30 keV) is accelerated from a field-emission gun (FEG) or W or $LaB_6$ filament and focused onto the sample. FEGs produce higher current density than W or $LaB_6$ and, thus, can produce smaller beam sizes and excitation volumes. The secondary electron and backscattered electron (SE and BSE, respectively) images are formed by the interaction of the primary electron beam with the atomic nuclei and electrons that surround them in the sample.

Inelastic interactions between the primary electron beam and electrons in the valence or conduction bands of the atoms in the sample can result in the ejection of relatively low-energy (<50 eV) electrons. Such secondary electrons are weakly bound near the sample surface and are not associated with specific atoms. So, they do not provide compositional information but instead produce high-resolution topographic images of the sample when collected with a secondary electron detector.

Alternatively, if the primary incident electron penetrates the electron cloud and approaches the nucleus, it can become strongly attracted to it and may elastically scatter at an angle >90°, i.e., backscatter out of the bulk specimen. The angle of the backscattering is proportional to the atomic number (Z) of the element. The backscattered electrons can be collected by an annular detector that sits above the sample about the optic axis and integrates the signal over an angular distribution. Thus, BSE images can provide qualitative compositional information, such as the existence of elemental zoning within individual mineral crystals (e.g., Fe-Mg zoning in an olivine), and can distinguish high-density minerals (e.g., Fe-bearing olivine) from low-density minerals (e.g., Ca, Na, Al-rich plagioclase).

**Sample preparation:**

For analysis of flat and polished specimens, flatness is typically achieved through grinding and lapping, and a finely polished surface is achieved through polishing by increasingly fine abrasive materials, e.g., from 15 to 6 to 3 to 1 to 0.25 mm (this stepwise progression can vary slightly from lab to lab). The surface of the sample must be conductive to avoid charge buildup on the surface. Although thin metal coatings are sometimes used for electron microprobe analyses and typically used for ion microprobe analyses, metals efficiently backscatter electrons, reducing the utility of the technique described here. Therefore, carbon coating is used.

Analysis of irregular samples often involves stages that allow tilt or rotation for viewing different surfaces. Effective coating of irregular samples can be difficult and is sometimes undesirable as it compromises them for future analyses. Many SEMs are capable of variable pressure conditions under which the specimen does not need to be at high vacuum. Finally, many SEMs are capable of holding large, irregular, uncoated samples, which could prove useful for analyzing spacecraft hardware or contact pads, for example.

**Sample impact:**

SEM is generally a non-destructive technique, as the electron beam does not typically cause structural damage to crystalline or amorphous materials. However, some samples can be sensitive to the electron beam interaction. Beam-sensitive materials are generally characterized by abundant light elements or volatile compounds (e.g., halogens, alkalis, water) or have quasi-stable or amorphous crystallinity. Users must be careful to understand the potential for beam damage in such cases. Another possible impact is deposition of impurities from the beam column. Proper preparation of both the sample and the beam can minimize contamination.

**Data quality:**

The spatial resolution of SEM is controlled by the nature of the emitter and the electromagnetic lenses in the electron column. For W and $LaB_6$ filaments, the beam can vary from 1 mm in size to either a defocused beam of ~10 mm diameter or a rastered beam with a 150-mm square edge. For FEGs, it is possible to achieve sub-micrometer beam sizes. In addition to beam rastering, stage movement can be used to map entire polished thin sections or larger, irregular samples.

**Data products:**

The data products expected from SEM include:

- SE and BSE images
- Metadata for each image, including magnification/scale and analytical conditions
- Location data (x-y-z) for each image
- Composite images of larger areas up to the centimeter scale

The image products are typically used in conjunction with compositional analysis via, e.g., EDS (as described above). The image data connect the physical locations and compositional analyses on the sample, providing microtextural context. All products can be delivered as image files (.tif, .jpg, .png, .bmp) including x,y,z information as needed.

**Facility(ies):**

Astromaterials Research and Exploration Science (ARES) Division NASA Johnson Space Center (JSC), Houston, TX, USA

The JEOL 7600F SEM produces ultra-high resolution electron images using a thermal FEG. This type of gun is a significant advancement over earlier SEM electron guns because of the fine (~1–2 nm) electron beam delivered to the sample surface. The tightly focused beam makes it possible to record electron images with spatial resolutions of 2 to 3 nm.

The 7600F SEM includes two SE detectors and two BSE detectors. In the case of the two SE and two BSE detectors, one of each is in-lens and one of each is in the sample chamber. The in-lens detectors make it possible to position samples very close to the bottom of the electron column, which permits the acquisition of images with a low-energy beam. The advantage of a low-energy beam is that very near-surface features are emphasized.

The 7600F is also equipped with an Oxford 170 mm$^2$ active area SDD EDS spectrometer system. This type of detector is a significant advance over earlier Si(Li) detectors in that it can acquire and process >100,000 X-ray counts per second. This high count rate permits production of high quality X-ray maps of geological or planetary samples in reasonable timeframes.

The JEOL 7900LV FE-SEM field-emission scanning electron microscope is equipped with the Oxford Symmetry EBSD system and an Oxford 170 mm$^2$ active area SDD EDS spectrometer that will enable simultaneous EDS and EBSD mapping of samples. The 7900LV provides outstanding performance for both imaging and analytical conditions at very low accelerating voltages, operation in a low vacuum mode (150 Pa or lower) to enable imaging of non-conductive samples, and energy filtering combined with specimen bias applied voltages (from +500 to –2000 V) to enable high-resolution imaging of astromaterials and surfaces down to 10V (accelerating voltage) with image resolution better than 2 nm.

The Oxford EBSD can collect high-resolution Kikuchi diffraction patterns at ultra-fast speeds (3000 patterns per second). This technique can be used for structural identification of mineral phases, quantification of intracrystalline plastic strain within individual grains, and determination of deformation mechanisms.

Smithsonian Institution, Washington, D.C., USA

- The FEI Nova NanoSEM 600 is equipped with a FEG, ThermoFisher SDD EDS detector, and Gatan MonoCL spectrometer. The SDD allows rapid full spectra mapping. The Nova also has a large sample chamber that can accommodate specimens up to 15 cm in diameter and several centimeters thick and a variable pressure capability that allows imaging and analysis of a sample without a conductive coating.

University of Arizona, Tucson, AZ, USA

- The Hitachi S-4800 cold-field emission SEM is equipped with an Oxford Instruments SDD-EDS running Aztec software for compositional analysis of elements with Z ≥ 4 in one and two dimensions.

- The Hitachi S-3400 tungsten thermal emitter SEM with a variable-pressure chamber is equipped with a Thermo-Noran SDD EDS system operating NSS software for compositional analysis of elements with Z ≥ 5 in one and two dimensions.

- The ThermoScientific Helios NanoLab 660 G$^3$ focused-ion-beam SEM (FIB-SEM) is equipped with an EDAX SDD EDS system and an EBSD system for compositional and crystallographic analysis in two and three dimensions and for compositional analysis of elements with Z ≥ 4 in one and two dimensions.

University of California, Berkeley, CA, USA

- A Tescan large-stage, low-vacuum-capable analytical SEM equipped with an Everhart-Thornley detector for SE imaging and a multi-segment solid state detector for BSE imaging. The EDX system comprises Oxford Instruments' 80 mm2 X-Max SDD, a large area silicon drift detector, and the INCA software suite.

Field Museum, Chicago, IL, USA

- Field Emission Scanning Electron Microscopy Hitachi SU-7000 with Oxford Instruments XMax 50 energy dispersive X-ray spectroscopy system

# 5.9 Transmission Electron Microscopy (TEM)

**Overview:**

Transmission Electron Microscopy (TEM) is a technique utilizing an energetic electron beam (typically 60 to 400 keV) to image thin samples under high vacuum. The electron beam is transmitted through the sample to detection devices to record images of the sample as well as to generate crystallographic data utilizing electron diffraction. Owing to the short wavelength of the electrons, TEM instruments are capable of imaging samples down to the atomic-scale. Images are obtained in two major modes: (1) conventional parallel illumination with a static beam (TEM mode) or (2) converged illumination with a scanned beam (scanning or STEM mode). In addition to a variety of imaging modes, several analytical techniques are available for chemical analysis, e.g., EDX and EELS, in STEM instruments.

**How it works:**

TEM imaging and associated analytical techniques rely on interactions within the thin (typically <100 nm thick) samples. Electrons are accelerated at the sample at energies between 60 and 400 keV and experience elastic and inelastic scattering with other electrons and the nucleus during transmission. The type of scattering together with the sample composition and atomic structure can be used to form different types of images of the sample. The image contrast in non-crystalline materials is determined by sample thickness and average atomic number (Z-contrast). In crystalline materials, there is an additional contrast effect related to the crystallographic orientation of the sample, so-called diffraction contrast. When equipped with a secondary electron detector, the STEM can image detect secondary electrons to see surface details in samples (including thick samples).

A wide variety of information about the analytical volume is provided by STEM signals:

- Brightfield imaging – using the direct beam and diffracted electron to form the image.
- Darkfield imaging – using only diffracted/scattered electrons to form the image and excluding the direct beam.
- Annular darkfield imaging – true atomic number (Z) contrast imaging
- Secondary electron imaging – surface imaging of the sample
- Electron diffraction – produces patterns that represent the crystal structure of the sample
- Energy-dispersive X-ray spectrometry (EDX)
- Electron energy-loss spectroscopy (EELS)

**Sample preparation:**

STEM techniques require a thin sample that is transparent to the incident electron beam, typically <100nm. Samples with small enough grain sizes, which can occur for synthetically prepared samples such as nanoparticles, can be dropcast on TEM support grids without modification. However, nearly all samples of planetary materials require some degree of modification to make them thin enough for proper scattering and imaging in the TEM. There are different approaches to achieving electron transparency with different degrees of site specificity. Such approaches include: (i) crushing a bulk sample in a mortar and pestle followed by dropcasting of the material onto a TEM support grid; (ii) ultramicrotomy involving mechanical slicing of a particle embedded in a medium such as epoxy; and (iii) FIB milling of a site-specific sample. We expect that both ultramicrotomy and FIB milling will be used to prepare samples returned from Bennu. We note that for many STEM applications, particularly those requiring atomic-resolution imaging, thinner samples (<50 nm) are preferred and dramatically reduces multiple scattering events that affect imaging and analysis.

**Sample impact:**

Aside from the modification during sample preparation noted above, the major concern with STEM imaging and analysis is damage to the specimen from the energy and flux of the incident beam. Samples exposed to the high vacuum of the TEM can also lose weakly-bound species (mainly $H_2O$ absorbed on surfaces) – for example, the interlayer water in smectite group clays is lost in the TEM and the structure collapses to a 1 nm basal spacing (001). The incident beam can result in amorphization of beam-sensitive materials, radiolysis (loss of light elements), and changes in bonding and oxidation state. Users must be careful to understand the potential for beam damage in such cases.

**Data quality:**

- Image resolution depends on the imaging mode and is typically reported as point-to-point resolution as determined by experimental images
- Atomic-resolution imaging (pm scale) can be achieved on properly prepared samples in microscopes equipped with high field strength objective lenses or with spherical-aberration correctors

Atomically resolved structural information (pm scale) is also provided by electron-diffraction patterns. However, the interplanar spacing measurements that result from such patterns are not as precise as those obtained by X-ray diffractometers owing to the difficulty in determining the exact camera length for the electron-diffraction experiment due to variations in objective-lens focus in the TEM. However, the small wavelength of the electron beam used in the TEM, as compared to that of XRD, provides a much larger sampling of crystallographic (reciprocal) space at any given sample orientation. Diffraction data are typically calibrated using well-characterized standards (e.g., evaporated Al or sputtered Au).

**Data products:**

The data products expected from STEM analyses include:

- image data
- electron diffraction data

These products are delivered in various image formats (e.g., TIFF, JPEG).

If also equipped with EDS and EELS (discussed elsewhere in this section), the STEM can produce:

- 1D point analyses (individual spectra from a specific location; applicable to EDS or EELS)
- 2D line scans (individual spectra plotted as a function of distance; applicable to EDS or EELS)
- 2D Energy-filtered TEM (EFTEM) images in which each image is acquired at a specific electron energy loss (applicable only to EELS)
- 3D (hyperspectral) spectrum images (2D maps of a sample in which every pixel contains a spectrum in the third dimension, also known as an image cube; ; applicable to EDS or EELS)
- 3D (hyper spectral) image stacks in which images of the sampler acquired at a specific energy loss over a range of energies (applicable only to EELS)

These products can be used qualitatively to identify what elements are present and where they occur. They can be used quantitatively to determine percent composition at the local scale (atomic to nanometer) and for larger regions of interest (micrometer to centimeter scale). All products can be delivered as various image formats (e.g., TIFF, JPEG).

**Facilities:**

NASA Johnson Space Center

- The JEOL 2500SE 200 keV field emission scanning transmission electron microscope is equipped with a JEOL 60 mm$^2$ silicon drift detector for EDX analyses quantitative X-ray mapping and a Gatan Tridiem imaging filter (GIF) for EELS analyses using Gatan GMS software. Brightfield and high-angle annular darkfield images are recorded using a Gatan Digiscan camera. High resolution TEM images and diffraction patterns are collected using a 2k X 2k Gatan Ultrascan CCD camera. The Tridiem GIF is equipped with a Gatan Ultrascan 2k X 2k CCD camera for EELS applications and imaging. The EELS energy resolution measured at the zero-loss peak is 0.7 eV. The point-to-point resolution in STEM mode is 0.16 nm.

- JEOL NeoARM (installation in 2023) 60-200 kV aberration-corrected field emission, scanning and transmission electron microscope (STEM) equipped with spherical aberration correctors to enable a finely focused probe scanning across the specimen, allowing images to be obtained in several STEM imaging modes. The STEM is equipped with dual, large-area, energy-dispersive x-ray (EDX) detectors for rapid acquisition of spatially resolved elemental composition maps, a cold field emission electron source for high-resolution spectroscopic measurements, and a direct-detection system for electron energy-loss spectroscopy (EELS) studies. The NeoARM utilizes a cold field emission (CFEG) source with a minimum energy resolution of 0.3 eV as measured by full width half maximum of the zero-loss peak. The microscope is equipped with a 5th-order Cs aberration corrector for STEM imaging. The STEM is equipped with detectors and cameras to enable bright field, annular bright field, low or medium angle annular dark field, high angle annular dark field, and secondary electron/backscattered electron (SE/BSE) imaging. The STEM high-angle annular darkfield imaging resolution is 0.083 nm at 200 kV and 0.136 nm at 80 kV. The EDX system consists of two large area (158 mm$^2$) JEOL silicon drift detectors (SDD), positioned for efficient EDX tomography (offset 90° from each other), each with a high takeoff angle (> 30°) to minimize shadowing. An Oxford Aztec system is used for pulse processing. The microscope has a large solid angle of 2.2 steradians. The EELS system utilizes a Gatan Continuum K3 imaging filter system with direct detection that enables EELS and energy-filtered transmission electron microscopy (EFTEM) measurements at low doses. The system has dual-EELS capabilities and high-speed spectroscopy, with simultaneous EELS/EDS acquisition.

University of Arizona

- The Hitachi HF5000 scanning transmission electron microscope is equipped with a cold-field emission gun enabling a temporal (energy) resolution of ±0.3 eV. It has electron-optical alignments at 60 keV and 200 keV and is also equipped with a Hitachi 3$^{rd}$-order spherical-aberration corrector for the scanning TEM (STEM) probe, as well as STEM bright-field, high-angle annular-dark-field (HAADF), and annular-bright-field detectors. Spectroscopic capabilities include: (1) Oxford Instruments twin silicon-drift detectors (SDD) for EDS providing a total solid angle of 2.0 sr; and (2) Gatan Quantum Imaging Filter (GIF) for electron energy-loss spectroscopy (EELS). The GIF is equipped with a 2k × 2k CCD camera for acquisition of EELS spectra and energy-filtered images, dual EELS for the simultaneous acquisition of EELS spectra in multiple energy ranges, and dual EDS and EELS capabilities for the simultaneous acquisition of X-ray and electron energy-loss spectra. The Gatan Microscopy Suite (GMS) is the software interface with the EELS system including spectrometer calibration and operation and data acquisition. The HF5000 is capable of ≤0.4 eV energy resolution for EELS and 78 pm point-to-point resolution in STEM mode for atomic-resolution imaging.

- Molecular Foundry at Lawrence Berkeley National Laboratory
  The FEI TitanX (custom analytical Titan) FEG TEM is used for HRTEM, imaging, diffraction and STEM/EDX. It has a Bruker 4-element EDS detector having a solid angle of 0.7 sr. This allows maps of minor elements with high accuracy as spectra often have on the order of $10^6$ counts/spectrum even when extracted from maps. Long duration spectra have the ability to quantify elements down to ~100 ppm depending on the phase and element – e.g. we have measured Se in sulfide and Ti, Ni in olivine. The high tension is configurable from 40-300 keV allowing for dose optimization. The STEM resolution is still sub-nm at 80 keV allowing for quantitative analysis of very small phases without destroying them with the electron beam. With a Hummingbird tomographic holder we can achieve tilts up to +/- 70° to reach zone axes and acquire tomographic stacks. ALCHEMI can be used to obtain site occupancies of regular crystals > 200 nm in size.
- Tecnai F20 FEG with a Gatan imaging filter is capable of EELS and EFTEM imaging.
- A Zeiss Libra is available for high quality diffraction and low magnification imaging work. The Libra is a cutting-edge instrument with an in-column Ω energy filter and Kohler illumination system that enables a monochromated parallel beam (as opposed to convergent beam) diffraction patterns from regions < 100 nm in size. CBED, EELS and EFTEM are also available. The Libra sample holder also can tilt by ±70◦ in α and ±30◦ in β allowing greater freedom to locate zone axes than is typical in TEMs.

Dr. Rolf M. Schwiete Cosmochemistry Laboratory at Goethe University Frankfurt – Frankfurt, Germany

- The Talos X-FEG F200X G2 (S)TEM is equipped with the new ChemiSTEM Super-X EDS detector system and four large windowless SDD detectors (120 mm$^2$) with a solid angle of 0.9 sr, a Fiori number of more than 4000, and spurious peaks well below 1%. An output count rate of more than 800 kcps can be obtained. It will enable super-fast quantitative 2D as well 3D element distribution measurements down to trace-element contents of about 100 ppm for almost all elements excluding hydrogen. It is also equipped with a Gatan Continuum S/1077 EELS system for the acquisition of electron energy-loss spectra including 3d metals and oxidation states. The instrument is specially designed to allow for TEM-, STEM- and EDX-tomography. A low-dose mode combined with the use of a nitrogen-cooled holder substantially decreases even small beam-damage effects. Selected area electron diffraction and high-resolution imaging will add detailed structural information.

# 5.5 Focused-Ion-Beam Scanning-Electron Microscopy (FIB-SEM)

**Overview:**

Focused ion beam–scanning electron microscopy (FIB-SEM) integrates a scanning electron microscope with a scanning ion microscope. It is a highly versatile technique that can be used for materials imaging and analysis as well as to prepare site-specific samples for higher-scale investigation using transmission electron microscopy (discussed elsewhere in this section).

**How it works:**

The SEM part of the instrument is a typical field emission gun–equipped instrument capable of secondary or backscattered electron imaging and energy-dispersive X-ray (EDS) analysis (see SEM and EDS descriptions elsewhere in this section). The ion column generates a primary beam of $Ga^+$ ions from a liquid metal ion source to provide the nanometer-precision milling of a sample through ion sputtering of the target specimen. The sample surface geometry can be modified by the ion beam into a wide variety of patterns and shapes depending on the specific research needs.

**Sample preparation:**

The FIB-SEM can accommodate nearly any type of sample or sample geometry with little to no preparation. Bulk unpolished samples, powders (deposited on conductive adhesives so they are fixed to the sample mount), or polished sections can all be imaged or cross sectioned with the FIB-SEM. The only requirement is that they be conductive to dissipate charge buildup during imaging or milling. If the sample is not intrinsically conductive, then a thin conductive coating, e.g. C, can be deposited via sputter coating prior to imaging or lift out.

FIB sample preparation begins with the identification of a region of interest (ROI) in the sample. A protective capping layer (or "strap"), typically 1 um wide and ≥10 um long (in the plane of the sample surface), is deposited on top of the ROI using electron- or ion-assisted deposition of a specific material, with carbon or platinum being the most common. The capping layer thickness (out of the plane of the sample surface) is ≥1 um thick to protect the surface ROI from ion implantation damage. The ion beam is then used to cut a trench on either side of the ROI typically to a depth of 5 to 10 um and a length that is 1.5× to 2× the depth. Thus, a section that is 5 um deep requires cutting back into the sample 7.5 um. This back cut is typically done in a pattern that resembles a set of stairs (stairstep profile) because it enables undercutting of the section to detach it from the substrate while minimizing the amount of material that is consumed in the creation of the FIB section. The FIB section is attached (welded) to a micromanipulator needle by ion-assisted deposition of Pt, and the ion beam is used to cut the bottom and sides of the section to free it from the substrate. Once detached from the substrate, the thick section is transferred to a TEM half grid using the micromanipulator and attached to the grid with a Pt weld. The ion beam is then used to cut the weld holding the thick section to the micromanipulator tip. The thick section, i.e., electron opaque (>100 nm in thickness), is then ion milled with successively lower voltages and currents to polish it and attain a final thickness of ≤100 nm.

In the context of planetary materials, the FIB-SEM is frequently used to prepare site-specific electron-transparent cross sections for TEM analysis (Lee et al., 2003; Wirth et al., 2004; Zega et al. 2007; Graham et al., 2008). The energy and orientation of the ion beam are controlled to produce such "FIB sections" from a specific site within a sample. The FIB section is extracted using a micromanipulator and attached to a TEM grid via ion-assisted deposition for final thinning to electron transparency (≤100 nm thickness). The FIB can also be operated in such a manner that specific patterns can be etched in samples, allowing for other types of analyses to be performed (e.g. NanoSIMS; see elsewhere in this section and Nguyen et al. 2016).

**Sample impact:**

FIB sample preparation is destructive. Ion milling of the sample to create the electron-transparent cross section consumes a volume of material (~50 to 100 $mm^3$ depending on the width and depth of slice) that is lost to the vacuum of the FIB-SEM and cannot be retrieved. In addition, there are impacts to the sample with the major concerns related to ion implantation damage ("side-wall damage") that can modify beam-sensitive materials through partial amorphization or implantation of Ga. On porous samples, material sputtered from elsewhere on the sample can redeposit in holes and cavities (typically nanophase Pt metal, but also carbon if that was used for the protective strap). If minerals with different sputtering rates are in the same section, then differential sputtering effects can develop, which can result in surface topology of the section referred to as "curtaining". Such curtaining can be removed through additional passes varying the voltage and the milling direction. Despite such potential for damage, FIB has revolutionized site-specific sample preparation for TEM and other types of analyses as well as enabled measurements with multiple analytical techniques to be coordinated ("coordinated analysis").

**Data quality:**

The image resolution achieved using the electron column is the same as for field-emission scanning electron microscopes described elsewhere in this section. Secondary-electron images can also be acquired with the FIB, which can form a focused 2.5-nm probe, but such imaging is used sparingly during preparation of FIB sections because of the ion beam's ability to damage the sample surface.

**Data products:**

The data products expected from FIB-SEM analyses include:

- image data
- energy dispersive x-ray data
- electron backscatter diffraction data

These products (discussed elsewhere in this section) are delivered in various formats (e.g., .tiff, .jpg, .txt).

**Facility(ies):**

Molecular Foundry at Lawrence Berkeley National Laboratory

An FEI Helios G4 Focused Ion Beam (FIB) is available at the Molecular Foundry in Berkeley. This instrument combines a FEG SEM for imaging with a Ga focused ion beam for milling and material deposition (Pt and carbon). The Helios has a 2 nm ion beam spot, and can achieve ion beam energies as low as 500V, which enables very gentle ion milling with essentially no amorphous layer. In addition, there is a Strata 235 DualBeam FIB with an EDX and EBSD for in situ analysis or orientation of samples during preparation. A Fischione Nanomill is also available for post-processing to mill any residual amorphous surfaces to negligible levels with a <1 keV Ar ion beam. A Gatan SEM mill is available for polishing of large areas using Ar ions down to 100 eV and can produce incredible surfaces suitable for NanoIR, EBSD, and other surface sensitive methods.

NASA Johnson Space Center, Houston, TX, USA

FEI Quanta 3D 600 Dual-beam Focused Ion Beam: This is a combined field-emission SEM and ion beam instrument for high spatial resolution ion beam cross-sectioning of samples for TEM study. A Gallium ion source is used for ion beam sectioning of samples using beam voltages of 30 kV for major milling down to a few kV for low voltage cleaning. Gas injector guns are used to deposit Pt and C for protective coatings and sample attachment to a micro-manipulator that extracts sections for TEM study from 10–20 µm regions of rock and mineral samples. The FIB is equipped with an Omniprobe micro-manipulator and a 70 mm$^2$ Thermo-Noran silicon drift detector for energy-dispersive X-ray analyses.

University of Arizona, Tucson, AZ, USA

FEI Helios NanoLab 660 G$^3$: This instrument in the Kuiper Materials Imaging and Characterization Facility at the Lunar and Planetary Laboratory is equipped with an Elstar electron gun and monochromator, and is capable of electron beam resolution down to 0.6 nm from 15 kV to 2 kV. Its Tomahawk Ga$^+$ ion column can be operated between 65 nA and 500 V for, respectively, removal of large volumes of material and final sample polishing. Under standard operating conditions, an ion beam resolution 2.5 nm at 30 kV is achievable. The Helios is equipped with several secondary electron detectors as well as a retractable backscatter electron detector. The Helios is also equipped with *in situ* micromanipulation for creation and transfer of lamellae for TEM analysis. Both an EDAX EDS system and electron backscatter diffraction (EBSD) analysis system are integrated for compositional and crystallographic analysis in two and three dimensions. Multiple polygons are supported for device patterning as well as the ability to directly import customized shapes for patterning or deposition. The Helios is equipped with C and Pt gas-injection systems.

# 5.6 Electron backscatter diffraction (EBSD) and transmission Kikuchi diffraction (TKD)

**Overview:**

Electron backscatter diffraction (EBSD) and related transmission Kikuchi diffraction (TKD) are non-destructive mapping techniques for microstructural characterization of a sample within a scanning electron microscope (SEM). They involve automated collection, processing, and indexing of electron diffraction patterns from regions of interest on polished samples, yielding maps of spatially resolved crystalline mineral phases and their crystallographic orientations. The small interaction volume of these techniques means that microstructures can be resolved to high spatial (~<50 nm) and angular (~<1°) resolution. EBSD data can be acquired over large areas (e.g., entire petrographic slides), whereas TKD data are collected from foils. These techniques enable quantification of phase distribution, grain shape and size distributions, and crystallographic preferred orientations. EBSD and TKD can also discriminate among polymorphs and reveal deformation microstructures. EBSD is a fundamental tool for characterization and targeting of specific phases or microstructures by other high-spatial-resolution analytical techniques.

**How it works:**

In EBSD, an incident electron beam interacting with the lattice structure of a crystalline solid results in the diffraction of electrons. When the beam is oblique to the sample surface (optimally, 20°), some diffracted electrons are emitted from the sample as backscattered electrons that can be collected on a phosphor screen, producing patterns of bands (Kikuchi bands) that represent the lattice planes within the sample. EBSD patterns are unique to the phase and crystallographic orientation of the analyzed sample. Kikuchi bands detected in the patterns can be compared to theoretical solutions from crystal structure files that are housed in databases covering thousands of crystalline solids.

Phase and orientation maps are produced by indexing gridded points over scanned areas. Map data are routinely processed to define grains based on phase and orientation, which can be used for visualization and/or to derive statistical quantities (grain size and shape distributions). Crystallographic orientation data from any crystalline phase can be plotted as pole figures and/or on thematic maps to assess the strength and patterns of preferred orientations, or the crystallographic relationships between phases, of any crystallographic axes or planes. The difference between different crystal orientations, referred to as misorientation analysis, can reveal a range of deformation and/or growth microstructures, such as twins, crystal plasticity, and reaction textures. Because an EBSD pattern is linked to crystallinity of the activated volume, maps of EBSD pattern quality can depict damage (e.g., fractures) and/or structural state (e.g., amorphous vs. crystalline).

In high-resolution EBSD (HR-EBSD), cross correlation analysis is performed for all of the diffraction patterns collected. This is only possible for small (<150 × 150 mm) areas of materials, but allows high enough precision in angular resolution (<0.01°) to detect low-angle grain boundaries and analyze elastic and plastic strains within the mapped area.

The TKD approach involves the collection of diffraction patterns generated by the electron passing through a crystalline material. The material must be thin (<100 nm thickness), so this technique is often used for the analysis of thin foils manufactured for TEM or needle-shaped atom probe specimens.

EBSD and TKD maps can be used to visualize and quantitatively analyze data in a range of ways, including as thematic maps; stereographic projections (pole figures or inverse pole figures); misorientation and angle/axis distribution plots; grain size, shape, and orientation statistics; crystallographic preferred orientation strength and pattern statistics; and phase distribution statistics. They can also be used in modeling of phase transformations or as inputs for numerical simulations of deformation. Spatially resolved quantification of grains in polycrystalline microstructures via EBSD and/or TKD analysis has advantages over other techniques because grains can be defined on the basis of phase and crystallographic orientation, which can also be linked with chemistry for combined EDS–EBSD/TKD acquisition systems.

Outputs from EBSD and TKD analysis can also be used to quantitatively characterize microstructures and textures, facilitating interpretations of mechanisms and history of crystal growth, deformation, metamorphic reactions, and phase transformations. EBSD and TKD data can be used to distinguish between polymorphs, aid the discovery of new mineral phases, and qualitatively assess crystallinity.

**Sample preparation:**

EBSD and HR-EBSD analyses are sensitive to mechanical damage of polished surfaces; special preparation of the sample is thus required beyond standard thin section polishing. The surface of the sample must be polished flat, with the layer of mechanical damage from grinding and lapping removed by fine polishing with colloidal silica, alumina, or ion milling. This additional polishing step typically removes a few tens of nanometers of material from the sample surface. The surface of the sample must be conductive to avoid charge buildup on the surface. Typically, a thin (few nanometers) layer of carbon is applied using an evaporative carbon-coater.

The preparation of <100-nm thin foils for TKD analysis of geological specimens involves milling the sample surface over a region of ~10 mm × 10 mm × 10 mm. The foils are prepared in a focused ion beam scanning electron microscope (FIB-SEM).

**Sample impact:**

EBSD is generally a non-destructive technique. However, some samples can be sensitive to interrogating radiation (electron beam with 20 kV acceleration voltage), which may induce loss of more volatile elements, changes to functional chemistry, and breakdown of long-range crystalline order. Beam-sensitive materials are generally characterized by abundant light elements or volatile compounds (e.g., halogens, alkalis, water) or have quasi-stable or amorphous crystallinity. Users must be careful to understand the potential for beam damage in such cases. Some products for fine polishing in preparation for EBSD analysis (e.g., colloidal silica in NaOH) can preferentially corrode or dissolve some phases. Therefore, users must choose polish methods for EBSD analysis carefully.

TKD requires that thin samples are created by ion-beam milling. Thus, TKD sample preparation is destructive. However, the thin samples created for TKD can also be used for TEM analysis, which is generally non-destructive.

**Data quality:**

The three main aspects of EBSD and TKD data quality are spatial resolution, angular resolution, and indexing accuracy.

Spatial resolution of EBSD is determined by the type, quality, and operating conditions (energy and size) of the electron beam, and is in part dependent on the material analyzed (which controls the activation volume). Nevertheless, the activation volume for EBSD is much less than for, e.g., energy-dispersive X-ray spectrometry (EDS) for equivalent beam conditions. For typical field emission source electron beams, operational acceleration voltage and beam current, and most rock-forming minerals, achievable spatial resolution is ~50 nm. However, newer SEMs have optimized field emission beams that can yield increased resolution to ~15 to 20 nm. Users can define step size for mapping to suit the requirements of the scientific problem. Modern EBSD systems also allow users to collect large-area maps composed of a mosaic of scans, which can encompass the entire area of a petrographic slide or epoxy grain mount.

For TKD, diffraction patterns are sourced from a very small volume (a few nanometers) on the exiting surface as the electron beam is transmitted through the sample, which results in a substantial increase in spatial resolution over EBSD (~2 nm compared with ~20 to 50 nm).

Angular precision is measured as mean angular deviation (MAD) of the indexing solution to the EBSD or TKD pattern at a given point. MAD depends on indexing parameters (such as the number of detected Kikuchi bands), and is sensitive to the quality of the EBSD or TKD patterns. Pattern quality can be phase-specific and is influenced by sample surface polish quality and EBSD camera and pattern acquisition settings. Pattern quality for TKD is sensitive to the analyzed material, sample thickness, and electron-beam character (voltage, beam current conditions). Therefore, users must choose sample preparation methods and data acquisition settings carefully. Typical MAD values for routine EBSD and TKD analyses are ~0.3° to ~1°. For HR-EBSD analysis, angular precision is typically less than 0.01°.

Indexing accuracy is phase-specific and influenced by EBSD and TKD pattern quality and indexing parameters. Errors in indexing can be systematic or non-systematic, resulting in misindexing of phase and/or orientation. Isolated non-systematic misindexed points and non-indexed points can be eliminated or reduced during data acquisition and subsequent processing. It is also possible to correct for systematic phase misindexing and systematic orientation misindexing via data processing algorithms. However, mitigating misindexing issues is preferred to correcting them. Therefore, users must carefully choose sample preparation, data acquisition, and processing parameters for EBSD analysis of their materials.

**Data products:**

The data products expected from EBSD and TKD analysis include:

- 1D point analyses (phase and crystallographic orientation from a specific point on a sample).
- Maps of regions of interest, in which every point contains information about phase, crystallographic orientation, and EBSD and TKD pattern quality. These are the most common type of product.

For most EBSD and TKD acquisition systems, data can be processed and visualized in proprietary software (e.g., Oxford Instruments AZtec and AZtec Crystal), exported for processing and visualization in other proprietary software packages (e.g., .cpr and .crc files, which can be read by Oxford Instruments Channel 5 software), or exported as text files (e.g., .CTF), which can be read by third-party or open-source data processing, visualization, and modeling software (e.g., MTEX, ELLE, ARPGE). Maps can be prepared with perceptually uniform and color-vision-deficiency–friendly color scales.

Final output files are typically image files in .bmp, .png, .jpg, or .tif formats or as text files such as .csv. Raw EBSD data file sizes can range from a few megabytes to tens of gigabytes, depending on the file type, number of analysis points, and whether EBSD patterns are included. For HR-EBSD, diffraction patterns must be saved, and typical projects yield ~0.5 Tb of data.

**Facility(ies):**

Curtin University, Perth, Australia

The Microscopy and Microanalysis Facility at the John de Laeter Centre houses three SEMs that are fitted with Oxford Instruments EBSD acquisition systems, each with different capabilities. All have capabilities for large-area mapping and simultaneous collection of EBSD and EDS data. The EBSD facilities include:

- TESCAN CLARA field emission SEM with Oxford Instruments Symmetry/AZTec combined EBSD/EDS acquisition system with Ultim Max 170 analytical silicon drift detector (SDD). The latest-generation electron column in this instrument enables beam currents up to 200 nA, but with a highly focused beam. This allows EBSD mapping at rapid (up to 3000 patterns per second) indexing rates and high spatial resolution. At typical EBSD operating conditions, spatial resolution is < 20 nm. Low-voltage operation facilitates analysis of beam-sensitive materials. The high-resolution beam enables enhanced EBSD mapping of TEM foils and atom probe needles. This instrument is state-of-the-art and is best suited for analysis of mission return samples.

- TESCAN MIRA3 field emission SEM with Oxford Instruments Symmetry/AZTec combined EBSD/EDS acquisition system, XMax 150 SDD. This instrument is a workhorse and well suited for most EBSD applications, with the exception of mapping at very high spatial resolution (>50 nm).

- TESCAN LYRA dual beam field emission FIB-SEM with Oxford Instruments NordleysNano EBSD detector and X-max 80 SDD EDS detector and Tofwerks time-of-flight mass spectrometer. This instrument has capabilities to prepare TEM foils, atom probe needles, and 3D slice-and-view reconstructions. This instrument is capable of EBSD and TKD mapping of TEM foils and atom probe needles, but is best suited to 3D EBSD analysis via integrated FIB milling.

Curtin University has multiple-user site licenses for Oxford Instruments EBSD acquisition and processing software AZtec, Channel 5, and AZtecCrystal.

NASA Johnson Space Center, Houston, TX, USA

- The JEOL 7900LV FE-SEM field-emission scanning electron microscope is equipped with the Oxford Symmetry EBSD system and an Oxford 170 mm$^2$ active area SDD EDS spectrometer that will enable simultaneous EDS and EBSD mapping of samples. The 7900LV provides outstanding performance for both imaging and analytical conditions at very low accelerating voltages, operation in a low vacuum mode (150 Pa or lower) to enable imaging of non-conductive samples, and energy filtering combined with specimen bias applied voltages (from +500 to –2000 V) to enable high-resolution imaging of astromaterials and surfaces down to 10V (accelerating voltage) with image resolution better than 2 nm. The Oxford EBSD can collect high-resolution Kikuchi diffraction patterns at ultra-fast speeds (3000 patterns per second). This technique can be used for structural identification of mineral phases, quantification of intracrystalline plastic strain within individual grains, and determination of deformation mechanisms.

University of Arizona, Tucson, AZ, USA

- The Kuiper Materials Imaging and Characterization Facility at the Lunar and Planetary Laboratory houses a ThermoScientific Helios NanoLab 660 G$^3$ FIB-SEM equipped with an EDAX SDD EDS and EBSD system for simultaneous compositional and crystallographic analysis in two and three dimensions. The Helios is equipped with a Schottky field emitter, acceleration voltage range from 20 V to 30 keV, beam currents from 0.8 pA to 100 nA, and resolution down to 0.6 nm at 15kV. EBSD patterns can be acquired on polished sections in two dimensions or used in conjunction with slice-and-view imaging for three-dimensional data sets. The EDAX system on the Helios is currently running the EDAX APEX platform. The EDAX Hikari XP EBSD camera is capable of data collection rates up to 1000 indexed points per second at 5 nA. It can be operated down to 5 kV acceleration voltage and 100 pA beam current.

# 5.7 Electron microprobe analysis (EMPA)

**Overview:**

Electron microprobe analysis (EMPA) is a scanning X-ray microanalytical technique used for the chemical analysis of solids (minerals, crystals, glasses). EMPA utilizes wavelength-dispersive spectrometry (WDS) to measure characteristic wavelength of X-rays emitted from a sample that has been subjected to an electron beam. Emitted X-rays have wavelengths characteristics of individual chemical elements, allowing for unique elemental identification and quantification. Electron microprobes are generally equipped with multiple WDS spectrometers to allow simultaneous collection of X-rays for the analysis of a specific phase. The electron beam interacts with a very small volume (a few $um^3$ or less) of material but larger volumes (up to a few hundred $um^3$ if necessary) can be analyzed either through stage scanning or beam rastering.

**How it works:**

EMPA is a scanning X-ray microanalytical technique for routine analysis of major, minor, and trace elements, where minor elements are <1.0 wt% and trace elements can be measured to ~ 100 ppm. The electron microprobe is similar in electron-optical design to an SEM but differs in that it is equipped with WDS spectrometers.

To stimulate the emission of characteristic X-rays from a specimen, a high-energy beam of electrons is accelerated (≤30 keV) from a field-emission gun (FEG) or W or $LaB_6$ filament and focused onto the sample being studied. FEGs have sharper tips and produce higher current density than W or $LaB_6$. Thus, FEGs can produce smaller beam sizes and excitation volumes from which X-rays are emitted. EMPA is based on the same atomic ionization model as EDS. At rest, an atom within the sample contains ground state (or unexcited) electrons in quantized energy shells bound to the nucleus. The incident beam may excite an electron from an inner shell and promote it to a higher-energy orbital, leaving behind a shell with a hole. For the atom to return to the ground state, an electron from an outer, higher-energy shell will fill the hole and in so doing, release an X-ray whose wavelength is proportional to the energy difference between the shells. Thus, the X-rays are characteristic of the atoms in the sample and, if measured with a wavelength-dispersive X-ray spectrometer, can be used to determine chemical composition.

Each WDS is equipped with a crystal that can be rotated into specific orientations in order to diffract, according to Bragg's Law, X-rays of specific wavelength. The X-rays are diffracted into a proportional counter filled with an inert gas mixture (typically 90% Ar, 10% $CH_4$) that becomes ionized when struck by the X-ray. The resulting charge pulse is proportional to the energy of the X-ray, enabling element identification in one (point) or two (map) dimensions. Quantification is achieved by comparison of that pulse to a sample of known composition under standard experimental conditions.

Typical EMPA instruments are equipped with four to six spectrometers, thus allowing simultaneous collection of x-rays for up to six different elements at a time. Each spectrometer can be moved to multiple elements, enabling measurement of numerous elements at one sample location by counting over a several minute timeframe. This analytical approach allows for a complete major and minor element analysis of a single sample location.

**Sample preparation:**

EMPA analysis requires a flat sample and fine polish preparation, to allow conduction of current across the sample, after it is coated with carbon or another conductive material, e.g., Al, Au, Pt. Flatness is typically achieved through grinding and lapping, whereas a finely polished surface is achieved by a series of polishing steps from 15 to 6 to 3 to 1 to 0.25 micron (typically, but can also vary slightly from lab to lab). The surface of the sample must be conductive to avoid charge buildup that would otherwise interfere with imaging and analysis. Typically, a thin (few nm) layer of conductive material (C, Al, Au, or Pt) is applied using a sputter-coater. EMPA can be carried out on a standard 30 micron thickness thin section, on a thick section, or on a thick mount (several mm to ~ 1 cm thick).

**Sample impact:**

EMPA is generally a non-destructive technique, as the electron beam and X-ray emission do not typically cause structural damage to crystalline or amorphous materials. However, some samples can be sensitive to the electron beam interaction; beam-sensitive materials are generally characterized by abundant light elements or volatile compounds, e.g., halogens, alkalis, water, or have quasi-stable or amorphous crystallinity. Users must be careful to understand the potential for beam damage in such cases. Loss of more volatile elements (alkali elements and S, P) can usually be avoided by use of accelerating voltage <10 kV, sample current <10 nA, and count times <20 seconds. Trace elements can be measured to levels as low as 100 ppm and even lower in some cases, but this is typically done by using high current (~300 nA) and long counting times (minutes). Such conditions can lead to beam damage on the samples.

**Data quality:**

EMPA detection limits in favorable cases can extend to as low as ~100 parts per million, but as mentioned above this depends on sample current, count times, and each element. The precision of EMPA is typically ~1 % relative for Ti, Fe, Mn, Ca, K, and 2 to 5% for Si, Al, Mg, Na, and P.

EMPA data quality is also dependent upon having very well-characterized standards, especially for elements such as Si and Al, that are in high concentrations in minerals and glass, and for which the analytical precision is slightly lower than other elements. Widely used microprobe standards have been distributed by the National Institute of Science and Technology (NIST), which produced standard glasses, and the Smithsonian Institution, who have characterized mineral standards.

EMPA is used to obtain a high quality high precision chemical analysis of mineral or amorphous solid (glass). The quality is high enough to determine the composition of, e.g, the olivine solid solution as $Fa_{25}$ or $Fa_{27}$ which would allow distinction between an L6 and LL6 chondrite. In addition, EMPA analyses can determine stoichiometry of elements in various mineral formulae, thus allowing detailed mineralogic determination (e.g., labradorite versus bytownite for plagioclase feldspars).

Matrix corrections must be applied to account for differences in atomic number (Z), absorption of X-rays (A), and production of secondary x-rays, or x-ray fluorescence (F) (abbreviated as ZAF corrections). Some approaches use PAP corrections, where PAP refers to the authors Pouchou and Pichoir (1991) and are phi-rho-Z corrections that have the same goal as ZAF but using a slightly different approach. PAP is a general model for calculating X-ray intensities and can be used for a wide range of X-ray energies (100 eV to greater than 10 keV) and accelerating voltages (1-40 keV).

The spatial resolution of EMPA is controlled by the nature of the emitter and the electromagnetic lenses in the electron column. For W and LaB$_6$ filaments, the beam can vary from 1 um in size to either a defocused beam of ~10 um diameter, or a rastered beam with 150 um square edge. An electron beam measuring 1 um in diameter interacting with an in situ sample, such as a polished petrographic thin section, at 30 keV generates an excitation volume with a diameter of approximately 3 um. For field emission guns, it is possible to achieve sub-micrometer beam sizes and analyses volumes, particularly with decreased beam currents and the use of Lα lines for transition metals (e.g., Fe), rather than the more commonly used Kα lines.

The advantage of WDS over EDS is its spectral resolution. WDS offers a spectral resolution of ~5 eV compared to ~130 eV for EDS. The better spectral resolution of WDS generally results in better X-ray peak separation and peak-to-background ratios than EDS, making it more precise with lower detection limits.

**Data products:**

The data products expected from EMPA measurements include:

- Elemental spectra acquired from standard calibrations
- Quantitative chemical analyses of standards to verify calibrations
- Elemental spectra acquired from unknowns
- Quantitative chemical analyses from the unknown spectra
- Back scattered electron images of sample locations that allow a x-y-z location of each analytical point from the analysis session, or a line scan or grid of analyses on any particular sample
- X-ray maps of any particular element for which chemical mapping was done on an unknown

These products would all be used to identify quantitatively what elements are present in minerals and amorphous phases (glass) and their concentration to the precision discussed above. The images document the physical locations of the analyses on the sample and provide microtextural context. All products can be delivered as text (asc, .qnt, .cor, .cnd) or image files (tif, jpg, png, bmp).

**Facility(ies):**

NASA-JSC has two electron microprobes: a JEOL FE 8530F Hyperprobe and a Cameca SX100. Either instrument can be used and their characteristics are summarized here:

The JEOL JXA-8530F electron probe, installed in early 2014 at NASA-JSC, is equipped with a FEG, 5 WDS spectrometers, as well as a ThermoElectron SDD EDS system.

- Highly focused electron beam yields better spatial resolution (~3 nm at 35 kV compared to 100-200 nm for standard EMPA) for imaging.
- Two high-count rate WD spectrometers permit trace element analysis with better detection limits produced with shorter count times (as low as 10 ppm at 100 s or some elements).
- SDD-type EDS detector yields very high count rates, permitting x-ray mapping (not limited to five elements as with WDS mapping).
- Highly focused beam permits development of low-kilovolt X-ray data collection and improved spatial resolution for chemical analysis (<100 nm).
- Several light-element-analysis diffracting crystals allow direct analysis of C, N, and O in astromaterials.
- Oil-free vacuum pumps enhance carbon analysis in chondritic meteorites and interplanetary dust particles.

Cameca SX-100

- 5 wavelength dispersive spectrometers (WDS)
- 1 Energy Dispersive Spectrometer (EDS) by iXRF
- Tungsten filament
- Automated stage down to < 3 micrometer precision
- Analysis down to several 100 ppm weight for selected elements
- Analysis of Boron to Bismuth
- Large variety of natural and synthetic standards

Smithsonian Institution

The JEOL JXA-8530F+ electron probe, installed mid-2017, is equipped with a FEG, 5 WDS spectrometers, as well as an integrated SDD EDS system and panchromatic cathodolumiscence detector.

- Vastly improved imaging (highly focused electron beam yields better spatial resolution for imaging)
- Five large-area crystal WD spectrometers will permit trace element analysis with better detection limits produced with shorter count times
- SDD type EDS detector yields very high count rates permitting x-ray mapping (not limited to 5 elements as with WDS mapping)
- Highly focused beam permitting development of low kV x-ray data collection, and improved spatial resolution for chemical analysis
- Several light-element-analysis diffracting crystals that allow direct analysis of C, N, and O in astromaterials.
- Oil-free vacuum pumps that enhance carbon analysis in chondritic meteorites and IDPs (interplanetary dust particles)
- Probe for EPMA (PFE) control software and ProbeImage software. These new software packages permit enhanced control of quantitative chemical analysis and WDS x-ray mapping with full quantification of maps. PFE software permits fitting of curved backgrounds in the automated mode, as well as "mean atomic number" background corrections and measurements of x-ray intensities extrapolated to time-zero for elements whose signal is altered due to damage induced in beam-sensitive materials.

Stanford University

- The Stanford Microchemical Analysis Facility houses a JEOL JXA-8230 SuperProbe electron microprobe. The microprobe is equipped with the following:
- Five WDS spectrometers
- Two large format crystals that permit high sensitivity trace element concentrations
- Seven unique diffraction crystals allowing for a wide range of X-ray wavelengths to be measured. These include: PETJ, PETL, LIF, LIFL, LDE1, LDE2, TAP
- EDS spectrometer for rapid X-ray mapping and qualitative analyses
- xCLent V hyperspectral cathodoluminescence (CL) system
- Computers are equipped with both the standard JEOL electron probe software as well as Probe for Windows software

University of Arizona

The KMICF at the Lunar and Planetary Laboratory is equipped with two electron microprobes.

- The Cameca SX-50 (installed in 1991 and in continuous service for 29 years) is equipped with a 30 keV W thermal emission gun, four wavelength dispersive X-ray spectrometers (WDS), 12 diffracting crystals, and a Princeton-Gamma Tech Si(Li) EDS system, allowing analysis of elements with Z≥4.
- The Cameca SX-100 Ultra (installed in 2011) is equipped with a 30 keV $LaB_6$ filament, five WDS, 14 diffracting crystals, and an SDD-EDS system, allowing analysis of elements with Z≥5.

# 5.8 High-resolution cathodoluminescence (HR-CL)

**Technique name: <u>High-resolution cathodoluminescence (HR-CL)</u>**

**Overview:**

Cathodoluminescence (CL) is an optical and electromagnetic phenomenon in which electrons impacting a luminescent material cause the emission of photons, which can have wavelengths in the visible spectrum. Cathodoluminescence in a scanning electron microscope can be used to characterize the composition and optical and electronic properties of materials and then correlate them with morphology, microstructure, composition, and chemistry at the micro- and subnanoscale.

Advantages of CL include that it allows investigation of optical properties at a spatial resolution better than the diffraction limit of light; enables geochemical processes to be reconstructed by revealing trace element distributions; measures dislocation densities; fully describes a sample by correlating shape, size, crystallinity, or composition with optical properties; and is non-destructive with minimal sample preparation requirements.

**How it works:**

CL occurs when high-voltage electrons impact a sample surface and promote valence electrons to the conduction band. Such ionization leads to the emission of photons that can have wavelengths in the visible spectrum. Theory predicts that the intrinsic CL of minerals is enhanced by defects and structural imperfections in the lattice and/or by substitutional or interstitial elements, which distort the lattice. Such lattice perturbations are referred to as "CL activators."

Some elements called "CL quenchers," with $Fe^{2+}$ being the most common, can play the opposite role in modifying the electron energy-level arrangement so that the CL process does not operate or is diminished. As long as CL quenchers are low in concentration, high-resolution CL (HR-CL) can detect very subtle changes in the concentrations of CL activators, for example, few parts per million.

For HR-CL, the setup has to be mounted on a high-resolution (usually a field-emission gun (FEG) scanning electron microscope (SEM). The CL setup includes detectors suitable both for hyperspectral analyses (CCD) and for panchromatic and monochromatic imaging (high-sensitivity photomultiplier). Integrated luminescence is extracted from the sample by using a parabolic mirror placed above the sample. The electron beam is directed perpendicularly to the sample surface through a hole in the center of the mirror. The electron beam current ranges typically from hundreds of picoamps to several nanoamps, and the beam voltage is set between 1 and 20 kV according to the probing range targeted (between some tens of nanometers to microns beneath the surface).

**Sample preparation:**

This technique is designed to study the luminescence characteristics of polished thin sections of solids irradiated by an electron beam. As the technique is implemented within a SEM (discussed elsewhere in this section), samples are required to be conductive or must be covered by a thin conductive layer (carbon or gold). Such coating is routinely performed and does not induce any sample modifications or changes.

**Sample impact:**

Similar to EDS, the technique is usually non-destructive, especially for minerals. CL does not require high electron-beam current and is fully efficient for beam currents in the range of 1 nA. Such values (except for in quantum-sized systems) do not provide enough incoming electrons for saturating energetic levels involved in the luminescence mechanism. Thus, these conditions allow extended observation without any sample degradation.

**Data quality:**

For minerals, CL reveals activation centers whose concentration can be as low as one part per million. Spatial resolution is given by the diffusion length of electron hole pairs produced by the excited activation centers. For some semiconducting materials, the diffusion length can be large, e.g., hundreds of microns for diamond. The diffusion length for most minerals is in the sub-micron range, allowing a precise correlation between optical properties and specific sample morphology. Chemical zoning, due to the fluctuation of activation center concentration, can therefore be detailed with a lateral resolution of tens of nanometers if using a FEG-SEM. HR-CL imaging should be considered as an indispensable prescreening tool for secondary ion mass spectrometry measurements (and other in situ analyses).

**Data products:**

CL can be operated both for imaging and for spectroscopic analysis in the UV/VIS/IR range. For imaging, a direct map of luminescence can be acquired simultaneously with SEM imaging (either secondary or backscatter), allowing a straightforward comparison between morphology and luminescence (e.g., Fig. 1). The imaging can be performed either in panchromatic mode, namely by recording the complete wavelength emission from the zone investigated on the sample, or in monochromatic mode, by selecting a specific wavelength of interest. For spectroscopic analysis, CL can provide 1D information by integrating the luminescence signal from a scan area or point. Hyperspectral analysis (3D) can be also performed (Fig. 2).

Data output from Digital Micrograph 2.3 or higher (Gatan) is in the proprietary formats .dm3 and .dm4. Conversion to more universal formats (.jpg, .tiff, .txt) is possible using a free offline version of Digital Micrograph available on the Gatan website. Metadata are included with hyperspectral data sets (not covered by free offline version).

**Facility:**

<u>Observatoire de la Côte d'Azur, Laboratoire J.-L. Lagrange, Nice, and Centre de Recherches sur l'Hétéro-Epitaxie et ses Applications (CRHEA), Valbonne, France</u>

The JEOL JSM7000F FEG-SEM has accelerating voltages from 1 to 30 kV and electron beam currents from 30 pA to 30 nA. The JSM 7000F is equipped with a MONOCL4 CL system from Gatan, which has a high-sensitivity photomultiplier detector (200 to 900 nm) and UV-enhanced CCD (215 to 1100 nm). The MONOCL4 is also equipped with a monochromator with gratings of 150 and 2100 lines/mm, allowing spectral analysis either under wide wavelength range (low spectral resolution) or narrow wavelength range (high spectral resolution). The JSM7000F is also equipped with a cooling/heating stage for analysis under variable temperatures between 77 and 373 K.

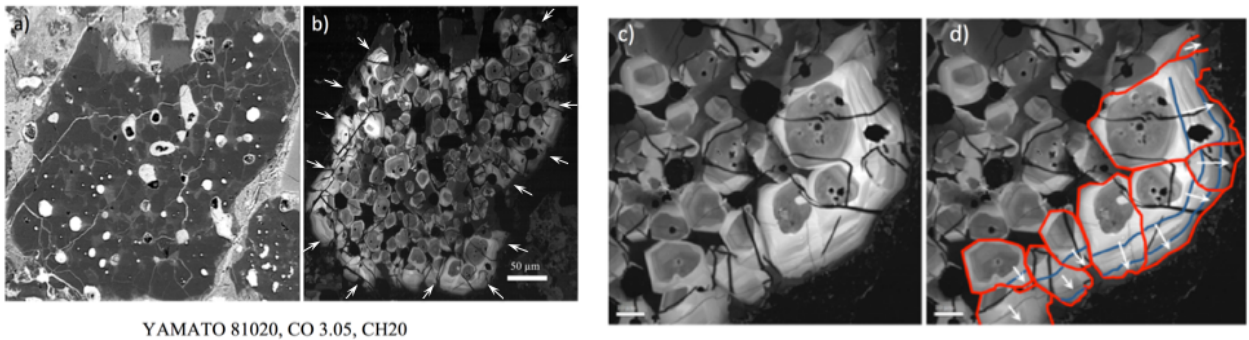

YAMATO 81020, CO 3.05, CH20

**Figure 1:** Examples of BSE and HR-CL panchromatic images. CH20 – Type I (FeO-poor) porphyritic olivine-pyroxene chondrule from Y-81020, showing an almost continuous multi-layer Mg-rich olivine shell at the chondrule margin. a) BSE image, b) CL panchromatic image reveals the occurrence of an almost continuous shell of olivine (white arrows) surrounding the chondrule at its outer edge. c) and d), image showing enlargement of the bottom right corner of the chondrule (Libourel and Portail, 2018). Asymmetric grain growth towards the free surface of the chondrule (white arrows) is particularly clear in this chondrule, leading to columnar-like textures, triple junctions and granoblastic textures (red outlines). From one crystal to another, sub-parallel epitaxial layers of Mg-rich olivine (blue outlines) being uninterrupted by grain boundary between two neighboring olivines delineate a more or less continuous olivine shell surrounding the whole chondrule. Metal blebs and melt inclusions are numerous in the CL revealed Mg-rich core of these olivines.

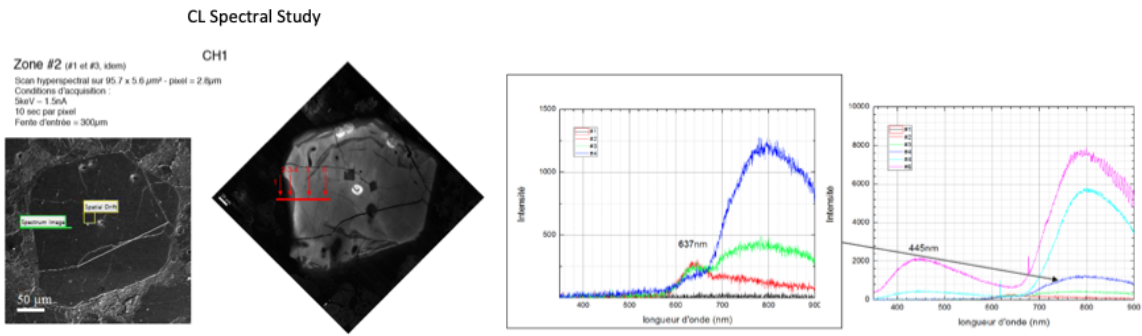

**Figure 2:** Example of a CL spectral analysis of chondrule Mg-rich olivines. Hyperspectral scan analyses and location of the corresponding spectra on a large, euhedral Mg-rich olivine from Yamato 81020. The two upper panels represent the BSE (left) and the FEG-SEM CL panchromatic (right) images of the euhedral Mg-rich olivine with the location of the hyperspectral scan analyses and location of the corresponding spectra (#1-#6). Going from the outside to the inside, spectra 2-3-4 are located in the low CL edge of the grain, while spectra 5 (light blue) and 6 (purple) are located in the higher CL intensity inner part of the grain. Notice that intensity is different and scaled to 10,000 on the bottom right panel. The FEG-SEM CL spectra collected on characteristic Mg-rich olivines contain a maximum of four broad emissions located at around 410, 640, 710 and 800 nm, and whose intensity varies according to the location of the analysis in the olivine. The red emission near 640 nm is located mainly in the edge zone and the far edge zone of the grains in contact with the glassy mesostasis, and clearly declines towards the olivine core. It is very often associated with the 710 nm emission in this edge region that occurs as a shoulder of a broader peak centered close to IR region at around 800 nm. The core region of grains is dominated by the positively correlated 410 and 800 nm emission bands, which increase in intensity from the edge to the core on average. Previous work on CL in forsterite and chondrule olivine has shown that i) the broad 410 nm blue luminescence corresponds to an intrinsic defect center related to lattice defect associated with $Al^{3+}$ substitution for $Si^{4+}$ ions in the Si-O tetrahedra, ii) the red emission band at around 650 nm is assigned to $Mn^{2+}$ impurity center in an octahedral coordination and preferred in the M2 site of forsterite, iii) The red-IR emission band at around 720 nm might be attributed to $Cr^{3+}$ in the M1 and/or M2 site, which acts as an activator and possibly associated with structural defect caused by interstitial Cr ions in IR region (around 800 nm), according to the result of CL spectroscopy for Cr-doped forsterite and meteorite samples (see more details in Libourel and Portail, 2018).

**References**

Libourel G. and Portail M. (2018) Chondrules as direct thermochemical sensors of solar protoplanetary disk gas. *Science Advances* 4, 1-12 DOI: 10.1126/sciadv.aar3321.

# 5.10 Energy-dispersive X-ray Spectroscopy (EDS)

**Overview:**

Energy-dispersive X-ray spectroscopy (EDS) is an analytical technique used for the chemical analysis of a sample. It requires the inelastic interaction of some source of interrogating radiation, such as electrons, ions, or X-rays, to produce X-ray excitation from a specimen. Emitted X-rays are characteristics of individual chemical elements, allowing for unique elemental identification. EDS spectrometers are generally coupled to electron, ion, or X-ray microscopes, as these instruments provide not only an interrogating radiation source but also high-resolution imaging—making a powerful combination for spatially resolved chemical analysis of heterogeneous materials.

**How it works:**

EDS is a suitable technique for routine analysis of elements that are heavier than or equal to beryllium in atomic weight. Minor elements, defined as 100 ppm to 1.0 %, can be analyzed but with lower precision (≥2% relative composition) compared with analyses of elements present in greater concentration. To stimulate the emission of characteristic X-rays from a specimen, a high-energy beam of particles or photons is accelerated and focused onto the sample being studied. At rest, an atom within the sample contains ground state (or unexcited) electrons in quantized energy shells bound to the nucleus. The incident beam may excite an electron from an inner shell and promote it to a higher-energy orbital, leaving behind a shell with a hole. For the atom to return to the ground state, an electron from an outer, higher-energy shell will fill the hole and in so doing, release an X-ray whose energy is proportional to the energy difference between the shells. Thus, the X-rays are characteristic of the atoms in the sample and, if measured with an energy-dispersive X-ray spectrometer, can be used to determine chemical composition.

**Sample preparation:**

EDS analysis may require special preparation of the sample. Bulk unpolished samples can be analyzed in an environmental scanning electron microscope. In comparison, in scanning electron microscopes and ion microprobes, the surface of the sample must be flat, relative to the size of the interrogating beam. This flatness is typically achieved through grinding, lapping, and mechanical polishing. The surface of the sample must be conductive to avoid charge buildup on the surface. Typically, a thin (few nm) layer of carbon or gold is applied using a sputter-coater. For transmission electron microscopy, samples must be electron-transparent (≤ 100 nm in projected thickness). This preparation is achieved using a variety of techniques such as ion milling, electropolishing, ultramicrotomy, or simple crushing and grinding.

**Sample impact:**

EDS is generally a non-destructive technique. However, some samples can be sensitive to interrogating radiation including loss of more volatile elements, changes to functional chemistry, and breakdown of long-range crystalline order. Beam-sensitive materials are generally characterized by abundant light elements or volatile compounds, e.g., halogens, alkalis, water, or have quasi-stable or amorphous crystallinity. Users must be careful to understand the potential for beam damage in such cases.

**Data quality:**

EDS detection limits in favorable cases can extend to hundreds of parts per million. The precision of EDS is 1 to 2% relative composition for major and minor elements.

EDS is often used qualitatively to determine what elements occur in the sample and where they occur. The mineral phase can be nominally inferred from such a first-look assessment. In comparison, quantitative analysis can be performed with or without standards. Standard-based analyses compares spectra from the sample to those from well-defined chemical and mineralogical standards, acquired under the same operating conditions. In all cases, matrix (ZAF) corrections must be applied to account for differences in atomic number (Z), absorption of X-rays (A), and production of secondary x-rays, or x-ray fluorescence (F). In thin electron-transparent samples used in TEM, A and F can be ignored but Z is still accounted for through the determination correction (k) factors specific to the instrument and detector being used.

The spatial resolution of EDS is controlled by the energy and size of the interrogating radiation and the nature of the sample. For example, an electron beam measuring 1 um in diameter interacting with an in situ sample, such as a polished petrographic thin section, at 30 keV generates an excitation volume with a diameter of approximately 3 um. In comparison, the ultimate spatial resolution of EDS is the atomic scale when coupled to a transmission electron microscope (TEM). The TEM accelerates an electron beam with a diameter of an atomic length scale (50 to 100 pm) at ≤ 300 keV that interacts with an electron-transparent sample (≤ 100 nm in projected thickness). The spectral resolution of EDS is approximately ~135 ±10 eV (depending on the system), and so is poorer than other types of spectroscopies used in electron microscopes, e.g., WDS (~5 eV), and EELS (down to 0.01 eV in monochromated TEMs), both of which are discussed elsewhere in this section. The advantage of EDS over WDS and EELS in the electron microscope is that it can analyze a sample for most elements in the periodic table simultaneously.

**Data products:**

The data products expected from EDS measurements include:

- 0D point analyses (individual spectra from a specific location)
- 1D line scans (individual spectra plotted as a function of distance)
- 2D elemental maps (images of X-ray intensity at a specific energy or range of energies)
- 3D (hyperspectral) spectrum images (2D maps of a sample in which every pixel contains a spectrum in the third dimension, also known as an image cube)

These products can be used qualitatively to identify what elements are present and where they occur. They can be used quantitatively to determine percent composition at the local scale (atomic to nanometer) and for larger regions of interest (micrometer to centimeter scale). All products can be delivered as text files.

**Facilites:**

NASA Johnson Space Center, Houston, TX, USA

NASA JSC has EDS systems incorporated into its electron microprobe, scanning electron microscope, and transmission electron microscope. These include:

- The JEOL JXA-8530F electron microprobe, installed in early 2014 at NASA-JSC, is equipped with a ThermoElectron silicon-drift detector (SDD) EDS system
- The Cameca SX-100 electron microprobe is equipped with an iXRF Lithium-drifted silicon or Si(Li) EDS system
- The JEOL 7600F scanning-electron microscope is equipped with an Oxford Instruments SDD EDS system
- The JEOL 2500SE S/TEM is equipped with a JEOL SDD EDS system

Smithsonian Institution, Washington, D.C., USA

The Smithsonian Institution has EDS systems incorporated into its electron microprobe and scanning electron microscope. These include:

- The JEOL JXA-8350 electron microprobe is equipped with a JEOL SDD EDS system
- The FEI Nova NanoSEM 600 is equipped with a ThermoFisher SDD EDS system

University of Arizona, Tucson, AZ, USA

Each electron microscope in the Kuiper Materials Imaging and Characterization Facility (https://kmicf.lpl.arizona.edu) at the Lunar and Planetary Laboratory is equipped with an EDS system. These include:

- The Hitachi S-4800 cold-field emission scanning electron microscope (SEM) is equipped with with an Oxford Aztec EDS SDD system for compositional analysis of elements with $Z \geq 5$ in one and two dimensions.
- The Hitachi S-3400 tungsten thermal emitter SEM with a variable-pressure chamber is equipped with a Thermo-Noran silicon drift detector (SDD) EDS system operating NSS software for compositional analysis of elements with $Z \geq 5$ in one and two dimensions.
- The ThermoScientific Helios NanoLab 660 $G^3$ focused-ion-beam scanning electron microscope (FIB-SEM) is equipped with an EDAX SDD EDS system and electron backscatter diffraction analysis system for compositional and crystallographic analysis in two and three dimensions for compositional analysis of elements with $Z \geq 5$ in one and two dimensions.
- The Hitachi HF5000 scanning transmission electron microscope is equipped with an Oxford Instruments X-Max N 100 TLE EDS system with dual 100 $mm^2$ windowless silicon-drift detectors, providing a total solid angle of 2.0 sr for rapid compositional mapping of TEM samples of elements with $Z \geq 5$ in one and two dimensions.
- The Cameca SX-50 electron microprobe is equipped with a Princeton-Gamma Tech Si(Li) EDS system, allowing analysis of elements with $Z \geq 4$.
- The Cameca SX-100 electron microprobe is equipped with a Bruker SDD EDS system, allowing analysis of elements with $Z \geq 5$.

# 5.11 Electron Energy-Loss Spectroscopy (EELS)

**Overview:**

Electron energy-loss spectroscopy (EELS) is a technique for routine analysis of elements heavier than hydrogen. It is performed in an electron microscope and used for chemical analysis and chemical speciation studies of bonding environment, oxidation state, and light-element detection, among others. The EELS signal in the electron microscope probes primary interaction events between the incident beam and the sample electrons. EELS can be performed in spot analysis mode or mapping mode (spectrum imaging), either by itself or in combination with other analytical techniques such as energy-dispersive X-ray (EDS) spectroscopy.

**How it works:**

The EELS spectrometer measures the energy losses experienced by the incident electron that result from inelastic scattering during transmission through the analytical volume. Incident electrons can ionize those electrons in the sample from an inner shell to a higher-energy orbital, in turn losing a characteristic amount of energy (core-loss spectroscopy) depending on elemental composition. Thus, EELS can be used for quantitative chemical analysis, among other applications.

A variety of information about the analytical volume is provided by the EELS signal. The spectrum is generally divided up into energy regions. These regions include:

**Zero loss peak (ZLP):** contains elastically forward-scattered electrons including those that have not lost energy in transmission as well as electrons that have experienced small energy loss. The ZLP is used to measure the energy resolution of the microscope at full-width and half of the maximum height, but otherwise does not contain useful information.

**Low-loss region:** Contains the first 50 eV of energy loss electrons that have interacted with the weakly bound outer-shell electrons of the atoms in the sample, i.e., those in the valence and conduction bands, and so the low-loss region contains information on electrical properties including the dielectric function.

**High-loss region:** Occurs at energies above 50 eV and contains ionization edges and core-loss edges that result from the ionization of a core electron, i.e., those close to the nucleus in in K, L, and M shells. Core-loss spectroscopy provides both qualitative and quantitative information on elemental composition. Core-loss edges are divided into the electron energy-loss near-edge structure (ELNES) and the extended energy-loss fine structure (EXELFS). The ELNES is a histogram of unoccupied electronic states, i.e., the density of states, and contains chemical information on the local nearest-neighbor bonding environment including oxidation state. In comparison, the EXELFS contains radial distribution information on nearest atomic neighbors.

**Sample preparation:**

EELS requires a thin sample that is transparent to the incident electron beam, typically <100 nm; for some applications, samples <50 nm are preferred and result in higher-quality data. Samples prepared using focused ion beam and ultramicrotomy techniques, as well as powdered samples, can also be suitable for EELS studies.

**Sample impact:**

As with EDS analyses and electron microscope imaging, the main concern with EELS analysis is damage to the specimen from the energy and flux of the incident beam. The incident beam can result in amorphization of beam-sensitive materials, radiolysis (loss of light elements), and changes in bonding and oxidation state. Users must be careful to understand the potential for beam damage in such cases. For beam-sensitive samples, broad-beam (>10 nm) techniques such as energy-filtered imaging or spot-mode spectroscopy may be used.

**Data quality:**

EELS detection limits are variable. The spatial resolution of EELS can extend down to the atomic scale (the ultimate detection limit) but detection of individual atoms depends on the analytical conditions, the electronic structure of the element in question, and the quality of the sample preparation (thickness and damage layer). For some elements with strong core-loss features (e.g., Ca), detection limits are <100 ppm.

Quantitative chemical analysis is possible with EELS but is not as precise as EDS analysis and requires detailed knowledge of the absorption cross sections for the analyzed elements, the convergence angle of the incident beam, and the collection semi-angle into the entrance to the EELS spectrometer. Unlike EDS, absorption and fluorescence effects do not effect EELS analyses.

For ELNES studies, well-characterized standards in terms of structure, bonding, composition, oxidation state, etc., are required to interpret the near-edge structure (fingerprint). A critical variable in ELNES studies is the energy spread of the incident probe. For most field-emission instruments, the energy spread (as measured by the full width at half maximum of the zero-loss peak) is < 1 eV. Cold field-emission gun sources can achieve energy resolutions of ~0.3 eV. TEM instruments utilizing monochromators can achieve energy resolutions <0.1 eV.

**Data products:**

The data products expected from EELS measurements include:

- 1D point analyses (individual spectra from a specific location)
- 2D line scans (individual spectra plotted as a function of distance)
- 3D (hyperspectral) spectrum images (2D maps of a sample in which every pixel contains a spectrum in the third dimension, also known as an image cube)

These products can be used qualitatively to identify what elements are present and where they occur. They can be used quantitatively to determine percent composition at the local scale (atomic to nanometer) and larger regions of interest (micrometer to centimeter scale). All products can be delivered as text files or in image formats, e.g., *.TXT, *.TIF.

**Facility(ies):**

NASA Johnson Space Center, Houston, TX, USA

- The JEOL 2500SE field emission scanning transmission electron microscope is equipped with a JEOL 60 mm$^2$ silicon drift detector for EDS analyses and a Gatan Tridiem imaging filter (GIF) for EELS analyses using Gatan Microscopy Suite The Tridiem GIF is equipped with a Gatan Ultrascan 2k × 2k CCD camera for EELS applications and imaging.

University of Arizona, Tucson, AZ, USA

- In the Kuiper Materials Imaging and Characterization Facility at the Lunar and Planetary Laboratory, the Hitachi HF5000 scanning transmission electron microscope is equipped with a Gatan GIF Quantum ER (Model 965) EELS spectrometer. The Quantum has 5.0- and 2.5-mm entrance apertures for, respectively, collection of spectra with high signal-to-noise ratio (SNR) but lower than instrument-specified energy resolution, and spectra with low SNR but the highest instrument-specified energy resolution. It is also equipped with a 2k × 2k CCD camera for acquisition of EELS spectra and energy-filtered images, dual EELS for the simultaneous acquisition of EELS spectra in multiple energy ranges, and dual EDS and EELS capabilities for the simultaneous acquisition of X-ray and electron energy-loss spectra. The Gatan Microscopy Suite is the software interface with the EELS system for spectrometer calibration and operation and data acquisition.

# 5.12 X-ray Absorption Near-Edge Structure (XANES) Spectroscopy

**Overview:**

X-ray absorption near edge structure (XANES) spectroscopy uses monochromatized soft X-rays to provide information regarding the distribution of bonding environments for all elements in period 2 and greater. The XANES spectrum arises from absorption of monochromatized light that, when tuned to specific energies, has the capability of exciting core-level electrons (those "closest" to the nucleus) to higher-energy electronic states. XANES spectra for C, N, and O organics are well calibrated to known standards, making it possible to acquire quantitative information about bonding environments at spatial resolutions down to 30 nm and detection limits approaching 1 ppm.

**How it works:**

The basis for XANES is molecular orbital (MO) theory. In the simplest case, atoms with a single bond, e.g., C-C or C-O, each provide one electron to the bond so that it is formed by two paired electrons. In MO theory, for every electron-occupied single bond, there is an unoccupied anti-bond at an energy higher than the single bond. For elements in period 2 (Li to F, where Ne is excluded only because it does not form bonds), which are relevant to the measurement of CHON compounds, only the 1s electrons are core level. For heavier elements, core electrons occupy higher-level orbitals, e.g., for 3d metals, the 2p electrons can be considered core-level electrons. For light elements of importance to organic molecular structure such as C, N, and O, the electrons involved in bonding occur in 2s and 2p atomic orbitals.

The energy required to excite electrons of any element is referred to as the critical ionization energy ($E_c$). When 1s electrons of any element are excited above $E_c$, they produce a characteristic peak in an energy absorption spectrum referred to as the K-edge. The photon energies required for exciting a 1s (core) electron into the various unoccupied anti-bonding orbitals of C, N, and O are slightly less than their respective 1s ionization energies (K-edges). Thus, by sweeping the energy of an incident X-ray beam from below the energy of the K-edge to energies higher than $E_c$, the 1s (core electrons) can be temporarily excited (on the order of femtoseconds) into anti-bonds when the photon energy equals the energy of the anti-bond. Depending on the range of anti-bond types in a given organic molecular structure, the energy absorption spectrum that emerges is therefore diagnostic of the types of anti-bonds, which in turn represents the actual range of the types of bonds.

The spectral features that lie at energies below the ionization (or K-) edge are referred to as near- edge structure" spectral features. The energies of the anti-bond states are strongly affected by the electronegativity of the bonding neighbor. For example, XANES features for C shift to higher energies in order of C, N, and O bonding partners. Similarly, the near-edge structure shifts based on the functional chemistry. For example, for carbonyls, the ketones (1O) XANES features are at a lower energy than those of the carboxyls (2O), and carboxyls (2O) XANES features are at a lower energy that those of carbonate (3O).

**Sample preparation:**

The absorption coefficients for soft X-rays is very large, so samples for XANES spectroscopy must be on the order of 100 to 200 nm thick. Samples must therefore be amenable to modification of their geometries. Solid organics are typically prepared using either a diamond knife and an ultramicrotome or a focused ion beam. Preparation using ultramicrotomy involves embedding the sample in some kind of medium. For planetary materials, thermally activated epoxies and sulfur have been used.

**Sample impact:**

XANES with soft X-rays generally does not damage the sample. Ideally, XANES spectroscopy with scanning transmission X-ray microscopy (STXM) would precede analysis with transmission electron microscopy and be followed by analysis with nanoscale secondary ion mass spectrometry.

**Data quality:**

C, N, and O XANES spectroscopy of extraterrestrial organic solids is exclusively performed using STXM that affords spatial resolution down to 30 nm. XANES spectroscopy detection limits can approach 1 ppm depending on several variables, including, e.g., the absorption edge, sample thickness, and signal.

**Data products:**

The data products expected from XANES spectroscopy using STXM are:

- 1D point analyses (individual spectra from a specific location)
- 2D line scans (individual spectra plotted as a function of distance)
- 3D (hyperspectral) spectrum image stacks

The 3D image stacks are the typical data product for STXM. Such stacks are composed of multiple images each acquired at specific energy intervals over some range of energy. For example, an image stack of the C, K edge might be acquired over an energy range of 50 to 60 eV between the energies of 270 and 320 eV at intervals ranging from 0.1 eV to 1.0 eV, the lattermost depending on optimizing the acquisition time and resolving the near-edge structure. Image stacks yield a data set of $x$, $y$, and electron-volt information for each pixel, i.e., an image cube. The number of pixels typically ranges from 100 to 2000. All products can be delivered in the as-acquired (.xim, .hdr) or more universal formats (.txt, .tiff, .jpg).

**Facilities:**

Currently, the only sources for the monochromatized soft x-rays needed for XANES spectroscopy are X-ray synchrotron facilities. STXM capabilities are available at the Advanced Light Source at Lawrence Berkeley Laboratory (California), the Canadian Light Source (Saskatchewan, Canada), the Diamond Light Source (UK), SOLEIL (Italy), and the Swiss Light Source (Switzerland). Another STXM facility recently opened in Japan.

# 5.13 micro-X-ray computed tomography (XCT)

**Overview:**

X-ray computed tomography (XCT) is a 3D imaging technique that can be used to image meteorites, Apollo samples, and other planetary materials of all scales from interplanetary dust particles (25-100 microns) to hand specimen sized rocks (10-15 cm). With this technique, X-ray attenuation in a sample volume (dependent upon composition and density) allows interior components and textures to be observed and quantified. This technique can give details about the 3D petrography, petrofabric, porosity, and chemical composition of a sample, and is non-destructive (see exceptions below). The material below draws heavily on the review paper by Hanna and Ketcham (2017).

**How it works:**

An XCT dataset represents the attenuation of X-rays at each point within an object. The general attenuation of X-rays in a material is governed by Lambert-Beer's Law $I = I_0 \exp(-ux)$ for homogeneous material, or $I = I_0 \exp S(-u_i x_i)$ for heterogeneous material with more than one mineral or phase, where $I$ is the recorded X-ray intensity, $I_0$ is the initial X-ray intensity, m or $u_i$ is the linear attenuation coefficient of the material, and $x$ or $x_i$ is the path length of the X-ray through the material.

X-ray energies commonly used for tomography of planetary materials are $<\sim 450$ keV. The attenuation of X-rays through materials is the result of three processes whose relative influence varies depending on the incident energy and material properties: photoelectric absorption, incoherent (Compton) scattering, and coherent (Rayleigh) scattering.

- *Photoelectric absorption*: an incoming X-ray photon frees an inner-shell electron which leaves the atom ionized. The original X-ray photon is absorbed, and the probability of photon absorption (and associated ionization) depends on the photon energy and the atomic number (Z) (roughly to the fourth power $\sim Z^4$)
- *Compton scattering*: the incident X-ray photon frees an outer-shell electron (ionization), retains some of its energy, and thus is scattered as a lower energy photon (i.e., inelastic scattering), with the probability of this effect dependent on electron density of the material. Compton scattering is less dependent on the material composition than the photoelectric effect.
- *Coherent scattering*: an incoming X-ray photon causes an electron to vibrate at the same frequency as the photon, and thus emit an X-ray photon of the same energy (elastic), so that the incident X-ray is scattered but no energy loss occurs and the affected atom is unchanged (not ionized). The intensity of this effect is roughly proportional to $Z^2$.

Quantification of these three effects is possible and allows this technique to use attenuation to provide the detailed information about petrography, petrofabric, and composition.

System components: There are four primary components common among most XCT cone-beam systems: 1) an X-ray source that generates X-rays (typically polychromatic); 2) a rotating stage upon which the sample is mounted; 3) a detector that records the X-ray signal intensity after passing through the sample; and 4) a computer that drives the system and records the collected data; #1 and #3 are the most critical components of a system.

Spatial resolution: The spatial resolution is a function of several factors, including the imaging geometry, source, detector, and magnification optics. Generally, spatial resolution is a function of the width of the beam as it passes from the focal spot through the object to a detector element. Spatial resolution can vary from tens of microns at low resolution to sub-micron resolution in a smaller volume high resolution scan.

X-ray energy: The energy used depends on which materials are of interest and the research goals, but most planetary materials are typically imaged at energies <450 keV. Lower-energy X-rays can resolve contrast between most minerals, but they are also less penetrating, and can cause data contamination by noise or artifacts. A compromise is to image at the highest energy that will still allow sufficient attenuation contrast between the phases of interest. In some cases, imaging the sample at two different energies (dual energy or absorption edge imaging) may be useful. Because the X-ray attenuation of a material is a well-defined function of its density, atomic number Z, and the X-ray energy, imaging the same material at dual energies allows an estimation of its density and effective (average) Z.

Calibration: A series of careful calibrations must be performed prior to, during, and/or after XCT imaging to ensure data quality. A dark-field calibration (or 'offset') is a projection image taken while the X-ray beam is off. This helps to correct for detector bias present while the X-rays are off. The bright field calibration (or 'gain') is taken with the X-rays on and the sample outside of the FOV, and is essential to correct for the differential response between detector elements, falloff (due to spherical beam), and X-ray source intensity drift.

**Sample preparation:**

XCT requires little to no sample preparation, and any that is done involves how the sample is encapsulated and/or held in position during imaging. For example, precious astromaterials are usually kept in packaging that will protect them from contamination or being compromised in an instrumental environment. This can easily be done with packaging that offers no contamination threat, and that is transparent to X-rays. Such materials can be Teflon, or other polymers or compounds that will allow X-ray transmission. Otherwise, the ideal rock sample geometry is a cylindrical shape, where the X-rays are passing through a consistent amount of material throughout the rotation. However, samples of all sizes and shapes can be imaged, but those with equidimensional or rounded edges offer the least amount of potential for aberration or imaging artifacts compare to angular shapes or sharp edges.

**Sample impact:**

XCT is frequently described as non-destructive, and this is largely true, but because the X-ray energy is interacting with material at the atomic level, ionization occurs and can cause heating. Some effort has been expended studying the effects of XCT on properties of meteorites and other planetary materials. Studies have shown that in general the power deposited into 1 cm$^2$ sample cross section during a synchrotron imaging experiment at 10 keV would be $\sim 4.2 \times 10^{-4}$ W, which is many of orders of magnitude lower than the absorbed power from an electron microprobe (Ebel et al., 2009). Studies of basalt magnetization revealed that XCT scanning imparts changes of only ~1% on the remnant magnetization measured in samples (Hanna and Ketcham, 2017). Amino acid studies have shown that neither abundances nor enantiomeric ratios were affected when XCT measurements are made at high energies (> ~25 keV) that are commonly used with laboratory-based

XCT instruments (Friedrich et al., 2019). Damage to organics can occur when using low X-ray energies (< 25 keV) or focused X-ray geometries that are common at synchrotron facilities (Hanna and Ketcham, 2017). Thermoluminescence measurements, however, have been shown to be sensitive to XCT measurements. The radiation dose experienced by a sample during an XCT analysis, is of similar magnitude to natural levels that can be recorded in a meteorite due to cosmic ray exposure and internal radioactivity (Sears et al., 2018). Thus, samples that might be used to measure cosmic ray exposure ages should avoid XCT and not be used in combination with XCT. In general, the XCT user should consider carefully the conditions being used to characterize any given sample and what other properties of that sample might be measured in the future.

**Data quality:**

XCT scanning is capable of imaging a wide variety of sample types ranging in size from ~0.5 mm to 20 cm. The spatial resolution of an individual scan depends on the size of the sample; smaller samples are imaged at higher resolution. The best achievable resolution can be determined by dividing the sample width by the detector width (which is 2000 pixels for the Nikon XTH 320, see below). For example, the maximum resolution for a 4 cm wide sample would be 20 μm/voxel. The optimal sample shape for XCT scanning is a cylinder whose height is the same as its diameter. However, the XCT scanning systems allow tall samples to be scanned in smaller sections then stacked so that the width of the sample can fill the field-of-view and still achieve maximum resolution. XCT data resolution is reported in terms of voxel size, but the effective resolution (i.e. the ability to resolve phases) is usually 2 to 3 times the voxels size.

The ability to differentiate materials depends not only on the effective resolution, but also on their respective linear attenuation coefficients, as well as other machine parameters. X-ray attenuation is a function of a material's density and atomic number (Z), as well as X-ray energy. Materials with very divergent densities and/or compositions are easy to differentiate. For geological purposes, most broad groups of silicate phases (i.e. olivine, pyroxenes, feldspars, sheet silicates) are usually distinguishable from one another and are easily distinguishable from more dense groups like oxides, sulfides, and native metals. Beyond material properties, machine parameters also affect the ability to differentiate materials. Larger samples require higher energy X-rays for adequate penetration. Unfortunately, the X-ray attenuation of most silicates begins to converge at very high (>200 keV) energies. For larger samples, where we need a high signal-to-noise ratio but want to avoid scanning at > 200 keV, a higher X-ray flux can be created by increasing the current in the X-ray tube. However, this leads to a larger X-ray focal spot size, thereby reducing the effective resolution. Several machine parameters can be tuned to optimize the signal-to-noise ratio while still being able to effectively differentiate materials.

**Data products:**

The data products expected from XCT imaging include:

- 2D X-ray radiographs (16-bit grayscale TIFFs taken at small rotational increments)
    - Up to 24 GB/scan
- 3D volume mapping X-ray attenuation
    - When stacked together, the 2D slices (TIFF) make up a 3D volume composed of a grid of cubic voxels (a pixel with a third dimension). Each voxel is assigned a CT number reflecting the average X-ray attenuation of the material constituting that voxel. Higher CT numbers are visualized with brighter gray values on the 16-bit grayscale slices. Although stacks of 2D TIFF files are typically used, other uncompressed image formats (e.g., BMP, DCM) or a 3D-format TIFF file can also be used. Convenient summaries of the 3D TIFF stacks are also exported as MP4 videos.
    - Up to 13 GB/scan

The radiographs can be used as an easy way to identify large variations in grain sizes and distributions, as well as other structural features (i.e. fractures, layering, void space within sample container). The CT volume is much more detailed and is used to image internal features in 3D. This permits qualitative observations of rock and mineral textures, relative grain sizes and distributions, presence of void space, among others. Further quantitative measurements such as true grain size distributions, spatial distribution analyses, true modal abundances, petrofabric analyses, and others can be extracted.

**Facilites:**

<u>Astromaterials X-ray CT Lab at NASA-JSC</u>

The **Nikon XTH 320** has four interchangeable X-ray sources: 180 kV nano focus transmission source, 225 kV reflection source with multi-metal target (Mo, W, Ag, Cu), a 225 kV rotating target (W) reflection source, and a 320 kV reflection source. The system also has a 16-bit, 400 mm$^2$ (2000 x 2000 pixel) CCD detector, as well as a heavy-duty stage that will accommodate large (up to 30 cm) and heavy (up to 100 kg) samples. The multiple sources, high-resolution detector, and large stage allow us the flexibility to analyze a wide range of sample sizes. The 180 kV transmission source will allow for high resolution (submicron) scans on small samples (less than ~5 mm), whereas the 225 kV and 320 kV sources will allow scans of larger samples at resolutions on the order of 10s or 100s of microns per voxel depending on the sample size.

The lab supports several software packages for 3D visualization of XCT data, as well as packages for manual or machine-learning segmentation (e.g., VGStudio, Avizo, Dragonfly, Octopus, ImageJ, are a few examples). Software for extracting 3D quantitative data are also regularly used in this lab (Blob3D, Quant3D, PhaseQuant, Slice, are a few examples). All software are stored on one of two in-lab high-performance computers.

<u>Natural History Museum London</u>

**Zeiss XRadia Versa 520 3D μCT** High resolution μCT scanner used to collect 3D reconstructions of samples ranging from ~150 μm to ~3 cm in size. Using energies of 30 - 160 kV, samples can be imaged with excellent absorption contrast at a spatial resolution down to ~0.25 μm (20× lens with a 0.5 mm FOV). The built-in enhanced absorption contrast detectors maximize collection of low-energy X-ray photons and are ideal for studying complex rocks.

**Nikon Metrology HMX ST 225 μCT scanner** A μCT scanner that enables 3D reconstructions of geological materials at the μm-scale. It provides energies of up to 225 kV and currents of 200 μA for measurements in either reflection or transmission. Specimens (0.5 – 25 cm) can be analysed with excellent absorption contrast and a spatial resolution down to ~5 μm, whilst higher energies allow for penetration of samples 10's cm in diameter.

Sears, D. W., Sehlke, A., Friedrich, J. M., Rivers, M. L., and Ebel, D. S. (2018) X-ray computed tomography of extraterrestrial rocks eradicates their natural radiation record and the information it contains. *Meteoritics & Planetary Science 53*, 2624-2631.

# 5.14 X-ray Diffraction (XRD)

**Overview:**

X-ray diffraction (XRD) is a well-established, flexible tool for characterizing complex multiphase materials and dynamic properties. XRD routinely provides phase identification (ID), quantitative phase analysis, atomic-scale crystal structures, textural analysis (mineral fabrics, lattice preferred orientations), grain size and crystallinity, and dynamic analysis of phase transformations and elastic properties. It can be used to investigate compositional and geological properties of extraterrestrial materials including meteorites [1–3]; returned asteroid [4], cometary [5] and interstellar grains [6]; and planetary bodies in situ [7]. Although XRD is most commonly performed on single or polycrystalline phases, it can also be applied to non-crystalline materials such as powders. A strength of XRD is the ability to identify and quantify clay minerals and phyllosilicates in multiphase samples.

XRD is generally a laboratory technique. However, larger synchrotron facilities enable the acquisition of high-resolution diffraction data with beam sizes down to ~50 nm. Such facilities can also select specific X-ray wavelengths or use polychromatic beams and can carry out 2- and 3-dimensional diffraction imaging and ptychography experiments. Moreover, synchrotron-based diffraction experiments are often performed with other chemical or spectroscopic imaging, providing multifunctional and multidimensional suites of data.

**How it works:**

A standard laboratory-based X-ray diffractometer generates X-rays by bombarding a metal target (e.g. Cu, Fe, Co, Mo, Cr) with electrons. The X-rays, whose wavelength is characteristic of the target material, are monochromated and often focused before interacting with the sample. The X-rays are diffracted at specific angles from each set of lattice planes in a sample and interfere constructively when conditions satisfy Bragg's Law,

$$2d \sin \theta = n\lambda$$

where $d$ is atomic spacing, n is an integer number (of wavelengths), $\lambda$ is wavelength of incident X-rays, and $\theta$ is the scattering angle of diffraction. The diffracted X-rays produce an XRD pattern that can contain spots, rings, or diffuse intensity depending on the nature of the sample (crystallinity and orientation) and the conditions used in the experiment. In the case of crystalline materials, the specific positions and intensities of the spots or rings (referred to as reflections) are related to the structure (unit-cell size and symmetry) and chemistry of the material being analyzed. As each mineral has a unique combination of crystal structure and chemistry, an XRD pattern acts like a fingerprint, enabling quick and reliable phase ID by software-driven comparison with more than 400,000 patterns in reference databases (e.g., International Centre for Diffraction Data, Crystallography Open Database), as well as analysis of the compositional, crystal-structural, and textural characteristics of a sample. XRD experiments can be performed in transmission or reflection geometries to suit, for example, the available sample volume, instrument requirements, or analysis of very low-angle diffraction peaks. Quantitative mineralogy can be accomplished with reference intensity ratio, full pattern quantitative analysis, and Rietveld techniques. Rietveld analysis can also provide crystal micro-strain and crystallite size determinations along with other crystal structure information.

**Sample preparation:**

XRD is capable of characterizing a variety of sample types and sizes, from single crystals and unprepared materials to powders and polished sections. In general, minimal specialized sample preparation is required. Single crystals (~5 to 30 µm in size) are attached to the tip of a fiber that is mounted on a goniometer and aligned to the center of rotation. For polycrystals, a powder analysis can be performed. Powder XRD requires a quantity (~milligrams to hundreds of milligrams) of the sample to be crushed to a fine, even grain size (≤~10 µm) and either thinly smeared onto a zero-background substrate, packed into deep-well or capillary sample holders, or held between two thin Kapton films. Samples can be powdered dry or in ethanol to prevent phase change and to avoid mineral grain stress or strain. It is important to ensure random orientation of the grains and a smooth sample surface in order to acquire a pattern with abundant peaks and high signal-to-noise ratio. Microdiffraction of specific regions of interest and diffraction mapping can be performed on polished sections and raw chips by using primary X-ray beams with spot sizes of tens of microns and area detectors. For quantitative bulk XRD, samples are typically mixed with a known amount of a mineral standard, usually 20 wt.% corundum. To ensure that OSIRIS-REx samples are contamination-free, an external corundum standard from the National Institute of Standards and Technology (NIST) SRM 676a can be use.

**Sample impact:**

Laboratory-based XRD is essentially a non-destructive technique. Once the sample has been prepared (powdered, sectioned, or polished, possibly exposed to an organic solvent), it is very rare for the sample to be damaged by the X-ray beam. Even samples that contain materials such as organics or hydrous phases can be measured with XRD without being damaged. Although very few minerals are sensitive to the beam, some samples may dehydrate or oxidize in air, in which case they can be mounted in environmental chambers during analysis. Powder diffraction patterns can be obtained non-destructively from single grains or clusters (e.g., interplanetary dust particles or asteroid grains) using Gandolfi movement goniometers and microbeams.

Beam damage to samples does need to be considered for some synchrotron-based experiments, including the analysis of small single crystals with high energy or brightness beams, or diffraction imaging with very small beam sizes. Nevertheless, beam damage of mineralogical samples is not common.

**Data quality:**

XRD instruments are robust, and data are highly reproducible. XRD is a well-established technique for phase ID, with numerous reference databases available (see above and Data Products section below). Phase ID from a powder ideally requires the identification and indexing of a minimum of three diffraction peaks from the unknown phase. Detection limits depend on the X-ray wavelength and structural and compositional characteristics of the single phase and mixture. They are often quoted in general terms as ~1 wt.% [8], although in favorable circumstances, phases can be confidently identified and quantified at levels below 1 wt.% [9].

As the positions and intensities of diffraction peaks are a function of phase chemistry and atomic structure, XRD is commonly used to assess semi-quantitatively element ratios within solid solutions, such as Mg/Fe in olivine or Ca/Mg in calcite, with a precision of a few percent. Modal mineralogy from XRD either using profile-stripping or Rietveld methods can quantify mineral proportions with a precision better than 1% and where phase volumes are below 1% (depending on the phase and nature of the sample). Crystal structural analysis, either by single crystal or powder diffraction, produces very accurate and precise descriptions, with uncertainties quoted on the order of 1 part in 10,000 or better. These numbers vary depending on properties such as the atomic number, number of atoms within the unit cell, and the atomic-number contrast within the phase.

**Data products:**

The data products expected from powder XRD measurements include:

- 2D diffraction spectra (intensity of diffracted X-ray photons vs. angle) for instruments in reflection geometry (most common)
- 2D ring diffraction patterns for instruments in transmission geometry

In addition to instrument-specific data formats, raw data from laboratory-based XRD analyses are typically in .txt, .raw, .dat., asc., or csv. formats, as standard image files (e.g., .tif), or in formats tailored for specialized analytical software. Data from samples and standards can be handled in both specialist software and more common spreadsheet and graphical packages. Processed data are typically saved as standard image files (e.g., .tif, .jpg). Accurate phase ID often requires access to a database of standard reference patterns, such as the powder diffraction file (PDF) of the International Centre for Diffraction Data. Crystal structure databases such as the Inorganic Crystal Structure Database (ICSD) are also valuable tools for phase ID and crystal structure analysis, with all deposited data supported by crystallographic information files (CIFs).

**Facility(ies):**

NASA Johnson Space Center Astromaterials Research and Exploration Science Division

- Malvern PANalytical X'Pert Pro XRD: Powdered, homogenized samples are analyzed using a PANalytical X'Pert Pro MPD X-ray Diffractometer, with an X'celerator detector and Co Kα radiation, with data typically collected at a step size of 0.02° per minute step counting rate from 2 to 80° 2θ at 40 mA and 45 kV. Samples are prepared on zero-background mounts in aluminum holders, and the XRD instrument is operated under ambient conditions on a spinner stage and calibrated with a Novaculite standard, a NIST Silicon 640e standard, and a NIST lanthanum hexaboride standard, which is used to characterize the instrument's line broadening function. Three non-ambient stages allow XRD analysis to be performed at temperatures ranging from –190 to 900°C, from 0 to 100% relative humidity, and at pressures ranging between 0.01 and 10 bar.

- Malvern PANalytical Empyrean XRD: We anticipate the purchase of a Malvern PANalytical Empyrean XRD, in FY21 or FY22, along with optimal hardware such as a 50 µm focusing lens, a micro-diffraction stage, an upgraded detector, and a capillary spinner stage. This configuration will allow for higher resolution, better detection of mineral trace phases, and mineral mapping. Sample preparation with the capillary stage allows for XRD analysis on minute quantities of sample that can be pre-loaded and sealed in a $N_2$(g) glovebox prior to analysis.

Natural History Museum, London, UK

- Three Enraf-Nonius PDS120 powder diffractometers for phase identification, modal mineralogy, and real-time dynamic property analysis. These instruments are equipped with Co Kα$_1$ and Cu Kα$_1$ radiation and can be adapted to other X-ray wavelengths including Mo, Fe, and Cr. Rapid, real-time data acquisition is achieved by using an INEL 120º curved position sensitive detector.

- Panalytical Scanning Diffractometer X'Pert Pro MPD a1 for high-resolution measurements, Rietveld crystal structure analysis, and quantitative phase analysis. This instrument can use either Cu Kα$_1$ or Co Kα radiation.

- Rigaku Rapid 2 Micro-XRD with curved 2D detector with beam size of 5 to 50 µm. Ideal for spatially resolved diffraction mapping, measurement of specific regions of interest, or non-destructive analysis of individual grains and particles. Equipped with Gandolfi movement goniometer.

- X-ray microdiffractometer with an Enraf-Nonius PDS 120 diffraction system consisting of an INEL 120º curved PSD and a high brightness Cu Kα source generated by a GeniX system with a Xenocs FOX2D CU 10_30P mirror. Beam size is between 30 and 300 µm. This instrument is equipped with a Gandolfi movement goniometer and a controlled atmosphere heating stage (25 to 1000º C).

- Agilent X'Calibur Single Crystal Diffractometer with CCD area detector for structural analysis of single crystals. This instrument incorporates a cryo-heat system for low temperature analyses and phase transformation studies between 100 and 500 K.

- Two Philips PW1050 scanning diffractometers, with Cu Ka radiation and fixed and variable slits, optimized for the analysis of clays and phyllosilicates.

# 5.15 Raman Vibrational Spectroscopy

**Overview:**

Raman vibrational spectroscopy is an analytical technique used for the chemical and structural characterization of carbonaceous materials. It is based on the inelastic scattering of photons on molecules induced by light originating from lasers at different wavelengths. During Raman scattering, the excited molecules shift their vibrational state. Input of energy at specific frequencies, therefore, is linked to specific vibrational modes. The molecular vibrational modes can depend on the orientation of atoms and bonds, the atomic mass of the atoms, and bond order. The technique requires very little sample preparation.

**How it works:**

Raman vibrational spectroscopy has been extensively used to describe the structure and the degree of crystallinity for carbonaceous material. For example, the structure of organic residues can be assessed by the distribution of the vibrons attributed to the crystalline and disordered bands of carbonaceous materials at ~1580 $cm^{-1}$ (G) and ~1350 $cm^{-1}$ (D), respectively. The relative intensities and full-width at half height (FWHH) of the D and G bands are known to be modified by the thermal histories that carbonaceous materials experienced since their formation. Raman carbonaceous material geothermometers, therefore, have been developed based on the distribution of peak areas and the FWHH of G and D bands, while specifically focusing on how these vibrational characteristics reflect conditions of thermal alteration of the graphitic material.

**Sample preparation:**

Solid samples are placed in quartz cuvette/vials with no additional treatment. The measurements are conducted using a confocal microRaman spectrometer with a sample size requirement of nearly 1 mg for IOM-like material. The beam diameter on the sample is nearly ~1 mm with 10 mm focal depth and 2-3 mm penetration depth in the sample. The power output of the laser on the sample can operate at 0.3 – 1 mW to minimize small degradation. During the course of the analysis, we can vary both the acquisition time and the power of the laser on the sample.

**Sample impact:**

When operating at a power output of < 1mW, Raman vibrational spectroscopy has been shown to be a non-destructive technique for the analysis of extraterrestrial organic residues and bulk samples.

**Data quality:**

The frequency resolution of the Raman acquisitions are 3to 4 $cm^{-1}$ at 600 grooves/mm, < 2$cm^{-1}$ at 1200 grooves/mm, and <1 $cm^{-1}$ at 2400 grooves/mm.

Background subtraction is conducted by fitting third-order polynomials through portions of the spectra that do not exhibit any signal intensity. Curve-fitting (Lorentzian/Gaussian) and background subtraction of the Raman spectra are performed using the commercial software Igor from Wavemetrics. The integrated areas and structures (i.e., FWHH) of the Raman bands reflect the average values between Lorentzian and Gaussian fitting functions. In this way, the standard deviation on the integrated areas between the fitted Raman spectra in each acquisition reflects a 5% level of uncertainty, which is larger than the uncertainty estimated for the integrated area of each individual Raman peak. The peak positions for the D and G bands of the carbonaceous materials will be cross-calibrated with Raman spectra of IOM extracted from Murchison. This will also serve a cross-calibration between the Raman analyses conducted in different instruments and/institutions.

**Data products:**

The format of the raw data output is defined by JASCO's proprietary control and data collection/processing software. The files are in ".jws" format. The typical file size of the raw data file is 0.1–0.5 MB. A typical dataset contains 20–50 files. Reduced data are delivered as MS Excel files.

**Facility(ies):**

Earth and Planets Laboratory, Carnegie Institution of Washington, Washington, D.C., USA

- JASCO NRS-3100 confocal microRaman spectrometer, equipped with three continuous wave laser lines at 490.2, 532, and 785 nm, operating at maximum power of 32, 7, and 400 mW at the sample, respectively. Signal detection is accomplished through a series of objective lenses (e.g., 100X/0.90, 50X/0.42, 50X/0.35, 20X/0.45, 20X/0.35, 20X/0.28, 10X/0.28, 5X/0.15). Spectra are collected in 600, 1200 and 2400 gratings/mm on a Peltier-cooled CCD at –69 °C (Andor Model DV401-F1 1024x128 pixel with 25 µm pixel size). The system is equipped with a holographic notch filter. All the Raman spectra collected are unpolarized.

The University of Arizona

The Kuiper Materials Imaging and Characterization Facility (KMICF) houses a Hitachi S-3400 SEM (W thermal emitter) with a variable-pressure chamber and Renishaw InVia Raman system with the Structural and Chemical Analyser (SCA) interface and Reflex microscope.

Field Museum, Chicago, IL

- WITec alpha 300 R Raman Spectroscopy System with 532 nm laser

# 5.16 X-ray Photoelectron Spectroscopy

**Overview:** X-ray photoelectron spectroscopy (XPS) is a non-destructive near-surface (< 10 nm) analytical technique that provides quantitative elemental composition and chemical-bond / oxidation state characterization for species with Z> 2. Based on the photoelectric effect, XPS determines electron-atom binding energies – highly sensitive to the local chemical environment – utilizing a monoenergetic X-ray source in combination with a high-resolution electron-energy spectrometer. Elemental and/or chemical composition can be determined at points or over areas, scanned along a line, and mapped over regions. Moreover, three-dimensional concentrations can be derived if XPS is combined with a sputter gun for depth profiling. XPS is a fundamental tool for understanding surface composition, which often differs from the bulk material, reflecting the surface's interaction with the surrounding environment with a sensitivity of < 0.1 atomic% and a spatial resolution < 10 um.

**How it works:** In standard XPS, monoenergetic soft X-rays (Xray) are directed at a solid sample and subsequently absorbed by core-level electrons orbiting the target atoms; if the resulting photoelectron has sufficient energy, it will be ejected from the target into vacuum, where its kinetic energy (KE) is measured with a high-resolution electron energy spectrometer. The atom-specific electron-orbital binding energy (BE) can be extrapolated as: BE = Xray - KE - φ, provided the spectrometer work function (WF) is known. Al Kα X-rays with Xray = 1486.7 eV are most commonly used in laboratory instruments. WF is typically extrapolated from the BE of a standard photoelectron feature such as the Au-4$f_{7/2}$ peak. XPS spectra and the binding energies of photoelectron peaks are measured with respect to the Fermi level, the top of the valence band for conductors (e.g. troilite) and metal species, which is equivalent for sample and spectrometer due to contact. Insulating materials such as oxide, silicate, and phyllosilicate minerals lack a Fermi-edge and develop a positive surface potential with the ejection of photoelectrons from the surface during analysis, shifting spectra to lower kinetic energy. Thus, most geologic materials require a binding-energy referencing methodology prior to spectral interpretation. Surface charging can be mitigated with an electron flood gun and low-energy ion source to suppress the buildup of static charge during XPS acquisition; spectral binding energies are generally charge-referenced post-acquisition to photoelectron features of a known position. XPS instrumentation is run at ultra-high vacuum (UHV) to preserve the surface composition of samples, which can be strongly affected by atmospheric (or other) environments, and because the inelastic mean free path for low energy (< 5 keV) electrons is short (~1mm) at low vacuum (~1 Torr). For measurements of returned samples, UHV is a reasonable match to the surface pressures of Bennu.

In practice, XPS spectra are acquired by systematically varying the analyzer input lens voltages to allow electrons of a particular energy (or band of energies) to pass through the spectrometer with fixed transmission characteristics; electrons passing through the spectrometer are then counted for a period of time (~ 50 ms) at each energy by a position-sensitive microchannel-plate detector. The analyzer energy resolution can be adjusted from high-pass energy for identification and quantification of elemental species to low-pass energy to improve energy resolution for investigations into sample chemistry, which can be inferred from subtle shifts in binding-energy resulting from local chemical bonds. Spectra are plotted as intensity (counts per second) vs. binding energy (eV). Dwell time, energy steps, and number of spectra to be averaged can be adjusted to optimize the acquisition. 2D and 3D chemical image maps can be acquired by rastering the X-ray source or moving the sample stage. An *in situ* keV ion gun can be used with low-energy (~ 500 eV) Ar or Xe to remove layers in a systematic manner, to study a grain composition and chemistry (e.g. oxidation state) as a function of depth. Care must be taken to minimize changes in chemistry and phase initiated by ion-impact. Quantification of each species (*n*) is accomplished by subtraction of the inelastic background (e.g. Shirley background) with subsequent peak-area integration to determine ( ) and normalization using a transition-specific and instrument-specific sensitivity factor ( ); concentrations ( are presented in atomic percent (at-%) and calculated as:

$$C_x = [\ (\ I_x / S_x\ ) \div (\ \sum I_i / S_i\ )\ ]$$

Features in XPS spectra include all core-level photoelectron peaks from atoms present in the target surface, a background due to inelastically scattered photoelectrons from the bulk, Auger-electrons from the decay of ionized target atoms, and valence-band electronic features including band-gap energy. Auger electrons in conjunction with photoelectron features can provide additional insight into sample surface chemistry.

The surface sensitivity of XPS data can be enhanced by selecting only photoelectrons ejected at grazing angles from flat samples, and non-destructive surface specific (< 5 nm) depth profiles can be acquired by selection of photoelectrons with different take-off angles (ARXPS or ARPES). The electronic density of states in a material, the band-gap and work-function can be studied in the valence band region of an XPS spectrum or by using ultra-violet light (UPS) rather than X-rays. XPS at synchrotrons can be tuned to measure composition at varied depths by changing the wavelength of the incident X-ray; spectra acquired at high-energies (> 5 keV) with hard X-rays (HAXPES) can probe bulk composition. XPS is also known as electron spectroscopy for chemical analysis (ESCA) and photoemission spectroscopy (PES).

**Sample Preparation:** Due to the surface specificity of XPS analysis, samples must be clean and free of both contaminants and dust, and handled only with non-magnetic stainless-steel or other inert tools (e.g. tweezers). Gloves are required when handling any materials, mounts, masks, or screws that will be inserted in to the UHV XPS system. If a sample or tool has been handled or has hydrocarbon contamination, it should be cleaned with an organic solvent (e.g. methanol, ethanol, or isopropyl) in successive 15-minute ultrasonic baths, then blown dry with compressed $N_2$. Samples must be mounted on platens using non-volatile materials that will not outgas in UHV; standard mounting methods include the use of retaining screws, washers, masks, and wires made from molybdenum, 300-series stainless steel, aluminum, copper, ceramic, and copper-beryllium. UHV epoxies can be utilized if required. Powders can be pressed into indium, compressed into self-sustaining pellets, held in place by Van der Waals forces (for individual < 50 um grains), or sprinkled across double-sided permanent scotch tape or copper tape. C-tape used for electron microscopy is discouraged, as the adhesive is volatile – particularly under X-ray irradiation.

**Sample impact:** XPS is generally non-destructive and inorganic samples can be utilized for subsequent analysis. Organic and biologic samples must be analyzed with care, as high levels of X-ray radiation have the potential to alter sample chemistry; techniques such as sample cooling and beam rastering can be used to mitigate these effects. If XPS depth profiling is done, the morphology, phase, and composition of the outer ~100-nm may be changed by momentum transfer from incident ions; below this depth, the sample remains unchanged.

**Data quality:** XPS is sensitive to all species, except H and He, to better than 0.1 at-% (< 500 ppm in the PHI Versaprobe III). Energy resolution (dE) is a function of electron energy-analyzer type, pass energy, and X-ray source width. For compositional analysis, survey spectra are acquired with dE ~ 1-2 eV, while chemical-state (high-resolution) data is taken with dE ~ 0.1-0.2 eV. XPS information depth is typically ≤ 7 nm for 95% of the photoelectron signal using Al Kα X-rays; while depth profiles can provide 3D-composition slices to ~ 1-um.

**Data products:** Data products for both composition and chemical analysis will be produced with PHI SmartSoft data acquisition software and processed with either PHI Multipak 9.8 or CasaXPS as required. Spectral data output files will be provided as SmartSoft native *.SPE files (~ 20kB/spectrum) and ASC II *.CSV files (~10kB/spectrum); data output from depth profiles will be provided as SmartSoft native *.PRO files (~ 100kB/file) and ASC II *.CSV files (~50kB/file); and XPS composition/chemical maps will take the form of SmartSoft native *.MAP files (~ 100MB/file), ASC II *.CSV files (~1MB/file) and * .TIF (~1MB/file). Additional files include: Sample Platen photos in *.JPG format (~600 kB/photo) and *.BMP format (~15 MB/photo), as well as X-ray induced secondary-electron images in SmartSoft native *.SXI format (~300kB/image).

Both native SmartSoft files and other common data-file types (*.CSV, *.TIF, *.JPG) data files will be delivered to the SAMIS Database, along with the XPS instrument description (*.TXT and *.DOCX), sensitivity factors (*.CSV file), and logbook/analysis notes (*.JPG).

Specific XPS Expected Data Products

- XPS compositional spectra from 1-3 points/sample for ~16 Contact Pad Grains and ~6 TAGSAM Grains

- XPS chemical spectra from at least 1 point/sample for ~16 Contact Pad Grains and ~6 TAGSAM Grains

- Depth profiles for at least 1 point/sample for ~8 Contact Pad Grains and ~4 TAGSAM Grains

- Compositional/chemical maps for heterogeneous grains for ~2-4 Contact Pad Grains and ~1-2 TAGSAM Grains

**Facilities:**

Nanoscale Materials Characterization Facility, University of Virginia, Charlottesville, VA USA

The PHI Versaprobe III XPS is installed in the Nanoscale Materials Characterization Facility (NMCF) at the University of Virginia as a multi-user instrument. The state-of-the-art imaging-XPS chemical microprobe has a spatial resolution < 10 µm, optimized for high sensitivity (< 500 ppm) and energy resolution (FWHM < 0.5 eV for the Ag 3d5/2 peak). The mu-metal lined system is maintained at UHV (< 3E-10 Torr) with oil-free pumping. A floating-column ion gun and electron flood gun allow charge compensation during analysis. Precision sample positioning is facilitated with an optical camera image or via *in situ* secondary-electron X-ray imaging. Angle-resolved XPS measurements can be made on this instrument, which also has an *in situ* hot-cold stage (120 – 850 K). If required, in situ irradiations to 1) reactivate the Bennu grain surface or 2) to acquire compositional information with depth can be accomplished directly within the Versaprobe main analysis chamber. Ion beams are defocused and rastered to provide uniform flux across the irradiated surface. A Faraday cup can be transferred to the ion gun focal point for accurate current density measurements.

# 5.59 Nanoscale Infrared Mapping (NanoIR)

**Overview:**

NanoIR is an umbrella term used to describe super-resolution (~10 nm) infrared mapping and spectral acquisition. Such acquisition is done via atomic-force microscope-assisted methods, most commonly **S**cattering **S**canning **N**ear Field **O**ptical **M**icroscopy (**s-SNOM**) and **AFM-IR**.

Both techniques rely on using an external light source to provide electromagnetic (EM) radiation for illumination of the sample-tip region, while an AFM tip localizes the electric field to provide sensitivity below the diffraction limit. Both approaches can be used to create IR spectral maps of a surface. Infrared spectra can be acquired at a spatial resolution of tens of nanometers in this fashion, which makes it possible to differentiate functional groups of IR-sensitive materials at the native scale of heterogeneity.

**How it works:**

<u>s-SNOM</u>

s-SNOM-based IR mapping measures the real and imaginary components of the optical constant of a material by detecting and measuring the amount of IR light that is scattered as a function of frequency (McLeod et al., 2014). The nanometer-scale spatial resolution of the technique is derived from the fact that the scattered light is heavily influenced by the interaction between the metal-coated AFM tip and the sample surface directly underneath the AFM tip's apex (Figure 1). The surface of the AFM tip is a conductor with a small radius of curvature that concentrates the electric field in a local region. A piezoelectric crystal is used to oscillate the AFM tip so that it "taps" the surface. At the top of the oscillation, the tip is many nanometers from the surface, while at the bottom of the oscillation, the tip comes within 1 Å of the surface. A light sensor measures the optical signal at both positions and subtracts them to extract the signal due only to the small volume of sample near the tip when it is at closest approach. Maps of the surface optical properties are generated by moving the sample in X and Y directions while repeating this measurement. Spectra are acquired by changing the laser wavelength or using an FTIR interferometer.

**Figure 1**: Nano-IR (s-SNOM) (a) A sharp metalized AFM tip interacts with a sample region with dielectric constant $\varepsilon(\lambda)$. (b) Incident electromagnetic field ($E_{inc}$ induces a dipole moment in the tip, which is mirrored in the sample directly underneath the tip. The induced dipole strength is a function of the local permittivity ($\varepsilon$). This interaction affects the tip's z-component of its dipole moment ($p_z(\varepsilon)$), thereby modulating the amplitude and phase of the tip-scattered electromagnetic field, $E_{scat}$. (Figure from Dominguez et al., 2014)

## AFM-IR

AFM-IR measures the absorption spectrum (or imaginary component of the dielectric function) of the material directly underneath the tip by measuring the thermal expansion of the material as a function of frequency (Dazzi et al., 2010, Lu et al., 2014, Kebukawa et al. 2019, Mathurin et al., 2020). The electric field enhancement caused by the shape of the AFM tip results in stronger absorption of the incident photons when the tip is near the surface (Figure 2). Thus, a several nm wide region near the tip will heat up by about 10 degrees C. This is enough to cause an expansion of the material by as much as a pm, depending on the phase, and the AFM records the force the sample exerts on the tip as it expands. Maps of the surface absorption are generated by moving the sample in X-Y directions underneath the tip while repeating the measurement. Spectra are acquired by changing the laser wavelength.

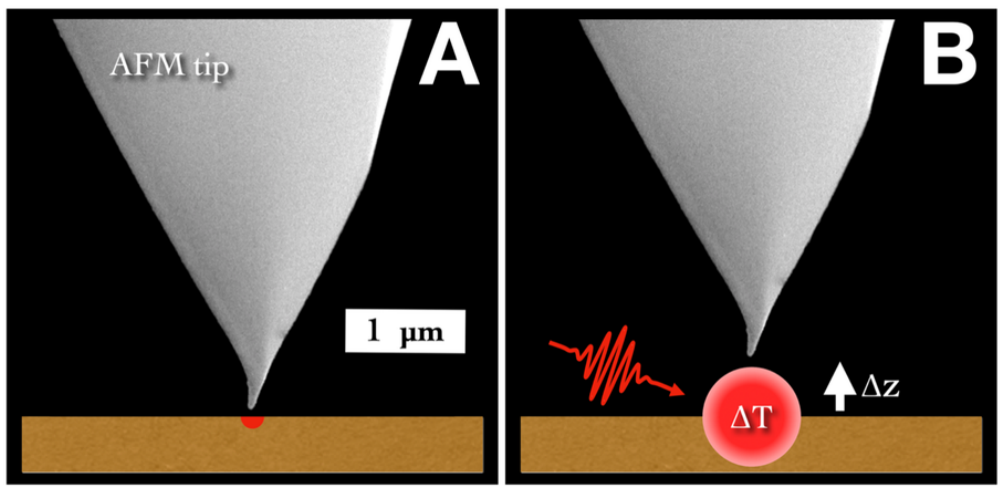

**Figure 2:** Mechanism of photothermal AFM IR spectroscopy. (a) An AFM tip approaches a nanophase material and records its height. This provides a baseline for thermal expansion. (b) An IR laser pulse is absorbed by the nanophase material, causing it to heat up on order 10 degrees ($\Delta T$). This causes it to expand, and the AFM probe records the height of the expansion, ($\Delta z$).

## Alignment of IR beam

Illumination of the sample is done by redirecting IR light from the source to the area where the AFM tip is interacting with a sample's surface using a set of mirrors that are controlled by the instrument software. This alignment must be done in two steps. In the first, coarse alignment step, a visible (green) laser whose optical path is co-aligned with the IR sources is positioned manually by a user. In the second step, fine alignment is done by measuring the amplitude of the AFM tip response and optimizing the amplitude by steering (via a computer) the IR beam in the X-Y and Z directions.

## Sample preparation:

For best results, samples should be smaller than 2.5 cm in diameter and relatively flat. Only the central ~1-cm-diameter region will be accessible to the AFM tip. For example, a 2.5-cm-thick section of Bennu could be measured so long as the rock was centered in the middle of the round, and only a few mm wide. Flattening of the sample can be achieved via standard techniques such as epoxy embedding followed by polishing. For very small samples, mounting the sample on an AFM disk (10, 12, or 20 mm) using carbon paint followed by FIB polishing can produce surfaces flat enough for high quality NanoIR imaging.

Air-free sample preparation is necessary for samples that can easily oxidize. Preparation of a flat surface using the FIB or cryogenic SEM ion mill begins with fixing the sample to an SEM stub or similar surface within an argon glove box. The stub is then transferred via airlock to the FIB or the ion mill, where it is processed in vacuum, and then vented to $N_2$ (FIB) or Ar (ion mill) and brought back to the Ar glove box.

FIB sections can be removed in the FIB and later transferred via air-free holder to the TEM. All ion milled surfaces can be transported to the NanoIR and placed into an air-free hood enclosing the AFM head (N$_2$) so IR spectra of the unmodified surfaces can be acquired.

Ion milling must be carried out at low energy. Initial rough shaping can be done, e.g., using Ga+ at 30 keV, but the damage layer left by the ion beam in such case can be tens of nanometers thick and will result in a low-quality NanoIR map if any map can be acquired at all. Energies below 2 keV (Ga) and 4 keV (Ar) have been successful, with improvement to be found at very low energies such as 1 keV (Ga) and 250 eV (Ar).

## Sample impact:

Typical use will result in no sample alteration/modification beyond the use of ion mills or mechanical polishing to prepare surfaces for study. NanoIR and IR preserve functional groups found on the sample including delicate organic species. The use of air-free/inert gas environments eliminates oxidation damage, which can alter the IR spectroscopy of air-sensitive samples.

When ion-milling porous materials (such as high-porosity aqueously altered material from asteroids), ion "sludge" consisting of the sputtered and redeposited ions from the sample can contaminate the pores. Various techniques can be used to mitigate sputtering and redeposition, such as low-incidence-angle milling with a low energy ion beam, and the use of metal shields to remove beam defocus and curtaining effects. The use of Ar tends to produce less deposition because it is an unreactive atom, but it cannot eliminate redeposition of ions already present in the sample (such as Al, Mg, Si, and O in the case of milling phyllosilicate).

## Data quality:

NanoIR spectra are highly reproducible but are sensitive to the laser tuning. As the technology is in its infancy, IR peak intensities of difficult samples are usually accurate within tens of a percent. Peak intensities of easy samples are reproducible within a percent for strong peaks. Weak peaks can have larger errors.

AFM-IR spectra can be compared directly to far-field infrared measurements such as Attenuated Total Reflectance and FTIR. They can be directly overplotted and will agree within experimental error of a few percent.

SNOM spectra require mathematical processing to compare against far-field IR spectra. Errors are similar to AFM-IR.

IR peaks can be calculated using density functional theory (DFT). Peaks identified in DFT-simulated phases have systematic shifts from experimental peaks due to incomplete handling of electron exchange-correlation within the DFT framework. However, these shifts are usually relatively small, and it is typically trivial to identify which phonon modes are contributing to which peaks in the experimental spectra.

## Data products:

AFM acquisitions include:

· Topography maps (the height of the sample surface in nanometers).

· Phase maps (the attractive force or "stickiness" of the surface to the AFM tip, which alters the phase of the tip's tapping.)

· IR absorption maps (AFM-IR, spatial resolution to 20 nm, and spectral resolution up to 1 cm$^{-1}$)

· IR optical constants, real and imaginary (SNOM) (spatial resolution up to 10 nm, and spectral resolution to 1 cm$^{-1}$)

AFM acquisitions are acquired in a proprietary Bruker specific format, .axz, which is processed using an in-house Python pipeline to produce various output formats. Typically, images are saved as .tif, .gsf, or .png files, hyperspectral stacks are saved as .tif stacks, and spectra are saved as .csv files. Other formats are possible if needed. Data are then further processed through linear algebra pipelines (e.g., PCA, ICA, NMF, peak fitting) and non-linear algebra pipelines (e.g., neural nets) to produce phase maps and quantify peaks. Outputs are typically in the form of images stored as artifact-free .tif or .png when artifacts are tolerable, such as scatter plots or 2D projections of a high-dimensional manifolds. Other types of outputs are guided by the data, e.g., ternary diagrams may be appropriate in some cases.

Data processing is done using a Python pipeline developed in-house and can be converted to other formats or processed to match other pipelines if desired.

A single experiment on a single sample will typically result in the acquisition of ~100 maps and ~100 spectra, which is ~1 GB of data.

Simulated spectra using DFT are stored as various text-formatted files produced by Quantum Espresso or Orca. These include dynamical matrix files (.dynmat) calculation overviews (scf.out, ph.out) and final processed spectra is stored as .csv. Vibrational movies can be stored as animated xyz files (.axyz) or CrystalMaker files (.cmdx). Final results can be displayed as MP4. Each mineral phase simulated will result in ~1 GB of data.

## Facilities:

### CSUSM NanoIR Facility

At California State University, San Marcos (**CSUSM**), we have two IR light sources available. The first is a pulsed Quantum Cascade (QC) IR pulsed laser that can illuminate a surface between 905 and 1950 cm$^{-1}$ to provide high-speed AFM-IR imaging. The other light source is an APE Carmina broadband continuous (CW, 670-4000 cm$^{-1}$) that is suitable for both s-SNOM imaging and AFM-IR mapping.

s-SNOM and AFM-IR are implemented on an Anasys/Bruker NanoIR3-s instrument at CSUSM (Fig. 3).

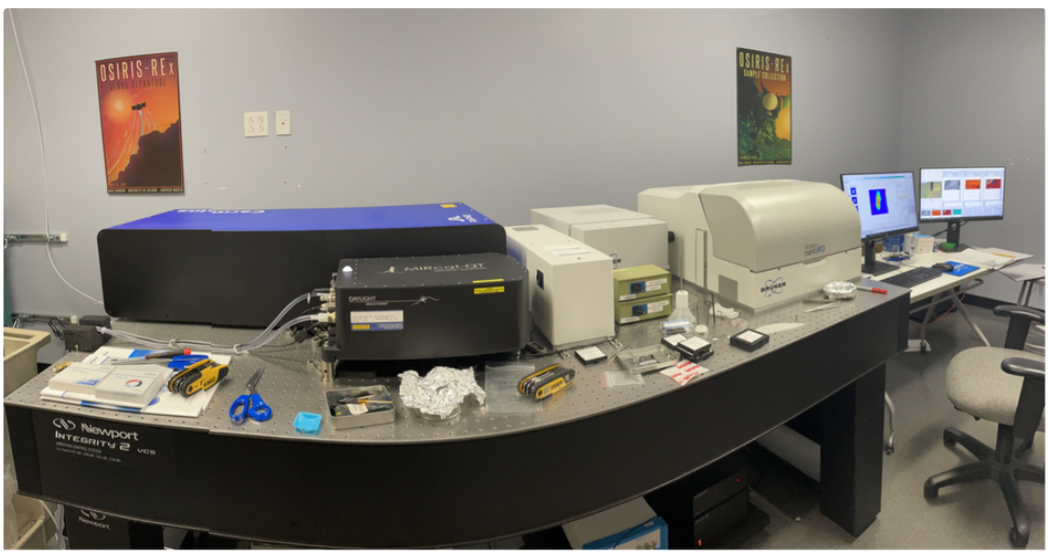

**Figure 3:** NanoIR Lab at CSUSM with SNOM and AFM-IR capabilities (fall 2022).

Instrument control is done via a Windows PC using an Anasys/Bruker Proprietary Data Acquisition environment.

### Lawrence Berkeley National Laboratory, Advanced Light Source

The Advanced Light Source (ALS) is a 1.9 GeV synchrotron light source located at the Lawrence Berkeley Laboratory (LBL). It has three infrared beamlines capable of both diffraction-limited infrared microspectroscopy and AFM tip-limited infrared nanospectroscopy.

Beamlines 2.4 and 5.4 have broadband s-SNOM capabilities for nearfield infrared spectroscopy. **Beamline 2.4** uses a Neaspec neaSNOM instrument with a spectral range between 330-3300 cm$^{-1}$, and **Beamline 5.4** is based on a modified Bruker Innova AFM and a Thermo Fischer Scientific 6700 FTIR spectrometer with a spectral range covering 450–5000 cm$^{-1}$. Beamline 2.4 also has nearfield nano-imaging capabilities with a QCL spanning 1160 cm– 1870 cm$^{-1}$.

## References:

**General SNOM and AFM-IR:**

**SNOM of planetary materials:**

AFM-IR of planetary materials:

# 5.60 Thermogravimetric Analysis (TGA)

**Technique name: <u>Thermogravimetric analysis (TGA)</u>**

**Overview:**

Thermogravimetric analysis (TGA) measures the mass of a sample as a function of temperature and can be used to understand phases transitions, particularly in conjunction with calorimetry. The technique has been applied to the study meteorites, often with a focus on examining the nature and abundance of hydrated components, providing insights into the aqueous and thermal alteration processes they have experienced. It can determine the fraction and composition of the volatile components of a sample by recording the weight change that occurs when the sample is heated over time. Using this information, it is possible to estimate the abundance of volatile species, for example $H_2O$, $CO_2$, $SO_2$, and some organics. TGA is a robust method that has previously been used to complement compositional studies of meteorites by deriving water content for primitive carbonaceous chondrites [1,2] and ordinary chondrites [3].

**How it works:**

A standard thermogravimetric analyser consists of a precision balance located within a furnace. The temperature of the furnace is increased at a constant rate that can be programmed. This process can occur in air, or, if the instrument has the capability of being connected to a gas cannister, under an inert gas atmosphere (e.g., $N_2$). Total weight change over the entire temperature range correlates to the volatile abundance of the sample. As the weight of the sample is constantly being monitored, the derivative of the weight change can be determined, showing specific 'breakdown' events at particular temperatures that correspond with specific mineral phases (determined by comparison with a set of standards also measured using TGA), and therefore with specific volatile species, giving their abundance.

**Sample preparation:**

TGA can be performed on powders or chips, with typically 10 to 15 mg of sample required. Powders are loaded into a crucible (usually alumina or platinum), which is gently tapped to flatten the surface of the powder, before being transferred onto the precision balance. Chips are also loaded into a crucible and require a geometry that allows them to sit below the top of it.

**Sample impact:**

TGA is a destructive technique, although it requires only very small masses. The process involves heating the sample, often to high enough temperature to completely de-volatilize the material (~1000°C). In some cases, this heating can induce phase changes in the material (for example, oxidation of minerals containing iron when heating in air). The heated samples are available for further characterisation once the experiment is concluded.

**Data quality:**

Mass loss is recorded over the whole temperature range with a mass resolution of 0.1 µg, meaning that a 10% mass loss from a 10-mg sample results in an error of ~0.01 % [2]. This error is lower for a 15-mg sample. The main source of uncertainty is from measuring the mass of the sample, which is done to a precision of ~0.01 mg. This gives an error of 0.1% for samples between 10 and 15 mg. The combined maximum error on the measured mass loss fraction is therefore ~0.1%.

**Data products:**

Typical output from TGA are .txt files with information about the experimental conditions (time, date, file location, temperature range, heating rate, and atmosphere) and eight columns containing the time (min), temperature (°C), weight (mg), heat flow (mW), temperature difference ($^oC$), temperature difference (µV), sample purge flow (mL/min), and the derivative of the weight (% per $^oC$) over the course of the experimental run (although these can be adjusted depending on the experiment). The files can be opened and manipulated using common spreadsheet and graphical packages. Additionally, the software used to operate the instrument can produce plots showing temperature on the x axis and weight (as a percentage of the total mass) on the y axis that can be saved as .pdf or .jpg files. A second y axis can be added to show the derivative of the weight (% per $^oC$) on the same plot.

**Facilities:**

Natural History Museum (NHM), London

The NHM has a TA Instruments SDT Q600, which provides simultaneous differential scanning calorimetry (DSC) and TGA, measuring both heat flow and weight changes in a sample as a function of temperature. The instrument has a horizontal balance that measures differential weight between the sample and a reference. It has Pt/Pt-Rh thermocouple pair for measuring temperature and a bifilar wound, horizontal furnace that is capable of heating samples up to 1500oC under an inert N2 atmosphere.

[1] Garenne et al. 2014, GCA, 137(93)

[2] King et al. 2015, EPSL, 67(198)

[3] Eschrig et al. 2021, Icarus 354(114034)

# 5.61 Synchrotron-based X-ray Fluorescence Spectroscopy (S-XRF)

**Overview:**

Synchrotron-based X-ray fluorescence spectroscopy (S-XRF) uses highly brilliant synchrotron X-ray radiation to induce X-ray fluorescence within a sample. Due to the unique characteristic energy emitted by each element, S-XRF allows detection and measurement of the major, minor, and trace elemental micro-distribution down to sub-ppm concentrations in a non-destructive manner. Depending on the incident X-ray beam energy, this technique can focus on medium-Z elements ($14 \leq Z \leq 42$) with a beam of typically around 20 keV, or high-Z elements ($18 \leq Z \leq 88$), including rare earth elements (REEs), with a high-energy incident beam of around 90 keV. Which elements are detected is further dependent upon the size of the sample and the depth of the element within the sample. For instance, with an X-ray energy of 90 keV, the K-line X-ray fluorescence of elements up to radium ($Z = 88$) can be detected from within millimeter-sized samples due to the deep penetration of highly energetic X-ray photons. Characteristic radiation from Ce ($Z = 58$), for example, can propagate through 1.55 cm of meteoritic material (CI mean composition at ~1.5 g/cm$^3$ density), whereas Zn propagates through only 340 µm of the same material.

**How it works:**

Synchrotron radiation X-rays are generated when relativistic charged particles (electrons or positrons) are inserted into a magnetic field and subsequently subjected to the Lorenz force, which causes the particles to make a curve in their trajectory. The very intense photon beams are almost 100% horizontally polarized and highly coherent. Aided by X-ray optics, the X-ray photon beam can be monochromatized, providing a well-defined and narrow range of photon energies, and focused to a nano- or micrometer-sized beam. When a sample is illuminated, a bound electron can be excited if the binding energy of that electron is lower than the incident X-ray energy. Subsequently, a vacancy in one of the core shells of the atom is created, rendering the atom ionized, which forces it to relax by filling the vacancy with a higher-shell electron. This relaxation in turn can result in the emission or fluorescence of an X-ray photon, the energy of which is characteristic for a specific element.

In a non-vacuum environment, i.e., atmospheric conditions, S-XRF is sensitive to elements with atomic numbers starting from silicon. Scanning a sample through the focused monochromatic X-ray beam in two or three dimensions and detecting the characteristic fluorescence photon energies in each pixel with an energy- or wavelength-dispersive X-ray spectrometer, results in a 2D or 3D elemental distribution image or volume, respectively. Alternatively, direct spatially resolved 3D information can be obtained from particular grains and volumes of interest from within the larger sample volume by limiting the detector view to a microscopic sub-volume(the confocal XRF approach). This method has the significant advantage that sample sub-volumes can be investigated with limited to no interference of the surrounding sample environment, which is ideal for quantification purposes, but comes at the cost of a slightly reduced sensitivity (ppm) and range of detectable elements ($Z \leq 40$).

**Sample preparation:**

S-XRF has the advantage that no sample preparation is necessary due to the penetrative character of the radiation. Samples can be measured as rock particles, slices, or thick or thin sections. Typically, the sample is positioned on a mount made of low-Z material for minimal interference with the measurement, e.g., ultra-pure polymer or quartz capillary. It is held in a fixed position to retain the coordinate system over multiple measurements. For ideal S-XRF investigation of rock particles and slices, a free-standing sample (not embedded in a polymer resin) is optimal to reduce spectral contribution due to scatter from the surrounding environment. For rock particles, sample sizes up to 1 mm for high-energy measurements, and up to a few 100 microns for medium-energy measurements, are preferred because of the physical constraints governing the energy-dependent information depth of the emitted radiation. For thick or thin sections, typically mounted on a glass substrate, a geometrically small substrate is preferred to allow close positioning of the detectors to the sample.

**Sample impact:**

S-XRF is a non-destructive and non-invasive technique, and the samples are neither damaged nor affected in any way. When exposed to high radiation doses, it is possible that organic phases (DNA, etc.) are altered or that temporary discoloration occurs.

**Data quality:**

The spatial resolution in the S-XRF application is directly related to the dimensions of the incoming X-ray beam and beam sizes. Depending on the size of the sample and the scanned area of interest, the time available, and the purpose of the scan, pixel sizes between 0.25 and 5 µm are usually selected, and minimal spatial resolutions between 200 and 1000 nm are typically achieved at the sample position. The typical energy or spectral resolution of an energy-dispersive S-XRF system is approximately 135 eV, allowing for a clear distinction between K-line emissions of the separate elements.

Detection limits for S-XRF data can reach down to sub-ppm levels for K-lines and between 10 and 100 ppm for L-lines. Detection limits are heavily influenced by the sample matrix and the measured background, as well as the dedicated measurement time. As such, the sample should be mounted and measured in an optimal form with minimal potentially interfering surroundings (e.g. without or with limited amount of encompassing resin, metallic supports, …), thus improving the signal-to-noise and signal-to-background ratios.

Post-processing of the XRF data may also include the application of dimensionality reduction methods such as K-means clustering and principal component analysis, as well as quantification of the results. Dedicated quantification algorithms result in quantitative concentration values of all detected elements within the sample. Typical Poisson statistics-governed relative errors for major and minor elements are 5–10%, whereas for trace elements, slightly larger errors may be obtained depending on the signal-to-noise ratio and the quantification method.

**Data products:**

Both qualitative and quantitative data on major, minor, and trace element distributions (including REEs) can be acquired with S-XRF. Specifically, the data products expected from S-XRF measurements include:

- 1D point analyses: individual high-quality (optimal signal-background ratio) elemental spectra from selected points of interest.

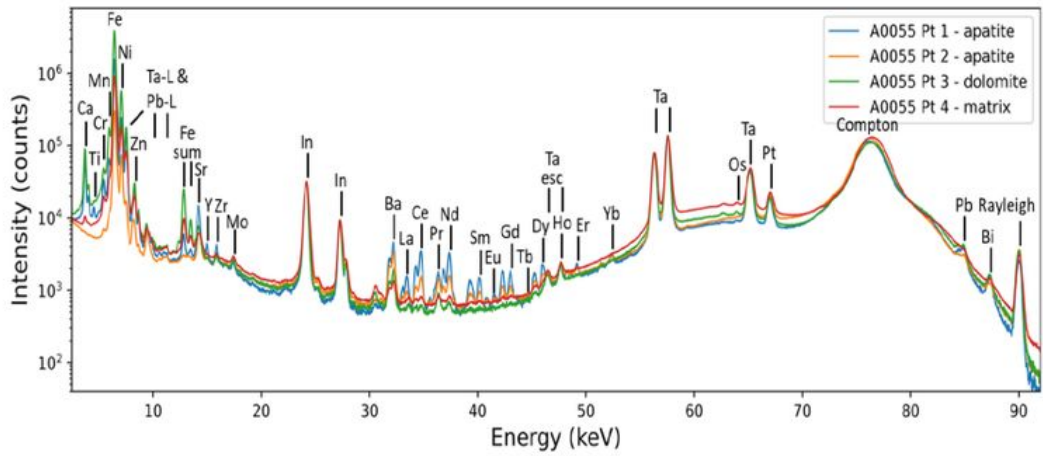

- 2D line scans: elemental distributions plotted as a function of distance or depth in the sample.

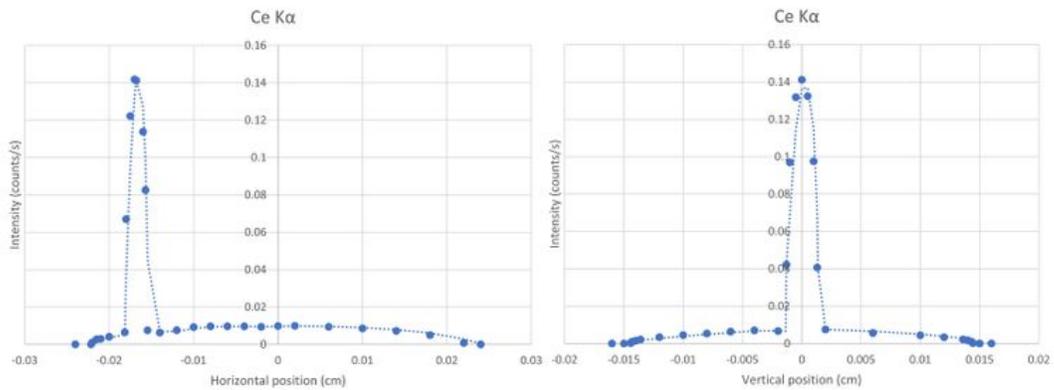

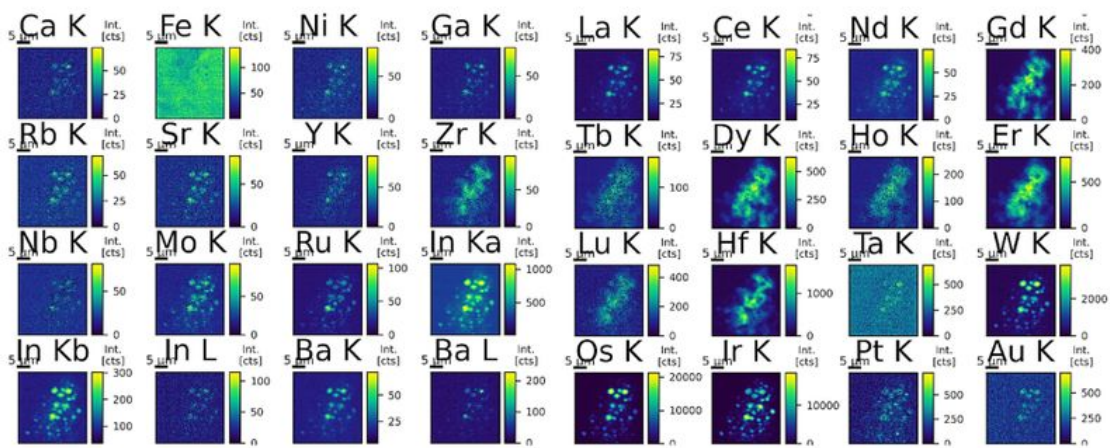

S-XRF 2D scan example

·   3D spectrum images: a spectrum for each pixel of the scanned area and, subsequently, a distribution image for all the detected elements. The latter is essentially a projection onto a 2D (YZ) plane of all the elements detected throughout the depth (XYZ) of the sample.

·   2D and 3D tomographic cross sections: virtual slices at specific selected heights, which can be combined to a full 3D volume.

·   3D confocal XRF measurements: direct 3D composition of selected microscopic areas of interest.

·   Quantified elemental concentration profiles obtained from the results of point analyses of selected points of interest.

·   Quantitative, spatially resolved distribution images of major and minor elements, obtained from 2D and 3D high-resolution maps.

All maps can be generated as single-element maps or as 3-element maps shown as 3-color RGB images to reveal more clearly the spatial coincidence of different elements.

Logbooks describing the S-XRF beamtime activity and linking unique measurement identifiers to the corresponding physical measurement description are stored as PDFs. Data for each specific measurement are stored in a compressed h5 (HDF5) file, supporting a defined structure where the raw data, fitted results, and further processed quantitative data are stored together. From this file, images can be produced and stored as a variety of image formats, for example, .png, .tiff, .bmp.

**Facilities:**

Include institutions, hardware, software, and any notable instrument-specific capabilities

Geologic/scientific team: Goethe University Frankfurt – Institute for Geosiences

Chemistry/instrumental team: Ghent University – X-ray Microspectroscopy and Imaging Group

S-XRF investigations require a synchrotron radiation facility, access to which is usually governed by a peer-reviewed proposal allocation procedure. In the scope of our collaboration, we currently have access to two synchrotron radiation facilities (in the context of long-term proposals ES-1018 and II-20210018 EC), providing two different beam energies and thereby covering the widest possible range of elements:

(1) Beamline ID15A at the European Synchrotron ESRF-EBS for high-energy measurements.

(2) Beamline P06 at PETRA III, the German Synchrotron DESY, for medium- to high-energy measurements.

ID15A – European Synchrotron Radiation Facility – ESRF – Grenoble, France

The excitation energy $E_0$ is selected around 90 keV, using a Laue crystal monochromator. A multilayer Kirkpatrick-Baez (KB) mirror system provides a 0.5×1 µm² beam spot at the sample position. X-rays are detected with a Canberra HPGe CMOS and a Vortex Si drift (SDD) detector. The setup is controlled with an ESRF inhouse-developed software called Bliss.

P06 – PETRA III – DESY: Deutsches Elektronen-Synchrotron – Hamburg, Germany

The excitation energy $E_0$ is mostly selected around 20 keV, using a Si(111) Bragg crystal monochromator. A KB mirror system provides a focused beam of around 0.2 x 0.2 µm² at the sample position. A Vortex 4-element Si drift detector is used in combination with a single-element Si-based Vortex detector equipped with a polycapillary X-ray optic for confocal detection. The setup is controlled with inhouse-developed software.

XRF data are processed using Python scripts and PyMca v5.6.2 and AXIL fitting routines at the home institutes.

# 5.62 Synchrotron-based Infrared Spectroscopy

**Overview:**

Infrared (IR) spectroscopy measures the abundance of chemical functional groups in samples by probing their vibrational states (Ventura et al 2014). Because of the high brightness and broadband spectral bandwidth of synchrotron IR radiation, synchrotron-based IR micro-spectroscopy (s-IR) enables diffraction-limited spatial resolution with high signal-to-noise ratios over the entire infrared spectral region. This allows spectral images to be acquired within hours instead of days, at a spatial resolution an order of magnitude higher than non-synchrotron sources (Martin et al. 1998, Levinson et al. 2006).

**How it works:**

For an IR micro-spectroscopy measurement, infrared light is coupled to an FTIR interferometer and IR microscope that focuses light onto a sample. The reflected or transmitted light is measured using a photo-sensor such as a liquid nitrogen cooled mercury-cadmium-telluride (MCT) detector. The intensity difference between the incident beam and the reflected/transmitted beam are caused by the absorption of light associated with vibrational modes of the sample, which can be related to different chemical functional groups.

The infrared light source for commercial IR micro-spectroscopy instruments is typically a globar, i.e., a blackbody radiation source, or a tungsten lamp. s-IR takes advantage of the high brightness (~1000 times brighter than a globar) and broad bandwidth of synchrotron radiation to enable high signal-to-noise measurements with diffraction-limited spot sizes of 2 to 10 microns. In a synchrotron, electrons are accelerated near the speed of light (1.9 GeV at the ALS, which is 99.999996% of the speed of light) in a polygonal evacuated pipe called the storage ring and deflected by magnetic fields. The lateral acceleration due to the Lorentz force causes the relativistic electrons to emit photons in the forward direction with energies spanning the infrared to X-ray region.

Photon energies used at the ALS IR beamlines range in energy from 200–11,000 cm$^{-1}$, with achievable spectral resolution down to 0.125 cm$^{-1}$, although spectral resolution of 4 cm$^{-1}$ is more common.

Figure 1 is an example of synchrotron IR spectra obtained from a Stardust cometary keystone showing the influence of space exposure on the surface of the aerogel. Several key chemical functional groups are labeled. In this configuration, dips in the spectra correspond to light that is absorbed by the sample because of the presence of a functional group. The space-exposed edge clearly contains SiH and SiH$_2$ functional groups that are not present in the interior of the aerogel tile, for example.

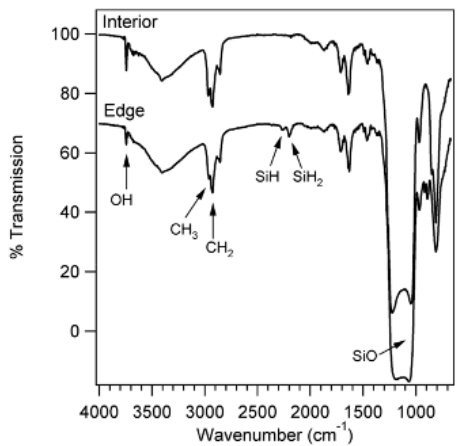

Figure 1) IR transmission spectra collected from the space-exposed edge and interior of Stardust cometary keystone C2005,5. The edge spectrum has been offset for clarity. Reproduced from Bechtel et al. (2012).

**Sample preparation:**

Samples can be prepared to measure in reflection mode (light reflects off the sample surface) or in transmission mode (light transmits through the sample).

For specular reflection measurements, samples should be as flat as possible on the length scale of the photons used to measure them. For example, vibrational modes in silicates respond to photons with a wavelength of 10 microns. The diffraction limit is thus about 5 microns, and the sample should be flatter than a micron over a width of 10 or more microns to produce high-quality spectra. Samples can be polished, cut, chipped, or prepared in any fashion that leaves them sufficiently flat. They cannot be coated with carbon or metal, as part of the sample preparation for SEM, because any coating can obscure the actual sample IR signatures.

Samples can be measured in the powder state or unprepared state, but artifacts will often be present from scattering and partial specular reflection. However, these types of spectra can be compared qualitatively, and in some cases quantitatively with spectra obtained by remote sensing instruments and are therefore still useful (Sultana et al. 2023).

Transmission mode samples should be flat both on the top and bottom faces. In addition, they need to be sufficiently thin so that a significant portion of the beam transmits through the sample, but not so thin that IR light transmits without appreciable absorption. For example, fused silica has an absorption length of several hundred nanometers (meaning ~67% of the incident light is absorbed in the absorption length). Therefore, a sample thickness up to about a micron would allow enough photons to pass through that a quality measurement can be obtained, but a 10-nm-thick coating on a transparent substrate (like KBr) would produce too little signal to measure.

**Sample impact:**

Infrared radiation is non-ionizing, and despite the high intensity of the synchrotron, temperatures in the sample do not increase markedly above room temperature. This is because the energy in each photon is relatively low, and the diffraction-limited nature of the incident radiation limits the total power delivery per volume. In most cases, the radiation will have no significant impact on the sample.

If the sample is left in a raw state or a powder, then there is little to no impact on the sample.

More typically, the sample can be prepared as part of a coordinated experiment involving SEM and other instruments. In this case, the sample can be polished but left uncoated. After the IR experiment it can then be coated and studied using SEM and other techniques. In this case, the impact on the sample is the same as with any standard polishing step.

**Data quality:**

Spectra are obtained from specific locations on the sample (often coordinated with an optical image for navigation). A grid of spectra can be obtained with spatial resolutions down to the diffraction limit for the wavelength of light used (typically a few microns). There is no point in reducing pixel sizes far below the diffraction limit, as neighboring spectra will be nearly identical.

A square array of spectra can be acquired. For example, 30x30 spectra would give spatially oriented information about the distribution of functional groups. This can be formed into a stack with a dimensions X, Y, and E, where X and Y are spatial coordinates and E is an energy coordinate.

**Data products:**

Synchrotron IR spectroscopy will produce reflectance and transmission spectra as:

- 1D spot spectra of interesting phases.
- 2D images of samples that show interesting functional groups.
- 3D (hyperspectral) spectrum images with dimension (X, Y, E).

Data will be recorded in OMNIC format and converted to CSV and HDF formats for delivery to the SAMIS system via a python pipeline.

**Facilities:**

Advanced Light Source (ALS) at Lawrence Berkeley National Laboratory

The Advanced Light Source (ALS) is a 1.9 GeV synchrotron light source located at the Lawrence Berkeley National Laboratory (LBNL). The [ALS Infrared Program](#) has far-field and near-field infrared microscopes available for users that have diffraction-limited (2–10 µm in the mid-infrared) and AFM tip-limited (< 25 nm for all wavelengths) spatial resolution, respectively. The far-field microscopes at Beamlines 1.4 and 5.4 are capable of rapid diffraction-limited (point) imaging with a standard frequency range of 600–11000 cm$^{-1}$ (extendable down to ~200 cm$^{-1}$ with the use of other detectors). Beamline 2.4 is equipped with an infrared microscope with a focal plane array detector, which enables rapid image acquisition with fields of views spanning 700 x 700 µm$^2$ (5.5 µm pixel size) to 84 x 84 µm$^2$ (0.66 µm pixel size), depending on the choice of objective and magnification optics. The near-field microscopes at Beamline 2.4 and 5.4 are capable of nanospectroscopy measurements between 300 and 4000 cm$^{-1}$ with the synchrotron source, as well as nano-imaging between 1160 and 1870 cm$^{-1}$ with quantum cascade laser sources.

# Elements and Isotopes

The Sample Elements and Isotopes Working Group (SEIWG) oversee element and inorganic isotope measurements. These analyses are aimed at measuring the elemental abundances and isotopic signatures of returned samples from Bennu to test OSIRIS-REx hypotheses.  This section describes the following techniques.

# 5.17 Secondary Ion Mass Spectrometry (SIMS)

**Overview:**

Secondary ion mass spectrometry (SIMS) is an analytical technique used for the isotope and trace-element analyses of solid samples. It requires a primary ion beam of keV energy level to produce secondary ions from a specimen. Emitted secondary ions can be separated according to their mass to charge (m/z) ratios, allowing for unique isotope identification. SIMS permits spatially resolved isotope and trace element analysis at the micrometer scale and allows for identification of heterogeneities within samples.

**How it works:**

SIMS can analyze all elements and isotopes from H to U in a solid. The detection limit is usually parts per billion level for areas of a few micrometers. Point, line, and image analysis can be obtained with a lateral resolution of sub-micrometer scale and depth resolution of 10 nm. Three-dimensional imaging can be possible if prolonged sputtering of the surface is allowed.

To stimulate the emission of secondary ions from a specimen, a high-energy beam of particles (usually oxygen or cesium) is accelerated and focused onto the sample being studied. Atoms of the sample are sputtered from the surface. Fewer than 1% of sputtered atoms are ionized as secondary ions. The secondary ions are then introduced to the mass spectrometer and separated along their m/z to make a mass spectrum. The intensity of ions with a given m/z is proportional to the abundance. The intensity of each m/z is transformed to chemical composition and isotopic composition by referencing to measurements of a standard material.

Standards should have a similar bulk composition to the unknown (should be matrix-matched) and the abundance of the elements of interest must be known. Standards and samples are measured under the same conditions. For isotope analysis, matrix-matched standards with known isotope abundance are necessary to correct for instrumental mass fractionation

**Sample preparation:**

SIMS analysis may require special preparation of the samples. The surface of the sample must be flat, relative to the size of the interrogating beam. Flatness is typically achieved through grinding, lapping, and mechanical polishing. The surface of the sample must be conductive to avoid charge buildup on the surface. Typically, a thin (few nanometers) layer of carbon or gold is applied using a vacuum evaporation method.

**Sample impact:**

SIMS is a destructive technique. The primary beam has lateral sizes from about 0.5 to 20 µm, depending on techniques and applications, and sputters the sample surface to make a crater (see Data Quality, below). The primary beam implants into the surface at a depth of < 100 nm, depending mainly on the impact energy of the primary beam. After the analysis, the sample surface (several hundreds of atomic layers deep) remains contaminated by the primary beam. The atomic arrangements are changed (destroyed) by the implanted particles and collisions of atoms.

**Data quality:**

SIMS detection limits in favorable cases can extend to less than parts per billion. The precision of SIMS is generally limited by statistics of secondary ion intensities, with instrumental precision typically 5 to 10 % for trace elements, and permil to 0.1 permil for isotope abundances.

The spatial resolution of SIMS for spot analysis and for scanning ion imaging is controlled by the size of the primary beam, which is controlled by the brightness of the source and the lenses of the primary ion column. The primary beam size is confirmed by observations of the sputter crater after measurements. For example, the smallest primary beam size of the Cameca IMS 1280 is somewhat less than 1 µm in diameter and for NanoSIMS is ~100 nm for both cesium and oxygen (with the Hyperion source).

The spatial resolution of SIMS can also be controlled by aberrations of secondary ion optics. For example, the smallest lateral spatial resolution achieved by the Cameca IMS 1280 with the Stacked CMOS-type Active Pixel Sensor (SCAPS) detector is about 0.1 µm for 2D imaging. SCAPS is a solid-state, imaging detector.

The mass resolution of SIMS depends on the system and is generally reported as M/ΔM. Resolving molecular interferences on transition metals of mass ~20-~30 (Ca to Zn) requires mass resolving powers of ~2500 to ~10,000. For the Cameca IMS 1280, the maximum useful mass resolving power is M/ΔM ~15,000.

The useful yield of SIMS refers to the fraction of sputtered isotopes that are ionized and detected. It is dependent on the element and the instrument and ranges from $10^{-4}$ to $10^{-2}$ for the Cameca IMS 1280.

**Data products:**

The data products expected from SIMS measurements include:

- 0D point analyses (individual data set from a specific location)
- 1D line scans (individual data set plotted as a function of distance)
- 2D images (2D maps of a sample in which every pixel contains a data set)
- 3D images (3D maps of a sample surface in which every voxel (3D pixel) contains a data set)

These products can be used qualitatively to identify what elements are present and where they occur. They can be used quantitatively to determine chemical and isotope compositions at the local scale and for regions of interest. All products can be delivered as text files.

**Facility(ies):**

Isotope Imaging Laboratory (IIL), Hokkaido University, Sapporo, Japan

The IIL has three SIMS systems for extraterrestrial material measurements:

- Isotope microscope I: Cameca IMS 1270 with SCAPS detector.
- Isotope microscope II: Cameca IMS 1280 with SCAPS detector.
- Cameca IMS 6f.
- See also the secondary neutral mass spectrometer (SNMS) section for a description of the IIL Isotope nanoscope the JEOL LIMAS.

Centre de Recherches Pétrographiques et Géochimiques – Centre national de la recherche scientifique (CRPG-CNRS), Nancy, France

The SIMS analytical platform at CRPG runs two SIMS:

- CAMECA IMS-1270E7 and IMS-1280HR2. These instruments are both equipped with the new Hyperion radio-frequency (RF) oxygen source and $10^{12}$ Ω resistors for Faraday cups developed in collaboration with CAMECA-AMETEK. These two upgrades allow significant ameliorations of the analytical precision and sensitivity with a primary beam smaller by a factor of 5 to 10 than obtained with a classical oxygen source.

M. Keck Cosmochemistry Laboratory, University of Hawaii at Manoa, HI, USA

A modified CAMECA IMS-1280 ion microprobe with Cs microbeam and RF oxygen sources. The monocollector system is equipped with electron multiplier and Faraday cup collectors. The multicollector system has five collectors, each of which can be operated with either an electron multiplier or a Faraday cup. The IMS-1280 is being upgraded so that the Faraday cups use either $10^{11}$ or $10^{12}$ Ω resistors. The instrument can be operated as either an ion microprobe to measure individual spots down to 0.5 to 1 µm in diameter or in scanning ion imaging mode, where the beam is scanned over the sample and images with the resolution of the primary ion beam are recorded. The IMS-1280 is also equipped with a SCAPS imaging detector that permits quantitative direct ion imaging of samples with a spatial resolution of <1 µm that is controlled by the ion optics of the mass spectrometer.

SIMS Laboratory, Department of Earth & Space Sciences, UCLA (accessed through PSP user status)

The CAMECA IMS-1290 ion microprobe has an identical mass spectrometer to that of the CAMECA IMS-1280HR, which has improved electronics and features much higher mass resolution (up to flat-top, M/ΔM = 40,000) compared to the UCLA IMS-1270. The UCLA IMS-1290 has a Cs microbeam and Hyperion-2 RF oxygen plasma source manufactured by Oregon Physics.

# 5.18 Nanoscale Secondary Ion Mass Spectrometry (NanoSIMS)

**Technique name:** <u>Nanoscale Secondary Ion Mass Spectrometry (NanoSIMS)</u>

**Overview:**

Nanoscale secondary ion mass spectrometry (NanoSIMS) is an analytical technique for the in situ isotopic and trace element analysis of ultra fine grains. A NanoSIMS instrument has a unique coaxial lens design that extends traditional SIMS analysis to extremely small spatial scales, achieving lateral resolutions down to 50 nm and small sputtering depths. NanoSIMS attains extremely high sensitivity at high mass resolution and can also measure up to seven masses in parallel. These features allow for detailed elemental and isotopic analysis of sub-micron grains and inclusions from astromaterial samples and the investigation of micron to sub-micron scale heterogeneities.

**How it works:**

A 16 keV focused ion beam bombards the sample surface. Energy from this primary ion beam is transferred to the surface and near-surface regions, causing elastic collisions between atoms. The energy generated by elastic collisions extends into the sample, initiating the ejection of atoms and molecules that have attained a kinetic energy greater than the surface binding energy. About 1 % of sputtered atoms are ionized. Negative secondary ions are generated by using $Cs^+$ primary ions, and positive secondary ions are generated by using $O^-$ primary ions. The generated secondary ions have a spread in kinetic energy, or velocity, and mass-to-charge ratio (m/z). The secondary ions are separated by a mass spectrometer consisting of an electrostatic analyzer that focuses the ions by kinetic energy (velocity) and a magnetic sector that separates the ions by m/z. Thus, ions of a given m/z that initially have a velocity spread are focused to the same spot. This beam focusing in energy and mass results in a higher mass resolution. In previous-generation ion microprobes, a high mass resolution was attained by sacrificing beam transmission. The improved optics and larger magnet of the NanoSIMS instrument allow for high sensitivity (beam transmission) at high mass resolution. The secondary ions are detected by miniature electron multipliers that operate by pulse counting. A Faraday cup can also be used for the detection of higher count rates ($10^6$ counts per second). A NanoSIMS instrument has up to six moveable detectors and one fixed detector to allow for multi-detection. Adjacent masses with a mass spread up to 58 can be measured at the parts per billion level.

NanoSIMS has the capability for spot analysis, depth profiling, line scan, and raster ion imaging. Spot analyses typically utilize larger primary beam currents than ion imaging and can thus acquire isotopic compositions with greater precision due to increased counts. In ion imaging, a focused primary ion beam is rastered across a sample's surface. The resultant chemical and isotopic maps can reveal variations across samples or highly anomalous "hotspots". For isotopic analyses, standard materials are used to calibrate the instrument and measurements. Because the secondary ion yield of an element can vary with the sample mineralogy and chemical composition, the mineralogy and chemistry of isotopic standards should ideally be the same as that of the sample.

**Sample preparation:**

NanoSIMS is routinely applied to analyze isotopes in nearly all components in carbonaceous chondrites, including fine-grained matrix material, organic matter, refractory inclusions (calcium-aluminum–rich inclusions), chondrules, presolar grains, carbonates, anhydrous, and hydrated minerals. The samples can be prepared as polished thin or thick sections or blocks, fine-grain dispersions on a substrate such as high-purity Au, samples pressed into Au or In, or electron-transparent sections produced through ultramicrotomy or focused-ion-beam preparation. High levels of sample flatness are required for the highest precision measurements, ideally approaching the diameter of the incident primary beam. For non-conductive samples (such as most meteorite samples), a conductive coating is necessary to avoid charge buildup on the sample surface. Typically, a ~10-nm-thick coating of C or Au is applied using a sputter-coater. A thicker coating is necessary for samples such as polished meteorite sections or large crystals that experience greater charge buildup. Because NanoSIMS operates under high vacuum ($10^{-9}$ to $10^{-10}$ torr), only resins (e.g., epoxy) with very low degassing properties should be used, and in minimal amounts. The amount of epoxy typically used in the preparation of thin sections is reasonable. However, the larger amount used in preparing thick sections can increase the vacuum to undesirable levels. Organic matter that will be analyzed for hydrogen isotopes should not be exposed to the transmission electron microscope (TEM) prior to analysis, as this technique is known to induce H isotopic fractionation in some organic materials. These sample preparation considerations also apply to standard or reference materials.

**Sample impact:**

NanoSIMS is a destructive technique, as primary ions bombard the sample and sputter away the surface atomic layers (approximately 1 nm depth). A crater is developed following prolonged sputtering, the depth of which is dependent on the primary ion beam current used and the duration of the measurement. This allows for depth profiling of a sample if that is desired. The implantation of primary ions enhances the production of secondary ions but contaminates the sample surface. The uppermost tens of nanometers become implanted with primary ions and are also amorphized. The sputtering rate and exact depth of beam damage depend on the sample phase and composition (e.g., mineral phase, glass, carbonaceous, organic), beam current, and analysis time. Samples are also impacted due to necessary sample preparation as described in the previous section.

**Data quality:**

The detection limit of NanoSIMS can extend to parts per billion levels. Precisions achieved are on the permil level when using electron multipliers. With Faraday cups, sub-permil precisions can be achieved. The precision of isotopic ratios is largely governed by Poisson counting statistics of secondary ions of the least abundant isotope. However, other instrumental factors can also influence measurement precision, including aging of the electron multiplier(s), quasi-simultaneous arrivals (QSA), sample topography, and charging. Aging of electron multipliers occurs most significantly with high count rates (> 300,000 counts per second) and leads to a shift in the peak height distribution and consequently a drift in the isotopic ratio during an analysis; e.g., the $^{18}O/^{16}O$ and $^{17}O/^{16}O$ ratios increase over time if the $^{16}O$ count rates continuously exceed 300,000 cps. However, there are methods to alleviate detector aging, such as applying a higher voltage on the detector. QSA is prominent in NanoSIMS due to its high transmission and results in undercounting of the major isotope. QSA occurs when a single primary ion produces multiple secondary ions that arrive at the detector at approximately the same time, causing these ions to be registered as a single pulse. The QSA effect varies with element and must be corrected for. Faraday cups do not suffer from aging or QSA, but they require higher count rates and beam currents compared to electron multipliers, both of which degrade spatial resolution. Moreover, ion imaging cannot be performed with a Faraday cup. Therefore, the choice to analyze with electron multipliers or Faraday cups depends on the sample size, spatial resolution required, precision required, mode of analysis (spot or imaging), and isotope systems of interest. Instrumental mass fractionation is very sensitive to both sample charging and sample topography (roughness), and these factors affect measurement reproducibility. To mitigate charging, an electron flood gun can be used for charge compensation.

NanoSIMS achieves a high lateral resolution with primary beam sizes down to 50 nm using low beam currents. The lateral resolution is approximately three times poorer for $O^-$ primary ion beams generated with a duoplasmatron compared to a radio-frequency plasma ion source. NanoSIMS instruments equipped with both of these sources will be available to the OSIRIS-REx sample-analysis team (see below). Mass resolving powers of at least several thousand can be achieved, enough to resolve isobaric interferences from most of the geochemically important elements or isotopes.

**Data products:**

NanoSIMS measurements can be performed as spot analyses, depth profiles, line scans, or raster ion images. Data files for spot, depth, and line analyses are saved in Cameca file formats (.stat, .statc) that can be read using MS Excel or Word. Ion images require specialized software to read and process. For these analyses, multi-layered image stacks are acquired for an area (generally <25 × 25 $\mu$m$^2$) and stored in native Cameca file format (.im). Analysis conditions are saved in the image header. Customized software is used to process and apply corrections to isotope data within the images. Processed isotopic ratio data of subregions (e.g., presolar grains or CAIs) within the images are saved to Excel spreadsheets. The software allows for visualization and export (in .tiff, .png, or .jpg formats) of isotope images. Metadata are saved in files (.chk_im) that are automatically generated for each image analysis and are readable as a Word document. For each type of measurement, the data products can yield qualitative or quantitative elemental and isotopic information.

**Facilities:**

NASA Johnson Space Center, Houston, TX, USA

- The Astromaterials Research and Exploration Science (ARES) Division houses a second-generation NanoSIMS instrument, the NanoSIMS 50L. This instrument has a larger magnet than the first generation NanoSIMS 50 and is capable of multi-detection of seven isotopes. The NanoSIMS 50L has duoplasmatron O and Cs primary ion sources, a secondary electron detector, and an electron flood gun. This instrument will receive an upgrade to a radio-frequency plasma O primary ion source in 2021/2022.

Stanford University, Stanford, CA, USA

- The Stanford Nano Shared Facilities houses a second-generation NanoSIMS 50L. The larger magnet allows for simultaneous multi-collection of seven isotopes. Each of the seven detectors can be swapped from electron multiplier (EM) to Faraday cup. The Stanford NanoSIMS is equipped with duoplasmatron O and Cs+ primary sources, a secondary electron detector, and an electron flood gun. In 2021/2022, the Stanford NanoSIMS acquired a custom-designed air-free glovebox for sample exchange. This glovebox allows for transfer of oxygen-sensitive materials into the sample airlock. The glovebox is back-filled using either Ar or N2 gas.

The Open University, Milton Keynes, UK

- The School of Physical Sciences houses a second-generation NanoSIMS instrument, the NanoSIMS 50L. This instrument has a larger magnet than the first generation NanoSIMS 50 and is capable of multi-detection of seven isotopes. The NanoSIMS 50L has duoplasmatron O and Cs primary ion sources, a secondary electron detector, and an electron flood gun.

The University of Arizona, Tucson, AZ, USA

- The Lunar and Planetary Laboratory will receive an upgraded NanoSIMS 50L instrument in 2022/2023. The tool will have the capability to detect up to seven isotopes simultaneously and will have Cs and radio-frequency plasma O primary ion sources, a secondary electron detector, and an electron flood gun.

Washington University, St. Louis, MO, USA

Cameca NanoSIMS 50: This is a novel type of ion microprobe whose characteristics make it the ideal workhorse instrument for isotopic and elemental studies of sub-micron presolar grains. The features of this instrument are:

- Cs+ primary beam of nominally 100 nm diameter, made possible by the unique coaxial design of the ion optics allowing for a normal incidence primary beam.
- High secondary ion yield (sensitivity) at high mass resolution due to the closeness of the secondary ion extraction optics to the sample.
- Multicollection of secondary electrons and up to 5 masses with movable detectors along the focal plane of the spectrometer.
- Automated particle definition in which discrete grains can be identified by image processing algorithms from ion images and then analyzed individually by rastering the primary beam over them for precise isotopic determination.
Measurement of TEM grids with minimal mechanical stress on the substrate and the ability to reanalyze in the TEM due to in-house development of unique sample holders for combined TEM-NanoSIMS studies.
Hyperion H201 RF Plasma Source: this new source replaces the duoplasmatron and greatly increases the capability of the NanoSIMS for measurements of positive secondary ions. The Hyperion primary ion beam of 16O- is brighter and more stable than the duoplasmatron, and can be run at small spot sizes with relatively high beam currents (e.g., 150-nm beam diameter at 10~pA).

# 5.19 Sensitive high mass resolution ion microprobe (SHRIMP)

**Overview:**

SHRIMP instruments are based on secondary ion mass spectrometry (SIMS). The SHRIMP behaves as an ion microprobe, i.e., it utilizes a static primary beam spot and collects secondary ions for transmission into the mass analyzer. However, it is designed to maximize the transmission of secondary ions through the mass spectrometer, yielding high sensitivity. Developments in the detector system allow isotope ratio measurements on Faraday cup electrometers in both current mode and charge mode; the latter having a very low inherent noise.

**How it works:**

The SHRIMP sputters a volume of the target to produce secondary ions. A negative molecular $O_2$ primary beam is typically used to sputter metals (left hand of the periodic table), and a positive Cs primary beam is used to sputter non-metals (right hand of periodic table) based on the propensity of elements to give up or accept an electron. Sputtering yields are related to several factors including the specific elements and the matrix. Electronegative oxygen and electropositive cesium enhance the ionization yields relative to inert species such as $Ar^+$ or $Ga^+$. For the $Cs^+$ primary ion beam, we use an electron beam with 2-keV electron energy at the target to neutralize charge buildup.

A feature of the SHRIMP configuration (Ireland et al. 2008) is the use of an optical microscope to place the sample plane in focus, thereby aligning the primary, secondary, and electron beam ion-optical axes. This minimizes the time sputtering the sample to otherwise optimize extraction conditions.

Much of the work done on SHRIMP instruments relates to the high-precision analysis of terrestrial materials. SHRIMP was originally developed to carry out a wide variety of measurements but found a niche in U-Pb analysis for geochronology, particularly from the mineral zircon (Froude et al. 1983). However, work was also carried out in stable isotopes (measurement of $S^+$ for $^{34}S/^{32}S$ analysis; Eldridge et al. 1991), cosmochemistry (isotope anomalies in Ti, Mg (Ireland et al. 1985, 1986), and in trace element analysis (Ireland and Wlotzka 1992). Recent cosmochemical applications have involved presolar grains (Ávila et al. 2013; Ireland et al. 2018) and Mn-Cr systematics (McKibbin et al. 2015). With the development of a negative ion capability on SHRIMP II and SHRIMP SI (see Facilities), a range of applications related to stable isotopes and water analyses is available (Ireland et al. 2006; Ávila et al. 2020; Turner et al. 2015).

**Sample preparation:**

As for all SIMS, SHRIMP analysis may require special preparation of the samples. All samples are solids. Samples can be embedded in epoxy or metal and require polishing and application of conductive metal coats (Al, Au, Cu, Ag, etc). Metal coats are applied in either a thermal evaporation unit or a sputter coater. We typically avoid C coats because of the poor conductivity and the poor vacuum quality in some C-coating apparatus, causing surface contamination. SHRIMP is relatively insensitive to some surface topography, and voids in the sample surface are only problematic in the horizontal plane of the incoming primary beam.

**Sample impact:**

SHRIMP analysis is a destructive technique. The main applications are for high-sensitivity isotopic analysis. Our preferred method is to use a large spot over a shallow depth so that context of the analysis is maintained. The typical spot diameter for minor isotope analysis is 25 µm with depth to a few microns. Smaller spot sizes are available, but these limit the number of secondary ions; halving the spot size usually leads to a factor of 4 reduction in secondary ions. The primary beam is typically configured to a Kohler mode of operation where the final lens is used to image a Kohler aperture, thereby defining the spot at the surface (Ireland 1995). This configuration produces well-defined primary beam spots with uniform intensity across the spot.

**Data quality:**

For a 25-µm spot, typical oxygen isotope $D^{17}O$ uncertainty can get down to 0.1‰ (1standard deviation). Grouped analyses can resolve the differences between martian shergottite, nakhlite, and chassignite (SNC) meteorites and terrestrial and main pallasite group olivines (Loiselle et al. 2019). The SHRIMP SI can operate with full resolution of the $OH^-$ signal from $^{17}O^-$ even in extremely hydrous minerals such as serpentine. The $OH^-/O^-$ signal can be used to determine water content and then to distinguish between different minerals sputtered during the course of the analysis.

For S isotope analysis, typical uncertainties for $δ^{34}S$ are around ±0.2‰; for $Δ^{33}S$, ±0.05‰; and for $Δ^{36}S$, 0.2‰, for a single spot analysis. Analyses of pyrite and troilite/pyrrhotite are routine (Li et al. 2020).

For U-Pb analysis, Pb sensitivity on the order of 25 c/s/ppm/na are routinely achievable. Oxygen flooding, as commonly used in CAMECA-brand methodologies, is not required. The reason for this is likely related to using a lower energy $O_2^-$ primary beam (the $O_2$ spits at the surface, depositing lower energy per $^{16}O$ ion, and therefore stays at the surface, enhancing secondary ionization).

**Data products:**

SHRIMP typically produces volume analyses of target materials. That is, the data are related to the integrated signal from a given spot. Data will show any variation of secondary ions with depth, allowing reconstruction of mineral variation during an analysis that can also be examined post analysis on a scanning electron microscope. Data are stored in .xml files with all instrument parameters recorded in the same file. Spot images from the optical microscope are taken after analysis to verify the location. Data are typically reduced using an in-house data reduction program (POXI).

**Facility:**

Australian National University, Canberra, Australia

There are three variants of SHRIMP instrument (the original SHRIMP I is now decommissioned):

- SHRIMP II is the commercial prototype capable of positive and secondary ion analysis. It is regularly used as the preferred instrument for U-Pb geochronology.
- SHRIMP RG (Reverse Geometry) provides positive secondary ion analysis only with an ultra-high-resolution single collector mass spectrometer capable of mass resolution over 30,000 (M/ΔM).
- SHRIMP SI (Stable Isotope) is a dedicated stable isotope instrument configured solely with a Cs gun primary beam. It utilizes charge mode electrometer collection to provide high-precision measurements of minor stable isotopes (e.g. $^{17}O$, $^{36}S$) at precision levels of 0.2‰ per spot.

Stanford University, Stanford, CA, USA

The Stanford-USGS Micro-Analytical Center (SUMAC) houses the SHRIMP-RG (reverse geometry) ion microprobe, and is cooperatively managed between Stanford University and the U.S. Geological Survey. It is a large format magnetic sector mass spectrometer that utilizes a single collector EPT discrete-dynode electron multiplier in pulse counting mode and an off-axis Faraday cup. The reverse geometry provides higher mass resolution than other large-format ion microprobes while retaining high transmission of secondary ions and less need for energy filtering. The Stanford SHRIMP-RG can be operated with a O-, O2-, or Cs+ primary beam, though it is primarily used for positive secondary ion analysis. Typical beam diameter ranges from 10 to 40 microns, depending on the primary current and instrument setup.

# 5.20 Thermal Ionization Mass Spectrometry (TIMS)

**Overview:**

Thermal ionization mass spectrometry (TIMS) is a technique for high-precision measurement of isotope ratios. After purification of the element of interest, the sample is loaded on a filament, which is heated to ionize the sample. Ions are then accelerated and focused by applying electrostatic potentials and mass-sorted using a magnet before hitting the array of multiple detectors for simultaneous detection of ions with different masses. TIMS yields data for isotope ratios in the parts per million range, which is only matched by multi-collector inductively coupled plasma mass spectrometry (MC-ICP-MS) analyses.

**How it works:**

The purified element of interest (see Sample preparation, below) is dissolved in an acidic medium and loaded onto a metal filament (made of, e.g., Re, W, or Pt) with an activator solution (a reagent that increases the ionization of the sample). The filament is heated under a high vacuum by an electric current. This volatilizes the loaded sample, and a portion of the desorbed atoms are ionized. This ionization can be described by the Saha-Langmuir equation. Elements with low first ionization potentials (e.g., Rb, Sr) are most easily ionized, producing +1 ions. However, elements with high ionization potentials (e.g., Zr) are not easily ionized, while others (e.g., Os) are best measured as negative ions ($OsO_3^-$).

After ionization, ions generated in the source are focused and accelerated by a series of electrically charged plates (lenses). They are then sorted by mass and energy with the help of a magnet before ion collection in a detector array (Faraday cups/electron multipliers/Daley detectors) to determine the intensity of ion beams of chosen isotopes simultaneously (multi-collection). In such a way, isotope ratio data with precisions as high as a few parts per million are achieved. However, such ultrahigh precision can only be obtained after careful purification of the target isotopes from the sample matrix, to avoid interferences and matrix effects. Moreover, due to the limited ionization of certain elements, the isotopes of only one element can be measured for each analysis. TIMS is the method of choice for specific elements (e.g., Sr, Nd, Pb).

**Sample preparation:**

The sample needs to be pristine and devoid of terrestrial contamination. The target sample material is physically separated from the main mass. The entire sample preparation procedure is carried out in a clean laboratory suite. The samples are crushed and weighed. This is followed by an acid digestion procedure using ultraclean labware to achieve low blanks and avoid contamination. The desired element is extracted from the sample matrix and purified using element-specific chromatographic separation procedures. These generally include several steps using carefully chosen chromatographic resins. After achieving a high degree of element purification, the sample is dissolved in an acid solution and a potential activator solution is added. The resulting sample is then carefully placed onto the filament, which is outgassed through short heating prior to loading into the mass spectrometer source.

**Sample impact:**

TIMS is a destructive technique. Samples are dissolved in an acidic medium and consumed by the mass spectrometric analyses.

**Data quality:**

TIMS provides high-precision isotopic ratio data in the range of a few to 100 parts per million (e.g. for Cr and Sr isotopes). Depending on the desired precision, 0.1 to 2 ug of the target element is needed. For Cr isotopes, when only a limited amount of Cr is available, TIMS achieves superior precision compared to MC-IMP-MS. This precision changes with larger samples and greater element concentrations.

**Data products:**

The data products expected include:

- High-precision isotope ratios (e.g. Cr, Sr, Ba, Pb) of bulk material or separated sub-components (chondrules, refractory inclusions, matrix)

These data can be used for:

- Determination of nucleosynthetic isotope compositions for comparison to known meteorite classes and identification of genetic relationships (e.g. carbonaceous versus non-carbonaceous material)
- Age determinations, such as Sm-Nd and Mn-Cr ages, depending on the exact nature of the sample material (suitable mineral occurrence and compositions)

TIMS data formats are determined by the company (Thermo Fisher Scientific) producing the commercial instruments. The software provides four files per run (one ASCII, one text file, and two data files that are read by the Triton software only). They are usually 3 to 4 MB together. The MS Excel reduction files (meta files) for data evaluation are typically about 2 MB per file. This amounts to a total of ca. 6 MB per measurement.

**Facility:**

Institute of Geochemistry and Petrology, ETH Zurich, Switzerland

The facility includes a dedicated suite of ultra-clean laboratories for the preparation of terrestrial and extraterrestrial samples. It features:

- Triton TIMS. The mass spectrometer is equipped with nine Faraday cups, three SEMs, and three compact discrete dynode detectors (Wotzlaw et al., 2017; Henshall et al., 2018).

Lawrence Livermore National Laboratory, CA, USA

The facility includes a dedicated suite of ultra-clean laboratories for the preparation of terrestrial and extraterrestrial samples. It features:

- Thermo Fischer Triton TIMS

# 5.21 High-resolution Inductively Coupled Plasma Mass Spectroscopy (HR-ICP-MS)

Write up for HR-ICP-MS and LA-ICP-MS

**Overview:**

High-resolution inductively coupled plasma mass spectrometry (HR-ICP-MS) is an analytical technique used for the quantification of bulk or mineral specific elemental abundances within a sample. It is a type of mass spectrometry that is capable of measuring the abundances of metals and non-metals at concentrations as low as parts per billion by weight (ng/g). This sensitivity is achieved by ionizing the sample with an inductively coupled plasma and then using a mass spectrometer to separate and quantify those ions. For bulk elemental abundances, samples are introduced to the plasma as a solution prepared using acid digestion. For spatially resolved analysis, material is ablated from the sample surface by a laser and is then introduced to the plasma.

**How it works:**

These analyses are performed by introducing the sample to the plasma, which generates ions for analysis by the mass spectrometer. The sample introduction method varies depending on the type of analysis (bulk or spot), which is discussed in more detail below. The plasma is a mixture of ions, electrons, and atoms that has an extremely high temperature (up to 10,000 K). At these temperatures, most elements transition from atoms to singly charged positive ions, making the mass spectra relatively simple to interpret.

The plasma is generated using a radio frequency (RF) generator. This system sends a high power (~1.5 kW) RF signal through a load coil wrapped around the ICP torch. A steady stream of Ar gas is introduced to the torch. The plasma is formed when electrons from an electric spark source cause Ar atoms to ionize within the torch gas flow. When these ions collide with other Ar atoms, they cause an ionization cascade that forms the plasma.

The sample interface is where the plasma is introduced to the high vacuum of the mass spectrometer system. This system is comprised of a pair of cones that allow ions from the sample to enter the high vacuum region. The sampler cone allows only the central channel of the plasma (where the sample ions are formed) into an intermediate vacuum region. In this region, the plasma undergoes supersonic expansion and the ions are pulled into the high vacuum region by passing through a second, skimmer cone. A negatively charged extraction lens is placed behind the cone to aid in the ion extraction through the skimmer cone.

Once they pass through the high-vacuum region, elemental masses of the sample ions are separated using a mass analyzer. There are two types of mass analyzers typically used in ICP-MS, quadrupole and magnetic sector. Quadrupole mass analyzers are sequential, so each element is measured in sequence. Magnetic sector analyzers can be either sequential or simultaneous depending on the geometry of the instrument components.

*Acid dissolution bulk analysis*

Acid dissolution HR-ICP-MS is a technique for analysis of bulk abundances of major elements (>1.0 wt%), minor elements (1000 ng/g – 1.0 wt%), and trace elements (0.1 ng/g – 1000 ng/g). The elements to be measured must meet the following criteria:

- Have relatively high ionization efficiencies in the plasma environment (true for most elements except for the halogens and noble gases).
- Are free of isotopic interferences or such interferences can be corrected for. Interferences arise from the Ar used to generate the plasma, molecules from air entrained in the plasma, isotopes of other elements present at higher abundances, oxides of higher abundance elements, and other sources. Most interferences caused by other elements in the solution can be resolved by careful selection of the resolution mode used.
- Are present in sufficient abundance in the sample.
- Are not lost during the acid dissolution process. For example, Si cannot be measured with this technique due to loss of this element as $SiF_4$ during preparation. Highly volatile elements can also be lost during this process.

The sample is introduced to the plasma as a liquid using a nebulizer and spray chamber. A peristaltic pump is used to sample the solution at a constant rate and deliver it to the nebulizer and spray chamber. The nebulizer uses supersonic expansion of the Ar carrier gas to turn the analyte solution into a fine mist, and the spray chamber then removes any droplets that are too large to be processed in the plasma. This occurs at the sample interface of the instrument.

*Laser ablation spot analysis*

The abundances of elements in individual mineral phases can be determined in situ by laser-ablation ICP-MS (LA-ICP-MS). Each analyzed spot is ablated by laser pulses at a high repetition rate, e.g., 10 Hz. Spatial resolution (see Sample impact) varies depending on the mineral under study and the abundance of the target elements. Larger spots produce more ions and hence more counts. Ideally, a single mineral phase is ablated. The mass spectrometer sweeps over the desired mass range while the ablated material is carried by Ar gas from the ablation chamber to the sample interface of the mass spectrometer. The signal from the transient laser ablation pulse is integrated over a period of time that begins before ablation and continues after the ablation ceases. This ensures sufficient time for the pulse to reach peak intensity and then decay to background levels.

**Sample preparation:**

For HR-ICP-MS bulk analyses, sample aliquots ranging from 50 to 100 mg must be dissolved using a series of acids such as concentrated $HF$-$HNO_3$-$HClO_4$. Once all solids are dissolved, the solution is evaporated at elevated temperature (typically 110 °C) and redissolved in 5% $HNO_3$ to produce a solution with a target amount of total dissolved solids, typically ~100 µg/ml. An analytical blank solution is prepared using the exact same procedure in parallel with the samples. Trace element standards are prepared that consist of successive dilutions of the trace elements to be quantified in a matrix solution. The matrix solution contains constant major element abundances in bulk proportions as close as possible to the sample. Successive dilutions of this matrix solution produce a standard series that are used to quantify major elements. Trace-element solution standards are targeted to cover six orders of magnitude (10 pg/ml to 1 µg/ml) in concentration to bracket expected elemental concentrations in the sample solution.

LA-ICP-MS analyses require a polished sample surface and the sample should be at least 100 µm thick. A typical preparation is a petrographic thick section, prepared in a similar manner to a thin section but ~3x thicker.

**Sample impact:**

HR-ICP-MS is a destructive technique. Samples are completely digested in acid solutions. Depending on the phases present in the sample, there may be acid insoluble residues left over from sample preparation. Sample mass for each aliquot should range from 50 to 100 mg. Assessment of sample heterogeneity requires repeated analysis of multiple aliquots.

LA-ICP-MS is a partially destructive technique. Each analysis will produce laser-ablated pits that can range from 25 to over 100 µm in diameter and are typically 10 to 25 µm deep.

**Data quality:**

HR-ICP-MS detection limits in favorable cases can extend down to parts per billion by weight (ng/g). The precision of HR-ICP-MS varies for each element as a function of its concentration in solution or the ablated mineral, the mass resolution required to resolve interferences (with higher resolutions yielding lower count rates), and the abundance of the element relative to the standards. Highest precision is obtained when the standard abundances bracket the elemental concentration in the sample, allowing the abundance to be determined using interpolation of the calibration curves.

**Data products:**

Quantification of bulk elemental abundances is achieved by comparison of the ion counts from the sample solution to that of a set of standard solutions, after subtracting signals from analytical blanks. Analysis of a range of standard solutions, with varying elemental concentrations, is used to produce a series of calibration curves, which plot ion counts vs. solution abundance for each element. The elemental abundance in the sample is determined by interpolating (ideally) or extrapolating the count rate from the sample against the calibration curve for each element of interest.

The data products expected from HR-ICP-MS bulk measurements include:

- Instrumental scan profiles, which indicate the count rate per ion as a function of time. These profiles are collected for each element in the sample, the standards, and the analytical blanks. If multiple mass resolutions are used, then mass resolution is indicated.
- Regions of stability in the scan profile. For each sample and standard, the user must identify a region of each scan profile where the signal was stable. This region is integrated to obtain a total ion count rate per element. A background subtraction is performed by subtracting similar regions for the analytical blanks from each sample and standard.
- Calibration curves for each element. These curves are generated by integrating ion counts over a specific time period in the scan profile and plotting the values vs. the known concentration of the element in each standard. These curves are typically fit using a linear regression.
- Sample elemental abundance tables. The abundance of each element is determined by plotting the ion counts from the sample on the relevant calibration curve, converting ion counts to solution abundances.
- Bulk elemental abundance tables. The concentration of each element in solution must be converted to bulk elemental abundances in the original sample. This simple calculation considers the total mass dissolved along with the final concentration of total dissolved solids in the sample solution.

Quantification in LA-ICP-MS is achieved by comparison of the ion counts (peak or integrated over a defined time) from the sample to that of a set of standards, after subtracting the blank signal. Instrumental sensitivity factors for each isotope are determined by measuring signal intensity from the solid standards which have known concentrations of the elements of interest. Elements of interest for LA-ICP-MS depend on the mineral phase under analysis. Typically sets of lithophile, siderophile, and chalcophile elements are defined for analysis of silicates, metals, and sulfides, respectively. Because of the transient nature of the signal, fewer elements are analyzed per spot compared to bulk techniques.

**Facilities:**

Pacific Centre for Isotopic and Geochemical Research (PCIGR), University of British Columbia, Vancouver, BC, Canada

This facility contains eleven mass spectrometers, together with sample preparation laboratories and the research and support teams.

The HR-ICP-MS systems at PCIGR include:

- A Thermo Scientific Element 2 sector-field double-focusing HR-ICP-MS. This instrument is capable of analyzing many elements in solution at concentrations as low as parts per trillion (ppt) and has a dynamic range of over nine orders of magnitude. Samples are introduced to the instrument using the direct nebulization system in the spray chamber.
- A Nu AttoM double-focusing, high-resolution magnetic sector mass spectrometer. The instrument is designed to provide precise and accurate isotope ratio analysis and elemental analysis of most elements in solution or in situ down to sub-ppt levels. The Nu AttoM can be connected to the CETAC Aridus II Desolvating Nebulizer System or the ESI Apex Desolvating Nebulizer System. The Nu AttoM is also connected to the Excimer Laser Ablation System RESOlution M-50-LR at PCIGR. Split-stream LA-ICP-MS is possible by coupling both the Nu AttoM and the Agilent 7700x to the same laser ablation system. This technique allows for the simultaneous measurement of isotope ratios and trace element concentrations on the same sample run.

Inst. für Geochemie und Petrologie, ETH Zurich, Switzerland

The facility includes a dedicated suite of ultra-clean laboratories for the preparation of terrestrial and extraterrestrial samples. Among other ICP-MS and TIMS instruments, it features:

- Thermo Scientific Element XR (HR-ICP-MS)

Lawrence Livermore National Laboratory, CA, USA

The facility includes a dedicated suite of ultra-clean laboratories for the preparation of terrestrial and extraterrestrial samples. Among other ICP-MS and TIMS instruments, it features:

- Thermo Scientific Element2 (HR-ICP-MS)

Natural History Museum, London, UK

The facility includes a dedicated suite of ultra-clean laboratories for the preparation of terrestrial and extraterrestrial samples.

- Agilent 7700x ICP-MS with laser (LA-ICP-MS)

# 5.22 Quadrupole and Multi-Collector (MC) Inductively Coupled Plasma Mass Spectrometry (ICP-MS)

See also HR-ICPMS

**Overview:**

Inductively coupled plasma mass spectrometry (ICP-MS) is a widely used geochemical tool for both elemental and isotope-ratio analysis. It enables element analyses at concentrations of 1 part per trillion or even lower depending on the instrument and application. In this technique, the sample is ionized in an inductively coupled plasma. The ions are extracted into a mass spectrometer that separates and quantifies ions of specific masses. Different types of mass spectrometers are combined with an ICP source, such as quadrupole, sector field, and multi-collector (MC) ICP-MS instruments. The MC-ICP-MS instrument is equipped with multiple detectors for the simultaneous detection of ions with different masses. The MC-ICP-MS analyses yield unprecedented precisions for isotope-ratio measurements in the parts per million range, which are only matched by thermal ionization mass spectrometry (TIMS) for certain elements.

**How it works:**

Samples are dissolved in an acidic medium and introduced as liquid samples, which are converted to an aerosol (fine liquid droplets) using a nebulizer. This primary aerosol is subsequently introduced into a spray chamber that filters out droplets with diameters >10 mm. The remaining aerosol is injected into the plasma, where the aerosol droplets evaporate and individual molecules are released and atomized before ionization. The extraction of ions from the plasma into the mass spectrometer requires bridging the pressure difference between the plasma source at atmospheric pressure and the mass spectrometer with a mass analyzer region at $10^{-5}$ to $10^{-9}$ mbar. This vacuum is necessary because collisions between sample ions and residual atoms or molecules present in the mass spectrometer severely affect the ability to separate ions of different masses.

*Quadrupole ICP-MS* is used for element concentration analyses and utilizes a quadrupole filter for mass separation. It is possible to scan continuously over the whole mass spectrum of the periodic table. However, shorter measurement times are achieved by rapid jumping from peak to peak (within a few milliseconds) instead of continuous scanning. The detector has a large linear dynamic range of nine to ten orders of magnitude to detect small to large ion beam intensities. This is typically achieved by combining two detection types (ion counting and analogue mode).

*MC-ICP-MS* is designed to achieve high-precision isotope ratio measurements. The precision achieved by quadrupole ICP-MS instruments is limited to about 2‰ mainly by the peak shape and the single-collector system. In MC-ICP-MS, these shortcomings are mitigated by combining a sector field mass spectrometer with a detector array (generally Faraday cups) that can simultaneously detect multiple isotopes. The instrument provides flat-topped peaks, which are important to ensure that small fluctuations in the ion beam position do not cause spurious results. MC-ICP-MS can provide isotope ratio data with precisions of a few parts per million. However, such ultrahigh precision can only be obtained after careful purification of the target isotopes from the sample matrix to avoid interferences and matrix effects. The purification procedure generally includes ion exchange chromatography.

**Sample preparation:**

The sample needs to be pristine and devoid of terrestrial contamination. The target sample is physically separated from the main mass. The entire sample preparation procedure is carried out in a clean laboratory suite. The samples are crushed and weighed. This is followed by an acid digestion procedure using ultraclean labware to keep the blanks low and avoid contamination. For concentration measurements, after digestion, the solutions are diluted to adequate concentrations with ultrapure acids and are then ready to be analyzed on an HR-ICP-MS. It is also possible to use the isotope dilution technique, which includes the addition of known amounts of an isotopically enriched solution to the analyzed sample. For high-precision isotope analyses by MC-ICP-MS, the desired element is extracted using element-specific chromatographic separation procedures, which generally involve several steps using different chromatographic resins. After achieving a high degree of element purification, the sample is dissolved in dilute acid and is ready for analysis.

**Sample impact:**

This technique is destructive. Samples are dissolved in an acidic medium and consumed by mass spectrometric analyses.

**Data quality:**

Depending on the element, concentrations can be measured at parts per trillion to parts per million with precisions of ca. 5% or better if isotope dilution is used. For isotopic analyses by MC-ICP-MS, the precision ranges from a few to 100 parts per million (e.g. for Cr, Ti and Zr isotopes). Ideally, 1 to 2 ug of each element is separated for high-precision analyses.

**Data products:**

The data products expected from HR-ICP-MS and MC-ICP-MS include:

- Element concentrations (excluding, e.g., noble gases, Hg)
- High-precision isotope ratios (e.g. Cr, Ti, Fe, Ni, Zr) of bulk material or separated sub-components (chondrules, matrix)

These data can be used for:

- Determination of nucleosynthetic isotope composition for comparison to known meteorite classes and identification of genetic relationships (e.g. carbonaceous versus non-carbonaceous material)
- Age determination such as Mn-Cr ages

MC-ICP-MS data formats are determined by the companies (Thermo Fisher Scientific & Nu Instruments) producing the commercial instruments (one binary, one ascii code [.txt] for Thermo Fisher Scientific or one text file [.txt] and one cvs file [.cvs] for Nu Instruments). Raw data files are on the order of 15 to 100 KB each. The MS Excel reduction files (meta files) are typically about 2 MB per file. For a given sample measurement, the total data volume is 6 to 10 MB.

**Facility:**

Institute for Geochemistry and Petrology, ETH Zurich, Switzerland

The facility includes a dedicated suite of ultra-clean laboratories for the preparation of terrestrial and extraterrestrial samples. It features:

- Four MC-ICP-MS (two NeptunePlus, a Nu Plasma II and a large-geometry Nu Plasma 1700)

- Agilent Triple Quad ICP-MS (8800 ICP-QQQ)
- See also TIMS section for Triton TIMS instrument

Pacific Centre for Isotopic and Geochemical Research (PCIGR), University of British Columbia, Vancouver, BC, Canada

This facility contains eleven mass spectrometers, together with sample preparation laboratories and the research and support teams.

The quadrupole ICP-MS to be used at the PCIGR is:

- The Agilent 7700x is bundled with a collision/reaction cell in helium mode that offers higher sensitivity, lower backgrounds, and more effective polyatomic interference removal in complex sample matrices. The 7700x also includes a frequency-matching RF generator. Our instrument, installed in 2010, is equipped with an ASX-500 autosampler. It is routinely used for the analyses of metals at ppb levels.

Lawrence Livermore National Laboratory, CA, USA

The facility includes a dedicated suite of ultra-clean laboratories for the preparation of terrestrial and extraterrestrial samples. It features:

- Thermo Fischer NeptunePLUS MC-ICP-MS
- Thermo Fischer Neoma MC-ICP-MS
- see also TIMS section for the Triton TIMS instrument

Washington University, St. Louis, MO, USA

lab website: https://sites.wustl.edu/cosmochemistry/facilities/

The facility includes a dedicated suite of ultra-clean laboratories for the preparation of terrestrial and extraterrestrial samples. It features:

- Thermo Fischer Scientific iCAP Qc for measuring major and trace elements
- Thermo Fischer Scientific Neptune Plus for measuring isotopes

# 5.23 Laser-ablation Inductively Coupled Plasma Mass Spectroscopy (LA-ICP-MS)

See write-ups for HR-ICP-MS and MC-ICP-MS

# 5.24 Inductively Coupled Plasma - Optical Emission Spectroscopy (ICP-OES)

**Overview:**

ICP-OES (Inductively coupled plasma - optical emission spectrometry) is a technique in which the composition of elements in (mostly water-dissolved) samples can be determined using plasma and a spectrometer.

**How it works:**

When plasma ionization excites core electrons in the component atoms into a higher "excited" state. When the electrons relax back to their ground level energy they emit photons. Each element has an own characteristic emission spectrum. The light intensity on each element's wavelength is measured and its concentration is determined.

**Sample preparation:**

Samples should be dissolved in acid (preferably with 1-5 % $HNO_3$ in order to keep metals in solution)

**Sample impact:**

Completely destructive - each analysis consumes ~10 mg

**Data quality:**

Varies with concentration. Target elements include Na, Mg, Al, P, K, Ca, Ti, V, Mn, Fe, and Ni

**Data products:**

Bulk abundances of major and minor elements.

**Facility:**

Pacific Centre for Isotopic and Geochemical Research (PCIGR), University of British Columbia, Vancouver, BC, Canada

This facility contains eleven mass spectrometers, together with sample preparation laboratories and the research and support teams.

The ICP-OES systems at PCIGR include: Varian 725-ES

# 5.25 X-ray Fluorescence Spectroscopy (XRF)

**Overview:**

X-ray Fluorescence spectroscopy is a non-destructive analytical technique used for the chemical and structural characterization of silicate minerals and rocks. This is a routine analysis. It is based on wavelength-dispersive spectroscopic principles that are similar to other methods such as electron microprobe, XRD, SEM-EDS. During X-ray Fluorescence molecules are excited by short-wavelength radiation (X-ray). The radiation emitted is of lower energy than the primary incident excitation. The fluorescent X-rays are characteristic for each particular element, and thus, they can be used to quantitatively determine the abundances of elements that are present in the sample The technique requires very little sample preparation. Our instrument is a Spectro-XEPOS X-ray Fluorescence EDS spectrometer with sample size requirements ≥ 0.2 gr.

**How it works:**

The instrument consists of a benchtop X-ray Fluorescence EDS spectrometer with a 10 mm$^2$ SDD detector with a spectral resolution of ≤ 155 eV at Mn Ka. Samples are excited by a 50 Watt end-window X-ray tube. The spectrometer is equipped with five secondary targets (Co, Pd, Mo, HOPG, $Al_2O_3$) to improve sensitivity for the analysis of elements from Na to U. The unit has a 12-position automatic sample changer (32/40 mm), and it can operate in He atmosphere. The system is equipped with Spectro's semi-quantitative program TurboQuant+. A method has been developed to support the analysis of ≥ 0.2 gr rock/mineral powders. The method is based on the TQ+ analysis of 72 rock reference samples for several elements (Si, Al, Mg, Ca, Na, K, Fe, Ti, Mn, P, Cr, Ni, Cu, Zn, Co).

**Sample preparation:**

Samples are loaded in special sample holders appropriate for small size powder samples (≥ 0.2 gr). The analysis is performed under He atmosphere.

**Sample impact:**

X-Ray Fluorescence spectroscopy is a non-destructive technique.

**Data quality:**

The method is for silicate minerals and rocks with a <15% uncertainty on major oxides (0.2 - 80 wt% concentrations). This level of uncertainty reflects <1 wt% deviation between samples and standards for oxides with less than 10 wt% abundance, and <3 wt% deviation in the case of major oxides (>20 wt% composition).

**Data products:**

The format of the raw data output is defined by Spectro's proprietary control and data collection/processing software. The extracted files containing the composition of major oxides are in ".txt" format with a typical file size of 2-5 kb. A typical dataset contains <10 files. Reduced data are delivered as MS Excel files.

**Facility:**

We will utilize the Spectro-XEPOS X-ray Fluorescence EDS spectrometer at the Earth and Planets Laboratory (CIW).

# 5.26 Elemental Analysis - Isotope Ratio Mass Spectrometry (EA-IRMS)

- Overview:
- How it works:
  - Standard EA-IRMS:
  - Nano-EA/IRMS:
- Sample preparation:
  - C-N Analyses:
  - H-O Analyses:
  - S Analyses:
- Sample impact:
- Data quality:
  - EA-IRMS
  - nanoEA/IRMS
- Data products:
- Facilities:

## Overview:

Elemental analysis - isotope ratio mass spectrometry (EA-IRMS) is an analytical technique used for the isotopic and chemical analysis of H, C, N, O, and S in a sample. These elements are released as gases ($H_2$, CO, $CO_2$, $N_2$, and $SO_2$) from solid and/or liquid samples by flash combustion/pyrolysis. The gases are separated in a gas chromatography (GC) column and introduced into the ion source of a mass spectrometer. Ions are accelerated by an electrostatic field and separated according to their masses by a magnetic field. The abundances of the elements and isotopes are determined by referencing the absolute and relative ion beam intensities to elementally and isotopically calibrated standards. With this technique, we can determine the elemental abundances of H, C, N, O and S, and the stable isotope ratios D/H, $^{13}C/^{12}C$, $^{15}N/^{14}N$, $^{18}O/^{16}O$, and $^{34}S/^{32}S$. Variations in isotopic compositions provides important insights on the physico-chemical-biological processes imprinted on the samples.

## How it works:

### Standard EA-IRMS:

The elemental analyzer (EA) operates based on flash combustion/pyrolysis of samples introduced into oxidation/reduction reactors at temperatures from 1020°C (C-N-S) to 1450°C (H-O). Flash combustion/pyrolysis generates gases ($N_2$, CO, $CO_2$, $H_2O$, and $SO_2$). Separation of these gases is achieved prior to their introduction into an isotope-ratio mass spectrometer (IRMS) via GC columns. Samples are carried from the EA through the GC columns to the IRMS by a high-purity He carrier gas. The EA is coupled to the IRMS via a series of regulators and valves (e.g., Conflo III) that control the influx of: (i) gases from the EA, and (ii) reference gases. Upon entering the IRMS, gases are ionized (e.g., $N_2^+$, $CO_2^+$) by the ion source and accelerate by extraction plates. The flowpath of the ionized gases is controlled by focusing plates and a deflecting magnetic field. Along the flight tube, the ions are separated according to their mass-to-charge ratio. At a constant accelerating voltage and magnet strength, lighter ions are deflected more than the heavier ones. The current of each ion beam is detected by Faraday cups placed at specific positions along the detector interface. The signals received by the Faraday cups are amplified and digitized for processing. The concentration and isotopic composition of each element is determined by comparison to a reference gas analysis. Isotopic abundances are expressed in delta notation defined as: d ‰ = [ ($Ratio_{sample}$-$Ratio_{standard}$)/($Ratio_{standard}$)]*1000.

## Nano-EA/IRMS:

The nano-EA/IRMS system is a modified elemental analyzer coupled to an isotope-ratio-mass spectrometer that permits high-precision isotopic analysis of 10s of nanomoles of carbon, nitrogen, and sulfur on a single sample without chemical or cryotrapping of gases. The nano-EA/IRMS system employs a Flash$^{TM}$ IRMS Elemental Analyzer (Thermo Fisher Scientific) that is coupled via a ConFlo IV Universal Interface to a Delta V Plus isotope ratio mass spectrometer (Thermo Fisher Scientific). Nano-EA modifications to the commercial system include a narrower-bore combination combustion/reduction reactor (10 mm i.d. quartz tube filled with tungsten oxide and reduced copper grains operated at 1020°C) and water trap (8 mm i.d. quartz tube filled with magnesium perchlorate), a carbonPLOT capillary GC column (0.32 mm i.d., 15 m, 1.5 µm film thickness) installed in a Trace 1310 GC oven for precise temperature control (held at 25°C for 6 minutes and ramped to 190°C at 50°C/min), and a custom open-split design to ensure optimal transfer of analyte to the ion source.

As sample size decreases, the ability to accurately and precisely characterize the carbon, nitrogen, and sulfur blank contribution becomes increasingly important. Any minor amounts of $N_2$, $CO_2$, and $SO_2$ consistently present in the carrier gas stream exiting the elemental analyzer can be accounted for by accurate baseline determinations, in contrast to trapping systems where background is collected with the sample. Atmospheric $N_2$ and $CO_2$ trapped in sample containers and autosampler cavities is reduced by isolating and purging the Costech zero blank with helium prior to the introduction of samples to the reactor. Replicate analyses of tin sample containers showed that nitrogen and sulfur blanks were too small to be measured directly but can be derived from regression. The carbon blank is large enough to be directly measured. All nano-EA/IRMS data is corrected for blank contribution.

## Sample preparation:

For analysis, the samples are weighed into tin (C-N analyses) or silver (H-O analyses) capsules of known weight. To minimize the amount of absorbed atmospheric water that the samples may contain, they are stored in a dry $N_2$ flushed oven at 50°C for at least 12 h. The samples are reweighed just prior to analysis.

### C-N Analyses:

The C and N elemental and isotopic compositions of solid residues are determined on the same samples (typically ~ 0.2–0.5 mg for pure organics) loaded into tin capsules. The analyses are conducted using a CE Instruments NA 2500 series elemental analyzer (EA) linked to a Thermo Delta V Plus IRMS. Samples are introduced directly from an autosampler (A2100) into the EA where they are combusted with ultra-pure $O_2$ at 1020°C in a quartz oxidation column containing chromium( III) oxide and silvered cobalt(II, III) oxide. The resulting gases ($CO_2$ and $N_2$), mixed with zero-grade He as the carrier gas, are purified and separated by a packed gas-chromatography column prior to entering a Finnigan Conflo III interface. Both $N_2$ and $CO_2$ samples are analyzed relative to reference gas standards (high purity $N_2$, $CO_2$). Acetanilide ($C_8H_9NO$) is analyzed at regular intervals to monitor the accuracy of the measured isotopic ratios and elemental compositions.

For the N abundance and isotope analysis of $NH_4^+$ enriched fluid samples, dissolved $NH_4^+$ is converted to $NH_{3(g)}$ using a pH-buffered $Na_2B_4O_7$-NaOH solution (pH = 12.7) and then diffused into so-called diffusion packets. Diffusion packets are made with 1-cm-diameter glass fiber grade GF/D filters, pre-acidified with 25 ml of 2 M sulfuric acid (4 N) and sandwiched between 2.5-cm-diameter 10-um-pore Teflon membranes. Samples are incubated for 7 days at room temperature. After removal of the diffusion packets, the GF/D disks are oven-dried at 55°C. The $NH_4$-bearing GF/D disks are then placed in tin capsules.

### H-O Analyses:

The H and O elemental and isotopic compositions of solid samples are determined on separate aliquots using a thermal conversion elemental analyzer (TC/EA) coupled to a Thermo Finnigan MAT DeltaplusXL mass spectrometer. The H and O samples and standards (~0.2–0.5 mg in silver capsules) are converted to $H_2$ and CO in a pyrolysis reactor (made of glassy C and graphite) held at 1450 °C. Helium (UHP 5.5 grade) is used as a carrier gas to introduce the $H_2$ and CO into the GC. We use a He-flushed zero-blank autosampler to reduce the amount of water adsorbed from the atmosphere. We are developing protocols to maintain the autosampler at 80 °C to further reduce atmospheric $H_2O$ contributions. Reference gases are injected via dual inlet ($H_2$) and via the Conflo III interface (CO) into the IRMS before and after the samples for the computation of the isotopic compositions of the samples. Standards of known elemental and isotopic composition are analyzed at regular intervals between samples.

The stable H isotope analysis of aqueous solutions is performed by direct injection of liquid sample into the EA.

**S Analyses:**

Sulfur ($\delta^{34}S_{VCDT}$) isotopic values (analytical errors ±0.2 ‰) and elemental analysis for solid residues are conducted on the Thermo Fisher Scientific™ Delta V™ Plus IRMS attached to an Elementar Vario Pyro Cube elemental analyzer. Time pulses of $SO_2$ reference gas of high purity are introduced to the IRMs via a GasBench II interface.

# Sample impact:

EA-IRMS is a destructive technique and samples are not recoverable.

# Data quality:

## EA-IRMS

The measurement precision for elemental abundances is typically of the order 1–3% of the reported values. Because C and N are measured in the same samples, the precision of measured N/C ratios is typically about 1% of the reported values. The precision of C and N isotope measurements is generally 0.1– 0.3 ‰. The uncertainties associated with internal standards are; i) $\delta^{13}C$ 0.1 ‰; ii) $\delta^{34}S$ ± 0.2 ‰; iii) $\delta^{15}N$ ± 0.2 ‰, and iv) $\delta^{18}O$ ± 0.3 ‰. The accuracy of the H isotope measurements is more difficult to assess because it decreases with increasing D enrichment, and because there is a small memory effect associated with the measurements. We run blanks between different samples to reduce the memory effects. Furthermore, memory effects were monitored by analyzing in-house standards of H-bearing solids (e.g, syn-IOM; 67-1183 ‰) and $H_2O$ of varying dD content (-447 ‰ to 3259 ‰). There is no memory effect for the C and N analyses.

International (SMOW, NBS-22, air, IAEA $Ag_2S$ S-1/S-2/S-3) and commercially certified standards from Isoanalytical, USGS, Costech and Oztech are used to calibrate and correct the data; internal working gas standards (reference $H_2$, $CO_2$, CO, $N_2$) are analyzed at regular intervals to monitor the accuracy of the measured isotopic ratios and elemental compositions throughout the run. Sample heterogeneity is also a potential source of uncertainty in all the measurements.

## nanoEA/IRMS

The sample size required by the nano-EA system is ~2 orders of magnitude less than conventional analyses. Measurement precision for elemental abundances is typically on the order of 5-10% of the reported values, depending on sample size. After blank correction, nano-EA data achieve precision better than 1‰ for isotopic analysis of 100 nanomoles or less of carbon, nitrogen, and sulfur in isotopic standards. There is no memory effect for the C and N analyses. There is a minor sulfur memory effect (~0.013‰ for every permil difference between sequential samples), but only between sequential samples with isotopic differences greater than 30‰. Accuracy and precision of isotopic measurements were evaluated using international and laboratory standards: CH-6, N-1, N-2, NO-3, S-1, S-2, and S-3 (International Atomic Energy Agency), US Geological Survey reference standards 24 and 40, Urea # 1 (Indiana University), and in-house standard Peru mud.

# Data products:

The format of the raw data output from EA-IRMS analyses is compatible with the ISODAT data acquisition and control *software. The files are ".cf" (Isodat NT) and in ".dxf" (ISODAT 2.5).* These products can be used qualitatively to identify the elemental and isotope HCNOS composition of solid samples. The typical file size of the raw data file is 0.1-0.5 Mb. A typical dataset contains 20-50 files. Reduced data are delivered as MS Excel files.

# Facilities:

The Stable Isotope Laboratory Dionysis Foustoukos at the Earth and Planets Laboratory (CIW) has the capability to provide high precision-high accuracy stable isotopic analysis of a broad range of materials including gasses, solutions, and solids and inorganics and organics. Currently we support a Thermo Scientific Delta V$^{Plus}$ mass spectrometer and CE Instruments NA 2500 series elemental analyzer (EA) for C and N analyses, and a Thermo Quest Finnigan Delta$^{Plus}$ XL mass spectrometer and Thermo Finnigan thermal conversion elemental analyzer (TC/EA) for H and O analyses. The Thermo Scientific Delta V$^{Plus}$ is also interfaced with an Elementar Vario Pyro Cube elemental analyzer to facilitate high-temperature pyrolysis (1500 °C). Our facility is capable of providing stable isotopic data for H, C, N, O, and S in both organic and inorganic solids.

The Deines Laboratory Kate Freeman at The Pennsylvania State University houses a nano-EA/IRMS for analyses of solid samples and isolated molecules that are too small for analysis by conventional methods. Highly precise isotopic measurements (1 sigma = 1‰) are possible for <100 nanomoles of carbon, nitrogen, and sulfur. The instrument is comprised of a Thermo Delta$^{Plus}$ with a modified and streamlined design and capillary gas separation and enables isotope analyses of nanomolar quantities of solid samples for $^{15}$N, $^{13}$C, and $^{34}$S isotope systems.

Brown Lab for Organic Geochemistry (BLOG), - Yongsong Huang at Brown University has a multi-Shot Pyrolyzer (EGA/PY-3030D) made by Frontier Laboratories, Ltd. is used to thermally extract volatile organic compounds and thermally break down macromolecules from carbonaceous meteorites or asteroid samples. The evolved compounds are first trapped using liquid nitrogen with the MicroJet trap system. Subsequently, the compounds are separated using GC columns containing appropriate stationary phases and analyzed by a quadrupole mass spectrometer for identification. The EGA/PY-3030D system can be easily mounted and dismounted from the GCMS. After compound identification using GCMS, EGA/PY-3030D can be mounted onto the GCIRMS system for compound specific carbon and hydrogen isotopic analyses.

# 5.27 Noble gas and nitrogen static mass spectrometry

**Overview:**

Noble gases (He, Ne, Ar, Kr, Xe) are chemically inert elements that mainly record physical processes such as radioactivity, diffusive loss or fractionation, irradiation, adsorption, and phase changes. They can provide information on the sources of matter that formed the solar system, including primordial organic matter and presolar grains that carry signatures of stellar nucleosynthesis, as well as on processes that affected the host phases. The abundance of noble gases in extraterrestrial materials is generally very low.

Noble gas mass spectrometry (NGMS) is an analytical technique that is used for the elemental and isotopic measurements of noble gases. Given the low abundance of these elements, mass spectrometers work in static mode, that is, with the pumping system shut down during analysis. This requires specifically designed mass spectrometers with very low analytical background. Some of these systems permit the analysis of nitrogen isotopes at very low levels together with the noble gases.

**How it works:**

Noble gases trapped within solid materials are extracted from samples by heating, UV laser ablation, crushing, or etching. The heating method of extraction occurs by means of an IR laser or in a crucible that is heated by, e.g., resistance, electron impact, or high-frequency electromagnetic waves. All extraction procedures need to be done under high vacuum (typically $10^{-9}$ mbar) to prevent contamination of sample gases by atmospheric gases. The extracted gases are purified using hot metals or alloys (and cold traps) that bind or condense chemically reactive species, leaving essentially only noble gases in the gas phase. Noble gas elements can be separated using cryogenic traps and sequentially analyzed. Most mass spectrometers have a classical Nier-type design in which atoms in the sample are ionized, in this case with an electron impact ionization source, and accelerated to the mass spectrometer. The mass spectrometer consists of a tunable sector-type magnet for separation of the ions according to their mass/charge ratio and a collection block consisting of single or multiple Faraday cups and/or electron multipliers. The whole system, including valves, is made of metal or glass, both of which have very low degassing rates. The system can be baked up to 300°C to lower the degassing rate during purification and lower the background during analysis in the static mode.

**Sample preparation:**

To reduce contamination, particularly of the heavy noble gases, samples ideally should have no contact with the terrestrial environment between recovery from the re-entry capsule and introduction into the analytical system. Samples can be accommodated as bulk rock, mineral separates, or petrographic sections. The sample needs to be as pristine as possible. Its preparation involves physical separation from the main lithology at room temperature in a dust-free environment and without using any liquid. If a sample needs to be crushed, for instance to achieve homogenization and splitting to allow simultaneous analyses of, e.g., major element chemistry and noble gas contents, this should be performed under vacuum to similarly avoid exposure to Earth's atmosphere. After physical separation of the targeted mineral phase, the sample is loaded in the extraction/purification line, and the whole system is pumped to very low pressure (Ultra High Vacuum, ≤$10^{-9}$ mbar) and baked for several hours or days typically at temperatures up to 110°C to remove potentially adsorbed atmospheric gases. Laser extraction can also be performed on thin and thick sections to achieve spatial information on the gas distribution and carrier minerals. In this case, the sections can be prepared following standard techniques, while avoiding the use of organics and halogens as much as possible.

**Sample impact:**

This technique is destructive: samples are heated to their melting point and ablated or dissolved in acidic solvents. Melted samples can be recovered for refractory element analysis. Laser spot analysis on thin or thick sections is much less destructive (typically a few hundred-micron craters on sections). The sample is ablated to a depth to which sufficient gas is released, while avoiding ablation of another mineral phase or epoxy below to sample of interest. Thus, the ablation depth can range from a few microns to a few hundred microns.

**Data quality:**

The detection limit is very low, on the order of a few hundreds of thousands of atoms in conventional systems and much lower (thousands of atoms) in special mass spectrometers equipped with, e.g., a compressor source or resonance ionization. These specialized systems are used to measure some of the least abundant isotopes. The precision of noble gas isotopic ratios depends on the ratio, typically 1% for helium isotopic ratios and down to 1 to 2‰ for most other isotopic ratios. Noble gas concentrations can typically be measured at the percent level.

**Data products:**

Data products include raw data ASCII files, log, and meta files (e.g., in MS Excel format). The data products expected from noble gas measurements include:

- The duration for which samples were exposed to cosmic rays through measurement of cosmogenic isotopes ($^3$He, $^{21}$Ne, $^{38}$Ar, $^{78,81,83}$Kr, $^{126}$Xe);
- Exposure to the solar wind;
- Absolute gas retention ages using $^{238}$U+$^{232}$Th-$^4$He, $^{40}$K-$^{40}$Ar, and $^{238}$U-$^{131\text{-}136}$Xe chronometers;
- Extinct radioactivity closure ages using $^{129}$I-$^{129}$Xe and $^{244}$Pu-$^{131\text{-}136}$Xe systems;
- Origin(s) of components present in the samples: pre-solar, solar, chondritic, cometary;
- Processes having affected the samples such as atmospheric escape, impacts, regolith gardening and irradiation, diffusion, thermal and aqueous alteration;
- Gas carrier minerals/phases, via stepwise (heating or etching) release.

**Facilities:**

Centre de Recherches Pétrographiques et Géochimiques–Centre national de la recherche scientifique (CRPG-CNRS) and Université de Lorriane, Nancy, France

- Nu Instruments Noblesse HR: Multi-collection (five collectors) noble gas mass spectrometer fitted to an extraction/purification line for the analysis of noble gases and nitrogen abundances and isotopes. Extraction from bulk rock or mineral separates is achieved by (i) heating the sample with a $CO_2$ laser or a resistance mini-furnace or (ii) crushing the sample under vacuum. Thermo Scientific Helix MC Plus: Multi-collection (five collectors) noble gas mass spectrometer fitted to an extraction/purification line for the analysis of noble gases and nitrogen abundances and isotopes. Extraction from bulk rock or mineral separates is achieved by (i) heating the sample with a resistance mini-furnace, (ii) crushing the sample under vacuum, or (iii) laser ablation (193 nm or 216 nm) on thin/thick sections.

Inst. für Geochemie und Petrologie, ETH Zurich, Switzerland

- Nu Instruments Noblesse multi-collection (5F5M configuration) mass spectrometer for He-Xe isotope analysis.
- Custom-built "Albatros" (1F1M) mass spectrometer for He-Xe isotope analysis.
- High-sensitivity compressor-source equipped custom-built "Tom Dooley" mass spectrometer for detection of He and Ne isotopes.
- Custom-built multi-collection mass spectrometer with nine fixed multipliers dedicated to the measurement of Kr and Xe isotopes in a single step each.
- Gas extraction devices include electron-impact heated Mo crucibles (up to ~1800 °C); a Nd:YAG IR laser system for (stepwise) heating; a programmable UV laser system with computer controlled x-y stage for ablation; closed system (stepwise) etching enabling gas release by applying various dissolving agents at low temperature in ultra-high vacuum; and various vacuum crushers.

# 5.28 Secondary neutral mass spectrometry (SNMS)

**Overview:**

Secondary neutral mass spectrometry (SNMS) uses sputtered neutrals instead of secondary ions to determine the isotopic composition of a sample. This technique was developed to overcome the issues of low secondary ionization yield and matrix- and element-dependent ionization that are characteristics of secondary ion mass spectrometry (SIMS). In recent years, a time-of-flight (TOF) SNMS was developed to measure depth profiles for noble gas isotopes (Bajo et al., 2015; Tonotani, 2017). The laser ionization mass nanoscope (LIMAS) a type of TOF-SNMS was applied to extraterrestrial materials to measure solar wind–implanted noble gases in Genesis samples (Tonotani, 2017) and provides a complementary analytical technique to static noble gas mass spectrometry.

**How it works:**

SNMS uses a 20-keV primary $Ga^+$ ion beam to sputter the sample. Both ionized and neutral species are generated. Whereas SIMS measures the ionized species, SNMS measures the neutral species by ionizing them selectively by laser beam and thus produces a greater useful yield than SIMS. The ions are then introduced into the mass spectrometer. In LIMAS, a femto-second laser is used to achieve a strong-field (tunneling) ionization to ionize all elements efficiently, including noble gases.

LIMAS consists of a Ga liquid metal ion source, a femto-second laser, and TOF mass spectrometer. The smallest primary beam size of the JEOL LIMAS, for example, is ~10 nm in diameter for Ga under the performance of an aberration corrector for spherical and chromatic aberrations in the primary column. The primary beam is pulsed and emitted on the sample surface. A neutral plume generated by the primary beam is ionized by the synchronized laser beam pulses. The ionization efficiency is 70% for helium and 100% for other elements. The ions are then introduced to the TOF mass spectrometer and separated by their mass-to-charge ratio (m/z) to make a mass spectrum. The intensity of ions with a given m/z is proportional to the abundance. The intensity of each m/z is transformed to chemical and isotopic compositions using standard materials.

**Sample preparation:**

SNMS analysis may require special preparation of the samples. The surface of the sample must be flat, relative to the size of the interrogating beam. Flatness is typically achieved through grinding, lapping, and mechanical polishing. The surface of the sample must be conductive to avoid charge buildup on the surface. Typically, a thin (few nanometers) layer of carbon or gold is applied using a vacuum evaporation method.

**Sample impact:**

SNMS is a destructive technique. The primary beam sputters the sample surface to make a crater with lateral sizes from about 10 nm to 1 μm depending on techniques and applications. The primary beam implants into the surface at a depth of <100 nm, mainly depending on the impact energy of the primary beam. After the analysis, the sample surface remains contaminated by the primary beam to a depth of several hundreds of atomic layers, or ~50 nm. The atomic arrangements are changed (destroyed) by the implanted particles and collisions of atoms.

**Data quality:**

SNMS detection limits in favorable cases can extend to less than parts per billion. The precision of SNMS is generally limited by statistics of secondary ion intensities, with instrumental precision typically 10 to 30% for compositions of major to trace elements, and several percent for isotope abundances.

Quantitative analysis of trace elements is performed by a calibration curve method. Matrix-matched and composition-known standards should be prepared for the calibration curve. For isotope analysis, matrix-matched standards with known isotope abundance are necessary to correct for instrumental mass fractionation.

The spatial resolution of SNMS is controlled by the size of the primary beam. The spatial distribution of the primary beam determines the primary beam size. The primary beam size is confirmed by observations of the sputter crater after measurements. The smallest primary beam size of JEOL LIMAS is about 10 nm in diameter, which is very close to the theoretical limit of sputtering theory.

The mass resolution of LIMAS depends on the flight time of an ion packet because the LIMAS is installed with a multi-turn TOF mass spectrometer. A mass resolution of $M/\Delta M = 1,000,000$ is achieved.

The useful yield of SNMS refers to the fraction of sputtered isotopes that are ionized and detected. It is ~0.4 % for the JEOL LIMAS.

**Data products:**

The data products expected from SNMS measurements include:

- 0D point analyses (individual data set from a specific location)
- 1D line scans (individual data set plotted as a function of distance)
- 2D images (2D maps of a sample in which every pixel contains a data set)
- 3D images (3D maps of a sample surface in which every voxel [3D pixel] contains a data set)

These products can be used qualitatively to identify what elements are present and where they occur. They can be used quantitatively to determine chemical and isotope compositions at the local scale and for regions of interest. All products can be delivered as text files.

**Facility:**

Isotope Imaging Laboratory, Hokkaido University, Sapporo, Japan

- JEOL LIMAS isotope nanoscope. This is a SNMS using strong-field (tunneling) post ionization by a femto-second laser. It can analyze all isotopes including noble gases with high sensitivity.

**References:**

Bajo K.-I. et al. (2016) High spatial resolution imaging of helium isotope by TOF-SNMS. *Surf. Interface Anal.* **48**, 1190–1193.

Tonotani, A. (2017) Development of depth profiling for solar wind noble gases implanted in Genesis diamond-like-carbon on silicon substrate targets by using isotope nanoscope. Doctoral Thesis, Hokkaido University, Japan.

# 5.29 Laser-assisted fluorination for bulk oxygen isotope ratio measurements

**Overview:**

Oxygen 3-isotope measurements ($^{17}O/^{16}O$ and $^{18}O/^{16}O$ ratios) of meteoritic and other planetary materials are one of the cornerstones in our understanding of early solar system reservoirs and processes. Bulk measurements of extraterrestrial samples provide a framework that allows identification of common origins and/or the nature of subsequent processes (e.g. gas/solid or water/rock interaction).

In the laser-assisted fluorination technique, the oxygen in bulk samples is extracted through reaction with a powerful oxidizing reagent while being heated by an IR laser. The purified oxygen gas is then analyzed on a high specification gas source mass spectrometer, providing good sensitivity and very high analytical precision.

**How it works:**

Aliquots of powdered samples, or one or more sample chips (see Sample preparation, below), are placed in wells in a nickel tray that is then mounted in a sample reaction cell attached to a vacuum system. Multiple samples (up to ~16 plus a number of standards to verify instrument performance throughout analyses) can be loaded into a single tray. The samples and sample reaction cell are then degassed at ~70°C (12 to 24 hours) under vacuum to remove adsorbed atmospheric moisture.

Oxygen gas is liberated from the samples by adding an aliquot of purified $BrF_5$ gas and heating the sample with a 10.6-µm laser (typically 25–50 W). Fluorine from the $BrF_5$ replaces oxygen in the silicate and oxide phases. Use of an IR laser allows the sample to be heated to very high temperatures, >1200°C, without heating the sample chamber walls, and driving the reaction to completion in a few minutes, both of which help keep system background levels very low. The laser can be focused or defocused (diameter nominally 400 mm to 3 mm) depending on the optimized reaction requirements to ensure quantitative extraction of oxygen from the sample.

The liberated gas is purified from excess $BrF_5$ and other reaction products (e.g., $SiF_4$, $F_2$, $Br_2$) by passing over a series of traps at liquid nitrogen temperature and reaction bed(s). The purified oxygen gas is then cryo-focused immediately before the entrance to the mass spectrometer inlet prior to being expanded into the inlet system. In the case of small samples (≤200 mg silicate), the gas is cryo-focused within the inlet system itself.

Isotopic measurements are ideally performed on a stable isotope gas source mass spectrometer with large radius and high mass resolution such as a Thermo MAT 253 to provide good stability and control of peak tailing effects on very low abundance $^{17}O/^{16}O$ peaks. The gas inlet system is a dual inlet configuration, allowing for carefully balanced flow of sample and calibrated reference gas that are repeatedly compared to provide high-precision isotope ratio measurement.

Most sample types are essentially unreactive to BrF5, allowing for multiple samples to be loaded in a tray. However, hydrated phyllosilicates and amorphous silicates display significant reaction with BrF5 at room temperature; therefore, for samples such as CM2s, CIs, and samples of Bennu, a modified technique can be used where only one sample and one standard are loaded in the system. This allows for successful measurement of such samples free from the effects of pre-reaction.

**Sample preparation:**

Samples can be analyzed with minimal preparation if required. The simplest preparation involves pieces of sample, picked or extracted from a larger sample. Samples can be analyzed as powders or chip(s). Typically, laser fluorination systems require 1–2 mg of silicate (or similar) material for each analysis, but sample sizes of about 100 µg can also be analyzed. As oxygen typically makes up ~40 wt% of a typical rock, there is an abundance of analyte available. One of the benefits of this abundance is that the effects of many other analytical techniques are negligible, and therefore oxygen isotope measurements can be performed on bulk samples following detailed characterization by other non-destructive techniques, e.g., X-ray CT, XRD, spectroscopy. The only requirement is that the samples are protected from exposure to water, and, if possible, atmospheric oxygen. Micro-sampling techniques (e.g., picking with a needle, micro drill) also allow samples to be extracted from polished blocks, and therefore techniques such as SEM, EPMA, and SIMS/NanoSIMS can also be applied prior to bulk oxygen isotope measurements.

**Sample impact:**

Laser-assisted fluorination for bulk oxygen isotope measurements is a completely destructive technique. All silicates and oxides are converted to an amorphous mix of metal fluorides, some of which may be volatile (e.g. $SiF_4$) and therefore lost from the sample well, making the residue of very little use for subsequent study.

**Data quality:**

Oxygen isotope compositions are reported in delta notation: the deviation in the $^{17}O/^{16}O$ or $^{18}O/^{16}O$ ratio from the same ratio in a reference material expressed in ‰ (parts per thousand). While mass fractionation effects usually result in co-variation of $^{17}O/^{16}O$ and $^{18}O/^{16}O$ in a mass-dependent manner, mass-independent effects are commonly observed in extraterrestrial samples, and therefore a particularly useful notation is $\Delta^{17}O$, which measures the deviation from a mass-dependent line defined by all terrestrial rocks (TFL in Fig. 1).

Analytical precision on stable (i.e., anhydrous), homogeneous (i.e., not carbonaceous chondrite) materials is very small compared to whole-rock variation in carbonaceous chondrites, i.e., ±0.09‰ for $\delta^{18}O$ and ±0.017‰ for $\Delta^{17}O$ (2 sigma). Samples smaller than 1 to 2 mg can be analyzed, which is useful in cases where there are very limited amounts of a specific lithology or where there is a need to analyze materials previously characterized by other destructive techniques. Sample sizes < 100 µg can be analyzed with some loss of analytical precision (±0.25‰ for $\delta^{18}O$ and ±0.035‰ for $\Delta^{17}O$).

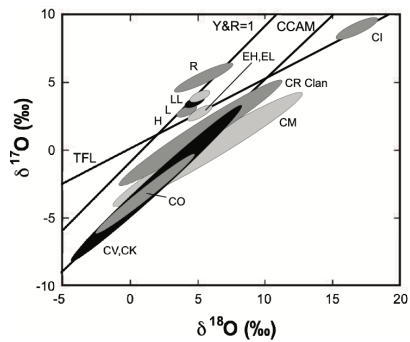

Figure 1. Oxygen isotopic composition of chondritic meteorites, with each meteorite group (CM, CV, CO, etc.) having a unique, or nearly unique, isotopic signature. $\delta^{17}O$ and $\delta^{18}O$ refer to variations in $^{17}O/^{16}O$ and $^{18}O/^{16}O$ in parts per thousand. All terrestrial rocks fall along the terrestrial fractionation line (TFL); other lines are examples of possible solar nebula reservoir mixing.

**Data products:**

The data files generated from each analysis are saved in Thermo file format. These files record the measured ion beam intensities for each sample or reference comparison. Corresponding metadata are recorded within the file. Summary export reports can be generated as text files.

**Facility:**

Open University, Milton Keynes, UK

- The laser-assisted fluorination system utilizes a Photon Machines 50W 10.6 mm laser. The sample chamber and sample gas purification vacuum line were designed and built in-house and are connected directly to the dual inlet of a Thermo MAT 253 fitted with a microvolume third inlet for very small samples.

# 5.30 Stepped heating carbon and nitrogen isotopic compositions

**Overview:**

Within samples containing an intimate mix of small phases, different components (e.g., organics, carbonates, silicates, diamonds, silicon carbide) decompose, oxidize, or degas volatiles over discrete and often distinct temperature ranges. A fully automated system called FINESSE was designed and built at the Open University and includes bespoke sample heating, gas processing, and three mass spectrometers. Small bulk samples are heated in a stepwise manner from room temperature up to 1400°C, and the abundances and isotopic compositions of carbon and nitrogen are analyzed at each temperature step. As different components combust or decompose at different temperatures, FINESSE can provide a measurement of the abundance and isotopic composition of the C and N associated with individual components within the bulk sample (e.g. organics, carbonates) (Figure 1).

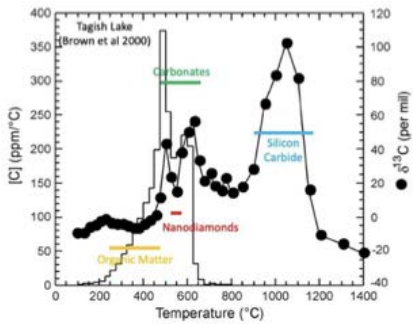

Figure 1. The release of carbon (histogram) and its associated isotopic composition (data points) from the Tagish Lake meteorite at each temperature step. The approximate temperature range of some of the principle components present in the sample are marked by colored horizontal bars. Deconvolution and integration of these features can provide the abundance and isotopic signature of each component.

One particularly important capability of stepped-heating extraction is to resolve contaminating species. Contamination is a major problem when studying meteorites, particularly finds; although handling and curation of Bennu samples may be exemplary, exposure to $N_2$ from TAGSAM and curation activities will likely add some N to the sample. Most gaseous contaminants are released at relatively low temperatures and therefore can be readily resolved from indigenous components in most cases.

**How it works:**

Small samples (typically a few milligrams) of powder or chips are heated stepwise from room temperature to ~1400°C, either in vacuo (pyrolysis) or in the presence of ~10 mbar pure $O_2$ (combustion). Step size is typically between 25 and 200°C, depending on, for example, the complexity of the release profile, availability of sample, and abundance of C and N. At the end of each step, volatile species of C and N liberated from the sample are cryo-focused into a gas purification system to convert all C-bearing species to $CO_2$ and all N-bearing species to $N_2$. $CO_2$ and $N_2$ are then cryo-separated, and other species (e.g., $H_2O$, $SO_2$) are removed, before these two species are admitted to the inlet system for each gas. Here the volume of gas is optimized for analysis in the inlet sections of dedicated mass spectrometers, which are optimized for the measurement of small amounts of each gas. A quadrupole mass spectrometer is used in this optimization process and, if sufficient gas is available, can also determine the abundance of some noble gases and their associated isotopic composition (usually $^4$He, $^{20}$Ne, $^{21}$Ne, $^{22}$Ne, $^{36}$Ar, $^{38}$Ar, and $^{40}$Ar). This information can be useful to correlate results with dedicated noble gas measurements.

The mass spectrometers operate in static mode, whereby the mass spectrometer volume acts as the sample reservoir, with a precision gas pipette system dispensing an exactly matched volume of reference gas immediately after analysis of the sample for sample-standard comparison. Given that many step analyses can require >36 hours, the system runs fully automatically.

**Sample preparation:**

Samples are loaded into small, pre-baked platinum buckets, either in the form of powders or chips. The sample mass required is a function of several parameters, not least of which is the abundance of C and N in the sample as well as the number of steps planned, the abundances of C and N present in minor components that will be characterized, and the analytical precision required. For CM2-like material, this translates into typical sample size requirements of 3 to 6 mg. This mass provides an abundance of gas from the major components (organics, carbonates) but also allows for determination of the abundance of presolar grains that are normally present in very low concentrations but have characteristic extreme isotopic signatures (e.g. presolar silicon carbides and nanodiamonds). Smaller sample mass (~1 mg) would be suitable for characterization of major components in CM2-like material.

**Sample impact:**

The sample is usually heated to ~1400°C, resulting in loss of many volatile elements and partial fusion of residual material within the platinum bucket. Normally, samples are considered destroyed after analysis.

**Data quality:**

Analytical precision for individual measurements at each step is typically <± 1‰ for $\delta^{13}$C ($^{13}$C/$^{12}$C ratio) and $\delta^{15}$N ($^{15}$N/$^{14}$N ratio). Replicate measurements of carbonaceous chondrites and comparison with elemental analyzer measurements indicate that accuracy is also ~± 1‰ for both $\delta^{13}$C and $\delta^{15}$N. Isotopic signatures of individual components may be poorer, depending on how well resolved they are from other components in the release profile. For minor components, analytical precision usually decreases. For very minor components such as presolar grains, where only ≤1 ng of gas may be liberated, precision may be ± 3‰.

Abundance measurements are conservatively estimated at <±10% of measured value.

System backgrounds are typically <<1 ng $N_2$ and <1 ng C as $CO_2$ at all steps. System blanks are usually performed before and/or after each sample analysis, and the results are automatically corrected for system contribution.

**Data products:**

Raw data are generated as tab-delimited .txt files capturing temperature, gas yield, isotopic ratio measurements, and optimization for analysis conditions. System blank files are stored separately.

Summary reports of the full stepped heating experiment are generated as MS Excel spreadsheets. These reports contain partially processed results. A summary plot of the analysis of the Tagish Lake meteorite (adapted from Brown et al. 2000) is shown in Figure 1.

**Facility:**

Open University, Milton Keynes, UK

- FINESSE is a bespoke instrument, incorporating a fully automated gas extraction and purification system coupled to two mass spectrometers for the measurement of nanogram quantities of $N_2$ and $CO_2$.

Reference:

Brown, P.G., Hildebrand, A.R., Zolensky, M.E., Grady, M., Clayton, R.N., Mayeda, T.K., Tagliaferri, E., Spalding, R., MacRae, N.D., Hoffman, E.L. and Mittlefehldt, D.W. (2000). The fall, recovery, orbit, and composition of the Tagish Lake meteorite: A new type of carbonaceous chondrite. *Science*, *290*(5490), pp.320-325.

# 5.32 Atom Probe Tomography (APT)

**Overview:**

Atom probe tomography (APT) is a nanoscale analytical technique capable of 3D imaging and chemical-isotopic identification of atoms within a small, targeted sample volume. It combines single-ion detection with time-of-flight mass spectrometry to produce a 3D spatial reconstruction of elemental distributions at the atomic level. The output consists of individual atoms positioned in 3D space and identified by the mass-to-charge ratio of the detected ion. Typical sample volumes are approximately 100 × 100 × 500 nm, with a spatial resolution of ~1 nm and chemical sensitivity down to 10 µmol/mol. Recent advances in APT hardware, in particular the use of ultraviolet laser pulsing, have allowed routine analysis of non-conductive materials and have opened the technique to geological applications.

**How it works:**

Sample atoms are removed from the tip of a needle-shaped specimen and ionized by a high electric field in a process of "field evaporation". The evaporation is initiated by a thermal pulse from a picosecond laser focussed on the specimen apex. The projection of the evaporated ions onto a position-sensitive detector, combined with the ion-detection sequence, allows the original location of each atom to be determined. The ion-detection time, relative to the laser pulse, is used to identify each chemical and isotopic species through time-of-flight mass spectrometry. Recent advances in APT hardware, in particular the use of ultra-violet laser pulsing, have allowed routine analysis of non-conductive materials and have opened the technique to geological applications.

**Sample preparation:**

Beginning with a clean, polished sample mount, such as a thin-section or epoxy mount, regions of interest are targeted for site-specific atom probe specimen preparation using a focused ion beam scanning electron microscope (FIB-SEM). Specimens are extracted in-situ from the surface using a 30 kV $Ga^+$ ion beam to mill out material that is subsequently attached to a pre-fabricated silicon post. The sample material and silicon post are then sharpened with the ion beam using several stages of annular milling to form atom probe needles of < 200 nm diameter at the apex. To reduce high-energy $Ga^+$ ion implantation, a final ion milling stage is performed at 2 kV to remove the damaged surface layers.

The orientation of the specimen needle axis is usually perpendicular to the original polished sample surface, with the needle apex close to the surface. The sample mounts must be vacuum compatible for SEM work and coated as necessary to reduce charging under the imaging electron beam. Chemical-vapor-deposition assisted by the ion beam (using an in situ gas injection system) is used throughout the preparation process for placing protective layers and for joining the lift-out sample material to the silicon posts.

During APT analysis a high voltage (up to 10 kV) is applied to the specimen, producing a strong electric field at the apex and a high axial stress on the needle. The specimen is actively cooled to below 100 K, but the laser pulse typically heats the apex region by several hundred degrees for a few hundred picoseconds. Data acquisition conditions must be selected to avoid fracturing of the specimen due to mechanical stress, or melting of the apex region due to the laser pulse heating.

**Sample impact:**

Atom probe tomography is a destructive technique that completely disassembles the targeted sample volume (100 × 100 × 500 nm) at the atomic level. Furthermore, the lift-out specimen preparation procedure involves $Ga^+$ ion milling of a larger volume around the region of interest; typically a 20 × 20 µm region of the polished surface, and around 5 µm in depth. A larger area of the sample surface (~50 × 50 µm) is often subject to $Ga^+$ ion bombardment during the preparation process in the FIB-SEM, though this may be minimized if particular care is taken during the procedure.

**Data quality:**

Due to the high ionization efficiency of the field-evaporation process, for most mineral phases APT provides quantitative results, with equal sensitivity to all elements in the periodic table, without the need for calibration using reference materials. In some cases, small differences in quantification can arise, particularly for oxygen, but usually, the precision is limited by the counting statistics related to the number of ions collected. The spatial resolution of APT is difficult to quantify for non-conductive materials but is usually taken to be ~1 nm for most mineral phases. Boundaries between different phases, or microstructural features in the sample, can produce local distortions in the reconstructed data that reduce the effective spatial resolution in these regions. The 'mass spectrum' generated from the ion mass-to-charge ratio values has a mass resolving power (M/DM) of approximately 1000 to 1200 for most acquisitions. The chemical sensitivity obtained from the mass spectrum is typically ~10 µmol/mol. Typical data volumes are approximately 100 × 100 × 500 nm, and the detection efficiency across all elements is close to 38% within this volume.

**Data products:**

The data products from APT include:

- the raw data file (.rhit)
- secondary file format (.apt)
- range files (.rrng)

The raw data file (.rhit) from an atom probe acquisition contains the acquisition metadata, ion detection coordinates, detection timing, and other relevant data, such as the specimen voltage, corresponding to each detector ion "hit" in the data acquisition. A proprietary file format is used for the raw data. Following reconstruction, the data are contained in a secondary file format (.apt) consisting of x,y,z spatial coordinates and the mass-to-charge ratio for each detected ion. Other acquisition parameters are also recorded for each ion, in addition to acquisition metadata. Range files (.rrng) contain user-defined mass ranges in the mass-to-charge spectrum corresponding to particular identified ion species. These are small text files that are combined with the mass-to-charge ratio values in the reconstructed data to identify the detected ions. The reconstructed data can be manipulated and analyzed in many ways to generate useful geochemical and isotopic data such as local compositions, chemical profiles, isotopic ratios, spatial information on microstructural features, and so on.

**Facility:**

Curtin University, Perth, Australia

- The Geoscience Atom Probe (GAP) facility within the John de Laeter Centre operates a Cameca LEAP 4000X HR atom probe microscope. The LEAP 4000X HR uses ultra-violet laser pulsing and contains a reflectron in the ion flight path that provides a higher mass resolving power.
- The Centre also houses a FIB-SEM (Tescan Lyra3) used for preparing APT specimens through the ion-beam lift-out method.

# 5.33 Neutron Irradiation Noble Gas Mass Spectrometry (NI-NGMS)

**Overview:**

Neutron irradiation noble gas mass spectrometry (NI-NGMS) is a technique used for the chemical analysis of a sample. It provides information on the abundances and ratios of the halogen (Cl, Br and I) and the noble gas (Ar, Kr and Xe) elements. The technique requires neutron-irradiation of the sample to produce noble gas isotopes from halogen isotopes within the sample based on (n, $\gamma$, $\beta$) reactions (e.g., $^{79}$Br to $^{80}$Kr). Irradiation requirements are similar to those required for $^{40}$Ar-$^{39}$Ar dating, except that the sample is unshielded for thermal neutrons. After irradiation, gases are released from the sample by heating, purified and measured by conventional noble gas mass spectrometry. Information on the K, Ca and Ba abundances in the sample is also provided by (n, p), (n, $\alpha$) and (n, $\gamma$, $\beta$) reactions. NI-NGMS requires only small sample masses (~1 mg) and is capable of detecting low halogen abundances (<1 ppb Br and I), making it a particularly useful technique for the analysis of planetary samples.

**How it works:**

NI-NGMS is a suitable technique for the analysis of the halogen elements (Cl, Br, and I) in bulk geologic samples. NI-NGMS is typically used when knowledge of halogen abundances is needed in (i) a sample where halogen abundances are low, e.g., ppm Cl and ppb Br and I and cannot be detected by more routine methods; (ii) in a sample where the material is limited and only a small amount (e.g., mg-sized) is available for analysis, or (iii) where obtaining both halogen and noble gas isotope ratios in a single analysis is desirable. Bulk NI-NGMS analysis is widely applicable to samples of varying mass (<1 mg to ~10's mg) and halogen abundances (<1 ppm to wt.% Cl and <1 ppb to 1000's ppm Br and I).

Samples are wrapped in Al-foil packets, encapsulated in a $SiO_2$ tube under vacuum, and neutron-irradiated with monitor minerals of known composition. Neutron irradiation produces noble gas isotopes following specific reactions, which are given in Table 1.

| Parent Isotope | Noble Gas Product | Reaction |
|---|---|---|
| $^{37}$Cl | $^{38}$Ar$_{Cl}$ | n, $\gamma$, $\beta$ |
| $^{79}$Br | $^{80}$Kr$_{Br}$ | n, $\gamma$, $\beta$ |
| $^{81}$Br | $^{82}$Kr$_{Br}$ | n, $\gamma$, $\beta$ |
| $^{127}$I | $^{128}$Xe$_{I}$ | n, $\gamma$, $\beta$ |
| $^{39}$K | $^{39}$Ar$_{K}$ | n, p |
| $^{40}$Ca | $^{37}$Ar$_{Ca}$ | n, $\alpha$ |
| $^{130}$Ba | $^{131}$Xe$_{Ba}$ | n, $\gamma$, $\beta$ |

The irradiation parameters and interference correction factors necessary to convert measured noble gas isotopes to halogen abundance are determined from analysis of the monitor minerals and salts that are included in the $SiO_2$ tube during irradiation. These parameters include the irradiation neutron fluxes (e.g., fast, thermal, and epi-thermal) and irradiation parameters J, $\alpha$, and $\beta$, which monitor the production of neutron-produced noble gases and determined from monitor minerals. After irradiation, samples are unpacked from the $SiO_2$ tubes, removed from their foil packets, and loaded separately into a drilled Al-sample holder. The sample holder is placed in a laser port, which is evacuated under vacuum. The sample port is pumped to UHV and heated to ~100°C for 12-24 h before analysis can begin.

Using a 10.6 µm 50W RF excited $CO_2$ laser the samples are either step-heated in increments of increasing laser power (W) until the point of fusion to resolve different potential halogen hosts, or fused in a single step for bulk analysis. Gas is purified and gettered in the extraction line until inlet in the mass spectrometer, where Ar, Kr, and Xe are measured over the course of 8-10 cycles using Faraday cup and compact discrete dynode (CDD) detectors. Instrument blank and air calibration standard analyses are interspersed between samples. After analysis, raw data (in fA) are corrected for instrument blank, mass discrimination, minor reactions producing interference isotopes during irradiation and decay of certain isotopes since the time of irradiation (e.g., $^{37}$Ar and $^{39}$Ar) to analysis.

**Sample preparation:**

There is minimal sample preparation required for NI-NGMS. The sample material, which may be present as bulk powder, bulk or separated mineral grains, or larger chips of sample are weighed and wrapped in small (~4 mm x 4 mm) individual Al-foil packets. Sieving of the sample to achieve a consistent size fraction or picking individual grains under a binocular microscope, is sometimes warranted.

**Sample impact:**

NI-NGMS is a destructive technique and the sample is entirely consumed during analysis. Thorough documentation and characterization of the sample to be analyzed should take place prior to analysis.

**Data quality:**

NI-NGMS is a sensitive method for the detection of halogens and can measure <1 ppb Br and I and <1 ppm Cl. The typical precision of NI-NGMS is ~2-10%, depending on the halogen element and abundance.

**Data products:**

The raw data products (all delivered as text or excel files) expected from NI-NGMS measurements include:

- Individual bulk sample measurements, presented as noble gas abundances in fA

- Standard analyses, presented as noble gas abundances in fA, for irradiation monitor minerals to determine irradiation neutron flux (and calculate halogen abundance)
- Standard air calibration analyses, presented as noble gas abundances in fA, for calculating mass discrimination and instrument sensitivity
- Instrument blank analyses, presented as noble gas abundances in fA, for subtraction from the sample gas

The processed data products (all delivered as Excel files) expected from NI-NGMS measurements include:

- Individual bulk sample measurements, presented as Cl, Br, I, Ca, K and Ba abundances in ppb/ppm concentrations and K/Ca, K/Cl, Br/Cl, I/Cl ratios of the sample
- Standard analyses, presented as noble gas abundances in $cm^3 \, g^{-1}$, for irradiation monitor minerals. This will include calculated irradiation neutron fluxes (e.g., fast, thermal and epi-thermal), irradiation parameters (J, $\alpha$, and $\beta$) determined from monitor minerals and any correction factors.
- Standard air calibration analyses, presented as noble gas abundances in $cm^3 \, g^{-1}$ and isotope ratios
- Instrument blank analyses, presented as noble gas abundances in $cm^3 \, g^{-1}$

These products can be used qualitatively to calculate the halogen, Ca, K and Ba abundances and the noble gas isotope ratios and abundances of the analyzed sample.

**Facility:**

The neutron-irradiation of samples is carried out offsite at a nuclear reactor facility by arrangement. There is flexibility in the selection of irradiation facility and the choice is typically dependent upon availability, schedule, shipping arrangements, and cost. Irradiation requirements and duration depend on the reactor power and resultant neutron fluence as well as the age and chemistry of the sample.

<u>University of Manchester</u>

The NI-NGMS facility at the University of Manchester includes:

- Fusions 10.6 µm 50W RF excited $CO_2$ laser (Teledyne-CETAC) for uniform stepped or single fusion heating of the target sample with a 150 µm to 5.5 mm spot size.
- New Wave UP 213 Nd:YAG UV laser ablation system for *in situ* analysis and depth-profiling with spot size down to 5 µm resolution.
- Thermo Scientific[TM] preparation bench with extraction line for the gettering and purification of sample gas prior to inlet in the mass spectrometer.

Thermo Scientific[TM] Argus VI[TM] static vacuum noble gas mass spectrometer with low internal volume (~700 $cm^3$) and high sensitivity (e.g., <1 × $10^{-3}$ Amp/Torr Ar) for the analysis of Ar, Kr and Xe on the Faraday cups (Ar) and a CDD multiplier (Kr and Xe

# 5.34 Resonance ionization time of flight noble gas mass spectrometry (RI-TOF-NGMS)

**Overview:**

Noble gases are extremely useful tracers of solar system evolution. They are highly depleted in rocky Solar System materials. Xenon and krypton are particularly important because they have enough isotopes to allow multiple contributing sources to be unambiguously identified. Distinct isotope signatures have been identified in the sun (solar wind), from primitive meteorites, and from the coma of a comet, allowing contributions from these sources to be traced in different materials. Several radioactive species decay with different characteristic timescales to produce distinct xenon isotopic signatures, in this context $^{129}$I and $^{244}$Pu are most significant since they form the basis of the I-Xe and Pu-Xe dating schemes. Artificial irradiation of several species also each produces a characteristic xenon isotopic signature. And both xenon and krypton isotopes can be produced by cosmic ray induced spallation reactions with target elements. Cosmic ray exposure ages can be calculated using the $^{81}$Kr-Kr system. In exploiting these processes to understand the provenance and history of extraterrestrial material, the analytical challenge is often to detect anomalies in small amounts of gas.

The RELAX (Refrigerator Enhanced Laser Analyser for Xenon) and RIMSKI (Resonant Ionization Mass Spectrometer for Krypton Isotopes) mass spectrometers at The University of Manchester are ultra sensitive, time of flight, resonance ionization mass spectrometers for measuring xenon and krypton isotope ratios respectively. Resonance ionization is a selective ionization technique; only atoms or molecules that have a resonance or resonances at the wavelength(s) generated by one or more lasers are efficiently ionized. Other species are only ionized by non-resonant processes, which are inherently less efficient. This allows for the detection of very small quantities of a given species in the presence of much larger quantities of other species, and thus measurement of isotopic ratios without isobaric interferences from other species is possible.

We anticipate that our focus is likely to be on:

- Exploiting the ability to make useful xenon/krypton isotopic measurements on smaller samples than conventional instruments. This potentially allows grain-to-grain variability to be studied, and consumes smaller aliquots of material from separates.
- I-Xe analysis after neutron irradiation of suitable samples.

Since the technique is destructive, samples should be exhaustively characterized before analysis.

**How it works:**

The RELAX and RIMSKI mass spectrometers both combine a resonance ionization ion source, a cryogenic sample concentrator and a time of flight mass analyser. Gas is extracted from samples, usually by laser step heating. Once admitted into the mass spectrometer, gas is continually condensed on a localised cold spot, and released every 0.1 s by pulses from an infra-red laser. The ionizing laser(s) are then fired through the resulting plume, selectively ionizing xenon or krypton. Resonant ionization of xenon is achieved using two 249.6 nm photons to excite the xenon atoms, which are then ionized from the excited state by a third photon of the same wavelength. Ionization of krypton is more complicated: the krypton atoms are initially excited by a photon at 116.5 nm, they are further excited by a photon at 558.0 nm, and then finally ionized from the second excited state by a 1064 nm photon. The vacuum ultra violet (vuv) light required from the first excitation steps is generated by nonlinear four-wave sum frequency mixing in a xenon/argon cell. Isotopes are separated according to their masses by the time of flight mass spectrometer, and microchannel plate detectors (MCPs) allow for the detection of all xenon or krypton isotopes from each laser pulse.

**Sample preparation:**

Samples can be single grains, chips up to about 3 mg, or mineral separates. They can be characterized by non-destructive techniques before analysis, but it is vitally important that samples are not impregnated with epoxy, crystal bond or similar for any characterization experiments prior to being analyzed in the RELAX or RIMSKI mass spectrometers.

I-Xe dating requires samples to be artificially neutron irradiated prior to analysis with fluences $\sim 10^{19}$ n cm$^{-2}$.

**Sample impact:**

These analyses are destructive. The samples are heated with an infra-red laser until they have melted and all the gas has been extracted. Fused glass beads often remain after analyses are complete.

**Data quality:**

The detection limits of both instruments are limited by variations in the blank, and are of the order of 1000 atoms of $^{84}$Kr and $^{132}$Xe. At maximum sensitivity samples are limited to $\sim 10^6$ atoms. Measurements are reproducible, and precise measurements of isotope ratios can be determined by averaging repeat analyses. Ratios for the major xenon and krypton isotopes can be determined to 1 % from a single analysis of 100,000 atoms per isotope.

The MCP detectors allow for all xenon or krypton isotopes to be recorded for each laser pulse, and the length of the flight tube is sufficient to ensure all isotopes are fully resolved from each other. There is, however, a non-resonantly ionized isobaric hydrocarbon that that is not resolved from $^{128}$Xe, and benzene is not resolved from $^{78}$Kr. Blank corrections can account for much of the contribution of these hydrocarbons to the measured peaks, but also add an additional contribution to the uncertainties of the corrected isotope ratios.

Date are collected over 5 minutes from each release, and typically 3 analyses (sample, air calibration or blank) can be performed per hour of operation. The instruments require around one hour to achieve operating temperature once they have been switched on, and are returned to room temperate at the end of each day's analyses.

**Data products:**

The data from both instruments are mass spectra as a function of time (samples at 1 ns intervals), over ranges covering around 20 units of mass-to-charge ratio. Diary files of metadata (date, sample, file number, etc) are automatically created alongside the mass spectra. The mass spectra are used to calculate isotope ratios and abundances with reference to calibration samples (derived from terrestrial atmosphere) and procedural blanks. The raw data are binary files (two bytes per time step), the diary files are text files, and the isotope ratios and abundances are usually calculated using Excel.

**Facility:**

Include institutions, hardware, software, and any notable instrument-specific capabilities

The University of Manchester

Both the RELAX and RIMSKI mass spectrometers are housed in the Isotope Geochemistry and Cosmochemistry Group (IGCG), in the Department of Earth and Environmental Sciences (DEES) at the University of Manchester. Both instruments were developed and built in house, and use custom software.

Both instruments comprise a mass spectrometer and a laser system.

Mass spectrometer (both instruments):

- Custom built time of flight mass spectrometer with associated gas extraction lines.
- Mass spectrometer incorporates Photonis MCP 18/12/10/12 D 40:1 Detection Quality Long-Life™ Microchannel Plate detectors.
- Mass spectrometer incorporates a sample concentrator in the ion source. A cryogenerator cools a localized cold spot, the temperature is monitored and maintained using thermocouples, a heating resistor, and a Eurotherm temperature controller.

RELAX lasers:

- Continuum Powerlight 8000 Nd:YAG laser, output 310 mJ at 355 nm
- Sirah Cobra Stretch dye laser with a frequency doubling unit, output 70 mJ at 500 nm using Coumarin 503 laser dye, ~10 mJ at 250 nm.
- Litron Minilase Nd:YAG laser, output 75 mJ at 1046 nm.
- JK Lasers JK50FL fiber laser, output 50 W at 1080 nm.

RIMSKI lasers and associated cells:

- Xenon/argon 4-wave mixing cell, and krypton and NOx detection cells, attached to the mass spectrometer, with associated gas filling lines
- Continuum Powerlight 9100 Nd:YAG laser with seeded output, output 410 mJ at 355 nm, 165 mJ at 532 nm, 170 mJ at 532 nm, 160 mJ at 1064 nm, and 290 mJ at 1064 nm.
- Sirah Cobra Stretch dye laser with a frequency doubling unit, output 84 mJ at 505 nm using Coumarin 503 laser dye, ~13 mJ at 252.5 nm.
- Sirah Cobra dye laser, output 18 mJ at 558 nm using Pyrromethene 580 laser dye.
- Sirah Cobra stretch dye laser with a Difference Frequency Mixing Near Infrared (DFMNIR) unit, output 39 mJ at 623 nm using DCM laser dye, 5 mJ at 1507 nm.
- Continuum Minilite I Nd:YAG laser, output 20 mJ at 1064 nm.

The group's facilities also include sample handing and preparation facilities, class 1000 and class 10000 clean rooms, and secure sample storage dedicated to analysis of extraterrestrial materials.

# 5.35 $^{40}$Ar/$^{39}$Ar geochronology and thermochronology

**Overview:**

The $^{40}$Ar/$^{39}$Ar technique is an improved derivative of the K/Ar method, which is based on the decay of $^{40}$K to $^{40}$Ar with a half-life of ca 1.25 Ga. This technique can be used to date essentially all minerals and rocks that contain potassium (e.g., plagioclase, groundmass and even pyroxene). The $^{40}$Ar/$^{39}$Ar geo- and thermo-chronometer is a very powerful tool to investigate the crystallization, igneous and/or cooling history of an extraterrestrial rock/particle (Jourdan et al., 2020). If the material is subsequently shocked after it cooled below its closure temperature, and owing to the relative sensitivity of the K/Ar system to post- crystallization perturbations, the $^{40}$Ar/$^{39}$Ar technique can instead record impact events which ages are not easily accessible by other techniques (Bogard, 2011; Bogard and Garrison, 1995). For instance, the $^{40}$Ar/$^{39}$Ar method can easily record major impact events and even disruption of asteroids by investigating the age of melt rocks and/or shock-reset minerals (Kennedy et al., 2013; Kennedy et al., 2019). $^{40}$Ar/$^{39}$Ar results can then be combined with diffusion modeling (e.g., Jourdan et al., 2014) to obtain the time-temperature history of a given meteorite and determine information such as the maximum temperature experienced by a given meteorite (Jourdan and Eroglu, 2017). Last, the $^{40}$Ar/$^{39}$Ar is particularly well placed to date hydrothermal alteration events by dating recrystallization products (e.g., sericite/albite after plagioclase; Verati and Jourdan, 2014).

**How it works:**

This technique requires that the samples are sent for irradiation in a nuclear reactor. $^{40}$Ar/$^{39}$Ar dating allows the derivation of age information from a single analysis of similarly behaving isotopes. This feature enables the degassing of samples and even single tiny chondritic particles (e.g. 91 µm; Jourdan et al., 2017) in multiple steps by laser (± furnace) incremental heating with temperatures ranging from 50°C to >1500°C. The step-heating degassing procedure allows construction of age spectra (Fig. 1) and inverse isochron diagrams, both of which presenting powerful statistical tests of the reproducibility of a sample age. A $^{40}$Ar/$^{39}$Ar age provides information on when a given mineral was raised above its Ar closure temperature and in most cases and when successful, either indicates the age of crystallization/cooling of the particle, or the age of an impact event. Furthermore, owing to the next generation of noble gas machines (cf. below), regolith particles can now be precisely analyzed on a single particle basis (Fig. 2; Jourdan et al. (2017)). Furthermore, from the same dataset and using the cosmochron approach, Ar isotopes can provide (1) a $^{38}$Ar Cosmic-ray exposure age which records the time span over which a given sample has been exposed unshielded to the bombardment by cosmic rays in the vacuum of space and (2) the $^{38}$Ar/$^{36}$Ar composition of the implanted solar wind at the surface of the grain (Levine et al., 2007).

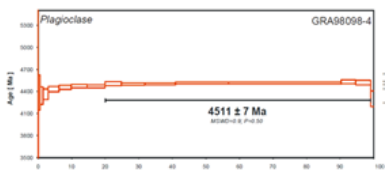

***Fig. 1***: Single-grain laser step heating $^{40}$Ar/$^{39}$Ar age spectrum of plagioclase from the metamorphic eucrites GRA98098 indicating crustal cooling at ca. 4511 Ma (Jourdan et al., 2020).

**Sample preparation:**

$^{40}$Ar/$^{39}$Ar sample preparation and analyses will be undertaken at Western Australian Argon Isotope Facility (WAAIF; lab director: F. Jourdan). Handling of ~70-100 µm-size grains is a routine procedure at the WAAIF since our initial work on Itokawa (Jourdan et al., 2017). Furthermore, we developed robust transfer procedures specific to return sample particles. All transfers will be carried out in The Advanced Ultra-Clean Environment (ACE) Facility located at Curtin University. The ACE consists of a series of class 10 ultra-clean laboratories enclosed within a ~400 m² class 1000 containment space and where the extremely low ultimate particle counts are achieved with successive 'spaces within spaces' and HEPA (99.999% high efficiency particle arresting) filtration at each stage. Particles from Bennu will be carefully transferred from the NASA containers into individual aluminium packages themselves loaded into large wells of one 1.9 cm diameter and 0.3 cm depth aluminium disc carefully packed and courier to the Oregon State University nuclear reactor (USA) where they will be irradiated for 90 hours.

Upon return from irradiation, the samples will be loaded and step-heated in a vacuum chamber (P = ~10$^{10}$ mbars) using a $CO_2$ (IR; 10,200 nm) laser rastered over the particle for 1 minute to ensure a homogenously distributed temperature. The gas will be purified in an ultra-low volume 200 cc stainless steel extraction line using a single SAES AP10 getters to minimize the volume of the line. Ar isotopes will be measured in static mode using an ARGUS VI mass spectrometer, the only mass spectrometer that has successfully handled *single particle* $^{40}$Ar/$^{39}$Ar analyses of regolith dust particles. We will follow the stringent procedures established during previous AOs related to asteroid Itokawa and carry out peak jumping analyses on the CDD detector. The ARGUS VI has been fully operational since 2014 and has successfully analyzed so far 6 particles returned from Asteroid Itokawa demonstrating the expertise of the team in this field (Fig. 2).

**Sample impact:**

Once the analyses have been completed, the samples will be completely fused (or even vaporized) and the residues will be lightly radioactive. Hence the residue will be disposed of in special containers for radioactive material at the WAAIF. Note that the residue can be made available for researchers that could make use of it, provided that regulations of handling radioactive material are complied with.

**Data quality:**

The success of a robust $^{40}$Ar/$^{39}$Ar analysis is ultimately constrained by the amount of radiogenic $^{40}$Ar measured which is itself constrained by the amount of $K_2O$ contained in the samples and its recorded age. At Curtin, we have experience with a large range of minerals and groundmass compositional types including chondrite and carbonaceous chondrites. Particularly relevant here, is our experience on tiny particles brought back from asteroid Itokawa (Jourdan et al., 2017 and in prep.). For reference, the particles analyzed contained 50-60% of plagioclase (each crystal with 0.8-1.0 wt% $K_2O$) and the total particle sizes range from 90 µm to 150 µm.

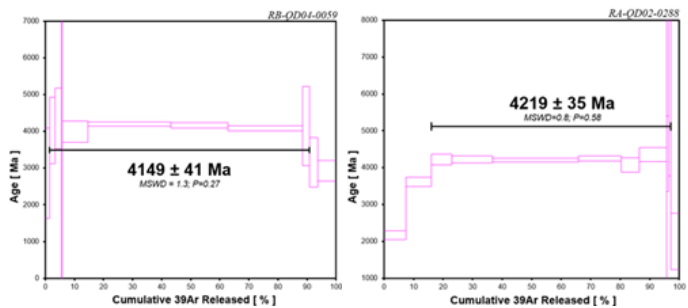

**Fig. 2**: Single-grain laser step-heating $^{40}$Ar/$^{39}$Ar age spectra of particles #0059 and #0288 recovered from Asteroid Itokawa and measured on the ARGUS VI mass spectrometer at Curtin University and showing plateau ages of 4149 ± 41 and 4219 ± 35 Ma (Jourdan et al., in prep.).

When larger amount of material becomes available, the overall precision of the age obtained increases. For example, recent analyses containing each 2 or 3 150-µm chondrule particles from the Murrilli meteorite and replicated on five aliquots yielded an ultra-precise $^{40}$Ar/$^{39}$Ar age of 4475.3 ± 2.3 Ma (2σ) and $^{38}$Ar Cosmic ray exposure age of 7.12 ± 0.41 Ma. (Anderson et al., in prep.).

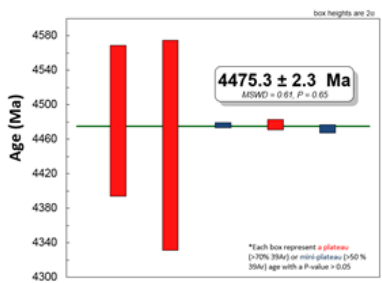

**Fig. 3**: $^{40}$Ar/$^{39}$Ar age of five aliquots of pyroxene-rich chondrule fragments analysed by laser step-heating at Curtin University (Anderson et al., in prep.)

One last example is given for small plagioclase crystals from a carbonaceous chondrite (PCA 02012) that yielded a young (impact) age of 166 ± 14 Ma (Anslem, in prep.) as per our recent focus on carbonaceous chondrite material in preparation for samples returned from asteroids Bennu and Ryugu.

The data accuracy is tested using the step-heating approach and essentially verifies that the system remained closed since the last recorded event (i.e. that it develops a concordant series of step ages that form a "plateau"; Fig. 1 &2).

**Data products:**

The raw data produced are under the form of beam intensities for each of the five isotopes measured (from $^{36}$Ar to $^{40}$Ar) for each heating step and delivered as a .txt document. Those are converted into ratios and age information using the ArArCALC software and ultimately produce Excel spreadsheets (Koppers, 2002) that are about 400 kb in size each.

**Facility:**

Curtin University: Western Australian Argon Isotope Facility

Ar isotopes are purified in an ultra-low volume extraction line designed for small (meteoritic) samples and measured in static mode using a low-volume (600 cm$^3$) ARGUS VI mass spectrometer from Thermo Fisher© (Jourdan et al., 2020) set with a permanent resolution of ~200. Measurements are carried out either in multi-collection mode using three Faraday cups equipped with three 10$^{12}$ ohm (masses 40; 38; and 37) and one 10$^{13}$ ohm (mass 39) resistor amplifiers and a low background compact discrete dynode (CDD) ion counter to measure mass 36 or, for ultra-small particles (e.g. Itokawa particles), isotopes are measured in peak hopping mode on the CDD. We measure the relative abundance of each mass during 10 cycles of peak-hopping and 16 seconds of integration time for each mass. Blanks (background analysis without firing the laser) are analyzed for every three to four incremental heating steps and typical $^{40}$Ar blanks range from 1x10$^{-16}$ to 2 x 10$^{-16}$ mol. Mass discrimination is monitored using an automatic air pipette. The raw data are processed using the ArArCALC software (Koppers, 2002).

# 5.63 Ion Chromatography (IC)

**Overview:**

IC is used for analysis of aqueous samples and can analyze inorganic anions (fluoride, chloride, nitrate, nitrite, sulfate, carbonate, bromide, iodide, phosphate, chlorate, perchlorate) and organic acids (acetate, formate). The team has access to both an ion chromatograph (IC) instrument and ion chromatography-orbitrap-mass spectrometer (IC-Orbitrap-MS). Cation analysis is also possible, but requires a different configuration.

**How it works:**

*Ion chromatography*

IC measures anion concentrations by separating them in a chromatographic column. 0.5 mL sample is placed in a sample vial and sealed with a cap. Sealed samples are placed in an autosampler, run through IC instrument equipped with chromatographic column. Sample solutions pass through a 4 x 250 mm AS-18 or AS-20 where ions are absorbed. As an ion extraction liquid (KOH solution of varying concentration) runs through the column, the absorbed ions are separating from the column. The retention time of different species and peak intensities are used to identify anion and its concentration.

*Ion chromatography with mass spectrometry*

The distribution of anions in extracts can be analyzed by using a Dionex ICS-6000 IC system (Thermo Fisher Scientific, Waltham, USA) equipped with an Orbitrap Exploris 480 mass spectrometer (Thermo Fisher Scientific). For the IC separation, a Dionex IonPac AS11-HC analytical column (2 x 250 mm, Thermo Fisher Scientific) with a guard column can be used at 35 °C. A conductivity detector combined with an AERS suppressor (Thermo Fisher Scientific) can be utilized to cross-check the detection of anions by the subsequent mass spectrometry.

The Orbitrap mass spectrometer is equipped with an electrospray ionization (ESI) source and operated in negative ion mode. The flow rates of nitrogen gas for desolvation is set to 40 arbitrary units (Arb) of the sheath gas, 5 Arb for the auxiliary gas, and 0 Arb for the sweep gas. The ion transfer capillary temperature and ESI spray voltage is set to 320 °C and 2.5 kV, respectively. Full scan mass spectra are acquired over a mass range of $m/z$ 50 to 750 with a mass resolution of 120,000 (at full-width-half-maximum for $m/z$ 200). Most ions are detected in the deprotonated form, $[M-H]^-$. The full scan measurements exhibit a general mass accuracy of less than 1 ppm, defined as [(measured $m/z$) − (calculated $m/z$)] / (calculated $m/z$) x $10^6$ (ppm). An exclusion list composed of the two largest background peaks, $m/z$ 112.9856 (corresponding to $CF_3COO^-$) and $m/z$ 68.9958 (corresponding to $CF_3^-$) can be implemented with a mass width of ±10 ppm to decrease the background signal.

**Sample preparation:**

IC analyzes liquid samples prepared in water-based matrix (acidic or alkaline solutions). Samples should be filtered to remove any residual particles. For analysis of the solid samples, samples should be dissolved in aqueous solution (water, acid, etc).

**Sample impact:**

The technique is destructive. However, impact the sample flow is designed to be a follow-on method that utilizes washes from SOAWG purification steps that would otherwise go unanalyzed.

**Data quality:**

This technique can obtain the ppm-level concentrations of several anions (e.g., perchlorate, sulfate, chloride) simultaneously in aqueous solutions.

**Data products:**

- Chromeleon Backup file (*.cmbx, ~5MB) The file contains the injection list, measurement method set up, the calibration set up, data processing set up, and measured concentrations

- Report file (*.xlsx, ~100kB) The file contains sample ID, retention time, peak area and height, and anion concentration in mg/l.
- Report file (*.csv, ~50kB) The file contains sample ID, retention time, peak area and height, and anion concentration in mg/l.

**Facility:**

NASA Johnson Space Center, Houston, TX, USA- The Astromaterials Research and Exploration Science (ARES) Division-Analytical Geochemistry Lab

- The Astromaterials Research and Exploration Science (ARES) Division houses an DIONEX AS-DV autosampler and DIONEX Integrion IC instrument equipped with 4 x 250mm AS-18 or AS-20 chromatographic column. Data outputs are analyzed with Chromeleon software.

Department of Earth and Planetary Sciences, Kyushu University, Fukuoka, Japan

- Thermo Fisher Scientific Dionex ICS-6000 IC system with Orbitrap Exploris 480 mass spectrometer (see for Orbitrap)

# 5.65 Time-of-flight secondary ion mass spectrometry (ToF-SIMS)

**Overview:**

Time-of-flight secondary ion mass spectrometry (ToF-SIMS) is a surface-sensitive analytical technique that provides detailed elemental, isotopic, and molecular information on surfaces, interfaces, and thin layers (depth profiling) with detection limits typically in the parts-per-million (ppm) range. The technique is known for its wide range of applications due to its universal applicability to all materials, as well as its ability to give direct molecular information (Benninghoven, 1994). ToF-SIMS allows parallel detection of organic and inorganic species in a sample and provides structural information about organic molecules that are intimately associated with minerals. One of the main advantages of ToF-SIMS is that ions are detected simultaneously over a wide mass range (0 to >10,000 m/z) and are not limited to counting over a predefined set of ion peaks. Other advantages of ToF-SIMS include nanoscale spatial and depth resolution (Fig. 1), identically small analytical volumes (~50 nm spot x 1 nm depth) for all ion species, extremely low detection limits, and capability of ionising and detecting large organic species with minimal fragmentation.

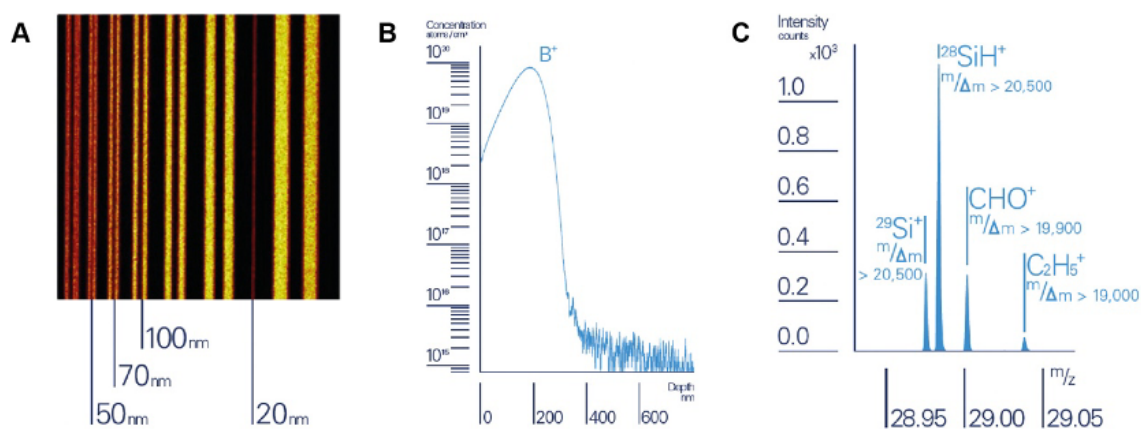

**Figure 1**. A. ToF-SIMS map showing spatial resolution to ~50 nm. B. Depth profile of a boron NIST implant standard. C. Example of a ToF-SIMS mass spectrum demonstrating high mass resolution (IONTOF M6 Brochure).

**How it works:**

A ToF-SIMS instrument detects the secondary ions that are emitted from a surface when it is bombarded with energetic primary ions (Benninghoven, 1994; Vickerman and Briggs, 2013). A liquid metal primary ion source generates primary ions, which interact with the sample surface to emit secondary ions. Secondary ions are extracted and focused into a narrow beam by ion optics. A time-of-flight analyzer consisting of an ion reflectron separates secondary ions based on their mass-to-charge ratio. The reflectron is an ion mirror that is used to reflect and focus secondary ions to the microchannel plate (MCP) detector. MCP detectors offer high sensitivity and fast response times, which are crucial for detecting low-abundance species and acquiring high-resolution mass spectra.

ToF-SIMS is a semi-quantitative approach for compositional abundances at the surface of a sample. Secondary ion yields depend on the ionization efficiencies of elements and molecules, which in turn are dependent on the primary ion sources. Instruments may have a range of different ion sources, such as $O_2$, Ar, Bi, and/or Xe. Multi-beam ToF-SIMS instruments can be run in positive or negative mode to suit the detection of ionic species of interest. Data can be collected in positive and negative mode sequentially to maximize detected species in a region of interest.

Mass ranges of ToF-SIMS instruments cover the periodic table from H to U, as well as large molecular ions (>10,000 m/z). Parallel detection of all masses allows all ionic and molecular species to be detected simultaneously if required. Options for collecting data at a reduced mass range (for example, collecting up to m/z of 100 instead of 1000) facilitates quicker measurement times when advantageous.

ToF-SIMS mapping is a useful targeting step in a correlative workflow for atom probe tomography (APT) and geochronology (Fig. 2A,B) (Rickard et al., 2016; 2020; Fougerouse et al., 2018; Timms et al., 2020). Multi-ion source instruments enable a wider range of ions to be analysed with increased sensitivity because the primary ion source that gives the greatest ionization efficiencies can be selected. Dual-beam analytical routines (i.e., separate ion sources for sputtering and analysis) for depth profiling and 3D analysis are also enabled by multi-ion source instruments (Fig. 2E).

Recently, gas cluster ion beams (GCIBs) with low kinetic energy per atom (E/n) have enabled the ionisation of large molecules with minimal fragmentation (Fig. 2C) (Shen et al., 2015). ToF-SIMS is one of the few in situ high-spatial-resolution techniques capable of identifying organic compounds and correlating their distribution with morphological features (Fig. 2F) (Naraoka et al., 2015; Siljeström et al., 2016). Some ToF-SIMS instruments (e.g., IONTOF M6) can operate in a low-energy mode suited to analyzing organic molecules. The use of large argon clusters as a sputter species in ToF-SIMS experiments allows depth profiling of organic materials to be carried out whilst retaining the intact molecular information. For example, organic biomarkers have been mapped in various materials including shale rock, fossils, and oils (e.g., Siljeström et al., 2009).

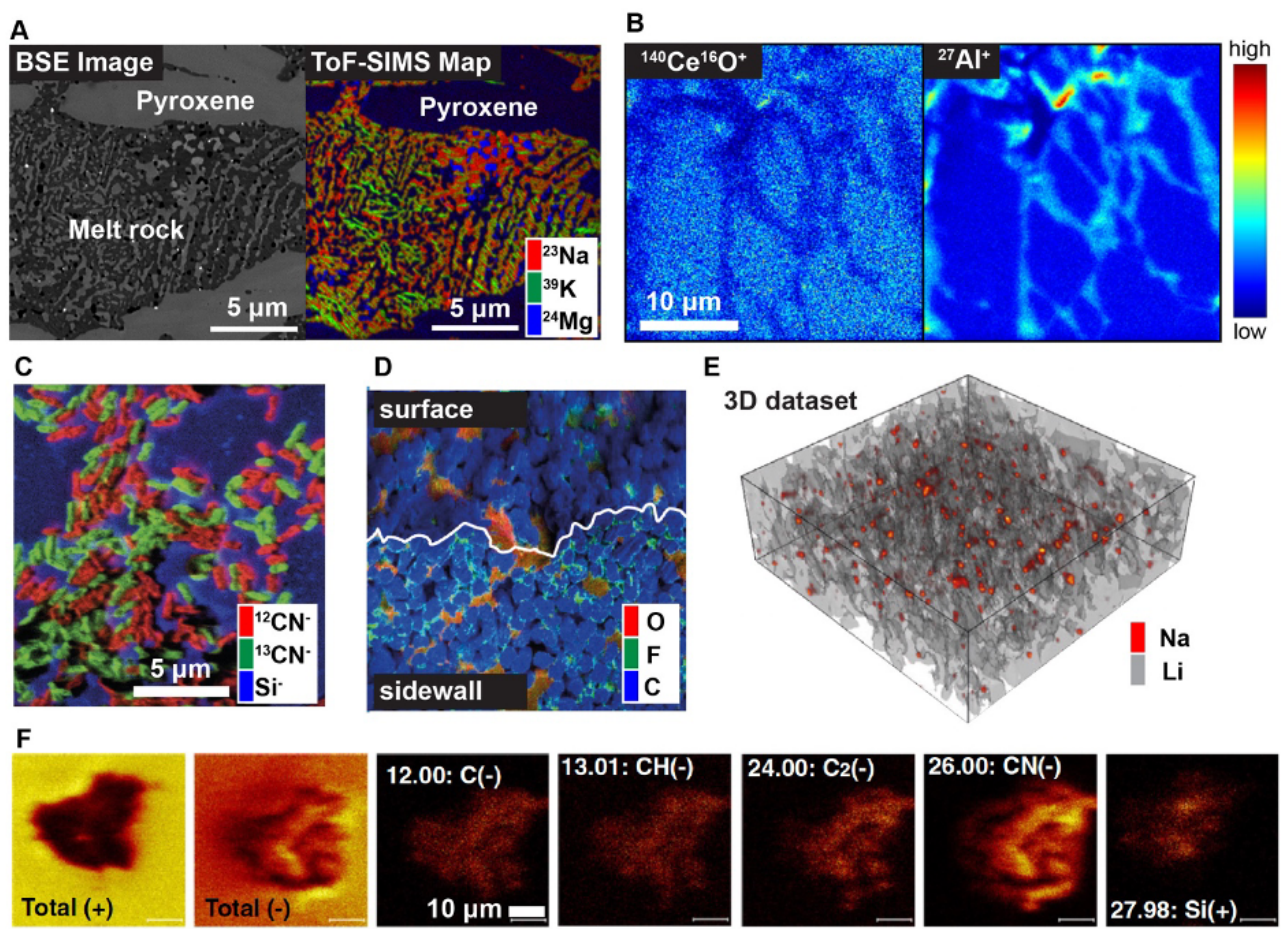

Figure 2. Examples of ToF-SIMS maps. A. Fine-scale major element variations in a silicate melt rock domain in a dust particle from asteroid Itokawa (Timms et al., in prep). B. Trace element variations in hydrothermally altered shocked titanite from the Chicxulub impact structure (Timms et al., 2020). C. Surface image of $^{12}$C and $^{13}$C labelled *Escherichia coli* cells on silicon (IONTOF M6 Brochure). D. Sidewall and surface image of a lithium-ion battery (IONTOF M6 Brochure). E. 3D tomographic image of a lithium-ion battery (IONTOF M6 Brochure). F. Images of positive (+) and negative (-) ions at specific m/z of a carbonaceous particle from asteroid Itokawa (Naraoka et al., 2015).

**Sample preparation:**

Solids, powders, and embedded samples (including geological thin sections) are common sample forms. Samples for ToF-SIMS analysis must be vacuum-compatible (dry, minimal outgassing). ToF-SIMS can be done on topographic samples; however, high-mass-resolution analysis requires flat surfaces with minimal topography. Surface topography reduces ion transmission, limits the achievable mass resolution, and can result in unwanted edge highlighting in ion maps. Thus, where possible, mechanically polished samples are preferred. Some ToF-SIMS instruments can analyze rough samples via the use of a delayed extraction mode and analyzer optics optimised for depth of field, significantly reducing any topographic contrast.

Matrix-matched standards are required for analyses where semi-quantitative concentration information is to be generated. Ion imaging analyses do not require standards.

For most geological samples, preparation methods analogous to those for scanning electron microscopy can be used. However, there is no requirement for a conductive coating, as the ToF-SIMS uses an electron flood gun for charge compensation.

**Sample impact:**

ToF-SIMS sputters ions from a sample surface and as such is a destructive technique. ToF-SIMS uses a pulsed primary ion beam with analysis currents commonly << 1 pA. For ion imaging, the typical primary current used is 0.02 pA, and <50 nm of material is generally removed from the surface of a small region of the sample. Users have control over the beam settings of ToF-SIMS instruments, and only the top few nanometers of material are removed during data collection in some instances. Primary ion beam spot sizes are 30–70 nm, which minimizes the area of the sample that is impacted. Ion implantation and damage to crystal structures from the primary ion beam can occur, but these phenomena are typically constrained to <50 nm depths below the sample surface and limited to the mapped regions. Localized ion redeposition during data collection is minimal.

A sputter beam (O2, Cs, or $Ar_x$) can be used for surface cleaning, ion yield enhancement, coating removal, depth profiling, or 3D analysis. In such instances, an ion current of >100 nA can be applied over a specific region to sputter away between a few nanometers or up to 10 μm, depending on the application. Sputter beam energies are low (<2 keV) to minimize sample damage below the sputter region.

**Data quality:**

Very high mass resolution (>30,000) can be achieved, which is sufficient to resolve molecular interferences on most inorganic ion peaks, such as $H_2O$ and $^{18}O$.

All ions in organic and inorganic samples can be detected in parallel with a very high mass range of >10,000 m/z.

The detection limit is typically <1 ppm (element dependent) and can reach the parts-per-billion level with seven orders of magnitude in dynamic range.

Mass accuracy (the difference between the measured m/z of an ion and the actual m/z) is typically <1 ppm, with seven orders of magnitude in dynamic range for (semi-)quantitative compositional analysis. Isotope abundances can be determined for a number of elements with an accuracy of <10 permil.

Spatial map (x-y) and depth (z) resolution is controlled by the primary beam source (e.g., Bi, Ga, and Xe ions) and energy. IONTOF M6 can achieve ultrahigh map (x-y) resolution of ~50 nm and depth (z) resolution of ~1 nm. This lateral spatial resolution is achievable with the Nanoprobe 50 liquid metal primary ion source, which has nine different aperture options.

**Data products:**

The data products expected from ToF-SIMS analysis include:

· 2D hyperspectral map data comprising full m/z spectra at each pixel on a map. This is the raw data file, usually in a proprietary format from the instruments that can be extracted in non-proprietary formats. Files contain a wide variety of metadata (e.g., sample ID, instrument settings, analysis conditions, x-y positions, sidecar optical images, total ion maps, ion induced secondary electron images, user-added descriptions).

· 2D images (maps) of relative abundances of ions, ion isotopes, and/or ion ratios. Images can be combined into tricolor (RGB) images with species assigned to each channel, or as single-species images. 2D images are the most common data types used for most applications of ToF-SIMS. Images can be exported in common formats (.tif, .jpeg, .png, etc.) with embedded metadata (sample ID, instrument settings, analysis conditions, x-y positions, user-added descriptions, etc).

· 3D hyperspectral tomographic datasets with full m/z spectra at each voxel in a data volume. As per 2D hyperspectral map data, but for depth profiles and 3D volumetric datasets.

· Mass spectra comprising ion counts per mass unit at a point or region of interest in a mapped area.

· Tabulated summaries (e.g., as .csv or .txt files) and plots (as common image file formats such as .jpeg, .tiff, or .png) of multivariate analysis of ToF-SIMS data. These can be formatted. These are more common data products for analyses of organic molecules.

**Facility:**

IONTOF M6 ToF-SIMS facility at the John de Laeter Centre at Curtin University, Perth, Western Australia

· Optimized for inorganic and organic molecular structure elucidation.

· Electron impact gas ion source; $O_2$, Ar, an Xe ion beam sources; and a thermal ionization caesium ion source.

· Ultrahigh imaging resolution: <50 nm lateral spatial resolution achievable with Nanoprobe 50 liquid metal primary ion source, which has nine different aperture options.

· Parallel detection of all ions in organic and inorganic samples.

· High mass resolution: >30,000.

· Very high mass range: >10,000 m/z.

· High mass accuracy (<1 ppm), with seven orders of magnitude in dynamic range for (semi-) quantitative compositional analysis.

· Intensities of more than 100 ions per pulse per mass with excellent linearity and reproducibility.

· Unique delayed extraction mode for high transmission with simultaneous high lateral and high mass resolution.

· Dual beam depth profiling for high-quality analysis over extended depths. Up to 50 Hz repetition rate resulting in up to three times faster imaging capabilities despite high sputter rates of 100 nm/min.

· Gas cluster ion source for molecular analyses.

· 3D analysis, including for organic materials.

· SurfaceLab 7 software package including fully integrated Multivariate Statistical Analysis (MVSA).

# 5.66 Accelerator Mass Spectrometry

**Overview:**

Accelerator mass spectrometry (AMS) is a technique for measuring long-lived cosmic-ray–produced (cosmogenic) radionuclides, with half-lives ranging from several thousand to several million years, in terrestrial and extraterrestrial samples. AMS uses a large particle accelerator in conjunction with ion sources, large magnets, and detectors to separate out molecular and isobaric interferences and to count individual atoms in the presence of up to $1\times10^{16}$ stable atoms per radioactive atom in the same sample. The AMS facility at PRIME Lab, Purdue University, routinely measures six different cosmogenic radionuclides; $^{14}$C, $^{10}$Be, $^{26}$Al, $^{36}$Cl, $^{41}$Ca and $^{129}$I. These nuclides are used for a wide variety of dating and tracing applications in the geological and planetary sciences. Measuring the concentrations of $^{10}$Be, $^{26}$Al, $^{36}$Cl and $^{41}$Ca in samples of asteroid Bennu will provide information on their cosmic-ray exposure history, including irradiation depth and time, which will constrain the formation age of Hokioi Crater and other recent surface processes.

**How it works:**

The AMS facility consists of six main components, including an ion source, an injector magnet, a tandem accelerator, analyzing and switching magnets, an electrostatic analyzer, and a gas ionization detector (Figure 1). These main elements are connected by a beam line, through which the ions travel. Several vacuum pumps remove the air from the beamline, so the beam particles have a free path from the ion source to the ionization detector. In addition, there are several beam-profile monitors and other electronic tools that help to guide (tune) the beam through the beamline by tuning each section of the AMS setup for the element and isotope of interest. We briefly describe each element below.

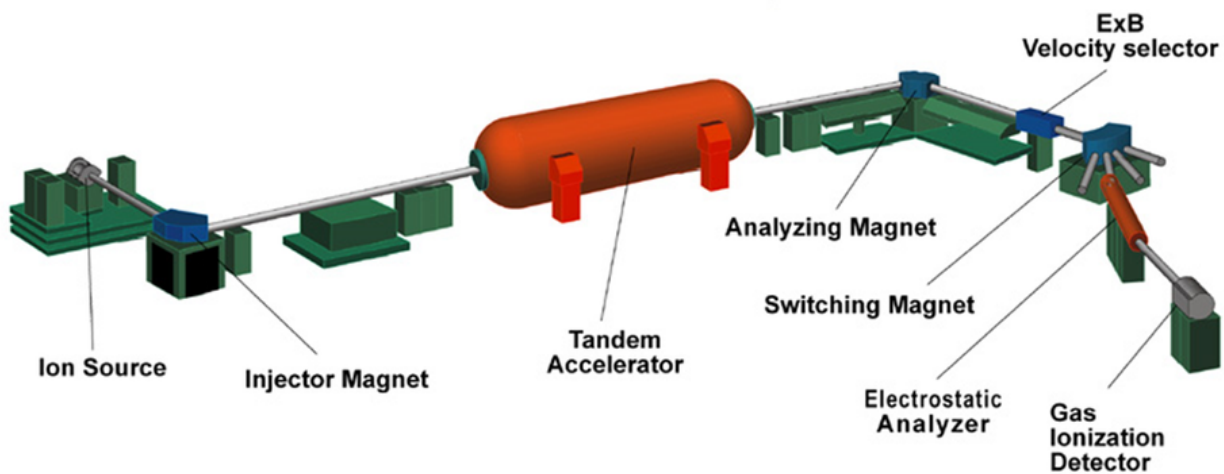

**Figure 1**. Schematic of PRIME Lab's AMS facility.

(1) The ion source produces a beam of ions (atoms that carry an electrical charge) from a few milligrams of solid material. The desired element is first chemically extracted from the sample (for example, a rock, a meteorite), then it is loaded into a metal holder and inserted into the ion source through an ultra-high vacuum sample changer. Atoms are sputtered from the sample by cesium ions, which are produced on a hot spherical surface called an ionizer and focused to a small spot on the sample. Negative ions produced on the surface of the sample are extracted (by electrostatic attraction) from the ion source and sent down the evacuated beam line towards the first magnet. At this point the beam is about 10 microamps, which corresponds to $10^{13}$ ions per second (mostly the stable isotopes).

(2) The injector magnet bends the negative ion beam by 90° to select the mass of interest, a radioisotope of the element inserted in the sample holder, and reject the much more intense stable isotopes of neighboring elements. Because this is only a first mass selection, at this point there are still numerous molecules and isobars (isotopes of neighboring elements having the same mass) that must be removed by magnets after the accelerator.

(3) The FN tandem accelerator consists of two accelerating gaps with a large positive voltage in the middle. Think of it as a bridge that spans the inside of a large pressure vessel containing $CO_2$ and $N_2$ insulating gas at a pressure of over 10 atmospheres. The bridge holds two long vacuum tubes with many glass (electrically insulating) sections. The center of the accelerator, called the terminal, is charged to a voltage of up to 10 million volts by two rotating chains. The negative ions traveling down the beam tube are attracted (accelerated) towards the positive terminal. At the terminal they pass through an electron stripper, either a gas or a very thin carbon foil, and emerge as positive ions. These are repelled from the positive terminal, further accelerating to ground potential at the far end. The name tandem accelerator comes from this dual acceleration concept. The final velocity is a few percent of the speed of light, or about 50 million miles per hour.

(4) The analyzing and switching magnets select the mass of the radionuclide of interest, further reducing the intensity of neighboring stable isotopes. In addition, they eliminate molecules completely by selecting only the highly charged ions that are produced in the terminal stripper. (Highly charged molecules are unstable because they are missing the electrons that bind the atoms together). Isotope ratios are measured by alternately selecting the stable and radioisotopes with the injector and analyzing magnets. The beams of each stable isotope of the element of interest are sent into separate Faraday cups where their currents are measured, thus allowing us to determine the ratio of the radioisotope to the stable isotope in each sample.

(5) The electrostatic analyzer is a pair of metal plates at high voltage that deflects the beam to the left by 20 degrees. This selects particles based on their energy and thus removes the ions that happen to receive the wrong energy from the accelerator, as they either have the wrong mass and/or the wrong charge state. Recently, a gas-filled magnet has been added to eliminate isobaric interferences from the radionuclide beam (e.g., $^{10}B$ from $^{10}Be$, $^{26}Mg$ from $^{26}Al$, $^{36}S$ from $^{36}Cl$).

(6) The gas ionization detector counts ions one at a time as they come down the beamline and enter the detector through a thin foil window. The ions are slowed down and come to rest in the gas. As each individual high-energy ion is slowed down to zero velocity, electrons are knocked off the gas atoms along the path of the slowing ion. These electrons are collected on several metal plates strategically placed within the detector, amplified, and read into the computer. The rate of energy loss as a function of distance traveled within the ionization detector determines the element's atomic number and distinguishes it from other stray ions entering the detector.

To correct the measured isotopic ratios of the samples for mass-dependent fractionation within the AMS system, which may vary slightly from one run to the next, AMS standards with known ratios of $^{10}Be/Be$, $^{26}Al/Al$, $^{36}Cl/Cl$, or $^{41}Ca/Ca$ are measured in the same run as the unknowns. We use internationally distributed AMS standards with well-known ratios (Table 1) so that the results can be verified in other AMS laboratories around the world.

**Table 1**. List of the cosmogenic radionuclides in samples of asteroid Bennu that will be measured by AMS, the main production mechanism and target elements for their production in chondritic samples, and the AMS standard used for calibration of the measurements.

| Nuclide | Half-life (yr) | Main Production* | Main Target Elements | AMS standard |
|---|---|---|---|---|
| $^{10}Be$ | $1.36 \times 10^6$ | GCR | O, Mg, Al, Si, Fe | Nishiizumi et al. (2007) |
| $^{26}Al$ | $7.05 \times 10^5$ | GCR + SCR | Al, Si | Nishiizumi (2004) |
| $^{36}Cl$ | $3.01 \times 10^5$ | GCR, (n,g) | K, Ca, Fe, Cl | Sharma et al. (1990) |
| $^{41}Ca$ | $1.04 \times 10^5$ | GCR, (n,g) | Fe, Ca | Nishiizumi et al. (2000) |

*GCR = galactic cosmic rays, SCR = solar cosmic rays, (n,g) = thermal neutron-capture.

**Sample preparation:**

Samples are dissolved, along with a solution containing a small (and well-known) amount of Be and Cl carrier, in concentrated $HF/HNO_3$. After dissolution, the Cl fraction is separated as AgCl. After removing Si as $SiF_4$, the residue is dissolved in dilute HCl, and a small aliquot of the dissolved sample is taken for chemical analysis by inductively coupled plasma optical emission spectrometry (ICP-OES). For samples with less than 1 mg of Al and Ca, a small amount (0.5–1 mg) of Al and Ca carrier is added. From the remaining solution, we separate Be, Al,

Ca, and Mn using ion exchange chromatography and acetyl-acetone extraction and purify each fraction (Welten et al. 2010; Nishiizumi et al. 2014). Be and Al are converted to BeO and $Al_2O_3$, respectively, mixed with high-purity Nb powder, and loaded into stainless-steel holders. AgCl is dried and loaded into a copper holder filled with AgBr. The Ca fraction is converted to $CaF_2$, dried, mixed with Re powder, and loaded into an aluminum holder. These holders are sent to PRIME Lab for AMS analysis. The Mn fraction is saved for future AMS analysis of $^{53}Mn$.

**Sample impact:**

The AMS measurements of cosmogenic radionuclides are destructive, as the samples are dissolved in $HF/HNO_3$.

**Data quality:**

Analytical uncertainties of the radionuclide measurements are generally dominated by counting statistics of the radionuclide ions in the ionization detector, and thus mainly depend on the concentration of the radionuclide in the sample and the amount of sample dissolved. For extraterrestrial samples of 10–100 mg with a few million years of cosmic-ray exposure, 1s uncertainties are generally 1–3% for measured $^{10}Be$, $^{26}Al$ and $^{36}Cl$ concentrations and 5–10% for $^{41}Ca$ concentrations. For smaller samples and/or samples with lower concentrations, the uncertainties are larger.

**Data products:**

The data products include concentrations of $^{10}Be$, $^{26}Al$, $^{36}Cl$, and $^{41}Ca$ reported in disintegrations per minute per kilogram (dpm/kg). These concentrations are calculated from the amount of sample dissolved; the amount of stable Be, Al, Cl, and Ca (which is the sum of the carrier amount added and — for Al and Ca — the amount in the sample itself); and the isotopic ratio of $^{10}Be/Be$, $^{26}Al/Al$, $^{36}Cl/Cl$, and $^{41}Ca/Ca$ measured by AMS. The data file also contains the chemical composition of the sample as determined by ICP-OES from dilutions of a small aliquot of the dissolved sample. The chemical analysis includes concentrations of Mg, Al, K, Ca, Ti, Mn, Fe, Co, and Ni.

The ICP-OES data for each sample and the AMS data for each isotope are included in separate .csv or .xlsx files.

**Facilities:**

Chemical separation and ICP-OES analysis: Space Sciences Laboratory, University of California, Berkeley, CA, USA

- The ICP-OES instrument at SSL's Cosmochemistry Lab is a Thermo Scientific iCAP 6300 duo and is used for measuring the chemical composition of all terrestrial and extraterrestrial samples studied in our lab.

AMS analysis: Purdue Rare Isotope Measurement (PRIME) Laboratory, Purdue University, West Lafayette, IN, USA

- The PRIME Lab is a dedicated research and service facility for AMS that measures up to ~10,000 samples per year. The AMS system is based on an upgraded FN (7 MV) tandem accelerator, as described in Sharma et al. (2000).

# Organic Chemistry

Organic chemistry is principally organized under the Sample Organic Analysis Working Group (SOAWG) with relevant techniques also employed by the Contamination Control/knowledge Working Group (CCWG) (CCWG). Organic analyses are concerned with both soluble and insoluble organic matter that will contribute to testing numerous OSIRIS-REx hypotheses.  This section describes the following techniques.

- 5.2 Visible, near-infrared, and mid-infrared (VNMIR) spectroscopy: μ-FTIR
- 5.12 X-ray Absorption Near-Edge Structure (XANES) Spectroscopy
- 5.15 Raman Vibrational Spectroscopy
- 5.18 Nanoscale Secondary Ion Mass Spectrometry (NanoSIMS)
- 5.26 Elemental Analysis-Isotopic Ratio Mass Spectrometry (EA-IRMS)
- 5.36 Gas Chromatography-Mass Spectrometry (GC-MS)
- 5.37 Liquid Chromatography-Mass Spectrometry (LC-MS)
- 5.38 Fourier Transform Ion Cyclotron Resonance Mass Spectrometry (FTICR-MS)
- 5.39 Microprobe Two Step Laser Mass Spectrometry (μL$^2$MS)
- 5.40 Nuclear Magnetic Resonance Spectroscopy (NMR)
- 5.41 Gas Chromatography–Isotopic Ratio Mass Spectrometry (GC-IRMS)
- 5.42 Solid-State Nuclear Magnetic Resonance Spectroscopy (SS-NMR)
- 5.56 Desorption Electrospray Ionization-Orbitrap Mass spectrometry (DESI-Orbitrap MS)
- 5.63 Ion Chromatography (IC)
- 5.64 Capillary Electrophoresis (CE)

## 5.2 Micro-Fourier Transform Infrared Spectroscopy (µ-FTIR)

See Spectroscopy section 5.2 Visible, near-infrared, and mid-infrared (VNMIR) spectroscopy

# 5.12 X-ray absorption near edge structure (XANES)

See 5.12 X-ray Absorption Near-Edge Structure (XANES) Spectroscopy

# 5.15 Raman Spectroscopy

See Mineralogy and Petrology section 5.15 Raman Vibrational Spectroscopy

# 5.18 Nanoscale Secondary Ion Mass Spectrometry (NanoSIMS)

See 5.18 Nanoscale Secondary Ion Mass Spectrometry (NanoSIMS)

# 5.26 Elemental Analysis-Isotopic Ratio Mass Spectrometry (EA-IRMS)

See Elements and Isotopes section 5.26 Elemental Analysis-Isotopic Ratio Mass Spectrometry (EA-IRMS)

# 5.36 Gas Chromatography-Mass Spectrometry (GC-MS)

- Overview:
- How it works:
  - Gas chromatography
  - Mass spectrometry
- Sample preparation, chemical derivatization and analytical methods:
- Sample impact:
- Data quality:
- Data products:
- Facilities:

## Overview:

Analysis of a suite of soluble organics in extraterrestrial materials requires a coordinated approach that uses (i) a variety of sample-preparation techniques (e.g., extraction, purification, derivatization), (ii) chromatographic separation, and (iii) detection and quantitation of individual compounds using a mass spectrometry detector. The choice of which sample preparation and extraction methods and instrumentation are used depends on whether or not specific organic compound classes are being targeted, the expected abundances of individual organic compounds, and whether enantiomeric and compound specific isotopic measurements can be made. Below is a description of Gas Chromatography-Mass Spectrometry and the sample preparation required for the analytes to which it is best suited:

- Abundance and distribution of primary and secondary aliphatic amines
- Abundance and distribution of aldehydes and ketones (method 1)
- Abundance and distribution of aldehydes and ketones (method 2)
- Abundance and distribution of carboxylic acids
- Abundance and distribution of hydroxy acids
- Abundance and distribution of alcohols
- Abundance and distribution of polyols and aldoses
- Abundance and distribution of volatile compounds (Method 1, GSFC)
- Abundance and distribution of volatile compounds (Method 2, Brown)
- Analysis of bulk organics (Method 1, GSFC)
- Analysis of bulk organics (Method 2, Brown)
- Analysis of IOM aliphatics

## How it works:

### Gas chromatography

Gas chromatography is a broad laboratory method for physically separating compounds from a mixture based on the preferential partitioning of different molecules across two different phases. The analyte is carried in the mobile phase (gas), then passes through a stationary phase (liquid or solid), and some analytes are retained more efficiently than others. Different types of chromatography are based on the geometry (through a stationary phase–filled tube or "column"), the type of mobile/stationary phase interaction (gas/liquid or gas/solid), and the nature of the analyte/stationary phase interaction (e.g., volatility, polarity, etc.). Column chemistries can also be combined to broaden analytes or improve difficult separations.

Column chromatography is well suited for real-time interrogation of the eluting compounds with detectors. In gas–liquid chromatography (often just called gas chromatography, GC), the mobile phase is a gas such as helium and the stationary phase is a high boiling point liquid adsorbed onto a solid. GC is well suited for any number of mass spectrometric detection techniques (GC-MS). Two-dimensional gas chromatography (GC×GC) is a technique that utilizes two different columns with two different stationary phases to produce a three-dimensional plot rather than a traditional chromatogram. GC×GC is commonly coupled to a mass spectrometer (GC×GC-MS) to study complex mixtures.

Preparative liquid chromatography (see LC-MS) is used to purify large quantities of a compound of interest for subsequent analyses and is employed in some sample preparations. Though solid-phase extraction (SPE) is also effective for concentrating analytes from the gas phase (e.g. sample headspace gas analysis) for subsequent GC-MS analysis. Methods for SPE gas analysis for possible use on Bennu samples and SART are in development.

## Mass spectrometry

Mass spectrometry is an analytical method used to identify and quantify chemical species based on their mass-to-charge ratio (m/z). It is also widely used to determine the isotopic abundances of various elements. The m/z ratio of ionized molecules is determined by their motion or traveling through electromagnetic fields. Element composition and chemical structure of molecules can be deduced with high-resolving-power instruments and/or via the interpretation of the mass spectra of molecular fragments. Mass spectrometers ionize, separate, then detect an analyte.

Of the two are two main methods of ionization available for GC-MS electron impact, (EI) is commonly used over chemical ionization (CI). In EI an analyte is bombarded with electrons (~100 eV) to form ions and these electrons typically fragment organic compounds in a characteristic pattern and produce the molecular ion ($M^+$) peaks in a mass spectrum. The ionization efficiency of a given technique is compound-dependent.

The separation of ions in a mass spectrometer is achieved by means of different technologies. A quadrupole mass spectrometer (QMS) selects ions by their movement through a sweeping electric and radio frequency field (video). A QMS provides low-resolution (nominal mass) spectra. The linear sweeping of a QMS means that the more ions being measured, the lower the dwell time, and thus sensitivity, of any particular ion. A QMS typically operates from 1 to 500 m/z, with the lower masses <20 m/z dominated by background. A QMS can also be used as a mass filter to permit only specific ions through to a different stage of a MS. A triple-quadrupole mass spectrometer (QqQ-MS) has a central collision cell to form fragments sandwiched between two QMS filters. A QqQ can operate as a QMS or filter a parent ion, fragment it, and observe a daughter ion or spectrum.

A time-of-flight mass spectrometer (ToF-MS) pulses ions from the ionizer to the detector. The time it takes for ions to pass down the flight tube indicates the m/z. ToF-MS provides rapid scans without loss of sensitivity and can provide high to ultrahigh resolutions (5,000 to 50,000 m/Δm) with mass ranges from 100 to many thousands m/z. Masses below 100 have lower mass accuracy and sensitivity.

Mass detectors are typically either microchannel plates (ToF) or electron multiplier tubes (QMS).

## Sample preparation, chemical derivatization and analytical methods:

All glassware, ceramics, and sample handling tools used in sample processing are rinsed with type 1 polished ultrapure water (18.2 MΩ, < 3 ppb total organic carbon, typically via a Millipore purification system), wrapped in aluminum foil, and then heated in a furnace at 500°C in air overnight to remove organic contamination. Only the two analysis of volatiles methods require intact individual sample fragments (bulk regolith sample or separate mineral lithologies); the rest of our analyses rely on sample fragments that are separately crushed to a fine powder (particle size typically < 150 mm grain size) using a ceramic mortar and pestle, transferred to glass vials, and homogenized using a vortexer inside a positive pressure ISO 5 high efficiency particulate air (HEPA)–filtered laminar flow hood housed in an ISO ≤8 white room. The specific sample extraction protocols, chemical derivatization techniques, and analytical methods for each soluble organic compound class in the current analysis plan are described in the individual sections below.

Chemical derivatization is often required to make analytes of interest compatible with chromatographic and detection methods. For example, in gas chromatography, analytes must be sufficiently volatile to be in the gas phase under the conditions used for separation. A derivatization reaction using trifluoracetic acid (TFA) can be used to lower the polarity of amino acids, thus increasing their volatility (Fig. 1). Each of the analytical methods sections below describes derivatization methods optimized for specific compounds.

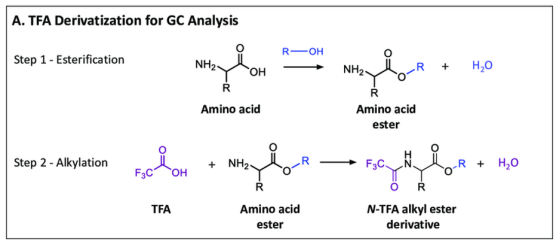

Figure 1. Schematic of the trifluoroacetic acid (TFA) derivatization reaction commonly used for analyses of meteoritic amino acids (Simkus *et al.* 2019a).

### Abundance and distribution of primary and secondary aliphatic amines

**Sample extraction and purification.** A portion of each powdered sample (~0.01 to 0.1 g) is transferred to the bottom of a borosilicate glass ampoule, 1 mL of ultrapure water is added to the ampoule, and the top of the ampoule is flame-sealed with an oxy-propane torch. The solid sample is not altered by heat during the rapid flame-sealing process. Water-soluble organic compounds are then extracted from the solids into the water supernatant by heating the ampoule at 100°C for 24 hours in an oven. After heating, the sealed ampoules are removed from the oven, cooled to room temperature and then opened. Optionally, half of the water supernatants are transferred to separate glass tubes and dried under vacuum, and then the water-extracted residues are subjected to a 6 M HCl vapor hydrolysis procedure at 150°C for 3 hours to determine total acid-hydrolyzable (free + bound) amine abundances and enantiomeric compositions. The remaining non-hydrolyzed water extracts of the samples are taken through the identical procedures in parallel with the acid-hydrolyzed extracts to determine the abundances and enantiomeric compositions of free amines. The water supernatants are transferred to separate glass tubes and stored at –20°C prior to derivatization and GC analyses. Prior to storage, optionally, As controls, a procedural water blank and/or crushed serpentine mineral heated to 500°C in air overnight is taken through the identical extraction procedure as the samples and processed in parallel.

***S*-TPC derivatization.** The derivatization method for the GC separation of primary and secondary aliphatic amines is carried as follows (Aponte et al. 2014; Aponte et al. 2016): the portions of the water extract allocated for analysis of amines are acidified with 1 mL 6 M HCl, dried under reduced pressure and subjected to acid-vapor hydrolysis using 1 mL of 6 M HCl at 150°C for 3 h. After hydrolysis, iron hydroxides are precipitated by the addition of 4 mL of 1 M NaOH; the aqueous layer is then separated by centrifugation and the solid residue rinsed with water (2 × 4 mL). The aqueous portions are combined, re-acidified using 1 mL of 6 M HCl, and dried under reduced pressure. The residues are re-dissolved in 2 mL of 1 M NaOH and extracted using dichloromethane (DCM) (3 × 1 mL). The combined DCM extracts are passed through a plug of anhydrous $Na_2SO_4$, rinsed once with 0.5 mL of DCM, and derivatized by stirring with 50 mg of Si-Gua and 50 µL of 0.1 M (*S*)-(–)-*N*-(trifluoroacetyl)pyrrolidine-2-carbonyl chloride (*S*-TPC) for 1 hour at room temperature. Next, 50 mg Si-$NH_2$ are added to the mixture, which is stirred for 30 min at room temperature. Finally, the slurry is filtered, rinsed with ~3 mL of DCM, dried under flowing nitrogen gas ($N_2$) then dissolved in 50 µL of ethyl acetate for analysis by GC-MS.

**GC-MS Analysis.** Separation and quantification of the derivatized amines are analyzed on a GC-MS instrument suite, consisting of a Thermo Trace GC and a Thermo DSQII electron-impact quadrupole mass spectrometer or Thermo Finnigan Trace 1310-TSQ8000 GC-triple-quadrupole MS. The GC-triple-quadrupole MS provides a slightly improved sensitivity, but one of the single quadrupole instruments available can divert 90% of the eluent to perform simultaneous combination GC-C-IRMS to permit $δ^{13}C$, $δ^{15}N$, or δD stable isotope analyses of individual compounds.

The derivatized amines are analyzed using a 5 m base-deactivated fused silica guard column (Restek, 0.25 mm ID), four Chirasil L-Val (25 m length × 0.25 mm I.D. × 0.25 µm film thickness; capillary columns connected in series using SilTite µ-union connectors, Restek), and the following temperature oven program: initial temperature 40°C, ramped at 10°C/min to 162°C, ramped at 2°C/min to 176°C, and ramped at 20°C/min to 200°C with a final hold of 20 min. The carrier gas used is ultrahigh purity (UHP) helium (5.0 grade) at 0.6 mL/min flow rate. Triplicate sample injections are made in splitless mode in aliquots of 1 µL; splitless mode is used to maximize sensitivity and minimize potential isotopic fractionation during injection. Analysis of the MS data for compound identification and quantification is performed with XCalibur software. The mass spectrum is used to identify and quantify amines in the samples by comparison to reference standards and application of calibration curves. Five-point external calibration curves are prepared for each individual hydroxy acid standard. Selected ion mass-to-charge ratio (m/z = 100+114+166+224+237) is used to identify and quantify compounds in the samples. Concentrations are calculated using quadratic equations derived from the calibration curves of each individual standard.

### Abundance and distribution of aldehydes and ketones (method 1)

This method shows superior compatibility with other hot water extraction methods, unlike method 2.

**Sample extraction and purification**. Same procedures as described above for amines except acidification of the extract is not required, but may be performed.

**PFBHA derivatization.** The derivatization method for the GC separation of aldehydes and ketones (collectively referred to as "carbonyl compounds") is an optimized Environmental Protection Agency (EPA) Method #556 using *O*-(2,3,4,5,6-pentafluorobenzyl) hydroxylamine hydrochloride (PFBHA) (Simkus et al., 2019b). The procedure is carried out as follows: 1 mL of 0.2 mg/mL PFBHA solution is added to the water extract. The solution is agitated for 5 minutes and left to react for 24 hours. 100 μL of 0.4 M HCl solution is added to each sample to quench the reaction and 2 mL of DCM is added to each solution to extract the derivatized carbonyl compounds. The solutions are shaken for 5 minutes and then left undisturbed for 30 minutes to allow the dichloromethane and water layers to separate and settle. The DCM layers are extracted and acid-washed (3 mL of 0.4 M HCl). The dichloromethane extraction of the derivatized carbonyl compounds and the subsequent acid-wash step are repeated for each sample with another 2 mL of dichloromethane. The 2 × 2 mL DCM extracts containing the aldehyde/ketone derivatives are concentrated down to small (~ 200 μL) volumes under a stream of nitrogen and then analyzed by GC-MS.

**GC-MS analysis.** Separation and quantification of the aldehyde and ketone derivatives are analyzed on a GC-MS instrument, consisting of a Thermo Trace GC, a Thermo DSQII electron-impact quadrupole mass spectrometer or Thermo Finnigan Trace 1310-TSQ8000 GC-triple-quadrupole MS. The GC-triple-quadrupole MS provides a slightly improved sensitivity, but one of the single quadrupole instruments available can divert 90% of the eluent to perform simultaneous combination GC-C-IRMS to permit $\delta^{13}$C or δD stable isotope analyses of individual compounds.

For the aldehyde and ketone measurements, GC separation uses a 5-m base-deactivated fused silica guard column (Restek, 0.25 mm ID) and three 30 m length × 0.25 mm I.D. × 0.5 μm film thickness Rxi-5ms capillary columns (Restek) connected using Press-Tight connectors (Restek) and the following temperature program: initial temperature 40°C, ramped at 10°C/min to 160°C, ramped at 5°C/min to 190°C, ramped at 10°C/min to 290°C and held for 7 min. The carrier gas used is UHP helium (5.0 grade) at a 2.6 mL/min flow rate. Sample injections are made in splitless mode in aliquots of 1 μL; splitless mode is used to maximize sensitivity and minimize potential isotopic fractionation during injection. Analysis of the DSQII MS data for compound identification and quantification is performed with XCalibur software. The mass spectrum is used to identify and quantify carbonyl compounds in the samples by comparison to reference standards and application of calibration curves. Five-point external calibration curves are prepared for each individual carbonyl standard. Selected ion mass-to-charge ratio (m/z = 181.0) is used to identify and quantify compounds in the samples. Concentrations are calculated using quadratic equations derived from the calibration curves of each individual standard.

### Abundance and distribution of aldehydes and ketones (method 2)

This method shows superior performance over method 1, however is incompatible with water extractions.

**Sample extraction and purification**. The powdered sample is transferred to a glass ampoule with 2 mL of dichloromethane, then frozen in liquid nitrogen and flame-sealed. The sample is extracted in the sealed ampoule at 50°C for 24 hours. After heating, the sealed ampoule is removed from the oven, cooled to room temperature and opened. The dichloromethane supernatant is transferred to a separate vial, and dissolved sulfur is removed by adding clean metallic copper beads. The solution is transferred to a separate vial and water is removed by adding clean $MgSO_4$. The solution is then filtered, transferred to a glass conical vial and stored at –20°C in preparation for derivatization.

**(*S*,*S*)-DMB-Diol derivatization.** The derivatization method for the GC separation of aldehydes and ketones is carried out (Aponte et al., 2019b; Aponte et al. 2020) using the reagent (*S*,*S*)-(–)-1,4-dimethoxy-2,3-butanediol, ≥99.0% ((*S*,*S*)-DMB-Diol; sum of enantiomers, GC) as follows: 20 μL of $BF_3$ and 50 μL of 0.1 mM (*S*,*S*)-DMB-Diol are added to the DCM sample extract, then the mixture is heated at 110°C for 1 hour in sealed glass vials with PFTE-lined screw caps. The reacted solution is concentrated to ~1 mL by flowing nitrogen gas ($N_2$), and extracted using 1 mL of water three times to remove the unreacted (*S*,*S*)-DMB-Diol and $BF_3$. After aqueous extraction, the organic layer is passed through a plug of anhydrous sodium sulfate ($Na_2SO_4$; 45 mm length × 5 mm I.D.), rinsed using 0.5 mL of DCM, and blown dry with $N_2$. Samples are dissolved in 70 or 100 μL of ethyl acetate for analyses by GC-MS.

**GC-MS analysis.** Separation and quantification of the aldehyde and ketone derivatives are analyzed on a GC-MS instrument, consisting of a Thermo Trace GC, a Thermo DSQII electron-impact quadrupole mass spectrometer or Thermo Finnigan Trace 1310-TSQ8000 GC-triple-quadrupole MS. The GC-triple-quadrupole MS provides a slightly improved sensitivity, but one of the single quadrupole instruments available can divert 90% of the eluent to perform simultaneous combination GC-C-IRMS to permit $\delta^{13}$C or δD stable isotope analyses of individual compounds.

GC separation uses a 5-m base-deactivated fused silica guard column (Restek, 0.25 mm ID) and four thickness CP-Chirasil Dex CB capillary columns (25 m length × 0.25 mm I.D. × 0.25 µm film, Agilent) connected using SilTite µ-union connectors (Restek) and the following temperature program: the initial temperature of 40°C is held for 4 min, ramped at 5°C/min to 100°C and held for 5 min, ramped at 2°C/min to 116°C and held for 10 min, ramped at 2°C/min to 120°C and held for 15 min, ramped at 2°C/min to 130°C and held for 7 min, ramped at 2°C/min to 144°C and held for 10 min, ramped at 2°C/min to 164°C and held for 3 min, and finally ramped at 15°C/min to 220°C with a final hold of 15 min. The carrier gas used is UHP helium (5.0 grade) at a 1.8 mL/min flow rate. Triplicate sample injections are made in splitless mode in aliquots of 1 µL; splitless mode is used to maximize sensitivity and minimize potential isotopic fractionation during injection. Analysis of the DSQII MS data for compound identification and quantification is performed with XCalibur software. The mass spectrum is used to identify and quantify the aldehydes and ketones in the samples by comparison to reference standards and application of calibration curves. Five-point external calibration curves are prepared for each individual aldehyde or ketone standard. Selected ion mass-to-charge ratio (m/z = 87+161+175+189) is used to identify and quantify compounds in the samples. Concentrations are calculated using quadratic equations derived from the calibration curves of each individual standard.

### Abundance and distribution of carboxylic acids

**Sample Extraction and Purification**. Same procedures as described above for amines, except acidification of the extract is not required, but may be performed.

**Esterification derivatization.** The derivatization method for the GC separation of carboxylic acids (Aponte et al. 2019a; Aponte et al. 2020a) is carried as follows: the portions of the water extract allocated for analysis of carboxylic acids are basified with 40 µL 2 M NaOH, dried under vacuum, and then derivatized with 2-pentanol using previously described methods (Aponte et al., 2019, 2020). The dry residues are suspended in 50 µL of 6 M HCl, 20 µL of 2-pentanol, 100 µL of DCM, and heated at 100°C for 15 hours in sealed PFTE-lined screw cap vials in a heating block. After cooling to room temperature, the derivatized samples are passed through a plug of aminopropyl silica gel (45 mm length × 5 mm I.D.), rinsed using ~3 mL of DCM, gently dried with flowing nitrogen gas ($N_2$), and dissolved in 100 µL of DCM for analysis by GC-MS.

**GC-MS analysis.** Separation and quantification of the carboxylic acid esters are analyzed on a GC-MS, consisting of a Thermo Trace GC, a Thermo DSQII electron-impact quadrupole mass spectrometer or Thermo Finnigan Trace 1310-TSQ8000 GC-triple-quadrupole MS. The GC-triple-quadrupole MS provides a slightly improved sensitivity, but one of the single quadrupole instruments available can divert 90% of the eluent to perform simultaneous combination GC-C-IRMS to permit $\delta^{13}$C or δD stable isotope analyses of individual compounds.

The derivatized acids are analyzed using a 5-m base-deactivated fused silica guard column (Restek, 0.25 mm ID), two Rxi-5ms (30 m length × 0.25 mm I.D. × 0.5 µm film thickness; capillary columns connected in series using SilTite µ-union connectors, Restek), and the following temperature program: the initial temperature is held at 40°C for 1 min, then ramped at 15°C/min to 110°C, ramped at 10°C/min to 140°C and held for 2 min, ramped at 10°C/min to 145°C, and finally ramped at 30°C/min to 300°C with a final hold time of 5 min. The carrier gas used is UHP helium (5.0 grade) at 4.8 mL/min flow rate. Triplicate injections of carboxylic acid derivatives are made in split mode (split flow: 5 mL/min, held for 1 min) in aliquots of 1 µL. Analysis of the DSQII MS data for compound identification and quantification is performed with XCalibur software. The mass spectrum is used to identify and quantify carboxylic acids in the samples by comparison to reference standards and application of calibration curves. Five-point external calibration curves are prepared for each individual carboxylic acid standard. Selected ion mass-to-charge ratio (m/z = 55+60+70+81+89+99+101+105+169) is used to identify and quantify compounds in the samples. Concentrations are calculated using quadratic equations derived from the calibration curves of each individual standard.

### Abundance and distribution of hydroxy acids

**Sample Extraction and Purification**. Same procedures as described above for amines, except acidification of the extract is not required, but may be performed.

**Esterification Derivatization.** The derivatization method for the GC separation of hydroxy acids (Aponte et al., 2020b) is carried as follows: the portions of the water extract allocated for analysis of hydroxy acids are basified with 40 µL 2 M NaOH, then dried under vacuum after which 200 µL of DCM, 10 µL of *iso*-butanol, and 20 µL of boron trifluoride diethyl etherate (≥46.5% $BF_3$ basis) are added and the mixture is heated at 80°C for 1 hour. After cooling to room temperature, 80 µL of acetyl chloride are added to the mixture, which is again heated to 80°C for 1 hour. The reaction mixture is passed through a plug of aminopropyl silica gel and rinsed with ~1 mL of DCM, then gently dried with flowing nitrogen gas, and dissolved in 100 μL of DCM for analysis by GC-MS.

**GC-MS Analysis.** Separation and quantification of the hydroxy acid derivatives are analyzed on a GC-MS instrument, consisting of a Thermo Trace GC, a Thermo DSQII electron-impact quadrupole mass spectrometer or Thermo Finnigan Trace 1310-TSQ8000 GC-triple-quadrupole MS. The GC-triple-quadrupole MS provides a slightly improved sensitivity, but one of the single quadrupole instruments available can divert 90% of the eluent to perform simultaneous combination GC-C-IRMS to permit $\delta^{13}$C or $\delta$D stable isotope analyses of individual compounds.

The derivatized hydroxy acids are analyzed using a 5-m base-deactivated fused silica guard column (Restek, 0.25 mm ID), four CP-Chirasil-Dex CB (Agilent, 25 m length × 0.25 mm I.D. × 0.25 μm film thickness; capillary columns connected in series using SilTite μ-union connectors, Restek), and the following temperature program: initial temperature 40°C, ramped at 10°C/min to 113°C, then ramped at 2°C/min to 134°C and held for 8 min, then ramped at 2°C/min to 165°C, and lastly ramped at 10°C/min to 210°C with a final hold of 12 min. The carrier gas used is UHP helium (5.0 grade) at 2.0 mL/min flow rate. Triplicate sample injections are made in splitless mode in aliquots of 1 μL; splitless mode is used to maximize sensitivity and minimize potential isotopic fractionation during injection. Analysis of the DSQII MS data for compound identification and quantification is performed with XCalibur software. The mass spectrum is used to identify and quantify hydroxy acids in the samples by comparison to reference standards and application of calibration curves. Five-point external calibration curves are prepared for each individual hydroxy acid standard. Selected ion mass-to-charge ratio (m/z = 101+115+129) is used to identify and quantify compounds in the samples. Concentrations are calculated using quadratic equations derived from the calibration curves of each individual standard.

**Abundance and distribution of alcohols**

**Sample extraction and purification**. The powdered sample is transferred to a glass ampoule with 1 mL of dichloromethane, then frozen in liquid nitrogen and flame-sealed. The sample is extracted in the sealed ampoule at 50°C for 24 h. After heating, the sealed ampoule is removed from the oven, cooled to room temperature and opened. The dichloromethane supernatant is transferred to a separate vial, and dissolved sulfur is removed by adding clean metallic copper beads. The solution is then passed through a glass pipette plugged with 200 mg of $MgSO_4$ in order to remove water, collected in a glass conical vial, and stored at –20 °C in preparation for derivatization.

**S-TPC Derivatization.** The derivatization method for the GC separation of alcohols is carried out using the reagent (S)-(–)-N-(trifluoroacetyl)pyrrolidine-2-carbonyl chloride (S-TPC) as follows: 1 μL of pyridine and 25 μL of S-TPC are added to each sample in a conical vial. The vials are capped, and the samples are heated at 90 °C for 20 minutes. After heating, the samples are cooled to room temperature, and then passed through glass pipettes plugged with 200 mg of aminopropyl silica gel in order to remove any excess S-TPC reagent. The dichloromethane solutions containing derivatized alcohols are concentrated down to small (~ 200 μL) volumes under a stream of nitrogen and then analyzed by GC-C-IRMS to permit $\delta^{13}$C or $\delta$D stable isotope analyses of individual compounds.

**GC-MS Analysis.** Separation and quantification of the alcohol derivatives is carried out on a GC-QqQ-MS instrument, consisting of a Thermo 13100 GC and a TSQ8000 triple quadrupole mass spectrometer. For the alcohol measurements, GC separation uses a 5 m base-deactivated fused silica guard column (Restek, 0.25 mm ID) and two 30 m length × 0.25 mm I.D. × 0.5 μm film thickness Rxi-5ms capillary columns (Restek) connected using Press-Tight® connectors (Restek) and the following temperature program: TBD. The carrier gas used is ultrahigh purity (UHP) grade helium (5.0 grade) at a TBD mL/min flow rate. Sample injections are made in splitless mode in aliquots of 1 μL; splitless mode is used to maximize sensitivity and minimize potential isotopic fractionation during injection. Analysis of the TSQ8000 QqQ-MS data for compound identification and quantification is performed with XCalibur software. The mass spectrum is used to identify and quantify the alcohols in the samples by comparison to reference standards and application of calibration curves. Five-point external calibration curves are prepared for each individual alcohol standard. Selected ion mass-to-charge ratio (m/z = 166.0) is used to identify and quantify compounds in the samples. Concentrations are calculated using calibration curves of each individual standard. Method development work for the analysis of alcohol derivatives is underway to focus on optimizing the purification of the samples to obtain sufficient peak separation and compound-specific isotope measurements.

### Abundance and distribution of polyols and aldoses

**Sample extraction and purification.** Crushing procedures as described above for primary amino acids, except powdered samples (approximately 1 g) are sonicated in 20 mL 2% HCl in a glass test tube held at room temperature with ice 4 times. The residue is further extracted with ultrapure water 2 times. Each time, the test tube was centrifuged at 2000 rpm for 30 min and supernatant was collected in a flask. The extract was dried. Then, sugars in the precipitates were extracted in 50 mL methanol and dried (Furukawa et al. 2019; Oba et al. 2019). The extracted sample is subjected to cation exchange chromatography (AG50W-X8 resin; Takano et al. 2010) and polyols are eluted with water. The eluate also contains uracil and thymine, so a fraction (~10%) is allocated to analysis of N-heterocycles (Method 2). The eluate is dried, dissolved in ultrapure water and filtered using a 0.20 μm PTFE cartridge filter.

**Aldonitoril acetate derivatization.** Aldonitoril acetate derivatization was conducted the same procedure as reported in Furukawa et al. (2019). This derivatization converts all the anomers of mono carbohydrates into their liner form (Figure Sugar 1). A dried sample after cation removal was added by 0.5 mL of 0.2 mol/L hydroxylammonium chloride in pyridine and heated at 90°C for 60 min. This was then added by 0.5 mL acetic anhydride and heated at 90°C for 60 min for acetylation. The sample was condensed with $N_2$ flow. Sugar derivatives in the dried sample was dissolved in 0.5 mL of DCM. The DCM solution was washed with 1 mol/L HCl and $H_2O$ repeatedly and dried. The dried sample was dissolved in a solvent (hexane:ethyl acetate=1:1) for GC-MS analysis.

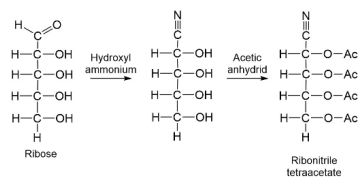

Figure Sugar 1. Derivatization reaction

**GC-MS Analysis.** The identification of sugars is conducted by GC-MS (Shimadzu QP2010 or Agilent 5977B) with a a DB-17ms column (60 m long, 0.25 μm thick, 0.25 mm ID). The carrier gas (He) flow rate was 0.8 mL/min. GC inlet, MS ion source, and MS quadrupole temperatures were 250°C and 230, and 150°C, respectively. The column oven temperature was programed as follows: initial temperature of 50°C for 2 min, then ramp up at 15°C/min to 120°C (hold 5 min), 5°C/min to 160°C, 3°C/min to 195°C (hold 5 min), and 3°C/min to 240°C (hold 10 min). The sample injection was conducted with solvent vent mode with 1/10 split ratio. Peak integration and quantification were conducted with m/z of either 103, 115, or 145 (Figure Sugar 2), depending on the effects of interfering peaks.

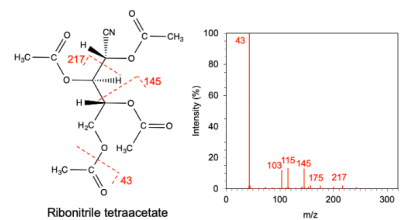

Figure Sugar 2. Electron impact ionization fragmentation pattern

### Non-targeted analyses of low polarity organic molecules

**Sample Extraction and Purification**. Same procedures as described above for Aldehydes and Ketones (Method 2).

**GC×GC/ToF-MS analysis of soluble organic compounds.** The separation, quantification, and compound identification of the DCM-soluble organic compounds are performed on a GC×GC-ToF-MS instrument suite, consisting of an Agilent 7890B GC and a LECO Pegasus HRT+ 4D. Method optimization is underway, for the free organics measurements, the GC×GC column set consists of a 5-m base-deactivated fused silica guard column (Restek, 0.25 mm ID) and two Rxi-5ms (30 m length × 0.25 mm I.D. × 0.5 μm film thickness, Restek) in the first dimension modulator-coupled to a Stabilwax secondary column (0.6 m × 0.25 mm ID, 0.25 μm film thickness, Restek) in series using SilTite μ-union connectors (Restek). The following temperature program is used: the initial temperature of 40°C is held for 2 min, ramped at 5°C/min to 100°C and held for 10 min, ramped at 10°C/min to 160°C and held for 10 min, ramped at 10°C/min to TBD°C, and finally ramped at 15°C/min to 220°C with a final hold of 15 min. The secondary oven is maintained at +5°C relative to the primary oven, and a modulation period of 4 seconds is applied. The carrier gas used is UHP helium (5.0 grade) at a 1.0 mL/min flow rate. Triplicate sample injections are made in splitless mode in aliquots of 1 μL; splitless mode is used to maximize sensitivity and minimize potential isotopic fractionation during injection. The ToF-MS operates at a storage rate of 150 Hz, with a 40–400 amu mass range and a detector voltage of 1.7 kV. Data are processed using the LECO Corp. ChromaTOF software. Compound identification is performed by comparison with the chromatographic retention in both dimensions and mass spectra of authentic standards where possible.

## Abundance and distribution of volatile compounds (Method 1, GSFC)

**Sample Extraction.** A sample chip (~0. 1 to 0.6 g) is transferred to the bottom of a borosilicate glass freeze-thaw apparatus (named "Cuernavaca-X" after the cow horn-shaped flask). The container is evacuated using a series of valves connected to a turbo pump and a pressure gauge, 4 mL of ultrapure water is added to one of the compartments of the freeze-thaw apparatus containing the sample chip and the is proceeded to three cycles of freezing to remove all gases dissolved in the water. Once water is degassed and the chamber containing the sample chip are under reduced pressure, the water is released and put in contact with the sample and the entire apparatus is kept under helium flow. The portion of the glass freeze-thaw apparatus containing the solid uncrushed sample in water is submerged in liquid nitrogen (-196 °C) for ten minutes and then let it warm to room temperature, completing once freeze-thaw cycle. The chip sample will be subjected to as many as necessary freeze-thaw cycles until the chip disaggregates into smaller pieces (10 to 40 cycles). Once the sample disaggregates, the freeze-thaw apparatus will be kept at room temperature and purged with helium gas (10 mL/min) evacuating the volatile compounds (including volatile organics) to a water removing filter and into a coil at liquid nitrogen temperature where the room temperature volatile molecules will be condensed and trapped. After helium purge for 10 minutes, the cryogenically trapped compounds will be in taken to room temperature and analyzed by GC-MS-IRMS.

**GC-MS-IRMS analysis.** Separation and quantification of the volatile compounds are performed on a GC-MS instrument, consisting of a Thermo Trace GC, a Thermo DSQII electron-impact single mass spectrometer in combination GC-C-IRMS to permit $\delta^{13}C$ or $\delta D$ stable isotope analyses of individual compounds.

The underivatized organic compounds are analyzed using a 3-m base-deactivated fused silica guard column (Restek, 0.25 mm ID), and one PoraBOND Q Column (25 m length × 0.25 mm I.D. × 3 μm film thickness; capillary columns connected in series using SilTite μ-union connectors, Restek), and the following temperature program: the initial temperature is held at 120°C for 1 min, then ramped at 20°C/min to 180°C held for 15 minutes, and ramped at 30°C/min to 300°C with a final hold time of 15 min. The carrier gas used is UHP helium (5.0 grade) at 5 mL/min flow rate in spitless mode. Analysis of the DSQII MS data for compound identification and quantification is performed with XCalibur software. The mass spectrum is used to identify and volatile compounds in the sample by comparison to reference standards and application of calibration curves where possible. Five-point external calibration curves are prepared for each individual standard. Selected ion mass-to-charge ratio may be used to identify and quantify compounds in the samples. Concentrations are calculated using quadratic equations derived from the calibration curves of each individual standard.

**Abundance and distribution of volatile compounds (Method 2, Brown)**

**Sample Extraction (Active).** The volatile organics released from sample crushing in a custom chamber will be purged with helium under gentle heating and possible vibration, to a system that will trap all evolved volatile organic compounds. The trapping system consists of three-stage trapping: 1) a cold finger that is liquid nitrogen or dry ice/isopropanol cooled, 2) followed by non-polar and polar organic traps commercially available, and 3) subsequently a capillary column with divinylbenzene (DVB) stationary phase submerged in liquid nitrogen.

> Stage 1 cold finger trap will permit initial trapping of most of the polar polar organic volatiles such as C1 to C5 monocarboxylic acids, ketones/aldehydes, amines, esters, alcohols etc. A septum on top of cold finger permits needle penetration and the use of solid phase micro-extraction (SPME) to remove desired amounts of these volatile organics to be delivered to GC-MS and GC-C-IRMS. Repeated loading (e.g., for compound specific H isotopic analysis using GC-C-IRMS that needs more mass) may be performed with the help of cryotrapping in the GC column front. The baseline technique is to use direct SPME to deliver the compounds with SPME on-fiber derivatization as a back up in the event that concentrations are too low for direct SPME. The efficiency of delivery for these two compound classes is enhanced by 200-1000 times when on-fiber derivatization methods are used. A polar SPME fiber (e.g. Carboxen) will permit delivery of alcohols into the instruments in Stage 1 as an alternative or augmentation to the more aggressive solvent extraction of alcohols, above. Direct SPME may also collect free cyanides, amides etc.

> Stage 2 trapping will be accomplished by using commercially available organic traps such as PDMS, Carbonxen, Tenax, molecular sieves or activated carbon (to be selected during SART tests). This stage will allow trapping of similar types of organic volatiles as in Stage 1. However, instead of using SPME to deliver different compound classes into GC-MS and GC-C-IRMS (which can be selective and partial depending on the filber used, and whether on-fiber derivatization is used), the traps will be placed into a Frontier Lab thermal desorption/pyrolysis device interface to desorb all absorbed compounds into GC-MS and GC-C-IRMS. An ultra-inert wax GC column will be used to separate all compounds with no derivatization. Based on published data so far, all common polar organic compound classes can be quite well separated chromatographically with no derivatization. In order to analyze sample on GC-MS as well as on GC-IRMS, the packing materials, after trapping, could be dug out, and divided into appropriate portions for mass allocations.

> Stage 3 trapping is dedicated for hyper volatile hydrocarbons such as methane, ethane etc. It is necessary in this case to use a DVB capillary column to trap these compounds with liquid nitrogen, as common trapping materials cannot retain these hyper volatiles. The trapped hyper volatiles will be directly flushed into GC-MS and GC-C-IRMS for analyses.

**Sample Extraction (Passive).** The JSC glove box GN2 glove box exhaust can be passed through a polar compound and nonpolar compound traps to collect gases released by sample manipulation and that would otherwise be lost. The traps will be selected to be of no impact to the operation of the glove boxes or sample integrity. Small portions of the traps are be placed in thermal desorbers for GC-MS and GC-C-IRMS analyses.  Such a collection could not only find released organics of scientific value, but also provide data to allow for curation decisions.

### Analysis of bulk organics (Method 1, GSFC)

**Sample Extraction.** Pyrolysis-GC-MS (py-GC-MS) requires minimal sample mass for the analysis of untreated organic-containing mineral samples and provided a chemical fingerprint of the compounds present. Typically, ~0.5 to 0.7 mg of sample is weighted inside of clean pyrolysis tube, where it is combined with calibration standards, and with the derivatization reagent (i.e., TMAH) as required. Prepared samples are subsequently loaded onto a pyroprobe autosampler interfaced with a GC-MS system. Samples are flash heated (10°C ms$^{-1}$) to ~600°C within the inert gas pyrolysis oven (CDS Pyroprobe 6000 series) and volatiles released are carried in a stream of He (35 mL min$^{-1}$) along a heated transfer line (300°C) directly into the GC-MS system. The pyrolyzed samples without pretreatment will characterize bulk organics (i.e., insoluble and soluble compounds). After the isolation of IOM, and analysis of SOM rinsed from solid samples prior to digestion by GC-MS to confirm a successful isolation protocol.   Instead of He, high pressure $H_2$ can be used to perform hydropyrolysis (hypy-GC-MS) with a slower heading rate than flash pyrolysis.  Hypy releases the organics with minimal structural rearrangement and less fragmentation than py.

**GC-MS analysis.** The GC is typically fitted with a nonpolar DB-5MS (5% phenyl) fused silica capillary column (though others can be used to achieve specific results), the GC oven is then programed to achieve the desired separation (e.g., 40 °C oven hold for 2 min, followed by a 4°C min$^{-1}$ ramp to 310°C, then a final isothermal hold at 310°C for 18 min), and the MS is operated in electron ionization (EI) mode at 70 eV scanning m/z 10 - 600. Thermally volatilized compounds are identified and confirmed by interpreting fragmentation patterns, using relative retention times, comparison to mass spectral libraries, and comparison to analytical reference standards where possible.

### Analysis of bulk organics (Method 2, Brown)

**Sample Extraction.** Sample residues after the removal of volatiles by cold-finger trapping  will be subjected to step wise thermal desorption and eventually pyrolysis analyses using a Frontier Lab device interfaced to a GC-MS and GC-C-IRMS. The Frontier Lab thermal desorption/pyrolysis device can be dismantled from one system and installed on another within 20 minutes. There is also a microjet (cryotrap) system to deliver liquid nitrogen for trapping analytes on the front of the column. This sharpen the chromatographic peaks especially for low molecular weight compounds to improve detections.  The sample consumption is very low - only 2 to 5 mg of sample is used for each analysis (split injector may be used as often too much, rather than too little, compounds are produced for instruments). The Frontier lab system also permits a rapid profiling of the evolved gas analysis (EGA), without using a GC column, to rapidly identify ranges of temperatures when most compounds or common gases are produced/evolved as temperature ramps typically from 50 to 700°C. One key regions of interest (i.e., high yield of compounds) are identified, the appropriate thermal desorption or pyrolysis temperature steps as set to screen compounds from individual zones (e.g., 100 to 150°C; 200 to 300°C; 600 to 800°C, and up to 1050°C), which not only reduce chromatographic coelution but also yield information of possible forms of these compounds exist in the samples.

Pyrolysis of insoluble organic matter, either directly or in the presence of TMAH, is followed by GC-MS and GC-C-IRMS analysis. The mass consumption is 1 to 2 mg for each run.

### Analysis of IOM aliphatics

**Sample Extraction.** Oxidation with $RuO_4$ of IOM samples permits the analysis of aliphatic linkages since $RuO_4$ oxidizes aromatic rings into carboxylic acids and leaves the aliphatic linkages intact. The resulting monocarboxylic acids are analyzed via SPME (Huang et al., 2007) and di- or multi-carboxylic acids (including those with additional substitutions like -OH) are analyzed using solvent extraction and derivatization (Remusat et al., 2005). All products are subjected to GC-MS and GC-C-IRMS analysis.

# Sample impact:

The analyses of soluble organic matter (SOM) is destructive and requires sample crushing and subsequent extraction in water or organic solvents by heating at elevated temperatures (up to 100°C for 24 hours). Soluble organic analyses are conducted on the solvent extracts;

however, the solvent insoluble residue can be used for other analyses, including acid dissolution of the solid residue and characterization of the IOM.

The SOM analyses described here are extremely sensitive to organic contamination, therefore sample handling and analysis of the samples prior to the extractions for organic analyses should be minimized. Exposure of the sample to some materials (e.g. latex gloves, epoxy cement, ink) can significantly compromise the sample quality. Even "clean" materials for inorganic analysis can be contaminated with organics, and vice versa. The glass and ceramics used in these analyses can leech inorganic ions.

Careful coordination of sample preparation to minimize sample consumption is also possible and is described here.

## Data quality:

GC-QMS detection limits are typically in the ~$10^{-12}$ mol range.

GC×GC-ToF-MS detection limits are typically ~$10^{-15}$ g (100 fg/μL octafluoronaphthalene (OFN)). The ToF-MS operates in High Resolution mode, R = 25,000 at m/z = 219 (mass range 40-600 m/z), or it can also operate in Ultra High Resolution mode with R = 50,000 at a mass range of 250 m/z. Typical analyses can provide accurate masses within ~0.8 ppm of the theoretical m/z value.

## Data products:

The data products expected from the GC-MS measurements include:

- 2D chromatogram (retention time vs. mass spectrum)

Data analysis consists of peak integrations vs. those of standards and mass spectrum of analyte vs. library. File types are proprietary conversion with Proteowizard is possible and will be tested prior to the SART.

The data products expected from the GC×GC-MS measurements include:

- 3D chromatogram (retention time A vs. retention time B vs. mass spectrum)

Data analysis consists of peak integrations vs. those of standards and mass spectrum of analyte vs. library. File types are proprietary; full automatic export method is not yet known.

## Facilities:

Astrobiology Analytical Laboratory, NASA Goddard, Greenbelt, MD, USA - Jamie Elsila

**Thermo Fisher Trace 1610 GC - Triple quadrupole 9610 MS system:** This system is equipped with a GC with dual split/splitless injector and pyrolysis configurations along with a subambient liquid $N_2$ cryogenic oven. The triple quadrupole MS with electron ionization source allows for full scans, single ion monitoring, and multiple reaction monitoring.

**Thermo Finnigan Trace 1310-TSQ8000 GC-triple quadrupole MS system**: This system is equipped with a GC with split/splitless injector and a subambient liquid $N_2$ cryogenic oven. The triple quadrupole MS with electron ionization source allows for full scans, single ion monitoring, and multiple reaction monitoring.

**LECO GC-HRT+ 4D system**: This GC×GC-ToF-MS system contains an Agilent 7890B gas chromatograph (GC), and LECO Dual Stage, Quad Jet Thermal Modulator serve as the inlet to high resolution time of flight mass spectrometer. The mass spectrometer has mass range of 10-1500 m/z, a resolving power of up to 50,000, and an acquisition of 200 Hz.

**Thermo Finnigan hybrid GC-MS/IRMS instrument with auxiliary TC/EA and EA inlets**: This is an instrument suite for both bulk and compound-specific isotope analysis. The core is a dual-inlet MAT 253 isotope ratio mass spectrometer (IRMS) with multiple inlets for sample introduction. The Trace GC is also connected to a DSQ II quadrupole mass spectrometer which can be used as the prime detector while looking for CSIA data of opportunity.

Brown Lab for Organic Geochemistry (BLOG), Brown University, Providence, RI, USA - Yongsong Huang

**Agilent GCMS** 7890B GC interfaced to 5977B inert plus single Quadrupole GCMS EI/CI Bundle

**Frontier Lab thermal desorption/pyrolysis device**

**Thermo Finnigan GC-IRMS instrument**: This instrument suite is for compound-specific isotope analysis. The core is a Delta Plus isotope ratio mass spectrometer (IRMS) with samples introduced by an Agilent 6890 plus GC.

Furukawa Lab Tohoku University, Sendai, Japan - Yoshihiro Furukawa

**Shimadzu QP2010 GC-MS:** This quadrupole GC-MS system is equipped with a split/splitless injector.

**Agilent 5977B GC-MS:** This quadrupole GC-MS system is equipped with a GESTEL PTV injector.

# 5.37 Liquid Chromatography-Mass Spectrometry (LC-MS)

- Overview:
- How it works:
    - Liquid chromatography
    - Mass spectrometry
    - Optical detectors
- Sample preparation, chemical derivatization and analytical methods:
    - Sample impact:
    - Data quality:
    - Data products:
    - Facilities:

## Overview:

Analysis of a suite of soluble organics in extraterrestrial materials requires a coordinated approach that uses (i) a variety of sample-preparation techniques (e.g., extraction, purification, derivatization), (ii) chromatographic separation, and (iii) detection and quantitation of individual compounds using one or two simultaneous detectors. Such detectors are of two kinds: mass spectrometers or mass spectrometers coupled with a UV optical (absorbance or fluorescence) detector.  The choice of which sample preparation and extraction methods and instrumentation are used depends on whether specific organic compound classes are being targeted, the expected abundances of individual organic compounds, and whether enantiomeric and compound specific isotopic measurements can be made. Below is a description of liquid chromatography-mass spectrometry (LC-MS) and the sample preparation required for the analytes to which it is best suited:

- Abundance, distribution, and enantiomeric composition of primary amino acids
- Abundance and distribution of primary and secondary amines/amino acids
- Abundance and distribution of peptides
- Abundance and distribution of free and acid-releasable cyanide
- Abundance and distribution of N-heterocycles (purines and pyrimidines) (Method 1)
- Abundance and distribution of N-heterocycles (purines and pyrimidines) (Method 2)
- Non-targeted analyses of polar organic molecules

## How it works:

### *Liquid chromatography*

Liquid chromatography is a broad laboratory method for physically separating compounds from a mixture based on the preferential partitioning of different molecules across two different phases. The analyte is carried in the mobile phase (liquid), then passes through a stationary phase (solid), and some analytes are retained more efficiently than others. Different types of chromatography are based on the geometry (through a stationary phase–filled tube or "column"), the type of mobile/stationary phase interaction (liquid/solid), and the nature of the analyte/stationary phase interaction (e.g., hydrophobic interactions, ion exchange, etc.). Column chemistries can also be combined to broaden analytes or improve difficult separations.

Column chromatography is well suited for real-time interrogation of the eluting compounds with detectors. In liquid–soild chromatography (often just called liquid chromatography, LC), the mobile phase is usually varied in composition over time to improve separation.  Mobile phases can vary, often aqueous buffers with increasing organic solvent over time ("reverse phase") is used. The pH, ionic strength, and organic component in the mobile phase can be varied to improve separation.  Numerous kind of stationary phase are available.  Common reverse phase columns are hydrophobic coated silica to separate analytes by polarity. LC can be on different scales or pressure domains.  Commonly HPLC or UHPLC:  High Precision or Ultra High Precision LC use higher pressures and finer particles in the column to achieve better and/or faster separations.  In hyphenated techniques they are all referred to as LC-.

Conversely preparative chromatography operates at ambient pressure and is used to purify large quantities of a compound of interest for subsequent analyses. This is necessary in sample preparation for many techniques, not just LC-MS. Though solid-phase extraction (SPE) is also effective for concentrating analytes from the gas phase (e.g., sample headspace gas analysis) for gas chromatography methods (see GC-MS).

Common detectors in LC are ultraviolet absorbance, fluorescence (FD), and various mass spectrometric methods (MS). Techniques that add more elaborate detectors with a chromatographic "front end" (or "inlet") are collectively referred to as hyphenated or hybrid techniques. Combinations of chromatography with mass spectrometry and optical spectroscopy increase the ability of an analytical chemist to study small quantities of complex mixtures.

### *Mass spectrometry*

Mass spectrometry is an analytical method used to identify and quantify chemical species based on their mass-to-charge ratio (m/z). It is also widely used to determine the isotopic abundances of various elements. The m/z of ionized molecules is determined by their motion or traveling through electromagnetic fields. Element composition and chemical structure of molecules can be deduced with high-resolving-power instruments and/or via the interpretation of the mass spectra of molecular fragments. Mass spectrometers ionize, separate, then detect an analyte.

There are several methods of ionization available, but atmospheric pressure electrospray ionization (ESI) is most commonly used with LC.  It is a soft ionization method that makes few but non-characteristic fragments by the addition or removal of $H^+$ from an analyte in an evaporating droplet of solvent under ~1.5 kV. It is most commonly commonly done in ESI+ mode to produce the $M+H^+$ parent ion.  In ESI, analytes are delivered in protic solution and may be derivatized to assist with chromatography.  ESI can suffer ion suppression, when excessive undesirable species are present and decrease the sensitivity of the analytes. Atmospheric pressure chemical ionization (APCI) or atmospheric pressure photoionization (APPI) are also available, and form $M+H^+$ or $M-H^+$ ions in a corona discharge or UV light, respectively.  They are appropriate for species that ionize poorly under ESI but tend to be less sensitive. An additional method that can be used to analyze surfaces is called Direct Analysis in Real Time (DART), which produces ions similar to those from ESI+ by shooting a stream of hot $N_2$ over a surface. It allows the desorption of intermediately volatile polar species from a solid and can allow some spatial resolution, but the ion efficiency and background effects can vary widely, making interpretation difficult. The ionization efficiency of a given technique is compound dependent. DART is employed instead of an LC inlet, but is used with an atmospheric inlet, as found on LC-MS.

The separation of ions in a mass spectrometer is achieved by means of different technologies. A quadrupole mass spectrometer (QMS) selects ions by their movement through a sweeping electric and radio frequency field. A QMS provides low-resolution (nominal mass) spectra. The linear sweeping of a QMS means that the more ions being measured, the lower the dwell time, and thus sensitivity, of any particular ion. A QMS typically operates from 1 to 500 m/z, with the lower masses <20 m/z dominated by background. A QMS can also be used as a mass filter to permit only specific ions through to a different stage of a MS. A triple-quadrupole mass spectrometer (QqQ-MS) has a central collision cell to form fragments sandwiched between two QMS filters. A QqQ can operate as a QMS or filter a parent ion, fragment it, and observe a daughter ion or spectrum. QqQ can selectively monitor all daughter of a specific parent, all parent of a specific daughter, or look for a specific parent/daughter combination. This technique is applied when a specific analyte with distinct fragments is to be measured over a large background.

A time-of-flight mass spectrometer (ToF-MS) pulses ions from the ionizer to the detector. The time it takes for ions to pass down the flight tube indicates the m/z. ToF-MS provides rapid scans without loss of sensitivity and can provide high to ultrahigh resolutions (5,000 to 50,000 m/Δm), with mass ranges from 100 to many thousands m/z. Masses below 100 have lower mass accuracy and sensitivity. A qToF-MS contains a QMS mass filter to permit increased selectivity or fragmentation before a ToF-MS.

An Orbitrap-MS is a type of Fourier transform MS, where a spindle-shaped ion trap captures ions in an orbit around the electromagnet spindle. The orbit is measured by radio frequency, and an ultrahigh-resolution spectrum is determined (30,000 to 100,000 m/Δm) for 20 to 2000 m/z. The spectral resolution is limited by the time an ion is allowed to reside in the Orbitrap. The Orbitrap can become saturated with undesired ions and decrease sensitivity. To solve this, a second ion trap can be placed before the Orbitrap as a filter. A method of fragmentation can also be employed in the ion beam to permit Orbitrap analysis of fragments.

Mass detectors are typically either microchannel plates (ToF), radio frequency (Orbitrap), or electron multiplier tubes (QqQ-MS).

### *Optical detectors*

LC systems can have additional optical detectors to assist in analyte quantitation. A UV/visible photodiode array (PDA) detector is used for the few analytes that significantly absorb light. The ability to absorb UV light by organic compounds is typically derived from conjugated ring systems. The absorbance is broad and relatively non-specific but can assist in verification of an analyte with the mass spectrum. The sensitivity is directly dependent on the molar absorptivity of the analyte and the solvent in which it is dissolved. A diode array detector is used in a flowing LC system to record the 1.2-nm resolution spectrum from 190 to 500 nm in real time. Photons from a $D_2$ lamp are focused on a ~0.5-μL flow cell and the absorbance is detected by the PDA. A fluorescence detector (FD) greatly increases sensitivity for the few analytes that are fluorescent (either natively or via derivatization). A specific wavelength of light from a Hg/Xe arc lamp is directed at a flow cell to excite the analyte and a specific wavelength of emission (fluorescence or phosphorescence) is detected orthogonal to the emission wavelength in real time. The detectors have an excitation wavelength range of 200 to 890 nm and an emission wavelength range of 210 to 900 nm with a 20-nm bandwidth.

## Sample preparation, chemical derivatization and analytical methods:

All glassware, ceramics, and sample handling tools used in sample processing are rinsed with type 1 polished ultrapure water (18.2 MΩ, < 3 ppb total organic carbon, typically via a Millipore purification system), wrapped in aluminum foil, and then heated in a furnace at 500°C in air overnight to remove organic contamination. Individual sample fragments (bulk regolith sample or separate mineral lithologies) are separately crushed to a fine powder (particle size typically < 150 mm grain size) using a ceramic mortar and pestle, transferred to glass vials, and homogenized using a vortexer inside a positive pressure ISO 5 high efficiency particulate air (HEPA)–filtered laminar flow hood housed in an ISO ≤8 white room. The specific sample extraction protocols, chemical derivatization techniques, and analytical methods for each soluble organic compound class in the current analysis plan are described in the individual sections below.

Chemical derivatization is often required to make analytes of interest compatible with chromatographic and detection methods. For example, in LC, reactions can be used to add a fluorescent tag such as *o*-phthaldialdehyde/*N*-acetyl-L-cysteine (OPA/NAC) to chiral primary amines to turn enantiomers (mirror-image molecules that are difficult to separate) into diastereomers that can be separated (Fig. 1). Each of the analytical methods sections below describes derivatization methods optimized for specific compounds.

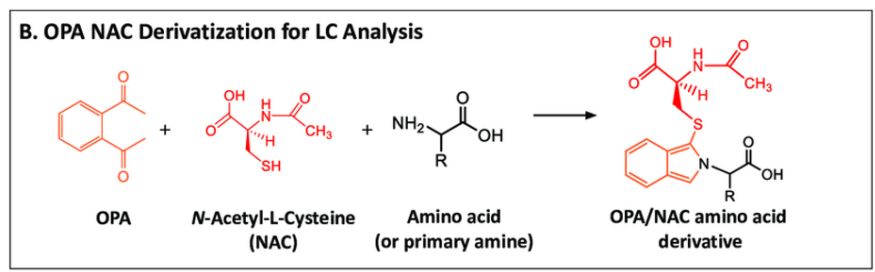

**Figure 1.** Schematic of the *o*-phthaldialdehyde/*N*-acetyl-L-cysteine (OPA/NAC) derivatization reaction commonly used for analyses of meteoritic amino acids (Simkus *et al.* 2019).

### *Abundance, distribution, and enantiomeric composition of primary amino acids*

**Sample extraction and purification.** A portion of each powdered sample (~0.01 to 0.1 g) is transferred to the bottom of a borosilicate glass ampoule, 1 mL of ultrapure water is added to the ampoule, and the top of the ampoule is flame-sealed with an oxy-propane torch. The solid sample is not altered by heat during the rapid flame-sealing process. Water-soluble organic compounds are then extracted from the solids into the water supernatant by heating the ampoule at 100°C for 24 hours in an oven. After heating, the sealed ampoules are removed from the oven, cooled to room temperature and then opened. Half of the water supernatants are transferred to separate glass tubes and dried under vacuum, and then the water-extracted residues are subjected to a 6 M HCl vapor hydrolysis procedure at 150°C for 3 hours to determine total acid-hydrolyzable (free + bound) amino acid abundances and enantiomeric compositions (Glavin et al. 2006). The remaining non-hydrolyzed water extracts of the samples are taken through the identical procedures in parallel with the acid-hydrolyzed extracts to determine the abundances and enantiomeric compositions of free amino acids. The non-hydrolyzed and HCl acid-hydrolyzed, hot-water extracts are desalted by using cation-exchange chromatography (BIO-RAD Poly-Prep ion exchange columns prepacked with 2 mL AG50W-X8 resin, 100-200 mesh, hydrogen form). After a series of column preparation steps, loading of the sample water extracts, and elution of the anionic species with water, the amino acids from the sample extracts are obtained by column elution with 7 mL aqueous 2M $NH_4OH$. The $NH_4OH$ eluates are each dried under vacuum, resuspended in 100 mL Millipore ultrapure water, transferred to clean glass autosampler vials, and stored at –20°C prior to OPA/NAC derivatization and liquid chromatography analyses. As controls, a procedural water blank and/or crushed serpentine mineral heated to 500°C in air overnight is taken through the identical extraction procedure as the samples and processed in parallel. Based on previous analyses of amino acid standards taken through the entire extraction, acid hydrolysis, and desalting procedure, there is no evidence of significant decomposition, racemization, or thermal degradation of the amino acids (Glavin et al. 2010).

OPA/NAC derivatization. Pre-column derivatization of samples prior to ultrahigh-performance liquid chromatography with UV fluorescence detection and time-of-flight mass spectrometry (LC-FD/ToF-MS) analysis involves the use of 0.1 M sodium borate, *o*-phthaldialdehyde/*N*-acetyl-L-cysteine (OPA/NAC), and 0.1 M hydrazine hydrate. Sodium borate is generated by heating solid sodium borate decahydrate at 500°C in air for 3 h, prior to dissolution in Milli-Q ultrapure water. The OPA/NAC derivatization reagent is prepared by firstly generating 0.1 M OPA by dissolving 0.1 g OPA in 7.5 mL methanol (Optima Grade), secondly generating 0.5 M NAC via dissolving 0.408 g NAC in 5 mL Milli-Q ultrapure water, and thirdly mixing 300 mL of 0.1 M OPA with 30 mL of 0.5 M NAC, and 670 mL of 0.1 M sodium borate. The 0.1 M hydrazine ($NH_2NH_2$) solution is prepared by vacuum distillation of concentrated anhydrous hydrazine (98% purity) and subsequent dilution in Milli-Q ultrapure water. Prior to analysis, 10 mL of each sample extract, procedural blank, or standard is added to 20 mL of 0.1 M sodium borate buffer and dried under vacuum. The dried residues are resuspended in 20 mL Millipore ultrapure water at room temperature and then derivatized by adding 5 mL of OPA/NAC. The OPA/NAC reaction is quenched after 15 min with 75 mL of 0.1 M hydrazine hydrate and 10 mL of the derivatized extract immediately then injected into the LC-FD/ToF-MS instrument for chromatographic separation and UV fluorescence and mass detection of the OPA/NAC amino acid derivatives.

**LC-FD/ToF-MS amino acid separation and quantification.** The amino acid concentrations, relative distributions, and enantiomeric ratios in the sample extracts are measured using an ACQUITY UHPLC with an UV fluorescence detector and Waters LCT Premier ToF-MS or Xevo G2-XS (qToF-MS) mass spectrometer in ESI+ mode for analysis of the OPA/NAC amino acid derivatives. Each derivatized extract is analyzed by three separate injections using our standard tandem LC column conditions for initial two- to ten-carbon ($C_2$ to $C_{10}$) amino acid separation and characterization (Glavin et al. 2010). The $C_2$ to $C_6$ amino amino acids can be chromatographically resolved using a Waters BEH C18 column (2.1 × 50 mm, 1.7-mm bead) and a Waters BEH phenyl column (2.1 × 150 mm, 1.7-mm bead) in series. Both columns are maintained at 30.0 ºC. During analyses, liquid chromatography is facilitated by the implementation of two solvents: solvent A (50 mM ammonium formate, 8% methanol, pH 8) and solvent B (Fisher Optima grade methanol). Solvent A is prepared by 2M $NH_4OH$ titration of a 50 mM formic acid solution to pH 8. The mobile phase conditions for amino acid separations are as follows: flow rate, 150 mL/min; gradient, time in minutes (%B): 0 (0), 35 (55), 45 (100). $C_5$ amino acid isomers and enantiomers are chromatographically separated using the same chromatography conditions as for the $C_2$ to $C_6$ amino acids but require the implementation of a different gradient. The gradient used for $C_5$ amino acid isomers and enantiomers was structured via time in minutes (%B): 0 (15), 25 (20), 25.06 (35), 44.5 (40), 45 (100).

For Xevo G2-XS analyses, a dual ESI system is be used for the purpose of implementing lockmass mass accuracy real-time corrections. The primary ESI+ source is operated using the following parameters: capillary voltage, 1.2 kV; sampling cone voltage, 40 V; cone gas ($N_2$) flow, 50 L/hour, source temperature, 120°C; desolvation gas ($N_2$) temperature, 500°C; desolvation gas flow rate, 300 L/hour. Due to the possibility that minor variations in the m/z scale may occur during the course of executing experimental runs after instrument calibration is performed, a reference ESI+ source is implemented to supply an independent leucine enkephalin standard signal. The reference ESI+ source is operated using identical parameters to those used for the primary ESI+ source, except the reference ESI+ source used a capillary voltage of 2.8 kV and a reference cone voltage of 30 V. The ToF analyzer is operated in Sensitivity mode, which used a reflectron to provide a full width at half maximum resolution of 22,000 based on the $[M+H]^+$ of leucine enkephalin, m/z 556.2771. The reference ESI+ source was operated with a scan time of 1 sec. at an interval of 120 sec. The detector voltage was set to 3325 V. The mass range over which mass data was acquired was 50–900 m/z (Glavin et al. 2020).

Amino acid concentrations (nmol $g^{-1}$) by bulk sample weight and their enantiomeric ratios in the sample extracts are determined by comparison of the peak areas generated from the UV fluorescence chromatograms (LC-FD, $\lambda_{ex}$ = 340 nm, $\lambda_{em}$ = 450 nm) of their OPA/NAC amino acid derivatives to the corresponding peak areas of standards run under the same chromatographic conditions on the same day. In addition to identifying the major fluorescent peaks present in the sample and procedural blank LC-FD chromatograms by retention time, we also search for peaks in the ToF-MS chromatograms corresponding to $C_2$ to $C_{10}$ amino acids by plotting the accurate mass to within 10 ppm of the theoretical m/z value of each OPA/NAC amino acid derivative over the elution time. Amino acid concentrations are determined from both UV fluorescence and single ion mass peak areas and include background level correction using a procedural blank and a comparison of the peak areas with those of an amino acid standard run on the same day. The reported uncertainties (δx) are based on the standard deviation of the average value of three to six separate measurements (n) with a standard error, $\delta x = \sigma_x \cdot (n)^{-\frac{1}{2}}$, and take into account peak resolution quality. The detection limit for LC-FD/ToF-MS is ~$10^{-15}$ to $10^{-16}$ moles of each amino acid derivative on column (Glavin et al. 2006; Glavin et al. 2010).

### *Abundance and distribution of primary and secondary amines/amino acids*

**Sample extraction and purification.** The same procedures are used for sample extraction as described above for primary amino acids, except desalting is not required.

**AccQ•Tag derivatization.** The derivatization method for the separation of primary and secondary amines and amino acids uses the Waters AccQ•Tag liquid chromatographic protocol described by Boogers et al. (2008). The portions of the sample water extracts that are allocated for amine or amino acid analyses are first acidified with 40 µL 6 M HCl, dried under vacuum, and then derivatized with the AccQ•Tag protocol. We use AccQ•Tag derivatization for these analyses because OPA/NAC does not react with secondary amines and therefore is unable to detect a variety of secondary amines and amino acids found in meteorites, including the protein amino acid proline (a secondary amino acid). The AccQ•Tag LC method is also well suited for the derivatization and analysis of more fragile protein amino acids such as glutamine and asparagine. In addition, the AccQ•Tag derivatization method is not sensitive to multivalent cation interferences like OPA/NAC; therefore, desalting of sample extracts using cation exchange chromatography is not required. The AccQ•Tag amine and amino acid derivatization products (Cohen & Michaud 1993) are also more stable than the OPA/NAC derivatives, which enables longer-term storage and automated sequential analyses. However, in contrast to OPA/NAC, AccQ•Tag derivatization does not facilitate chromatographic separation of chiral amines and amino acids into their corresponding enantiomers. Derivatives are prepared using 10 mL of standard or sample, 20 mL of commercial derivatizing agent, and 70 mL of borate buffer.

**LC-FD/ToF-MS amino acid and amine separation and quantification.** Separation and quantification of the AccQ•Tag amino acid and amine derivatives in the sample and procedural blank water extracts is achieved by ultrahigh performance liquid chromatography (UHPLC) with UV fluorescence and quadrupole time-of-flight mass spectrometry detection (LC-FD/Q-ToF-MS) using a Waters ACQUITY H Class UHPLC with fluorescence detector and Waters Xevo G2 XS. The amine derivatization and subsequent LC-FD/Q-ToF-MS analysis (Aponte et al. 2020) are modeled after the amino acid method used by Dworkin et al. (2018). $C_2$ to $C_6$ amino acids and $C_1$ to $C_6$ amines are chromatographically resolved using an AccQ•Tag Ultra RP column (2.1 × 100 mm, 1.7-µm bead), which is maintained at a constant temperature of 55 ºC. For LC analysis, a 250-µL syringe, 50-µL loop, and 30-µL needle are used, with a total injection volume of 1 μL. The mobile phase conditions for amine separations are as follows: flow rate, 700 μL/min; gradient, time in minutes (% B): 0.00 (0), 2.49 (10), 7.00 (20), 7.99 (50), 8.00 (100), 8.99 (100), 9.00 (0), 10.00 (0). Amino acid separation is identical except the gradient time in minutes is (% B): 0.00 (0.1), 0.54 (0.1), 5.74 (9.1), 7.74 (21.2), 8.04 (59.6), 8.70 (0.1), 10.00 (0.1). AccQ•Tag derivatives are monitored by LC-FD ($\lambda_{ex}$ = 340 nm, $\lambda_{em}$ = 450 nm), and peak identification confirmation is achieved by accurate mass using a match tolerance of 10 ppm. For amines, the Xevo ESI+ capillary voltage is set to 1.2 keV, the sampling cone to 40°C, the source temperature to 120°C, the cone gas flow to 50 L/hour, and the desolvation gas to 1000 L/hour. For amino acids, the Xevo ESI+ capillary voltage was set to 0.8 keV, the sampling cone to 30°C, the source temperature to 120°C, the cone gas flow to 70 L/hour, and the desolvation gas to 1000 L/hour. A set of amino acid and amine standards is prepared in water and analyzed at ten different concentrations to generate a linear least-square model fit for each analyte. The abundance of amino acids and amines is quantified from peak areas generated from the mass chromatogram of their Waters AccQ•Tag derivatives, which are given as the average of three separate measurements of the same extracted sample.

### *Abundance and distribution of peptides*

**Sample Extraction and Purification**. Same procedures as described above for primary and secondary amines and amino acids.

**AccQ•Tag Derivatization.** Three reagents are used for pre-column derivatization of samples and standards. All three reagents were purchased from Waters Corporation. These three reagents were: 1) AccQ·Tag Ultra borate buffer, 2) AccQ·Tag Ultra reagent powder, and 3) AccQ·Tag Ultra reagent diluent**.** The AccQ·Tag derivatization agent was prepared by mixing 1 mL of AccQ·Tag Ultra reagent diluent with AccQ·Tag Ultra reagent powder. The solution was then vortexed for 10 seconds and heated for 15 minutes on top of a heating block set at 55°C, until the powder fully dissolved. Pre-column derivatization was executed by mixing 10 mL of the sample or standard with 70 mL of AccQ·Tag Ultra borate buffer. This solution was then derivatized with 20 mL of the AccQ·Tag derivatization agent, prior to heating at 55°C for 10 minutes.

**LC-FD/ToF-MS peptide separation and quantification.** Two eluents are used to perform chromatography of derivatized peptides (Parker et al., 2020): (A) AccQ•Tag A buffer, (B) AccQ•Tag B buffer, (C) strong wash, and (D) weak wash. Eluent A is prepared by mixing 100 mL of AccQ·Tag Eluent A Concentrate (Waters Corporation) with 900 mL of ultrapure water. Eluent B is purchased directly from Waters Corporation. Chromatography is performed using a 2.1 x 150 mm, 1.7 mm particle sized Waters ACQUITY UHPLC BEH Phenyl column. Target analytes are eluted using the following gradient: 0 to 9.5 min, 0% B, 9.5 to 16 min, 0 to 8% B, 16 to 20 min, 8 to 18% B, 20 to 21 min, 18–0% B, 21 to 25 min, 0% B. The autosampler is maintained at 25°C, the injection volume is 2 mL, the eluent flow rate was held constant at 0.3 mL/min, and the column is maintained at 30°C. The FD is operated with an excitation wavelength of 266 nm and an emission wavelength of 473 nm. The ToF is operated as described for amino acids analysis.

### *Abundance and distribution of free and acid-releasable cyanide*

**Sample extraction and purification**. Cyanide ions ($CN^-$) are evaluated from the hot-water extract and from the mineral residues left after aqueous extraction such as for amino acids; both types of samples (liquid or solid) are acid-digested using a MicroDIST distillation apparatus (Lachat Instruments) (Smith et al. 2019; Aponte et al. 2020). Samples are placed into a micro-distillation tube with 5 mL of water, 750 μL of 9 M sulfuric acid used as a releasing agent for cyanide, and 1.5 mL of 0.08 M NaOH as a trapping solution for gaseous hydrogen cyanide (HCN); the acid digestion proceeds at 135°C for 30 min.

**Derivatization.** After distillation, 200 μL of 2 M NaOH is added to each distillate sample to raise the pH above 12; the total final volume of the distillate is measured by mass (3-4 g of distillates are typically collected). An aliquot of 50 μL of free and acid-releasable cyanide distillate solutions is derivatized using 50 μL of 10 mM sodium borate buffer (pH = 9), 50 μL of 1 mM naphthalene-2,3-dicarboxaldehyde (NDA) solution in methanol, and 50 μL of 0.1M glycine in water, reacted at room temperature for 30 min, and analyzed via LC-FD/ToF-MS. Free and acid-releasable cyanide species are quantified using a Waters ACQUITY Classic UPLC coupled to a fluorescence detector and a Waters LCT Premier time-of-flight mass spectrometer (LC-FD/ToF-MS). For LC analyses, a 250-μL syringe, 50-uL loop, and 15-μL needle were used, with a total injection volume of 1 μL. The NDA-cyanide derivative is chromatographically resolved using an ACQUITY UPLC CSH Phenyl Hexyl column (2.1 × 150 mm, 1.7-μm bead), an ACQUITY UPLC CSH C18 column (2.1 × 100 mm, 1.7-μm bead), and an ACQUITY UPLC CSH Phenyl Hexyl column (2.1 × 150 mm, 1.7-μm bead) attached in series. All columns are maintained at 30 ºC. Eluent A is an ammonium formate buffer prepared via $NH_4OH$ titration of a 50 mM formic acid solution to pH 9, and Eluent B is methanol. The mobile phase conditions for the separation of the NDA-cyanide derivative are as follows: flow rate, 100 μL/min; gradient, time in minutes (%B): 0.00 (70), 4.00 (90), 14.00 (90), 14.01 (100), 18.00 (100), 18.01 (70), 20.00 (70). The optical absorbance of NDA-derivative analytes is monitored by LC-FD ($\lambda_{ex}$ = 252 nm, $\lambda_{em}$ = 483 nm), and peak identification confirmation is provided by accurate mass using a match tolerance of 10 ppm (251.0815 Da). The ToF-MS ESI+ capillary voltage is set to 3.5 kV, the sampling cone to 50 V, the source temperature to 120°C, the cone gas ($N_2$) flow to 70 L/hour, and the desolvation gas ($N_2$) to 700 L/hour. Cyanide standards (potassium cyanide, KCN) are prepared in water and analyzed at 10 different concentrations to generate a linear least-square model fit for this analyte. The abundances of cyanide are quantified from peak areas generated from the mass chromatograms of the NDA-derivative. Abundances are given as the average of three separate measurements of the same extracted sample.

### *Abundance and distribution of N-heterocycles (purines and pyrimidines) (Method 1)*

**Sample extraction and purification**. Method 1 is a more aggressive extraction than method 2, providing increased yields of purines, but destroying more fragile species. Same crushing procedures as described above for primary amino acids, except powdered samples (approximately 50 to 500 mg) are extracted with 95% formic acid at 100°C for 24 hours in glass ampoules. Sample extracts are loaded onto a Waters Oasis Max SPE (solid phase extraction) cartridge and rinsed with 5% ammonia in water followed by methanol. Nucleobases are then eluted using 5% formic acid in methanol. Individual recoveries of nucleobases after the SPE procedure are determined using standards and calculated using an average of three runs. As a control, serpentine (a hydrated magnesium silicate) heated at 500°C overnight was carried through the identical extraction procedure. Pure standards carried through the same workup conditions are used to verify that no significant degradation of the canonical nucleobases occur, especially with respect to deamination of guanine to xanthine and adenine to hypoxanthine (which were both less than 0.04%).

**LC-PDA/QqQ-MS and LC- Orbitrap-MS analyses.** The procedure targets the five canonical RNA/DNA nucleobases (adenine, guanine, cytosine, thymine, and uracil) as well as 17 nucleobase analogs, which have been synthesized under plausible prebiotic conditions in the laboratory (with the exception of 3,7-dimethylxanthine; theobromine) (Callahan et al. 2011; Callahan et al., 2013) Sample analysis is carried out using either a Waters 2695 LC coupled to a Waters 2996 PDA detector and Waters Quattro Micro API triple-quadrupole mass spectrometer (QqQ-MS) operating in multiple reaction monitoring (MRM) mode or a Thermo Scientific Accela UHPLC coupled to a LTQ Orbitrap XL hybrid high-resolution mass spectrometer (Orbitrap-MS). Typically, initial screening and quantitation of compounds is performed by the LC-QqQ-MS in MRM mode while unambiguous structural confirmation was obtained using the LC-Orbitrap mass spectrometer, which permits high mass resolution (approximately 60,000 for our target masses) and excellent mass accuracy (≪5 ppm).

For LC-PDA/QqQ-MS analyses, nucleobase and nucleobase analog separation is achieved using two sequential Phenomenex Luna Phenyl-Hexyl columns (2.0 × 150 mm, 3-mm particle size) maintained at 50°C. Mobile phase (A) is composed of 10 mM ammonium formate buffer pH 3.9 with 2% methanol, and mobile phase (B) is 100% methanol. Mobile phase (B) is changed over a linear gradient as follows: 0–20 min 5–25% B, 20–21 min 25–95% B, 21–35 min 95% B, 35–36 min 95–5% B, and 36–46 min 5% B. A flow rate of 100 mL/min is used. In MRM mode, a specific parent-to-daughter ion transition for each nucleobase is acquired in addition to a specific retention time given by the LC separation. Typically, two MRM transitions are monitored for each nucleobase for identification and the most intense MRM transition is used for quantitation. The full fragmentation MS/MS spectrum, when the individual nucleobase concentration is sufficient for this type of acquisition is used for further verification. Molecular assignments based on the UV chromatograms is collected but is unlikely to be diagnostic due to an expected UV-absorbing background. This procedure is being updated for the newer UHPLC and QqQ systems.

LC-Orbitrap-MS analyses have the following parameters: ESI+ source voltage was 4.0 kV; sheath and auxiliary gases for desolvation of the electrospray were both nitrogen and set to 35 and 10, respectively (both arbitrary units); ion transfer capillary voltage and ion transfer capillary temperature are 21 V and 275°C, respectively; tube lens is 55 V; full scan spectra are acquired from 50 to 500 m/z; and to maintain a sufficient number of data points across chromatographic peaks, a mass resolution setting of 30,000 (at full-width-half-maximum for m/z 400) is used. Data-dependent MS/MS acquisition in both the linear ion trap (normalized CID energy 30% and isolation width of 2 Da) and the Orbitrap mass analyzer (HCD energy 60% and 100% and isolation width of 2 Da) is used. External calibration for ESI+ in the range of 120 to 2000 m/z is performed using a mixture of caffeine, MRFA (L-methionyl-arginyl-phenylalanyl-alanine acetate hydrate) peptide, and Ultramark 1621 in an acetonitrile-methanol-water solution containing 1% acetic acid, which resulted in a typical mass accuracy of <3 ppm. This procedure is being updated for the newer Orbitrap system.

### *Abundance and distribution of N-heterocycles (purines and pyrimidines) (Method 2)*

**Sample extraction and purification.** Method 2 is less efficient than method 1, but permits the recovery of more fragile compounds like cytosine. Crushing procedures as described above for primary amino acids, except powdered samples (approximately 1 g) are sonicated in 20 mL 2% HCl in a glass test tube held at room temperature with ice 4 times. The residue is further extracted with ultrapure water 2 times. Each time, the test tube was centrifuged at 2000 rpm for 30 min and supernatant was collected in a flask. The extract was dried by rotary evaporation. Then, the precipitates were extracted in 50 mL methanol and dried (Furukawa et al. 2019; Oba et al. 2019). The extracted sample is subjected to cation exchange chromatography (AG50W-X8 resin; Takano et al. 2010) and polyols (sugars), uracil and thymine are eluted with water. The majority of the water eluate (~90%) is allocated for polyol analysis by GC-MS. Most of the the N-heterocycles are then eluted using 3 bed volumes of 10% ammonia in ultrapure water. The eluate is dried, dissolved in ultrapure water and filtered using a 0.20 μm PTFE cartridge filter.

**LC-Orbitrap-MS analysis.** The procedure targets the five canonical RNA/DNA nucleobases (adenine, guanine, cytosine, thymine, and uracil) as well as 21 nucleobase analogs which possesses pyrimidine or purine structures. In addition, other kinds of N-heterocycles such as nicotinic acid and nicotinamide, approximately 50 species in total, are targeted. Separation and quantification of the purified N-heterocycles is achieved by HPLC with ESI Orbitrap using an Ultimate 3000 and Q-Exactive Plus. Purine nucleobases are chromatographically resolved using an InertSustain PFP separation column (1.0 × 250 mm, particle size 3 μm), which is maintained at a constant temperature of 40ºC. Other N-heterocyclic molecules including pyrimidine nucleobases are chromatographically resolved using a HyperCarb separation column (2.1 × 150 mm, particle size 5 μm), which is maintained at 40ºC. For LC analysis, a 10-μL syringe and a 5-μL loop are used, with a total injection volume of 1-5 μL, which varies with the target concentration. The eluent program for the HPLC setup with the PFP column was as follows: at $t$ = 0 to 5 min, solvent A (water), solvent B (acetonitrile + 0.1% formic acid) = 90:10, followed by a linear gradient of A:B = 50:50 in 20 min and maintained at this ratio for 25 min. The flow rate is 50 μL/min. The eluent program for the HPLC setup with the HyperCarb is as follows: solvent A (water + 0.1% formic acid) and solvent B (acetonitrile + 0.1% formic acid) = 99:1 at $t$ = 0 min, followed by a linear gradient of A:B = 70:30 in 20 min and maintained at this ratio for 20 min. The flow rate is 200 μL/min.

The mass spectra were recorded in the positive ESI mode with an *m/z* range of 60.5–400 (full scan) or 111–154 (targeted) and a spray voltage of 3.5 kV. Sheath and auxiliary gases for desolvation of the electrospray were both nitrogen and set to 25 and 5 (arbitrary units), respectively. The mass resolution is 140,000 at a *m/z* of 200. We confirmed the MS response linearity over a dynamic range of substrate concentrations on the order of ppt, ppb, and ppm scales. The capillary temperature of the ion transfer was 300°C. The injected samples were vaporized at 280°C. Peak identification confirmation is achieved by accurate mass using a match tolerance of better than 5 ppm (usually 3 ppm). The abundance of N-heterocycles is quantified by comparison between peak areas of each standard reagent with known concentration and those detected in the mass chromatogram of the sample.

MS/MS experiments were performed using a hybrid quadrupole-Orbitrap mass spectrometer (Q-Exactive Plus) with HPLC and ionization conditions identical to those used for the full-scan analyses. The extracted positive ion *m/z* (for example, for thymine, 127.05 ± 0.2) was reacted with high-energy collision $N_2$ gas to produce fragmented ions, and the mass range of *m/z* 50–160 was monitored using an Orbitrap MS with a mass resolution of ~140,000.

### *Non-targeted analyses of polar organic molecules*

**Sample extraction and purification**. Sequential solvent extractions are needed from polar to apolar.

**Tandem-LC-ToF-MS² molecular profiling.** Comprehensive non-targeted profiling of polar organic molecules requires two complementary chromatographic runs to achieve maximum compound coverage. A sensitive tandem-LC method, which combines the typical reverse phase (RP) and hydrophilic interaction chromatography (HILIC) separation in a single run is used (Hemmler, et al. 2018). In a first step, hydrophobic compounds are retained on a RP trap column while hydrophilic compounds are directly transferred onto a HILIC phase. Next, the pre-fractionated sample composition is analyzed by RP or HILIC chromatography, respectively. The presented setup allows individual and independent gradient elution as well as interfacing with MS. In total, thousands of chromatographic features are detected in meteoritic SOM samples. The simple and robust setup provides high flexibility in the selection of column combinations and does not require sophisticated instrumental setups or software. Tandem-LC is a valuable tool in the non-targeted screening of complex products, significantly increasing the covered polarity range compared to classical one-dimensional chromatography. The data is acquired in automatic tandem-MS mode by a Bruker maXis Q-ToF-MS as well to enable a structural correspondence and functional group annotation and will be integrated with the exact mass data as analyzed in direct-injection FTICR-MS.

Chromatographic analysis is carried out on a Thermo Scientific Dionex Ultimate 3000 system (Dreieich, Germany) equipped with two vacuum degassers, dual gradient pump, a temperature-controlled autosampler, a thermostat-controlled column oven containing two 10-port 2-position valves, and a variable wavelength detector (UV). The autosampler temperature is set to 5°C. Column temperature is maintained at 40°C. Connections were made using stainless-steel Viper capillaries (180 μm ID; Thermo Scientific, Dreieich, Germany) with the shortest possible lengths. A Kinetex C18 column (2.1 × 30 mm, 2.6 μm; Phenomenex, Aschaffenburg, Germany) is used for trapping hydrophobic analytes. For chromatographic separations, a ZIC-cHILIC (2.1 × 100 mm, 3 μm, Merck, Darmstadt, Germany) and a Kinetex C18 (2.1 × 100 mm, 2.6 μm, Phenomenex, Aschaffenburg, Germany) are used. Samples and blanks are injected in the LC flow stream via full-loop injection (20 μL). The LC system is coupled to a Bruker maXis qTOF-MS equipped with an APOLLO II electrospray ion source (Bruker Daltonics, Bremen, Germany).

## Sample impact:

The analyses of soluble organic matter (SOM) is destructive and requires sample crushing and subsequent extraction in water or organic solvents by heating at elevated temperatures (up to 100°C for 24 hours). Soluble organic analyses are conducted on the solvent extracts; however, the solvent insoluble residue can be used for other analyses, including acid dissolution of the solid residue and characterization of the IOM.

The SOM analyses described here are extremely sensitive to organic contamination, therefore sample handling and analysis of the samples prior to the extractions for organic analyses should be minimized. Exposure of the sample to some materials (e.g. latex gloves, epoxy cement, ink) can significantly compromise the sample quality. Even "clean" materials for inorganic analysis can be contaminated with organics, and vice versa. The glass and ceramics used in these analyses can leech inorganic ions.

Careful coordination of sample preparation to minimize sample consumption is also possible and is described here.

### Data quality:

LC-FD/ToF-MS detection limits are typically < $10^{-15}$ to $10^{-16}$ mol for OPA/NAC amino acid derivatives. The ToF-MS operated in Sensitivity mode can provide a full width at half maximum resolution of 22,000, and amino acids can be identified by accurate mass to within 10 ppm of the theoretical m/z value of each OPA/NAC amino acid derivative. Errors on the individual amino acid concentrations are typically <10% and depend on the specific amino acid, the amount of sample analyzed, and the number of runs. The uncertainties in the enantiomeric excesses of chiral amino acids that can be separated are typically 1–5%.

### Data products:

The data products expected from the LC-ToF-MS measurements include:

- 1D chromatogram if the FD is used (retention time vs. intensity)
- 2D chromatogram if the PDA is used (retention time vs. spectrum)
- 2D chromatogram (retention time vs. mass spectrum)

Data analysis consists of peak integrations vs. those of standards, outputted as .txt files. Unreduced data file types are proprietary; conversion with Proteowizard is possible and will be tested prior to the SART.

The data products expected from the LC-QqQ-MS measurements include:

- 1D chromatogram if the fluorescence detector is used (retention time vs. intensity)
- Either 2D chromatogram (retention time vs. mass spectrum) or 1D chromatogram for each parent or daughter ion monitored

Data analysis consists of peak integrations vs. those of standards. Unreduced data file types are proprietary; conversion with Proteowizard is possible and will be tested prior to the SART.

The data products expected from the LC-Orbitrap-MS measurements include:

- 1D chromatogram if the FD is used (retention time vs. intensity)
- 2D chromatogram if the PDA is used (retention time vs. spectrum)
- 2D chromatogram (retention time vs. mass spectrum)
- Fragmentation pattern for selected masses if $MS^n$ analysis is used

Data analysis consists of peak integrations vs. those of standards and $MS^n$ spectra, outputted as .txt files. Unreduced data file types are proprietary; conversion with Proteowizard is possible and will be tested prior to the SART.

The data products expected from Tandem-LC/MS² profiling include:

- The chromatogram and related extracted features in time (retention time vs. mass spectrum)
- The extracted mass difference abundances and network analysis for integration with FTICR-MS data
- The MS² database to infer functional information and abundance and possibly annotations from databases (Impurities, biogenic compounds, oligomers, prebiotic chemistry)

### Facilities:

Astrobiology Analytical Laboratory, NASA Goddard, Greenbelt, MD, USA - Jamie Elsila

**Waters ACQUITY UHPLC (3)**: Two UHPLCs are two channel LC that functions at high tolerances, one is a four channel LC. One has an ACQUITY Photodiode Array Detector, and all three ACQUITYs have Multi-Channel Fluorescence Detector (FD) for operations up to 15,000 psi for improved separations by allowing the use of columns packed with 1.7µm resin.

**Thermo Vanquish UHPLC**: This UHPLC (Ultra High Performance Liquid Chromatograph) is a four channel LC that functions at high tolerances. It has an Accela Photodiode Array Detector for operations up to 15,000 psi.

**Thermo Orbitrap Exactive**: This is a Fourier Transform (FT) mass spectrometer with a mass accuracy of <1 ppm below 400 m/z, a resolution of over 100,000 at masses below 400 m/z. Also contains an optional Direct Analysis in Real Time (DART) solid scanning inlet.

**Thermo Scientific UltiMate 3000 RSLCnano system**: This nanoflow UHPLC has a 900 bar maximum operating pressure, and stable flow rates from 0.05-1.5µL/minute. It is attached to a Picometrics ZETALIF laser induced fluorescence detector with a 355nm diode pulsed laser optimized for the flow rates and capillary tubes of the nanoflow UHPLC. Serves as an inlet to the Thermo Orbitrap Exactive.

**Thermo TSQ Quantum Ultra LC/MS/MS** This triple quadrupole (QqQ) mass spectrometer allows MS/MS for structural identification of compounds in a mass range of 30–1500 m/z. Sample is introduced via electrospray from a UHPLC.

**Waters Xevo TQ-S micro**: This triple quadrupole (QqQ) mass spectrometer excels at quantitation and allows MS/MS for structural identification of compounds in a mass range of 2–2048 m/z. Sample is introduced via electrospray (ESI) from a UHPLC.

**Waters Xevo G2 QToF**: This is a quadrupole time-of-flight (QToF) mass spectrometer with an electrospray (ESI) and corona ionization (APCI) source. It possesses a mass accuracy of 2 ppm, a resolution of 32,500, and has a mass range of 20 to 100,000 m/z. Sample is introduced via electrospray from a Waters ACQUITY.

Organic Geochem. & Cosmochem. Lab, Kyushu University, Fukuoka, Japan - Hiroshi Naraoka

**Thermo UltiMate 3000 RS UHPLC:** This UHPLC (Ultra High Performance Liquid Chromatograph) features a Rapid Separation pump that operates at up to 1000 bar.

**Thermo Q-Exactive Plus:** This is a Fourier Transform (FT) mass spectrometer with a mass accuracy <3 ppm and a maximum mass resolution of 280,000 at 200 m/z.

Helmholtz Zentrum Muenchen, Germany - Philippe Schmitt-Kopplin

**Thermo Scientific Dionex Ultimate 3000** system (Dreieich, Germany): Equipped with two vacuum degassers, dual gradient pump, a temperature-controlled autosampler, a thermostat controlled column oven containing two 10-port 2-position valves, a variable wavelength detector (UV), and coupled to the **Bruker maXis qToF-MS** equipped with an APOLLO II ESI source.

# 5.38 Fourier Transform Ion Cyclotron Resonance Mass Spectrometry (FTICR-MS)

- Overview:
- How it works:
- Sample preparation, chemical derivatization and analytical methods:
- Sample impact:
- Data quality:
- Data products:
- Facilities:

## Overview:

Untargeted analysis of polar/apolar solvent-soluble organics can be performed using extremely high resolution mass spectrometry to identify molecular formulas (but not structures) in a mixture. We showed SOM from Murchison CM2 being a complex mixture of over 15.000 elementary compositions as analyzed in electrospray ionization negative mode only; expending to positive mode electrospray and atmospheric pressure photoionization will even extend the coverage of the chemical space of SOM to more than 20.000 compositions reflecting hundred thousands of possible structures. The obtained compositional footprint can in a second step be compared to the compositional space database including meteorites of various classes, thermal- and shock-history and will enable us to replace the samples in a context of pressure and thermal stress as well as aqueous alteration.

## How it works:

Mass spectrometry is an analytical method used to identify and quantify chemical species based on their mass-to-charge ratio (m/z). It is also widely used to determine the isotopic abundances of various elements. The m/z of ionized molecules is determined by their motion or traveling through electromagnetic fields. Element composition and chemical structure of molecules can be deduced with high-resolving-power instruments and/or via the interpretation of the mass spectra of molecular fragments. Mass spectrometers ionize, separate, then detect an analyte

There are two methods of ionization available: (i) Atmospheric pressure electrospray ionization (ESI). It is a soft ionization method that makes few but non-characteristic fragments by the addition or removal of $H^+$ from an analyte in an evaporating droplet of solvent under ~1.5 kV. Most commonly done in ESI+ mode to produce the $M+H^+$ parent ion. (ii) Atmospheric pressure photoionization (APPI) forms $M+H^+$ or $M-H^+$ ions in UV light. It is appropriate for species that ionize poorly under ESI but tend to be less sensitive. In ESI, analytes are delivered in protic solution and may be derivatized to assist with chromatography. ESI can suffer ion suppression, when excessive undesirable species are present and thus, decrease the sensitivity of the analytes.

FTICR-MS (Ion cyclotron resonance) is a Fourier transform based MS (Schmitt-Kopplin and Kanawati 2019). A FTICR-MS uses superconducting magnetic field to trap ions for analysis. It has high mass precision (electron mass precision, half an AMU at m/z 400) and spectral resolution (higher 400,000 at m/z 400) enabling a direct conversion of the m/z space into the elemental compositional space in a wide mass range. The ultrahigh resolution and mass accuracy enables the FTICR-MS to be used in a direct injection mode in ESI or APPI with a direct information on the compositional space in the elements C,H,N,O,S,Mg and their corresponding isotopologues fine structure in natural abundance.

## Sample preparation, chemical derivatization and analytical methods:

**Sample extraction and purification**. Sequential solvent extractions of a crushed sample are needed from polar to apolar.

**FTICR-MS ultrahigh-resolution molecular profiling.** FTICR-MS analysis is destructive and involves a direct injection of the solvent extract of the crushed sample (as for tandem-LC/MS). To fully exploit the advantages of FTICR-MS, we routinely control the instrument performance by means of external calibration on arginine clusters prior to any analysis and internal calibration for each dataset. Relative m/z errors are usually < 100 ppb across a range of 150 < m/z < 1,500. The average mass resolution is near 400,000 at nominal mass 400. This exceptional mass accuracy and mass resolution makes it possible to assign the exact mass of molecular ions and the respective elemental composition directly out of mixtures with precision better than the mass of an electron (9.10938215 × 10$^{-31}$ kg or 1/1836.2 of the mass of a proton). Accordingly, two molecular compositions with mass differences smaller than that of a single electron mass can be differentiated at higher mass (500). Mass peak amplitudes grow with the square root of the number of acquired transients, similarly to NMR spectroscopy. The methanol extracts spectra will be measured in negative and positive electrospray ionisation modes [ESI– and ESI+] under conditions described previously (Schmitt-Kopplin et al., 2010) APPI if the sample amounts allows it (Hertzog et al., 2019); a maximum scan amount up to 3,000 scans will be accumulated with 4 million data points. The conversion of the exact masses into elementary composition will follow mass difference and network analysis to convert the exact massed into the compositional space in C, H, N, O, S, Mg (any other element will be able to be mined in the generated database). The representation of complex FTICR-MS datasets is possible in van Krevelen diagrams and other related representations (double bond equivalents/DBE, carbon oxidation state, aromaticity/aliphaticity, etc.) as needed to compare the sample to meteorites analyzed to date (150+).

## Sample impact:

The analyses of soluble organic matter (SOM) is destructive and requires sample crushing and subsequent extraction in water or organic solvents. Soluble organic analyses are conducted on the solvent extracts; however, the solvent insoluble residue can be used for other analyses, including acid dissolution of the solid residue and characterization of the IOM.

The SOM analyses described here are extremely sensitive to organic contamination, therefore sample handling and analysis of the samples prior to the extractions for organic analyses should be minimized.  Exposure of the sample to some materials (e.g. latex gloves, epoxy cement, ink) can significantly compromise the sample quality.  Even "clean" materials for inorganic analysis can be contaminated with organics, and vice versa.  The glass and ceramics used in these analyses can leech inorganic ions.

Careful coordination of sample preparation to minimize sample consumption is also possible and is described here.

## Data quality:

We will work with the FTICR-MS 12 Tesla Solarix Infinity system equipped with ESI and APPI ion sources (Bruker Daltonics, Bremen, Germany) at the Helmholtz center in Munich Germany. The injection will be followed in microspray (2 μl/min) or nanospray ionization source (ADVION, Nanomate).

The mass resolution we have in routine is around 500.000 at m/z 400 with a mass accuracy torards internal standards after calibration of 10ppb (0.1ppm). We analyze the mass with the precision of the mass of an electron and differenctiate to different masses with the mass of an electron at m/z 600.

## Data products:

The data products expected from the FTICR-MS profiling include:

- Mass spectra in comparable time of infusion and number of scans for all samples (relative abundance of the signals in the mass range 120 to 1000 amu)
- Data calibration and .dat files exported in m/z vs. peak intensity lists
- The abundance of elementary compositions of the experimental exact masses over the whole mass range (high resolution and accuracy)
- Plots of thousands of elementary compositions (H/C vs. O/C) in van Krevelen and related diagrams showing structural specificity relative abundances of the compositional CHNOSMg-space relative to that of meteorites

## Facilities:

Helmholtz Zentrum Muenchen, Germany - Philippe Schmitt-Kopplin

**FTICR-MS 12 Tesla Solarix Infinity system** equipped with an APOLLO II ESI source (Bruker Daltonics, Bremen, Germany): The system has an autosampler, but manual injection will be done for valuable samples with a syringe pump operating at 2 µL/min.

# 5.39 Microprobe Two-Step Laser Mass Spectrometry (µL²MS)

- Overview:
- How it works:
- Sample preparation:
- Sample impact:
- Data quality:
- Data products:
- Facility:

## Overview:

Microprobe two-step laser mass spectrometry (µL$^2$MS) is a technique that allows the detection and characterization of organic molecules with sub-femtomole (< 10$^{-18}$ mole) sensitivity and micrometer spatial resolution on microscopic and/or heterogeneous samples (Clemett and Zare, 1997). In brief, organic matter present on the surface of a sample is physically removed and ionized in two separate steps by laser pulses, and the ions are injected into a mass spectrometer, where the molecular weights of the species present are determined.

## How it works:

The sample to be analyzed is attached to a sample platter and placed within a high vacuum chamber. Analysis then proceeds in two steps; in the first step, an infrared (IR) laser pulse is focused down onto the sample, and the surface in the laser focus undergoes rapid a temperature excursion (≥ 10$^8$ K×s$^{-1}$) causing the desorption of surface-bound organic species without concomitant fragmentation. The magnitude of the temperature change is controlled through the fluence and duration of the laser pulse and is kept well below the plasma ignition threshold (~10$^6$ W×cm$^2$) to prevent ionization of desorbed species and minimize damage to the sample surface. Since this occurs in a vacuum, desorbed neutral species expand away from the sample surface in a plume and are allowed to enter the ion extraction region of a Wiley-McLaren source (Wiley and McLaren, 1955). Because the desorbed species are neutral, their trajectories remain unaffected by the electric field in extraction region. In the second step, a second UV laser is directed through the desorption plume, as it transits the extraction region, it effects the efficient "*soft*" photoionization (i.e., without concomitant molecular fragmentation) of any organic species present. Ions thus formed immediately interact with the electric field in the extraction region and can be injected into a reflection time-of-flight mass spectrometer to determine their mass. Because the desorption and ionization steps are spatially and temporally separated, each can be optimized independently and artifacts due to matrix ionization at the sample surface are avoided.

## Sample preparation:

No sample preparation for is required µL$^2$MS analysis. The two requirements for a sample are: (1) it can be physically attached to a sample platter; and (2) it maintains integrity when placed under vacuum.

## Sample impact:

Under nominal conditions, analysis by µL$^2$MS should be considered minimally destructive. Changes to the bulk organic abundance of a given sample, as a consequence of species desorbed during analysis, are generally negligible since only labile organic species present at the very uppermost surface of the sample are removed. The primary concern is the impact from the transient heating of the upper few microns of the sample by the IR desorption laser pulse. This can cause localized temperature spikes of up to a few hundred Kelvin over a period of several microseconds. Generally, this has no impact on the sample mineralogy, but could affect particularly thermally sensitive phases.

## Data quality:

Two-step laser mass spectrometry is a general technique that can be applied to detect a wide range of organic species on different samples depending on the application. Thus, the nature and quality of data produced is very much dependent on the specific μL$^2$MS instrument considered. For the purposes herein, discussion is focused on organic analysis using the μL$^2$MS instrument located at NASA Johnson Space Center.

The spatial resolution of the analysis is determined by how tightly the IR desorption laser beam can be focused on the sample surface. For desorption, the μL$^2$MS instrument uses a line-tuned $CO_2$ laser operating at 10.6 mm equipped with an inline spatial filter and plasma shutter. This laser is focused using a Cassegrain microscope objective to yield an effective, diffraction-limited analysis spot size of ~ 5 mm for single point analyses. This can be further reduced to ~ 2.5 mm during surface mapping by partially overlapping successively analysis spots.

The nature of organic species that can be measured depends on the particular photoionization scheme that is selected for a given analysis. In the current configuration of the μL$^2$MS instruments, two Nd:YAG laser systems (one with nanosecond and the other with picosecond pulses) are available as photoionization sources. Both are equipped with non-linear doubling crystals to allow harmonic generation of UV laser radiation 355 and 266 nm, and VUV laser radiation at 118 nm using a low-pressure phase-matched Xe-Ar 355-nm tripling cell. Two particular ionization schemes are most used; the first uses 266-nm UV laser pulses for the highly selective (1+1) resonance enhanced multiphoton ionization (REMPI) of polycyclic aromatic hydrocarbons (PAHs) and organic species incorporating an aromatic moiety. The second uses 118-nm VUV laser pulses for non-selective single photon ionization that allows the detection of virtually all desorbable organic species.

A linear time-of-flight mass spectrometer is used for determination of ion masses. It incorporates a two-stage reflectron for increased mass resolution (t/2×Dt ~ 2000) and uses and a large-area extended dynamic range dual microchannel plate detector assembly to maximize detection efficiency. Instrument detection limits are in the sub-femtomole (< 10$^{-18}$ mole) range (subject to specific species and operating parameters) with an effective mass detection range up to m/z ~ 2000.

## Data products:

The primary output from μL$^2$MS analyses are time-of-flight spectra that can be subsequently be converted into mass spectra using calibration spectra acquired both just prior to and after data acquisition from a given sample. These are used to create two data products:

- Single point datasets consisting of a mass spectrum formed from the addition of one or more time-of-flight spectra acquired from a specific known location on a sample. This may also be accompanied by an optical alignment image of the relevant sample surface.
- Hyperspectral datasets produced from the acquisition of mass spectra at defined points on the sample surface so as to create 2D image maps, where each pixel element can be associated with a specific sample location and mass spectrum.

These products can be used to identify the abundance and micrometer scale spatial distribution of organic species present on a sample surface that, along with the appropriated image documentation, enables subsequent correlated analysis of the same sample surface by other analytical techniques.

## Facility:

NASA Johnson Space Center, Houston, TX, USA

The μL$^2$MS instrument forms part of the suite of analytical instruments in the Astromaterials and Exploration Science Division of NASA Johnson Space Center. It is a one-of-a-kind instrument designed and built specifically to investigate organic species present in microscopic and/or heterogeneous astromaterial samples. It has already been applied to the analysis of organic matter in a wide variety of astromaterials (e.g., Clemett, et al., 1993; McKay et al. 1996, Clemett et al., 2010; Thompson et al., 2020), including meteoritic acid residues , carbonaceous & ordinary chondrites, Martian meteorites, Antarctic micrometeorites, interplanetary dust particles, interstellar graphite grains, laboratory synthesized interstellar ice residues, and Stardust cometary dust particles.

# 5.40 Nuclear Magnetic Resonance Spectroscopy (NMR)

- Overview:
- How it works:
  - Key variables in solution-state NMR spectroscopy
- Sample preparation and analytical methods:
- Sample impact:
- Data products:
- Data quality:
- Facilities:

## Overview:

Nuclear magnetic resonance (NMR) spectroscopy is a spectroscopic technique to observe local magnetic fields around atomic nuclei in molecules. The sample is placed in a static magnetic field $B_0$, and the NMR signal is produced by excitation of the sample with radio waves into nuclear magnetic resonance, which is detected with sensitive radio receivers. The intramolecular magnetic field within a molecule is atom-specific and is detected by high-precision frequency measurements (2p × n = g × B, where g is nucleus specific gyromagnetic ratio; B, local magnetic field). NMR spectroscopy provides unsurpassed accuracy in the determination of short-range atomic order in non-crystalline and dissolved materials, allowing reconstruction of unknown molecular structures without further information. In cases of complex mixtures, NMR spectroscopy provides identification and quantification of key substructures not available by any other method at present.

## How it works:

### Key variables in solution-state NMR spectroscopy

*Nuclear spin*

Nuclear spin is an intrinsic form of angular momentum carried by certain atomic nuclei with non-integer numbers of protons or neutrons, and a mandatory requirement for the observance of NMR spectra. Spherical spin ½ nuclei provide NMR spectra with narrow resonances, which are indispensable for NMR-based assignment of organic molecular structures. The most critical nuclei employed in the solution-state NMR–based analysis of soluble organic matter (SOM) are, in order of decreasing practical relevance, $^1$H > $^{13}$C > $^{15}$N ≈ $^{31}$P.

*Chemical shift d*

The chemical shift is the resonant frequency of a nucleus (e.g. $^1$H, $^{13}$C, $^{15}$N, $^{31}$P) relative to a standard in a magnetic field. Often, the position/number of chemical shifts is diagnostic of substructures within organic molecules. The overall magnetic field experienced by a nucleus includes local magnetic fields induced by currents of electrons in the molecular orbitals (electrons carry a magnetic moment themselves). These currents are affected by the local geometry (binding partners, bond lengths, angles between bonds, conformational flexibility, and others), which in turn determine the spin energy levels (and NMR resonance frequencies). The variations of NMR frequencies for given nuclei, caused by variations in the electron distribution, is called the chemical shift. The size of the chemical shift is given with respect to a reference frequency or reference sample, usually a molecule with a barely distorted electron distribution.

*Scalar coupling constant J*

Scalar or *J*-couplings (also called *indirect dipole–dipole couplings*) are mediated through chemical bonds that connect two or more spins. This indirect interaction between two nuclear spins arises from hyperfine interactions between the nuclei and local electrons. *J*-couplings contain information about relative bond distances and angles and about the connectivity of chemical bonds as visualized in 2D NMR spectra.

**Basic NMR characteristics of soluble extraterrestrial organic matter**

High-field NMR spectroscopy provides the capability for quantitative and non-destructive *de novo* determination of complex chemical environments from polydisperse and molecularly heterogeneous environmental samples. Quantitative relationships between number of spins and area (1D NMR) and volume (2D NMR) of NMR resonances operate in absence of differential NMR relaxation, which is more pronounced in 2D NMR than in 1D NMR experiments. This key feature implies the use of NMR spectroscopy as a quantitative reference for complementary structure-selective analytical methods, like mass spectrometry (which detects gas phase ions and is subject to ionization selectivity in the case of complex mixtures) and fluorescence spectroscopy (which selectively detects fluorescent chemical environments of $sp^2$-hybridized carbon).

NMR spectroscopy is particularly informative in the description of aliphatic chemical environments that are based on $sp^3$-hybridized carbon. NMR spectroscopy enables distinction of the size of aliphatic units but also allows for in-depth assessment of its intrinsic chemical environments, like open-chain and cyclic arrangements of carbon, speciation of oxygenated functional groups, and olefins and aromatic compounds.

## Sample preparation and analytical methods:

***Sample preparation for extraction of extraterrestrial soluble organic matter***

Extensive testing of a wide range of soluble organic matter across several classes of meteorites revealed that soft extraction with methanol at ambient temperature produced organic matter with the most meaningful, i.e., the best resolved and information-rich, NMR (Hertkorn et al., 2015; Schmitt-Kopplin et al., 2010). Methanolic extracts comprise the highest diversity of soluble molecules when compared with other solvent extracts, e.g., dichloromethane $CH_2Cl_2$ and cyclohexane $C_6H_{12}$. However, valuable pronounced extraction selectivity is observed, which can be used in the form of successive elution with (1) methanol $CD_3OD$, followed by (2) dichloromethane $CD_2Cl_2$, followed by (3) cyclohexane $C_6H_{12}$. Ultrasonic-assisted methanolic extraction of solid crushed meteorites under elevated temperatures (up to $45^oC$) resulted in enhanced mobilization of the mineral-organic continuum with concomitant degradation of NMR resolution for method-specific reasons. Metal complexation of organic ligands degrades NMR resolution because of accelerated transverse NMR relaxation in larger-sized molecules; in addition, multiple opportunities arise for chemical exchange of metal ions between different coordination sites in the case of polyfunctional molecules.

***Detailed procedure***

- Solid materials wil be washed with LCMS-grade methanol ($CH_3OH$), and the wash fluid (wash SOM) will be retained for NMR characterization.
- The washed solid material will be crushed under LCMS-grade methanol ($CH_3OH$); this fluid (crush SOM) will be used for FTICR-MS characterization.
- For NMR spectroscopy, $CH_3OH$ will be exchanged by $CD_3OD$ under vacuum line conditions.
- Transfer under $N_2$ into Bruker 1 mm (alternatively 1.7 mm)  MATCH NMR tune and seal.
- Solution-state $^1H$ detected NMR spectra will be acquired in sealed Bruker MATCH NMR tubes, according to the given table, depending on the amount of disposable SOM sample.
- Repeated extraction with $CD_2Cl_2$ and cyclohexane $C_6H_{12}$ can be done in analogous ways.

## Sample impact:

NMR is a non-destructive measurement, but sample preparation is destructive: as described above, the sample-preparation procedure involves progressive extraction of organic compounds with solvents, as well as crushing. The extract is finally dissolved in deuterated solvents for NMR analysis. Post NMR-analysis, the SOM samples can be reconstituted in any other solvent for further analysis if needed, e.g., mass spectrometry analysis.

## Data products:

**Possible NMR acquisition experiments and conditions specific for mass-limited sample return**

| NMR experiment / minimum realistic mass of dry extracted sample | Transfer mechanism and nominal time delay | General characteristics and key information |
|---|---|---|
| 1D $^1$H NMR  5-7 µg | | excellent overview of key functional groups at high sensitivity; differentiation and quantification of aliphatics, oxygenated functional groups, olefins and aromatic substructures; extensive acquisition time will apply for excellent definition of low amplitude NMR resonances at 1/500$^{th}$ of maximum amplitude |
| 2D $^1$H, $^1$H JRES  25 µg | $1/^nJ_{HH}$ ~ 100 ms | sensitive determination of methyl-related branching motifs and olefinic and aromatic substitution |
| 2D $^1$H, $^1$H TOCSY  30 µg | $1/^{2-5}J_{HH}$ ~ 100 ms | sensitive detection of connected spin systems at minor relative abundance, with excellent discrimination of aliphatic branching, heteroatom-containing functional groups as well as olefinic and aromatic substitution |
| 2D $^1$H, $^{13}$C DEPT HSQC  450 µg | $1/^1J_{CH}$ ~ 3.5 ms / | chemical environments of protonated carbon atoms; absorptive lineshape, large spectral dispersion in $^{13}$C frequency, multiplicity-editing according to $CH_3$, $CH_2$, CH groups |

All products can be delivered as images after data processing.

# Data quality:

**Practical sensitivity limits of 800-MHz solution-state NMR spectroscopy of SOM with cryogenic detection**

Given the availability of a 800-MHz NMR spectrometer at the the Helmholtz Center in Munich, Germany, with cryogenic probeheads for $^1$H and X-nuclei–based detection and lengthy acquisition times in excess of one week, the conditions are nearly ideal for recording of meaningful solution-state $^1$H NMR spectra of mass-limited SOM. From a year-long experience, the absolute *practical* limits of detection for extremely mass-limited samples are constrained rather by indirect effects and not by intrinsic NMR probehead sensitivity, as described below.

The use of small-diameter NMR tubes with minimal amounts of NMR solvent is advantageous from the standpoint of SNR alone, in particular for cryogenic probeheads. The commercially available Bruker MATCH system accommodates NMR tubes from 1 to 5 mm in diameter. For the proposed investigation of seriously mass-limited sample return SOM, we will most likely use 1-mm diameter tubes with 7-9 µL solution; alternatively, 1.7-mm diameter tubes with ~25 µL solution might be employed. The operation of 1-mm Bruker MATCH tubes is somewhat more challenging than that of 1.7-mm tubes but not overly difficult in case of methanol-$d_4$ solutions. The solution volume depends on the inner diameter of the NMR tube and dramatically shrinks with sample tube diameter. The use of 1-mm sample tubes carries the intrinsic risk of potential solubility limitation, which is difficult to assess by visual means (others are not practical).

A practical issue of concern is unavoidable impurities in NMR solvents. The overlay of solvent impurities with analyte NMR resonances of critical interest may happen (most often in the aliphatic section from $d_H$ ~0.9–1.3 ppm; in case of $CD_3OD$ also from OC**H** units, at $d_H$ ~3.3–3.6 ppm).

# Facilities:

Helmholtz Zentrum Muenchen, Germany - Philippe Schmitt-Kopplin

Two **Bruker Avance III NMR spectrometers** operating at 500.13 MHz ($B_0$ = 18.7 T) and 800.13 MHz ($B_0$ = 18.7 T) are available with cryogenic probeheads; excellent $^{13}$C NMR sensitivity at $B_0$ = 11.7 T is available from a cryogenic classical dual 5-mm z-gradient $^{13}$C/$^1$H 5-mm probe (~150% SNR of specification); three cryogenic probeheads are available for the 800-MHz NMR spectrometer: a 5-mm z-gradient $^1$H / $^{13}$C / $^{15}$N / $^{31}$P QCI cryogenic probe for $^1$H and $^1$H-detected 2D NMR spectroscopy, a 3-mm $^{13}$C / $^{15}$N / $^{31}$P QCO cryogenic probe for simultaneous detection of these three heteronuclei (multi receiver arrangement), and a selective 5-mm z-gradient $^2$H/$^1$H dual cryogenic probe, with $^2$H NMR spectroscopy.

# 5.41 Gas Chromatography–Isotopic Ratio Mass Spectrometry (GC-IRMS)

- Overview:
- How it works:
    - Gas Chromatography:
    - Combustion isotope ratio mass spectrometry (C-IRMS):
    - Picomolar-scale compound-specific isotope analysis (pico-CSIA):
    - Electron impact Orbitrap isotope ratio mass spectrometry (EI-Orbitrap-IRMS):
- Sample preparation, chemical derivatization and analytical methods:
- Sample impact:
- Data quality:
    - GC-C-IRMS
    - pico-CSIA
    - GC-EI-Orbitrap-IRMS
- Data products:
    - GC-MS/C-IRMS
    - GC-C-IRMS, pico-CSIA
    - GC-EI-Orbitrap-IRMS
- Facilities:

## Overview:

Stable isotopes have been used with great success in revealing the complex history of organic compounds in meteorites. Extraterrestrial organics frequently have exotic stable isotope compositions that are well outside of the range of terrestrial compounds. Many detected compounds in meteorites were regarded as suspect terrestrial contaminants until compound-specific isotopic analysis (CSIA) of these organic constituents helped resolve their extraterrestrial origin. The measurement of the $^{13}C/^{12}C$, $^{15}N/^{14}N$, and $^{2}H/^{1}H$ (measured in δ notation versus a standard) in a given molecule provides potential insight into the synthetic relationship of different compounds and their connection to the interstellar medium. CSIA of the following compound classes are described:

- Abundance, distribution, and stable isotopic composition of polycyclic aromatic hydrocarbons (PAHs)
- Stable isotopic composition of amino acids
- Stable isotopic composition of primary and secondary aliphatic amines
- Stable isotopic composition of aldehydes and ketones (method 1)
- Stable isotopic composition of aldehydes and ketones (method 2)
- Stable isotopic composition of carboxylic acids
- Stable isotopic composition of hydroxy acids
- Stable isotopic composition of alcohols
- Abundance and distribution of polyols and aldoses
- Stable isotopic composition of released volatiles (Method 1, GSFC)
- Stable isotopic composition of released volatiles (Method 2, Brown)
- Analysis of bulk organics (Method 2, Brown)
- Analysis of IOM aliphatics

## How it works:

### Gas Chromatography:

GC-IRMS relies on the chromatographic separation described in the GC-MS section.

### Combustion isotope ratio mass spectrometry (C-IRMS):

Once the analyte passes through the GC column, the flow of He and analyte split with 10% going to a quadrupole mass spectrometer for compound identity and purity verification and 90% going to the IRMS. The IRMS is a magnetic sector mass spectrometer uses strong magnetic fields to bend the path of an ion beam. A sector MS provides excellent precision and is used for stable isotopic ratio measurements. The instrument can be configured for one of the following: $^{13}CO_2$ and $^{12}CO_2$, $^{14}N_2$ and $^{14}N^{15}N$, $^{1}H_2$ and $^{1}H^{2}H$, or $^{12}C^{16}O$ and $^{12}C^{18}O$ ($^{18}O/^{16}O$ is typically not measured for CSIA) as detected by Faraday cups. To convert the analyte coming from the GC, it must be combusted into $CO_2$, or combusted then reduced into either $N_2$ or $H_2$ for detection. Each gas has different flow reactor combustion conditions and cannot be done simultaneously. Because isotopic precision is important, pulses of calibration gas are injected into the continuous flow of analyte at the start and end of the analysis. Because the IRMS is agnostic to the source of gas, interference from neighboring peaks or background must be minimized.

### Picomolar-scale compound-specific isotope analysis (pico-CSIA):

The pico-CSIA system is a modification of a GG inlet, where a custom streamlined GC-combustion interface coupled to a C-IRMS can measure as little as 50 picomoles of carbon for combusted organic molecules. The pico-CSIA system consists of a GC equipped with a programmable temperature vaporization (PTV) injector that is coupled to a C-IRMS. Pico-CSIA takes advantage of two key benefits of narrow-bore chromatography: improved resolving power and low volumetric flow rates. The improved resolving power significantly reduces total analysis time and improves separation of closely eluting compounds. Low volumetric flow rates improve sample transfer to the IRMS while maintaining narrow peak widths that improve signal to noise ratios. Modifications to the commercial system include using a microfluidic valve for solvent diversion, narrow-bore GC column (0.1 mm i.d., 30 m, 1.5 µm film thickness) and transfer lines, capillary combustion reactor, and cryogenic water trap. Pico-CSIA peak widths are about 500 ms (full width half maximum), or less than 25% the width of conventional peaks. To capture the full signal over such a short analytical window, the mass spectrometer must be fitted with collector amplifiers configured to 25 ms response times and a data logger board with firmware capable of rapid data acquisition.

### Electron impact Orbitrap isotope ratio mass spectrometry (EI-Orbitrap-IRMS):

Orbitrap-MS techniques are described in the LC-MS section since their high mass resolution and sensitivity are used for identifying and quantifying organic species. These instruments can also distinguish isotope substitutions of a single substitution of $^2H$, $^{13}C$, $^{15}N$, $^{17}O$ for ions smaller than about 250 m/z. Isotope ratios can be determined using isotopologue ion pairs. To account for fractionation within the mass analyzer, fragment isotope ratios are put on a delta scale with permil notation by evaluating the same ion fragments of an authentic standard. Delta values can also be calculated relative to the molecular average isotope value based on the isotopologue ion pairs of the intact molecular ion. Orbitrap methods have been used to determine isotope abundances for light elements, including $^{13}C/^{12}C$, $^{15}N/^{14}N$, $^{34}S/^{32}S$, $^2H/^1H$, and $^{18}O/^{16}O$.

Orbitrap MS has been optimized for both compound-specific (molecular averaged) and intramolecular isotopic analyses. The latter include analyses of isotopologues with a single rare isotope substitution, multiple rare isotope substitutions, and those with rare isotope substitutions in specific, chemically inequivalent positions within the molecule of interest. Multiple-substituted isotope ratios (e.g., $(2x^{13}C)/^{12}C$) are calculated relative to standards and assuming the ratios in standards follow a stochastic distribution set by their molecular-averaged $\delta^{13}C$ values. Differences between the measured ratios and the calculated, expected ratio of a standard are then used to compute corrected ratios for the samples. Finally, values for sample compounds are compared to their own stochastic reference frame set by their molecular-averaged $\delta^{13}C$ value.

Measured ion fragment ratios can be converted into position-specific ratios using an algebraic approach, and by correcting for instrument effects using standards, the ratios can be calculated on a delta scale. Isotope ratios for atom positions are determined algebraically from multiple ion fragments. Depending on fragmentation patterns, isotope ratios can be determined for individual positions, the molecular average (using the molecular ion), or for a sub-portion of a structure based on the averaged isotope value for two or more atom positions combined.

Orbitrap mass analyses are based on the specific angular frequencies ($\omega_z$) of ion movements along a central electrode. The angular frequency is related to the square root of an instrument constant ($k$) divided by the mass/charge ($m/z$) ratio:

$$\omega_z = (k(m/z)^{-1})^{\frac{1}{2}}$$

The angular frequency therefore conveys ion mass/charge, and the frequency amplitude (evaluated using Fourier transformation), represents the ion abundance. This is called a transient signal, and mass resolution increases with longer monitoring. Like mass resolution, isotope precision is also proportional to the acquisition time. As a result, molecules are monitored for an extended time relative to conventional GC-MS. Both GC-EI-Orbitrap instruments have modified inlet options that enable either direct injection into the instrument or injection into a peak trapping device that extends chromatographic peaks to 5-8 minutes, which is sufficient to yield precision in ion ratios of better than 1‰.

For GC-amenable organic molecules, ions are generated using electron ionization (EI) as described in the GC-MS section. Since ionization and fragmentation reactions can rearrange components of molecular structures, the atom position assignment to specific fragment ions is confirmed using labeled standards. Specific ion fragments can be selected for detailed study, and the instrument for this study has a mass filtering quadrupole that is tandem to and precedes the Orbitrap analyzer. The mass filter is used to restrict the range of ion masses permitted into the Orbitrap analyzer, which reduces space charge effects and improves statistics for low abundance ions, as in the case of rare isotope substitutions.

## Sample preparation, chemical derivatization and analytical methods:

**GC-C-IRMS.** Analytes have first been separated according to the methods described in the GC-MS section. Depending on the conditions used, the combustion interface (C) turns the carbon, nitrogen, and hydrogen in an eluting compound into $CO_2$, $N_2$, and $H_2$, whose $^{13}C/^{12}C$, $^{15}N/^{14}N$, and D/H ratios can be precisely measured in the IRMS, leading to compound-specific stable isotopic ratios. The instrument used at GSFC has a quadrupole MS added on a split before the combustion reactor to permit compound identification. This is referred to here as GC-MS/C-IRMS.

**GC-EI-Orbitrap-IRMS**. Analytes have first been separated according to the methods described in the GC-MS section. These compounds are then ionized and detected the Orbitrap and the isotope ratio is calculated based on a detailed understanding of the fragmentation pattern.

**Abundance, distribution, and stable isotopic composition of polycyclic aromatic hydrocarbons (PAHs)**

**Sample preparation and extraction:** Each powdered sample portion is transferred to a cleaned glass vial with Teflon cap and sonicated at room temperature with three 2-mL aliquots of methanol followed by two 2 mL aliquots of dichloromethane (DCM) or DCM extraction without methanol. Methanol and/or DCM extracts are combined and treated with elemental Cu beads to remove sulfur-bearing molecules that interfere with compound specific isotopic analyses. Cu beads that are pre-cleaned by treatment with HCl (5x) are then washed with 3 aliquots each of methanol and DCM, then stored under DCM to prevent oxidation. Extracts are concentrated using a custom-built spinning band distillation apparatus (Ace Glass) that gently evaporates low-molecular-weight solvent while returning higher-molecular-weight volatile species to a heated vessel. This concentration step is performed just at the boiling point of the methanol-DCM solvent mix (determined visually) to prevent loss of low molecular weight analytes of interest. While this process is time-consuming, it minimizes the risk of evaporating analytes if the extract is completely dried. Where necessary, HPLC purification is used to separate three- and four-ring species into fractions using a Restek Pinnacle II PAH column and methanol:water to acetonitrile gradient. These fractions are again concentrated by spinning band distillation, and residual water is removed by running extract through a pre-ashed sodium sulfate short column before isotopic analysis.

**GC-MS/C-IRMS analysis for D/H** uses a Thermo Scientific Trace GC connected both to a DSQ II quadrupole mass spectrometer and a MAT 253 isotope ratio mass spectrometer with a Thermo Scientific GC/TC high-temperature pyrolysis furnace held at 1400°C (Burgoyne & Hayes, 1998). The daily average $H_3$ factor is recorded throughout the analytical time frame. The isotopic composition is determined relative to reference gases calibrated to an international standard (Oztech) as well as by regular injections of a calibration mix (Restek 610 PAH calibration Mix B) and a previously characterized biphenyl standard (Arndt Schimmelmann, Indiana University). Bulk isotope analyses are performed on a high-temperature conversion elemental analyzer (TC/EA, Costech) coupled to the IRMS. The isotopic composition is determined relative to a reference gas (Oztech) as well as the international standard reference oil IAEA NBS-22.

**GC-EI-Orbitrap-IRMS analysis** uses a Thermo Fisher Q Exactive Orbitrap-based Fourier transform MS attached to a Trace 1310 GC with a modified sample inlet system. Samples will be injected in 1- or 2-µL aliquots using an autosampler with dedicated gas-tight syringes. The eluent off the GC column will either be measured directly ('direct injection') or 'trapped' in a passivated stainless steel reservoir, from which it 'drains' to the ion source, thereby producing a stable signal over several minutes. The number of species measured from a given injection is limited by the GC separation and the duration of signal needed for the desired precision. This method delivers high-precision analyses of extremely-low-concentration analytes (< 10 picomoles per injection) for which the molecular ion is the dominant peak. The choice of quadrupole mass window—which enables the isolation of multiply-substituted isotopologues and molecular fragment ions targeting specific sites of interest, effectively 'boosting' their signal to noise.

### Stable isotopic composition of amino acids

**Sample Extraction and Purification.** The distribution and stable carbon, nitrogen, or hydrogen isotopic composition of amino acids can be measured using GC-MS coupled to GC-C-IRMS (GC-MS/C-IRMS) from the same desalted hot–water extracted samples described in the LC-MS section. It is highly advisable to first analyze the amino acids by the vastly more sensitive LC-MS methods to determine whether sufficient material is available for GC-C-IRMS analysis. To conserve sample, for amino acid compound-specific isotope analyses, typically the entire sample hot-water extract is acid-hydrolyzed under HCl vapor and desalted, and 1% of the extract is analyzed by LC-MS to determine the total amino acid abundances and enantiomeric ratios. The remaining 99% of the acid-hydrolyzed water extract is used to measure the compound-specific isotope values of the individual amino acids. Based on our previous analyses of amino acid standards taken through the entire extraction, acid hydrolysis, and desalting procedure, there is no evidence of significant decomposition, racemization, thermal degradation, or carbon isotopic fractionation of the amino acids (Elsila et al., 2009; Elsila et al., 2012).

**IPA/TFAA derivatization.** To make the amino acids sufficiently volatile for GC analysis, extracts are first reacted with 70:30 isopropanol (IPA):acetyl chloride to form isopropyl esters; these esters are then reacted with trifluoroacetic anhydride (TFAA) to create the IPA-TFAA derivatives. Derivatives are dried using nitrogen gas and then dissolved in ethyl acetate for GC-MS/C-IRMS analysis.

**GC-MS/C-IRMS analysis.** Compound-specific carbon, nitrogen, and hydrogen stable isotopic values of the TFAA-isopropyl derivatives are analyzed on a gas chromatograph coupled to a GC-MS/C-IRMS instrument suite, which provides compound-specific structural and isotopic information from a single sample injection. A 1-µL aliquot of the derivatized amino acids in ethyl acetate is injected into the GC-MS/C-IRMS for each analysis. The GC-MS/C-IRMS suite consists of a Thermo Trace GC whose output is split, with approximately 10% directed into a Thermo DSQII electron-impact quadrupole mass spectrometer that provides mass and structural information for each eluting peak. The remaining 90% passes through a Thermo GC-C III interface, where eluting amino acids are oxidized to form $CO_2$, which is then passed into a Thermo MAT 253 isotope ratio mass spectrometer to measure the $^{13}C/^{12}C$ ratio of the amino acid–derived $CO_2$. Nitrogen isotopic values can be measured using the same instrumentation setup, as GC-C interface also converts nitrogen in amino acids into $N_2$ and the MAT 253 can be configured to measure the $^{15}N/^{14}N$ ratio of the eluting $N_2$. For measurement of hydrogen isotopes, a high-temperature GC-TC interface is used between the GC and the IRMS, allowing conversion of hydrogen in the amino acids into $H_2$, which can then be sent to the IRMS for D/H measurements.

GC separation uses a 5-m base-deactivated fused silica guard column (Restek) coupled with four 25-m Chirasil L-Val columns (Restek) and the following temperature program: initial oven temperature 50°C, ramped at 10°C/min to 85°C, ramped at 2°C/min to 120°C, ramped at 4°C/min to 200°C, and held for 10 min. Six pulses of high-purity reference gas ($CO_2$, $N_2$, or $H_2$, depending on the isotopes being measured) that are calibrated against commercial reference gases with known isotopic ratios are injected into the IRMS for computation of the isotopic ratios of the eluting derivatized amino acid standards and sample compounds. Analysis of the MAT 253 data is performed with Thermo Isodat 3.0 software.

### Stable isotopic composition of primary and secondary aliphatic

**amines Sample extraction and purification**. Procedures are described in the GC-

MS section.

***S*-TPC derivatization.** Procedures are described in the GC-MS section.

**GC-MS/C-IRMS analysis.** Separation, quantification, and compound-specific carbon stable isotopic values of the derivatized amines are analyzed on a GC-MS/C-IRMS instrument suite, consisting of a Thermo Trace GC, a Thermo DSQII electron-impact quadrupole mass spectrometer, a Thermo GC-C III interface, and a Thermo MAT 253 IRMS, as described above for stable isotopic compositions of amino acids. The derivatized amines are analyzed as described in the GC-MS section. Six pulses of high-purity reference gas ($CO_2$, $N_2$, or $H_2$, depending on the isotopes being measured) that are calibrated against commercial reference gases with known isotopic ratios are injected into the IRMS for computation of the isotopic ratios of the eluting derivatized hydroxy acid standards and sample compounds. Analysis of the MAT 253 data is performed with Thermo Isodat 3.0 software.

### Stable isotopic composition of aldehydes and ketones (method 1)

This method shows compatibility with other hot-water extraction methods, unlike method 2.

**Sample extraction and purification**. Procedures are described in the GC-MS section.

**PFBHA derivatization.** Procedures are described in the GC-MS section.

**GC-MS/C-IRMS analysis.** Separation, quantification, and compound-specific carbon stable isotopic values of the aldehyde and ketone derivatives are analyzed on a GC-MS/C-IRMS instrument suite, consisting of a Thermo Trace GC, a Thermo DSQII electron-impact quadrupole mass spectrometer, a Thermo GC-C III interface, and a Thermo MAT 253 IRMS, as described above for stable isotopic compositions of amino acids. For the aldehyde and ketone measurements, GC separation is described in the GC-MS section. Six pulses of high-purity reference gas ($CO_2$ or $H_2$, depending on the isotopes being measured) that are calibrated against commercial reference gases with known isotopic ratios are injected into the IRMS for computation of the isotopic ratios of the eluting derivatized carbonyl standards and sample compounds. Analysis of the MAT 253 data is performed with Thermo Isodat 3.0 software.

### Stable isotopic composition of aldehydes and ketones (method 2)

This method shows superior performance over Method 1; however, it is incompatible with water extractions.

**Sample extraction and purification**. Procedures are described in the GC-MS section.

**(*S*,*S*)-DMB-Diol derivatization.** Procedures are described in the GC-MS section.

**GC-MS/C-IRMS analysis.** Separation, quantification, and compound-specific carbon stable isotopic values of the aldehyde and ketone derivatives are analyzed on a GC-MS/C-IRMS instrument suite, consisting of a Thermo Trace GC, a Thermo DSQII electron-impact quadrupole mass spectrometer, a Thermo GC-C III interface, and a Thermo MAT 253 IRMS, as described above for stable isotopic compositions of amino acids. For the aldehyde and ketones measurements, GC separation is described in the GC-MS5.34 Gas Chromatography-Mass Spectrometry (GC-MS)section. Six pulses of high-purity reference gas ($CO_2$ or $H_2$, depending on the isotopes being measured) that are calibrated against commercial reference gases with known isotopic ratios are injected into the IRMS for computation of the isotopic ratios of the eluting derivatized aldehyde or ketone standards and sample compounds. Analysis of the MAT 253 data is performed with Thermo Isodat 3.0 software.

### Stable isotopic composition of carboxylic acids

**Sample extraction and purification**. Procedures are described in the GC-MS section.

**Esterification derivatization.** Procedures are described in the GC-MS5.34 Gas Chromatography-Mass Spectrometry (GC-MS)section.

**GC-MS/C-IRMS analysis.** Separation, quantification, and compound-specific carbon stable isotopic values of the carboxylic acid esters are analyzed on a GC-MS/C-IRMS instrument suite, consisting of a Thermo Trace GC, a Thermo DSQII electron-impact quadrupole mass spectrometer, a Thermo GC-C III interface, and a Thermo MAT 253 IRMS, as described above for stable isotopic compositions of amino acids. The derivatized acids are analyzed as described in the GC-MS5.34 Gas Chromatography-Mass Spectrometry (GC-MS)section. Six pulses of high-purity reference gas ($CO_2$ or $H_2$, depending on the isotopes being measured) that are calibrated against commercial reference gases with known isotopic ratios are injected into the IRMS for computation of the isotopic ratios of the eluting derivatized carboxylic acid standards and sample compounds. Analysis of the MAT 253 data is performed with Thermo Isodat 3.0 software.

**Stable isotopic composition of hydroxy acids**

**Sample extraction and purification**. Procedures are described in the GC-MS section.

**Esterification derivatization.** Procedures are described in the GC-MS section.

**GC-MS/C-IRMS analysis.** Separation, quantification, and compound-specific carbon stable isotopic values of the hydroxy acid derivatives are analyzed on a GC-MS/C-IRMS instrument suite, consisting of a Thermo Trace GC, a Thermo DSQII electron-impact quadrupole mass spectrometer, a Thermo GC-C III interface, and a Thermo MAT 253 IRMS, as described above for stable isotopic compositions of amino acids. The derivatized hydroxy acids are analyzed as described in the GC-MS section. Six pulses of high-purity reference gas ($CO_2$ or $H_2$, depending on the isotopes being measured) that are calibrated against commercial reference gases with known isotopic ratios are injected into the IRMS for computation of the isotopic ratios of the eluting derivatized hydroxy acid standards and sample compounds. Analysis of the MAT 253 data is performed with Thermo Isodat 3.0 software.

**Stable isotopic composition of alcohols**

**Sample extraction and purification**. Procedures are described in the GC-MS section.

**S-TPC Derivatization.** Procedures are described in the GC-MS section.

**GC-MS/C-IRMS analysis.** Separation, quantification, and compound-specific carbon stable isotopic values of the alcohol derivatives are analyzed on a GC-MS/C-IRMS instrument suite, consisting of a Thermo Trace GC, a Thermo DSQII electron-impact quadrupole mass spectrometer, a Thermo GC-C III interface, and a Thermo MAT 253 IRMS, as described above for stable isotopic compositions of amino acids. The derivatized alcohols are analyzed as described in the GC-MS section. Six pulses of high-purity reference gas ($CO_2$ or $H_2$, depending on the isotopes being measured) that are calibrated against commercial reference gases with known isotopic ratios are injected into the IRMS for computation of the isotopic ratios of the eluting derivatized hydroxy acid standards and sample compounds. Analysis of the MAT 253 data is performed with Thermo Isodat 3.0 software.

*Abundance and distribution of polyols and aldoses*

**Sample extraction and purification.** Procedures are described in the GC-MS section.

**GC-MS/C-IRMS analysis.** Separation, quantification, and compound-specific carbon stable isotopic values of the polyol derivatives are analyzed on a…

**Stable isotopic composition of released volatiles (Method 1, GSFC)**

**Sample extraction and purification**. Procedures are described in the GC-MS section.

**Aldonitoril acetate derivatization.** Procedures are described in the GC-MS section.

**GC-MS/C-IRMS analysis.** For compound specific carbon isotope analysis of polyols, the same extraction and derivatization procedures as the abundance and distribution analysis of polyols were used. The derivatives were analyzed by a 7890A gas chromatograph connected to a Delta plus XP isotope ratio mass spectrometer (ThermoFinnigan) via a GC IsoLink conversion unit and a ConFlo IV interface (Chikaraishi et al., 2004). Separation of sugars was achieved with a DB-17ms column (30 m long, 0.25 µm thick, 0.25 mm ID; Agilent). The calibration of isotopes was conducted with internal or external standards.

**Stable isotopic composition of released volatiles (Method 2, Brown)**

**Sample extraction and purification**. Procedures are described in the GC-MS section.

**GC-C-IRMS analysis.** Separation, quantification, and compound-specific carbon stable isotopic values of underivatized released volatiles are analyzed on a GC-C-IRMS, consisting of a XXX GC, Thermo XXX interface, and a Thermo Delta Plus IRMS. XXX pulses of ($CO_2$ or $H_2$, depending on the isotopes being measured) that are compared directly against the underivatized evolved gases. Analysis of the Delta Plus data is performed with XXX software.

**Analysis of bulk organics (Method 2, Brown)**

**Sample extraction and purification.** Procedures are described in the GC-MS section.

**GC-C-IRMS analysis.** Separation, quantification, and compound-specific carbon stable isotopic values of underivatized released volatiles are analyzed on a GC-C-IRMS, consisting of a XXX GC, Thermo XXX interface, and a Thermo Delta Plus IRMS. XXX pulses of ($CO_2$ or $H_2$, depending on the isotopes being measured) that are compared directly against the underivatized evolved gases. Analysis of the Delta Plus data is performed with XXX software.

**Analysis of IOM aliphatics**

**Sample extraction and purification.** Procedures are described in the GC-MS section.

**GC-C-IRMS analysis.** Separation, quantification, and compound-specific carbon stable isotopic values of underivatized released volatiles are analyzed on a GC-C-IRMS, consisting of a XXX GC, Thermo XXX interface, and a Thermo Delta Plus IRMS. XXX pulses of ($CO_2$ or $H_2$, depending on the isotopes being measured) that are compared directly against the underivatized evolved gases. Analysis of the Delta Plus data is performed with XXX software.

# Sample impact:

The analyses of soluble organic matter (SOM) is destructive and requires sample crushing and subsequent extraction in water or organic solvents by heating at elevated temperatures (up to 100°C for 24 hours). Soluble organic analyses are conducted on the solvent extracts; however, the solvent insoluble residue can be used for other analyses, including acid dissolution of the solid residue and characterization of the IOM.

The SOM analyses described here are extremely sensitive to organic contamination, therefore sample handling and analysis of the samples prior to the extractions for organic analyses should be minimized. Exposure of the sample to some materials (e.g. latex gloves, epoxy cement, ink) can significantly compromise the sample quality. Even "clean" materials for inorganic analysis can be contaminated with organics, and vice versa. The glass and ceramics used in these analyses can leech inorganic ions.

# Data quality:

## GC-C-IRMS

In the case of amino acids, other species are similar based on total moles of C, GC-C-IRMS typically requires ~1 nmol of an individual compound is required for triplicate $^{13}C/^{12}C$ measurements (typical accuracy of ±3 to 9‰, depending on abundances and complexity of background), while ~3 nmol is needed for $^{15}N/^{14}N$ with ~10‰ accuracy and ~6 nmol for D/H (50 to 200‰ accuracy).

## pico-CSIA

Highly precise (1 sigma = <1‰) isotopic measurements are possible for 100 picomoles of carbon, and robust isotopic measurements can be achieved on as little as 50 picomoles of carbon at slightly reduced precision (1 sigma = 1.5‰). For 780 pmol of carbon on column, the accuracy (±0.2‰) and precision (±0.3‰) approach that of conventional analyses. For 100 pmol and 50 pmol carbon on column, the accuracy is ±0.3‰ and ±0.4‰ and the analytical precision is ±0.9‰ and ±1.5‰, respectively. Accuracy and precision of isotopic measurements were evaluated using a mixture of *n*-alkanes with independently verified $\delta^{13}C$ ($C_{16}$-$C_{30}$, A. Schimmelmann, Indiana University). The instrument achieves <1‰ precision on natural abundance $^{13}C/^{12}C$ on 5-10 picomoles $CO_2$

## GC-EI-Orbitrap-IRMS

The Orbitrap has been optimized to require very small quantities of analyte (e.g., ~10s of picomoles per injection for PAHs) while still achieving high-precision isotopic analyses.

# Data products:

## GC-MS/C-IRMS

- 2D chromatogram (retention time vs. mass spectrum)
- Mass spectrum of analyte vs. library
- 1D chromatogram (retention time vs. isotopic ratio)
- Peak integrations vs. those of standards and mass spectrum of analyte vs. library and isotopic ratio of analytes

Data analysis includes peak integrations vs. those of standards, mass spectrum of analyte vs. library, and isotopic ratio of analytes. File types are proprietary; conversion with Proteowizard may be possible but not yet tested. Export of isotopic data as x,y pairs is possible.

## GC-C-IRMS, pico-CSIA

- 1D chromatogram (retention time vs. isotopic ratio)
- Peak integrations vs. those of standards

Data analysis includes peak integrations vs. those of standards and isotopic ratio of analytes. File types are proprietary; conversion with Proteowizard may be possible but not yet tested. Export of isotopic data as x,y pairs is possible.

## GC-EI-Orbitrap-IRMS

- Chromatograms showing separated elution peaks and their relative signal abundances for different analytes
- Mass spectra for different mass windows of different widths to isolate specific molecular ion and ion fragment isotopologues
- Text files containing all operating parameters, analytical parameters, instrument component settings, and Orbitrap-specific output (e.g., measured mass, absolute intensities, relative intensities, total ion current, injection time, peak noise, and peak resolution). These files are generated using the Thermo Fisher proprietary software FTStatistic.

# Facilities:

Astrobiology Analytical Laboratory (AAL), NASA Goddard, Greenbelt, MD, USA - Jamie Elsila

**Thermo Finnigan hybrid GC-MS/C-IRMS**: This instrument suite is for both bulk and compound-specific isotope analysis. The core is a dual-inlet MAT 253 isotope ratio mass spectrometer with multiple inlets for sample introduction including a Themo Trace GC via a ConFlo III interface or with auxiliary inlets from a Thermo TC/EA or Costech ECS 4010 EA. The Trace GC is also connected to a DSQ II quadrupole mass spectrometer.

Brown Lab for Organic Geochemistry (BLOG), Brown University, Providence, RI, USA - Yongsong Huang

**Thermo Finnigan GC-C-IRMS**: This instrument suite is for compound-specific isotope analysis. The core is a Delta Plus isotope ratio mass spectrometer (IRMS) with sample introduced by an Agilent 6890 plus model, with a split/splitless injector.

The Deines Laboratory, The Pennsylvania State University, University Park, PA, USA - Kate Freeman

**Thermo Fisher GC Q-Exactive Orbitrap MS:** For high-resolution, accurate-mass analysis of GC amenable compounds. The GC inlet system can be set up for position-specific isotope analyses, and isotope calculation subroutines and related software tools are available through collaboration with colleagues at Caltech and Themo Scientific. A modified inlet enables delayed and extended peak delivery necessary for high precision isotope analyses.

**pico-CSIA**: This instrument suite consists of a Trace 1310 gas chromatograph (Thermo Fisher Scientific) equipped with a programmable temperature vaporization (PTV) injector that is coupled via a GC Isolink and ConFlo IV to a MAT 253 isotope ratio mass spectrometer (Thermo Fisher Scientific). This prototype instrument introduced a new generation interface based on "fast GC" and microfluidic components, and fast electronics (i.e., 40 Hz sampling), tailored for picomolar quantities of compound-specific isotope analyses.

Laboratories for Stable Isotope Geochemistry, California Institute of Technology, Pasadena, CA, USA - Amy Hofmann

**Thermo Fisher GC Q-Exactive Orbitrap MS**: an Orbitrap-based Fourier-transform mass spectrometer, integrated with a gas ion source, quadrupole mass pre-selection, and GC sample introduction interface and a ThermoScientific TriPlus RSH autosampler and LabView-automated programmable valve-switching for custom analyte transfer methods between the GC and the ion source.

Chikaraishi-lab, Hokkaido University, Sapporo, Japan - Yoshihiro Furukawa

**Thermo Finnigan GC-C-IRMS**: This instrument suite is for compound-specific isotope analysis. The core is a Delta Plus XP isotope ratio mass spectrometer (IRMS) with sample introduced by an Aglient 7890A GC via a GC IsoLink conversion unit and a ConFlo IV interface.

# 5.42 Solid-State Nuclear Magnetic Resonance Spectroscopy (SS-NMR)

- Overview:
- How it works:
- Sample preparation:
- Sample impact:
- Data quality:
- Data products:
- Facilities:

## Overview:

Solid-state $^1$H and $^{13}$C nuclear magnetic resonance (SS-NMR) spectroscopy is an analytical technique that is used to determine quantitatively the distribution of organic functional groups (e.g., aromatic, olefinic, alcohol, methyl, methylene, methine, carboxyl, and ketone/aldehdyes in the cases of $^{13}$C SS-NMR, and aromatic, olefinic, aliphatic, alcohol, and carboxyl groups in the case of $^1$H SS-NMR).

## How it works:

SS-NMR works through the property of nuclear spin, i.e., angular momentum that is intrinsic to certain nuclei. Not all isotopes have the property of spin; for example, the most abundant isotope of carbon, $^{12}$C, does not have spin and therefore is not suitable for SS-NMR. The less abundant carbon isotope, $^{13}$C (at ~1% abundance), has spin. Nuclei that have the property of nuclear spin (in particular when the spin is I = ½, which is the case for $^{13}$C and $^1$H), when placed in the presence of a large magnetic field, will polarize into two different spin states with a small energy difference between them. If the nuclei, in the magnetic field, are irradiated with a high-power pulse of energy equal to the energy difference between the polarized spin states, this energy will be absorbed and create a new state defined as a single quantum coherence.

In this new state, the nuclei come under the influence of the electrons surrounding them, where the influence is a function of electron density, and hence, chemical bonding. The influence of surrounding electrons induces what is referred to as a "chemical shift", wherein nuclei in different bonding environments resonate at different frequencies. As the nuclei resonate, they emit energy near that of the initial energy that was used to excite them (typically energies corresponding to radio frequencies, e.g., tens to hundreds of MHz). These emitted frequencies are detected as an oscillating signal in the time domain. Applying a Fourier transform converts the time domain signal into a spectrum as a function of energy. Each organic functional group has a characteristic frequency, allowing quantification based on the intensity of spectral bands at those characteristic frequencies. Signal averaging is required to obtain suitable signal-to-noise ratios (SNRs), in some cases requiring from 1 to 7 days signal averaging, depending on the quantity of sample available.

## Sample preparation:

SS-NMR spectroscopy requires powdered samples of isolated insoluble organic matter (IOM; generally fine-grained as received after isolation) ranging in mass from 0.5 to 100 mg. The samples must not have abundant (no more than 5%, ideally less than 1%) paramagnetic elements, e.g., Fe-bearing oxides like chromite.

## Sample impact:

The isolation of IOM requires the dissolution or inorganic silicates, which is destructive to the majority mineral matrix of the sample. Once IOM is isolated, SS-NMR spectroscopy is non-destructive to the sample being analyzed. After analysis, the sample is available for other analytical techniques, such as IRMS, that are destructive.

## Data quality:

The quality of the data is related to the SNR of the NMR spectrum, which is proportional to the amount of sample analyzed and also scales as the square root of the number of acquisitions that add together to produce the spectrum. For the purposes of quantitating the relative abundance of various organic functional groups, we anticipate an uncertainty no greater than 5%.

## Data products:

The data products derived from SS-NMR spectroscopy are 1D spectra that report signal intensity as a function of frequency. These can be provided as .txt, ASCII, and/or MS Excel files.

## Facilities:

Earth and Planets Laboratory, Carnegie Institution for Science, Washington, D.C., USA – George Cody

The W. M. Keck Solid State NMR facility employs a Chemagnetics Infinity SS-NMR system. This includes a wide bore superconducting solenoid magnet with a field of 7 Tesla. High-power RF (1500 amp) amplifiers (for $^{13}$C and $^{1}$H, respectively) are used to excite the nuclei and to induce the spectral emission. For $^{13}$C SS-NMR, a double resonance probe capable of spinning the sample up to 12 KHz rotation (at the "magic angle") is employed. A fast digitizer (2 MHz) is used to capture and convert the analog signal to digital for subsequent data processing. For $^{1}$H SS-NMR, a double resonance probe capable of sample spinning up to 30 KHz is employed.

# 5.56 Desorption Electrospray Ionization-Orbitrap Mass spectrometry (DESI-Orbitrap MS)

## Overview:

High-resolution mass spectrometry imaging by desorption electrospray ionization coupled with Orbitrap mass spectrometry (DESI-Orbitrap MS) is an analytical technique used for organic compounds on the sample surface. The surface of solid sample is sprayed using an electrically charged polar solvent at an ambient condition. The desorbed organic ions are introduced into the ion source of an Orbitrap mass spectrometer to be analyzed according to their mass to charge (*m/z*) ratios. DESI-Orbitrap MS permits spatially resolved organic compound analysis at the sub-millimeter scale and allows for identification of heterogeneities within a sample.

## How it works:

An Orbitrap MS is a type of Fourier transform MS, discussed in more detail in the LC-MS section.

DESI-Orbitrap MS can analyze soluble polar organic compounds on a solid or dried liquid samples. The detection limit is usually sub ppm level for areas of about 50 x 50 µm depending on the ionization efficiency of molecules. The DESI-Orbitrap MS imaging is performed using a motorized x–y stage by continuously scanning the sample surface with a scanning rate of about 50 mm/s.

The spray solvent is a polar solvent such as methanol at a rate of 1-2 mL/min with an electrospray voltage of 1-3 kV. The DESI emitter (fused silica capillary with ~50 mm inner diameter) was mounted < 0.5 mm from the surface to be analyzed and was arranged at a 55° angle with respect to the sample surface. The nebulizer $N_2$ gas pressure was ~100 psi. The positive and negative ions are measured over *m/z* 70–750 range in a full scan mode with a mass resolution of 70,000-140,000 (m/Δm at *m/z* = 200).

The mass spectral data file is converted into the cube format using the FireFly software (Prosolia Inc.). A DESI/Orbitrap MS spectra image is reconstructed using BioMap (a free software package available at www.maldi-msi.org). A digital image processing is performed using a moving average with 3 x 3 pixels (1 pixel ~20 mm) on DESI images. The apparent mass resolution of the constructed DESI images is 0.001 Da, although mass spectra of higher resolution (~0.0001 Da) could be obtained from the raw spectral data.

## Sample preparation:

A sub-millimeter to several mm sample is generally embedded or pressed in a soft metal such as indium. The flat surface of sample is preferred with or without prior polishing, which is mounted on the moving stage. Incomplete fixed or fragile samples may be blown away by the nebulizer $N_2$ gas. The boundary between the surrounding metal and sample surface is to help identify and register the area of the sample surface analyzed to obtain DESI images (Hashiguchi and Naraoka 2019).

## Sample impact:

DESI-Orbitrap MS is a semi-nondestructive technique. The sample surface is sprayed only by a solvent such as methanol. After the DESI-Orbitrap MS analysis, the same sample surface can be used for other analyses including spectroscopy (e.g. FTIR or Raman), time of flight-secondary ion mass spectrometry (ToF-SIMS) or scanning electron microscopy (SEM).

## Data quality:

DESI-Orbitrap MS detection limits in favorable cases can extend to less than parts per million using a full scan mode. The analytical sensitivity strongly depends on the ionization efficiency of molecules and intensity of background signals. If specific ions are targeted for analysis using a single ion monitoring mode, the detection limit is down to parts per billion level.

The spatial resolution of DESI-Orbitrap MS for scanning ion imaging is controlled by the size of the splaying solvent, which is mainly controlled by the inner diameter of the emitter (fused silica capillary) and the scanning rate of the sample stage. These factors have influence on the sensitivity of detection.

The mass resolution of DESI-Orbitrap MS depends on the setting for the Fourier-transform calculation. The mass resolving power is chosen from 35,000, 70,000, 140,000 or 280,000 (at full-width-half-maximum for *m/z* 200).

## Data products:

The data products expected from DESI-Orbitrap MS measurements include ion intensities vs. time (distance):

- 1D line scans (individual data set plotted as a function of distance)
- 2D images (2D maps of a sample in which every pixel contains a data set)

All data products can be delivered as text files.

## Facility:

Department of Earth and Planetary Sciences, Kyushu University, Fukuoka, Japan - Hiroshi Naraoka

**A DESI-Orbitrap MS system for extraterrestrial material measurements:** 2-D DESI ion source (Omni Spray Source 2D, Prosolia Inc.) equipped with a hybrid quadrupole-Orbitrap mass spectrometer (Q-Exactive Plus, Thermo Scientific) in a clean room (ISO 6, Class 1000).

Laboratories for Stable Isotope Geochemistry, California Institute of Technology, Pasadena, CA, USA - Amy Hofmann

**Thermo Scientific Orbitrap Fusion Lumos Tribrid MS**: an Orbitrap-based Fourier-transform mass spectrometer, may also be equipped with a DESI or related scanning atmospheric ionization technique. The details of the interface are being investigated for implementation late during sample analysis.

# 5.63 Ion Chromatography (IC)

See Elements and Isotopes section 5.63 Ion Chromatography (IC)

## 5.64 Capillary Electrophoresis (CE)

### Overview:

Capillary electrophoresis (CE) is an analytical technique used for chemical analysis of soluble organic molecules in aqueous solutions. The CE instruments at the Institute for Advanced Biosciences, Keio University (Sasaki et al. 2019), and the Japan Agency for Marine-Earth Science and Technology (JAMSTEC) (Oba et al. 2023) are equipped to analyze water-soluble organic molecules in natural products (e.g., nucleobases, vitamins, nitrogen-containing imino acids, organic acids, sugars, and some group compounds)

### How it works:

CE measures soluble ions based on their electrophoretic mobility with the use of an applied voltage. The electrophoretic mobility is dependent on the charge of the molecule and the viscosity on the migration for organic cation, organic anion, and neutral species. If two ions are the same size, the one with greater charge will move the fastest. For ions of the same charge, the smaller particle has less friction and an overall faster migration rate. Each ion produces a peak on the chromatogram, and the peak height and area depend on the ion concentration in the sample. We can use internal and/or external standard reagents for quantification and qualification on the measurement.is 0.001 Da, although mass spectra of higher resolution (~0.0001 Da) could be obtained from the raw spectral data.

### Sample preparation:

CE analyzes liquid samples prepared in water-based (acidic, neutral, and alkaline) solutions. To avoid analysis of extreme-pH samples, pH measurements are taken prior to CE analysis. Samples should be filtered to remove any insoluble matters and precipitates. Samples, authentic standards, and process blanks are prepared for data quality assurance. To quantify ion concentrations in the sample, several standards with known concentrations are analyzed in the same sequential run.

### Sample impact:

CE is a destructive technique requiring at least 90 ml of sample for triplicate analysis. Depending on the expected sample concentration, the sample volume can be adjusted arbitrarily.

### Data quality:

The CE technique gives faster results and provides higher-resolution separation for soluble molecules than LC. CE separation can be connected with an online mass spectrometer (CE-MS), like liquid chromatography (LC)–MS. The data quality of mass spectrometry depends on the capability of the detector. In this context, both techniques are useful because there is a cross-validation between CE-MS and LC-MS for the target and non-target analysis of organic molecules.

### Data products:

The ion concentrations are the expected products from CE measurements. Based on the result summary determined by the instrumental software, the product can be delivered as Excel files.

### Facilities:

The CE instruments are located at (1) the Institute for Advanced Biosciences, Keio University

JAMSTEC

**The CE instruments include:**

- CE System G1600AX (Agilent Technologies Inc.)
- LC-30 AD pump (Shimadzu Corporation, Kyoto)
- Fused silica capillary (50 μm i.d. × 80 cm; Polymicro Technologies, Inc.)
- ZipChip Interface (Thermo Fisher Scientific, Inc.)
- Systems controller of ChemStation software (Agilent Technologies, Inc.)
- Orbitrap mass spectrometer of QExactive Plus (Thermo Fisher Scientific, Inc.)

# Physical Testing

# 5.43 Gas pycnometry

## Overview:

A gas pycnometer is a laboratory device used for measuring the volume of solids. Any sample geometry can be measured, including regularly or irregularly shaped, porous or non-porous, and monolithic, powdered, granular, or in some way comminuted. A pycnometer employs gas displacement and Boyle's Law. The utility of these measurements is to estimate quickly and easily sample grain density and porosity. Furthermore, some pycnometers have the ability to control the temperature of the sample cell, which allows for the measurement of the temperature-dependent change in sample volume and thus the determination of the coefficient of thermal expansion.

## How it works:

Gas pycnometry is suitable for the rapid estimation of sample volume. When coupled with measurements of mass, bulk sample shape, and volume (e.g., via 3D laser scan) gas pycnometry can be used to estimate sample bulk density, porosity, and the mean grain density (e.g. Consolmagno et al., 2008; Macke et al., 2011). The volume measured in a gas pycnometer amounts to the three-dimensional space into which the gas cannot penetrate. Ultimately, the volume within the sample chamber from which the gas is excluded is determined. Therefore, the completeness limit of the measurement (i.e., minimum size of roughness features and pores that are captured) depends on the atomic or molecular size of the gas. Helium gas offers the greatest advantage because the He atom is so small that it will quickly penetrate any cracks in the sample, and because He is chemically inert. Completely closed pores, i.e., those that do not connect with the surface of the solid, may not necessarily be penetrated by the gas, so they would be excluded from the microporosity values determined from this measurement.

The following method description is from Consolmagno et al. (2008):

> Typically, two sealed chambers of known volumes are connected through a valve. The meteorite is placed in one chamber, which then has He introduced into it while the connecting valve is closed. The second chamber is held at a different pressure from the first chamber. The simplest procedure is to raise the pressure of the meteorite's chamber to well above ambient room pressure (typically a pressure near 1.5 atm is used); then the second chamber may either be evacuated or simply be purged with He and maintained at the local ambient pressure. After pressure in the first chamber is measured, the valve is opened, allowing the gas to flow into the adjacent chamber until pressure in both chambers has equalized. In the case of previously evacuated chambers, the ideal gas law dictates the relationship between pressure and volume:

> where $P_i$ and $P_f$ are initial and final pressure, respectively; $V_A$ and $V_B$ are volumes of the two chambers, respectively (where chamber A is the one containing the sample); and $V_{gm}$ is the grain volume of the sample. Solving for the grain volume of the sample yields:

> For the case where the second chamber is initially at atmospheric pressure rather than evacuated, local atmospheric pressure must be taken into account, but this can be easily accommodated by measuring all pressures as overpressures relative to atmospheric rather than as absolute pressures. With pressures measured this way, the calculations are identical to that for an evacuated chamber.

> For hand-sized samples, evacuation of the second chamber is not necessary in order to obtain reliable measurements. A typical commercial pycnometer (Quantachrome) yields grain volume measurements repeatable to within ~0.02 cm$^3$ without employing a vacuum pump.

If the sample is smaller than the manufacturer recommendation for a given sample cell, or if variability between repeated measurements is unacceptably large, other objects may be placed inside of the cell to occupy some of the volume (e.g. Goodrich et al., 2019 Methods Supplement).

## Sample preparation:

Gas pycnometry does not require any particular sample preparation, although extraterrestrial samples may be placed in a vacuum to remove any absorbed terrestrial water before measurement. The sample may simply be placed inside of the measurement chamber.

## Sample impact:

Gas pycnometer is a non-destructive technique. The sample is subjected to slightly elevated pressures when helium is injected (~1.5 atm). The sample also comes into contact with the weighing paper (or another non-contaminating staging material) upon which it is placed in the pycnometer chamber. The entire apparatus could feasibly be operated in a nitrogen- or dry-air–purged environment to minimized atmospheric contamination risks.

## Data quality:

The accuracy of the result is in part related to the relative size of the sample to the sample cell in the pycnometer. Most commercial pycnometers come with sample cells of various sizes. For example, a few commercial pycnometers report the following minimum sample sizes: 0.025 cm$^3$ for the Ultrapyc 5000 Micro by Anton Paar (formerly the Ultrapycnometer by Quantachrome) and 0.01 cm$^3$ for the QPI-GP5800 by Qualtech Products Industry. Accuracy and repeatability for the latter device are reported as ±0.04% and ±0.02%.

Goodrich et al. (2019) recently measured two small meteorite samples at the same time with a total mass of ~200 mg. The resulting bulk density values were reported with uncertainties in the range of ±10–60 kg/m$^3$. The two meteorites were too small to measure individually (130 mg and 65.5 mg, respectively). Thus, we can to expect similar uncertainty values if we follow their methods, use a similar-sized sample cell, and utilize sample masses of at least ~150–200 mg for any given measurement.

## Data products:

The data products expected from gas pycnometry for each measured sample include pycnometer pressure readouts vs. time and calculated sample volumes. Each measurement will be repeated multiple times to assess variability. The final, averaged sample result will be reported with uncertainty based on variability between measurements. These results can be combined with 3D shape measurement and mass measurement data to determine sample bulk density, grain density, and microporosity values with associated uncertainties. All products can be delivered as text files.

The determined porosity values can be used to inform decisions for which samples should be used in thermal property measurements.

## Facility(ies):

NASA Johnson Space Center, Houston, TX, USA

- A custom pycnometer is currently being developed at the University of Arizona for eventual installation in the OSIRIS-REx curation facility at Johnson Space Center.

University of Calgary (Hildebrand Lab), Calgary, AB, Canada

- Quantachrome Instruments model MVP-D160-E

## References:

Consolmagno, G.J., Britt, D.T., Macke, R.J. (2008) The significance of meteorite density and porosity. *Geochemistry 68*(1), 1–29.

Goodrich, C.A., Zolensky, M.E., Fioretti, A.M., Shaddad, M.H., Downes, H., Hiroi, T., Kohl, I., Young, E.D., Kita, N.T., Hamilton, V.E., Riebe, M.E. (2019). The first samples from Almahata Sitta showing contacts between ureilitic and chondritic lithologies: Implications for the structure and composition of asteroid 2008 TC 3. *Meteoritics & Planetary Science 54*(11), 2769–2813. Methods supplement.

Macke, R. J, Consolmagno, G. J., and Britt, D. T. (2011) Density, porosity, and magnetic susceptibility of carbonaceous chondrites. *Meteoritics & Planetary Science* 46, 1842–1862.

# 5.44 Structured Light Scanning (SLS)

## Overview:

A benchtop structured light scanner (SLS) is used to create a detailed three-dimensional shape models of a sample (approximately 3 mm to 4 cm in size). The external sample volume calculated from the resulting 3d scan may then be used with grain volume measurements (via gas pycnometry) and mass measurements to determine sample bulk density, grain density, and porosity. Sample shape models are also available for sample shape and roughness analysis. They are additionally used in the thermal modeling of samples in the Spherical Cell Bulk Thermal Conductivity Analysis.

## How it works:

A blue-light LED projector casts parallel stripes of light on the sample. Two cameras are positioned on either side of the projector in order to view the sample from different perspectives. The blue-light lines are distorted by the 3D shape of the sample when viewed off-axis by the cameras. The data from the cameras are used to reconstruct the surface shape of the sample via the FlexScan3D software. The structured light scanner is also able to capture a grayscale texture map of the sample surface.

The scanner head and sample must first be positioned so that the sample is near the intersection of the field of view of the two cameras. The exposure of the cameras is then adjusted to minimize over-/underexposure pixels that fall on the sample. Following these steps, the sample is scanned repeatedly, changing the orientation of the sample between scans. An individual scan takes approximately 5 seconds and produces a partial 3D model of the sample face directly opposing the projector and cameras. The sample is typically rotated 10–20 times to obtain a partial 3D model of the top half. The sample is then flipped over so that the process of scanning and rotating can be repeated for the other half. The resulting partial scan files that are collected from each perspective are then merged to produce a full 3D model of the object. Subsequent processing may be performed to smooth the 3D model (noise reduction) and to fill any voids in order to make the model water-tight.

## Sample preparation:

No sample preparation is necessary. The sample is placed on a sample stage made of anodized aluminum or Teflon. Appropriate sample sizes are in the approximate range of 3 mm to 4 cm in width.

## Sample impact:

The blue light projector in the scanner is not expected to affect the sample.

Some manual manipulation of the sample is required in order to position the sample in the cameras' field of view and to rotate and flip it between scans. The impact of rotating the sample can be minimized by rotating the actual stage or surface upon which the sample rests, rather than manipulating the sample itself. The sample must be handled at least once when it is flipped over to obtain data of the obscured underside. For some samples, more than one flip is required to collect data of all faces.

In order to maintain sample pristinity, the scanner head can be positioned outside of a purged-air glove box, observing the sample through the glove box window. The scanner head has an IP 67 rating, meaning that it is dust- and water-tight. Its housing is constructed of anodized aluminum, stainless steel, and glass, with a single, rubber-coated power/data cable. The scanner head must be attached to a camera tripod or to a custom fixture. It must also be connected to a computer (with the FlexScan3D software) via ethernet data cable in order to collect data.

## Data quality:

Manufacturer-reported dimensional accuracy and point-to-point distance (i.e., facet size) in the resulting shape model are up to 12 µm and 20–25 µm, respectively. The sample volume measurement error is approximately 0.2% for larger samples (> ~1 cm) and approximately 1% for small samples (~5 mm), based on test measurements of samples with known volume. When scanning through a glove box window pane (Schott Amiran® glass, 3/8" thick), the measurement errors are approximately doubled.

## Data products:

3D triangular mesh shape models of scanned samples will be delivered in .stl and/or .obj format with corresponding .mtl texture map files. Scan files will be delivered as partial scans (from a single perspective) and combined scans. The combined scans will include both unprocessed and processed shape models where additional smoothing and hole-filling steps were taken by the user.

The determined sample volumes and associated uncertainty values can be provided in text files. They are calculated from the water-tight sample mesh files (i.e. shape models) using a tetrahedron signed volume method.

## Facility(ies):

University of Arizona, Tucson, AZ, USA; and Johnson Space Center, Houston, TX, USA

- Polyga Compact C506 structured light scanner (manufactured by LMI Technologies as Gocator 3506)
- FlexScan3D software license
- Lab jack for sample staging
- Tripods and aluminum rail system for mounting the SLS

## References:

# 5.45 Spherical Cell Bulk Thermal Conductivity Analysis

## Overview:

The spherical cell thermal conductivity analysis is used to measure the bulk thermal conductivity of an uncut regolith particle ranging in size from approximate 5 mm to 2 cm diameter. A sample is placed in between two hemispherical shells. A steady-state heat flux is applied to one hemisphere while the other serves as a heat sink at a constant temperature. The sample exchanges heat via infrared radiation with the two hemispheres. Once steady-state temperature distribution is achieved, the temperature of the non–temperature-fixed hemisphere is measured and related to the sample thermal conductivity via an iterative finite element model (FEM) simulation of the experiment. This is a novel experimental technique.

## How it works:

Note: This technique is under development; some design details are subject to change.

The technique is based on the commonly used guarded hot plate method. In a typical guarded hot plate configuration, the hot plate (constant heat flux) and cold plate (constant temperature) are flat and touching either side of a flat sample. Here, the hot and cold plates are replaced by two hemispheres that surround an irregularly-shaped geologic sample. Rather than contacting the sample directly, the hemispheres exchange heat radiatively with the sample. A steady heat flux is applied to one hemisphere while the opposite hemisphere serves as a cold sink and is held at a constant temperature. The two hemispheres are sized and shielded so as to minimize direct radiative heat transfer between each other. Rather, heat flow is directed from the hot hemisphere into the sample, and from the sample into the cold hemisphere. High-precision thermometers are placed in the hot hemisphere to measure the final, steady-state temperature that is achieved for a given prescribed heat flux. The simulation is ultimately reproduced in an FEM where the sample thermal conductivity is tuned to match the measured experimental temperature. The experiment is repeated with different values for the prescribed heat flux on the hot hemisphere and fixed temperature values on the cold hemisphere to achieve different mean sample temperatures, so that the temperature dependence of the sample thermal conductivity may be investigated.

For measurement of Bennu material, an uncut regolith sample will be suspended in the center of the cell by a thin, low-conductivity mesh basket that is designed to minimize conductive heat transfer between the sample and hemisphere wall while minimally affecting radiative view factors.

The measurement will be conducted in a high-vacuum chamber to eliminate gas conduction within the sample pores and between the sample and the cell. A residual gas analyzer will be included in the vacuum chamber to measure and characterize any volatile species that may be released by the sample during vacuum pump-down and temperature cycling.

## Sample preparation:

The technique is specifically designed to work with an uncut, as-is sample between 5 mm and 2 cm in diameter, so no direct sample preparation is necessarily required. However, measurements of the approximate shape, mass, and emissivity of the sample are needed to properly model the experimental results in the FEM. The required accuracy of those needed values is TBD.

## Sample impact:

This method is nominally no more mechanically destructive than any task that requires moving the sample from one container to another. However, the sample will be in a high-vacuum environment and will undergo user-specified temperature excursions, which could trigger phase changes, thermal fracturing, and/or loss of volatiles species.

There is a small risk that the sample could be mechanically disaggregated during handling and from contact with the sample basket (e.g. surface grains breaking off). The material of the basket is chosen to be non-reactive with the sample and not highly thermally conductive, such as a polymer material like Teflon.

The absolute temperature range of the experiment can be chosen, for example, to not surpass the modeled temperature range at the sample site at Bennu's present orbit. The values of the temperature and heat flux set-points can be constrained to limit the thermal gradient and/or the rate of heating in the sample in order to prevent thermal fracturing. The precise thermal gradient and heating rate limits are to be defined based on results from a thermal stress numerical study that will be performed later in the experimental design process.

## Data quality:

This technique is currently in the design phase, so data quality and accuracy are TBD. However, our objective is to determine sample thermal conductivity to within 10% error and uncertainty.

## Data products:

Raw experimental data can be prepared in ascii csv format. Data collected periodically throughout each measurement include:

- Raw and calibrated temperature sensor values
- Vacuum system pressure values
- Heater setpoints and heater output values
- Photographs of the sample positioned within the sample cell before and after the experiment

A numerical model (e.g., ANSYS, COMSOL) will need to be used to extract the sample material properties. All model files in their native formats for each specific sample run would be included as data products.

The final analytical solutions for thermal conductivity can be provided in a sample ascii format with supporting documents describing how the values were calculated from the experimental data.

## Facility(ies):

University of Arizona, Tucson, AZ, USA

The experiment will be housed in a thermal vacuum chamber, which will be moved to an unused lab space with an adjacent unused clean room for sample storage and prep in the Drake Building, under the management of Andrew Ryan.

# 5.46 Differential Scanning Calorimetry

## Overview:

A calorimeter is used for determining the specific heat ($c_p$) of a material. Knowledge of the heat capacity for the samples from Bennu is critical for comparing them to thermal inertia values determined from orbit, given that thermal inertia is a function of heat capacity, density, and thermal conductivity. It measures the energy difference between a sample and a reference material following a controlled temperature program. DSCs are also commonly used to measure the heat absorbed or released during various chemical reactions and/or phase changes. $c_p$ denotes specific heat at constant pressure and can be considered equivalent to the specific heat at constant volume $c_v$ for solid materials at low temperatures (i.e. temperatures on Bennu). The two might not be equivalent for solids only at very high temperature, e.g., in Earth's mantle. $c_p$ depends strongly on temperature, especially below room temperature, going to a value of 0 J kg$^{-1}$ K$^{-1}$ at 0 K. It also depends on mineral composition, varying roughly by a factor of 2 between the various (meteoritic analog) compositions (Biele et al., 2022).

## How it works:

For the measurement of specific heat at relatively high temperatures (≳300 K), a differential scanning calorimeter (DSC) is recommended. A DSC is used to measure heat flux *difference* between a sample and a reference material following a controlled temperature program. DSCs are also commonly used to measure the heat absorbed or released during various chemical reactions and/or phase changes. Specifically, for accurate specific heat measurements, a Power Compensated DSC is preferred because it will provide more accurate absolute values as compared to a Heat Flux DSC.

A power-compensated DSC consists of two identical sample cup holders with their heaters housed inside of a very stable heat sink (a calorimetric block). One cup contains the sample and the other is usually empty to act as a reference. As the two cells are gradually heated, their temperatures are monitored by thermocouples; The instrument electronics regulates the power input for the cells such that their temperature is always identical. The signal is the differential power between the two cells as a function of time, which can be used to determine the heat flow into the sample. A linear temperature heating ramp can be employed to determine heat capacity as a function of temperature. The obtained dataset consists of a series of temperatures and heat flow values. The specific heat ($c_p$) can be evaluated, after three runs (empty, with reference substance, and with the sample: "3-curve method") as:

$$c_{P,sample} = \frac{(dq/dt)_{sample} - (dq/dt)_{blank}}{m_s \, dT/dt} \, c_{P,ref} \Bigg/ \frac{(dq/dt)_{ref} - (dq/dt)_{blank}}{m_{ref} \, dT/dt}$$

where *dq/dt* is sample heat flow at time *t*, *dT/dt* is the temperature change at time *t*, *m* is the sample mass, *ref* stands for the calorimetric reference substance (usually corundum), *blank* for the empty run (Höhne, Hemminger et al. 2003). A heating rate of 10 K/min is typical.

For low-temperature specific heat measurements (~5–300 K), the hybrid adiabatic relaxation heat capacity method is more suitable. Quantum Design® constructed a fully automated commercial relaxation calorimeter (Lashley, Hundley et al. 2003), implemented as the heat capacity option of the Physical Properties Measurement System (PPMS). Often just called "PPMS", it used either powdered (wrapped and pressed into thin and light Al-foil holders weighing~5.5 mg) or cut to a cuboid of ~3.5 x 3.5 x h mm³. Accuracy can be better 1% except at very low temperatures (<60 K). More details on this technique may be found in Opeil et al. (2020) and references therein.

## Sample preparation:

The sample must either be powdered or cut to have a flat, polished surface and fit within the sample cup (<~ 5 mm diameter). As little as 3 mg of powdered sample can be used, though it is recommended to use 15–50 mg in order to obtain results with higher accuracy (as described in Data Quality, below). A cut and polished flat sample will make good thermal contact with the bottom of the sample cup. A powder can be more evenly spread in the pan, but as a tradeoff will not have as good of a thermal contact, which will have the effect of increasing temperature uncertainty in the resulting heat capacity values due to thermal lag. This may be countered by proper temperature calibration or by reducing the sample heating rate, which will increase the duration of a measurement run.

## Sample impact:

The sample is powdered (recommended for Bennu sample analysis) or cut as described above. During the measurement, the sample will be subjected to the user-prescribed temperature range that is of interest for measuring heat capacity. The maximum and minimum temperatures can be chosen based on model-predicted temperature range for the site on Bennu where the sample was collected in order to minimize the occurrence of thermally driven phase changes, mechanical breakdown, etc. that was not already occurring continuously on Bennu. If a bonding agent is used, such as thermal grease in the PPMS system, the lower portion of the sample (at minimum) will be contaminated by the grease. Depending on the agent, it could be dissolved with a solvent, at the possible expense of further contaminating the sample and/or altering the chemistry/mineralogy.

## Data quality:

Power-compensated DSCs using the 3-curve technique (with a reference substance) and the so-called "step-scan method" (Dachs, Benisek et al. 2018) can routinely achieve accuracies of <~2%.

The PPMS system can be used to obtain measurements with error <~2.5% at 300 K (Opeil et al., 2020).

## Data products:

Results from individual calorimetry runs will be reported in tables of calculated heat capacity versus temperature. Raw and intermediate sensor readouts (e.g., raw and calibrated sample cup heat flux values) can also be provided in these tables.

The results of repeat measurements of a particular sample will be combined by averaging to produce a final table of heat capacity versus temperature, with reported uncertainty based on expected instrument uncertainty and on variability between measurement runs.

All data products can be provided as ascii text files.

## Facility(ies):

<u>Nagoya University, Nagoya, Japan</u>

PerkinsElmer DSC 8000

- Power-compensated double furnace DSC
- Mass requirements: 5–20 mg powder per sample
- Temperature range: ~100–1000 K
- Two cooling systems available:
  - Intracooler II– Nitrogen purge (~200 – 1000 K)
  - Liquid Nitrogen coolant CLN2 – Helium purge (~100 – 470 K)

<u>Boston College, Chestnut Hill, MA, USA</u>

Quantum Design Physical Property Measurement System (QD-PPMS).

- Hybrid adiabatic relaxation heat capacity method
- Mass requirements: 20–60 mg powder per sample
- Temperature range: ~5–400 K

# 5.47 Scanning Thermal Microscopy (SThM) with AFM

## Overview:

Atomic force microscopy (AFM) is an analytical technique used to measure the physical properties of samples at the nanoscale. The AFM scans a sample in the X-Y-Z directions, resulting in three-dimensional image data. The AFM is equipped with a high-resolution scanning thermal microscope (SThM), which operates in two analytical modes: Temperature Contrast Mode (TCM) and Conductivity Contrast Mode (CCM). TCM is used to measure temperature variations on the surface of a sample, while CCM is used to measure the thermal conductivity of the surface. Topographic information about a sample's surface is also obtained with both analytical modes.

## How it works:

AFM analysis with the SThM is achieved by the interaction of a thermal probe (cantilever) with a sample's surface. The thermal probe is in direct contact with the sample during the measurement scans. Attached to the end of the thermal probe is a resistive element that serves as a thermometer in TCM and a heater in CCM. Scanning the sample in the Z direction measures the morphology of the surface, which allows users to create three-dimensional SThM contrast images, as well as topography images. All measurements are completed under ambient pressure and temperature conditions. Because the data obtained from TCM and CCM are contrast images, these analytical options are best for heterogeneous samples.

### TCM

In TCM, the probe scans the surface and measures its temperature. Essentially, a current is applied to the probe, and the AFM measures changes in resistivity over a sample's surface; the current is small enough to avoid heating the probe. Prior to scanning, thermal equilibrium is established between the probe and sample. As the thermal probe scans over the surface of the sample, changes in resistivity—or temperature—result in a voltage change of the system's circuit bridge. This change in voltage is represented by thermal contrast images.

### CCM

In CCM, a current is applied to the sample to heat the thermal probe to a temperature of approximately 55°C/131°F to create a thermal contrast to the sample surface, which is near ambient temperature. As the probe interacts with the sample and heat flows to the surface, the probe cools. The higher the surface thermal conductivity, the more the probe cools. The measured change in the resistance (temperature) of the probe is translated to thermal conductivity contrast maps.

## Sample preparation:

Sample preparation for the AFM is relatively simple. The surface of the sample needs to be cut flat. A standard geologic thin section (without the glass cover) could be used, for example. For best results, a well-polished surface is recommended, especially when conducting thermal measurements. Samples can be up to 50 x 50 mm in size and up to 20 mm thick. The weight of the sample should not exceed 500 g. Non–thin section samples are typically attached to the sample chuck with an adhesive double-sided tape. A different mounting method that does not involve adhesives could be developed for cut samples.

## Sample impact:

AFM is typically considered a non-destructive technique. However, the measurement probe (cantilever) is in direct contact with the sample's surface as it scans the sample and thus could produce nanoscale scratches and/or indentations.

## Data quality:

The resolution of the AFM depends both on the vertical (Z) and lateral (X-Y) scanning directions. The lateral resolution is determined by dividing the scan range size by the pixel size. For example, if a 10-μm scan range has been set for an image with 256 x 256 pixels, the lateral resolution is 10 μm/256 = 39.1 nm. The maximum lateral scanning range of the AFM is 100 μm and the pixel range is 4096 x 4096 pixels. Increasing the image pixel count increases the resolution, but also increases the scan time. Alternatively, decreasing the scan size can produce high-resolution data. The motion of the scanning probe can also influence the lateral resolution. Depending on the amount of voltage applied, the scanner proportionally expands or shrinks. Specifically, the scanner's lateral movement is controlled by a 20-bit DAC (digital-to-analog converter). For example, if the 20-bit DAC is used to move the scanner 50 μm, the lateral resolution is 50 μm /$2^{20}$ = 0.47 Å.

The vertical resolution is defined by the range of the vertical scanner motion, which is proportional to the vertical scanner's movable range. The maximum movable range of the scanner is 15 μm, under standard operating conditions, which corresponds to a scanner range of 1. If the scanner range is decreased to 0.5, the maximum movable range becomes 7.5 μm. Essentially, the smaller the scanner range, the smaller the movable range. By decreasing the scanner's maximum movable range, the vertical resolution is effectively increased. The height variation (topography) of a sample's surface must be considered when determining the appropriate vertical scanner range; the height variation must not exceed 15 μm.

## Data products:

The image data is processed as a TIFF file. The TIFF format allows the metadata of the image (e.g., width and height) to be stored within the file, and the files can be viewed with any image viewer application (e.g., Windows Photo Viewer). TIFF files can be compressed while preserving the original image quality. Raw TIFF files contain both the scan data and the image data. The TIFF files can be exported from the AFM software as either a text file or image file (.jpg, .png, .bmp, or .emf). If the file is exported as a text file, the file will contain both metadata pertaining to the original TIFF file and the data array of the image scan. Alternatively, if the TIFF file is exported as an image file, there will be no information saved regarding the scan data; only the image will be saved.

## Facility(ies):

Planetary Exploration Instrumentation Laboratory (PIL), York University, Toronto, ON, Canada

- The PIL at York University has a Park Systems NX10 AFM. The system is connected to a controller (NX Control Electronics), which is considered the core of the AFM; it controls the scanner and data acquisition and communicates with the computer connected to the AFM. The computer runs the necessary software to operate the AFM (XEP, data acquisition program) and process the data (XEI, image processing program). The XEI software allows the user to convert the AFM data to an image for analysis.
- Beyond thermal and topographical analysis, the AFM is also capable of analyzing the electrical, chemical, and magnetic properties of a sample. The type of cantilever used for these analyses vary and the measurement techniques differ compared to the SThM mode.

# 5.49 Particle cohesion determination with AFM

## Overview:

An atomic force microscope (AFM) is adapted to determine the magnitude of the cohesive force between dust-sized particles that are close or in direct contact. The force required to pull two contacting dust particles apart is measured to determine the cohesive force. Measurements of the local radii of the particles and/or their surface roughness can be used in conjunction with the measured cohesion values to calculate the Hamaker constants, which measure the magnitude of the attractive potential between the particles, and the value of the particles' surface energy. These quantities are used in models of rubble-pile asteroid cohesion, deformation, and disruption (e.g. Sánchez and Scheeres, 2014), as well in models of regolith thermal properties where heat transfer occurs due to phonon conduction (e.g. Sakatani et al., 2017).

## How it works:

Here, two tests are proposed. The first is to measure the cohesive force between a particle and a surface with a known surface energy. The second is to determine the particle-to-particle cohesive forces. In both tests, a single dust particle is affixed to the pin on the cantilever arm of an AFM. For the second test, a second particle is then fixed to a substrate. The cantilever arm is gently lowered to bring the particles in close proximity with either the free surface, or the other particle. If there is an attraction, a "pulling force" bending the AFM tip is detected by the machine. The particle is brought closer and closer to the substrate (either reference surface or a second particle) to characterize the force versus separation distance behavior, which helps determine the attraction energy potential. The particle attached to the AFM tip is then purposely brought into in contact with either the substrate or the second particle, then retracted to measure a resistance force to the retraction, which is the cohesive attraction force. The cantilever arm is then raised until the particle is sufficiently separated from the contact. For either test, the maximum force just before the particle separates is the cohesive force, which is assumed to be primarily due to van der Waals force. The Hamaker constant, which is related to the magnitude of the cohesive force, and r1 and r$_2$, the local radii of curvature where the two particles meet, can then be used to find the cohesive attraction potential (Sanchez and Scheeres, 2014). Alternatively, cohesion may be determined in terms of surface energy using the JKR theory (Johnson, Kendall, and Roberts, 1971), where the pull-off force is modeled as:

$$F_{JKR} = \frac{3}{2} \pi \gamma \frac{r_1 r_2}{r_1 + r_2}$$

where γ is surface energy in J/m$^2$. To determine the Hamaker constant and surface energy, the shape of the particles (and thus their local radii) will be measured by optical microscopy, but more precise measurements can be made through SEM. It is highly desirable to repeat this test with multiple particles to test the dependence of adhesion on their local radii at point of contact.

## Sample preparation:

For both tests, a particle is epoxied to the cantilever tip. For the second test, another particle is also epoxied to the substrate. Once the particles are epoxied to the tip or substrate, they will be difficult to remove. It is also likely that the particles would be damaged if detached from the epoxy. All particles should be in the size range of 10–100 microns in diameter. The maximum particle size that can be used is related to the size of the AFM tip. Larger particles will require larger and stronger AFM tips.

It is important that the samples are kept as clean from contamination as possible. Thin layers of adhering volatiles can affect the cohesive properties.

For comparison, the samples should be tested both as-is and after heating and outgassing to expel possible volatile surface contaminants. This would also provide enough data to compare the results to numerical and theoretical predictions.

## Sample impact:

Heating and outgassing of the sample to remove surface contaminants may alter the chemistry of the sample. The particles used in this analysis would likely be broken apart and almost (or mostly) destroyed in the process of removing them from the AFM arm and substrate. Some material may be lost in this process or contaminated with material from the AFM needle and substrate. Some portion of the particles may be recoverable without significant contamination, though it is not likely that this portion will be significant. It is possible to chemically remove the particles from the epoxy using organic alcohols, such as acetone, which may have the outcome of preserving better the complete physical shape of the particles but at the sacrifice of potentially altering the chemistry.

## Data quality:

The uncertainty in this test will be primarily related to (i) attaching the particles to the AFM arm and substrate; (ii) errors in measuring the particle diameter; and especially (iii) sample surface contamination. Uncertainty in determined values of the Hamaker constant and surface energy will be related to estimates of the local radii of curvature at the point where the particles contact. Knowledge of the sample contamination history (and, if possible, some measure of the thickness of adsorbed volatile layers on the sample surface) could be used to estimate how applicable the measured values are to the adhesive properties of the samples when they were still on Bennu. This is why the approach heating and off-gassing of volatiles is proposed.

## Data products:

Derived values of the force versus proximity (distance), as well as the Hamaker constants and inter-particle and inter-surface potentials, and their associated uncertainties will be stored in simple ascii tables and/or MS Excel sheets. A series of optical images will also be provided to qualitatively show the morphology (angularity) of the particles and show how the equivalent radii were calculated. These products should include a reference to SEM images of the particles that were investigated as well, to better understand the morphology and to find the local radii of curvature where the particles were brought into contact.

## Facility(ies):

Arizona State University, Tempe, AZ, USA

At the Experimental Multiscale Mechanical Materials Laboratory (EM$^3$ Lab, PI Christian Hoover):

- Nanosurf Atomic Force Microscope, manufactured by Anton Paar, for determination of cohesive forces and topology. The AFM is encased inside an acoustic sound isolation chamber as well as supported by a vibration dampening system consisting of a piezo table, which dampens vibrations similarly to how noise-canceling headphones work, and a large granite slab to attract any waves that move through the table away from the testing head. The chamber is also equipped with humidity controls ranging from less than 1% to over 65% relative humidity.
- The laboratory personnel also have access to a SEM and optical microscopes with total objective zoom to 1000× magnification for characterizing the shape and surface roughness of the sample particles.

# 5.50 Angle of Repose Measurement

## Overview:

Angle of repose measurements primarily provide the means to calculate the angle of internal friction, and secondarily provide information about the bulk cohesive properties of particle assemblages. The angle of internal friction is a measure of the ability of a particulate assemblage to withstand a shear stress; it is a key parameter in models of rubble-pile deformation (e.g., Scheeres et al., 2019), surface mass wasting (Barnouin et al. 2022), and TAGSAM surface penetration (Ballouz et al. 2021, Walsh & Ballouz et al. 2022). The internal angle of friction is generally equivalent to the angle of repose, i.e., the maximum slope angle formed by a pile of particles, when cohesive forces can be neglected (larger particle sizes). When cohesive forces are still prominent (sand and smaller particles), angle of repose measurements also provide the means to validate with a much larger system of particles the results of planned cohesion measurement between individual particles.

## How it works:

The angle of internal friction is often assumed to be equal to the angle of repose for cohesionless, densely packed particles (≥~1 mm size; Metcalf, 1965). Although there are variations in the definition of angle of repose, the static angle of repose can generally be described as the steepest possible slope of an unconfined material, measured from the horizontal plane on which the material can be heaped without collapsing (Al-Hashemi and Al-Amoudi, 2018). The dynamic angle of repose can be measured after the pile has collapsed. A test measurement for angle of friction should not be done with very small particles (<100 μm), where yet undetermined cohesion would compete with gravity, to avoid the uncertainty.

Several methods exist for creating an angle of repose slope (e.g. Carrigy, 1970; Calle and Buhler, 2020)—for example, placing a collection of grains in a transparent rectangular container and tilting it until mass wasting event occurs, or pouring grains onto a sphere to form a cone-shaped heap. The angle can be measured manually with a protractor or similar device, or digitally by photographing the heap to create a 3D reconstruction.

The friction properties of the Bennu sample can be determined during the removal and processing of the returned sample at the curation facility by monitoring the mechanical behavior of particles as they are physically manipulated. These methods involve passive photo-documentation of curation activities during the preliminary examination of the sample. There are four distinct measurements that can be made: i.Monitor edges of tray after perf-ring removal. ii.Monitor on-set of granular flow during tilt-induced motions. iii.Monitor pile-up in sub-trays. iv. Monitor particle dynamics (collisions, sliding, rolling, twisting motions) during pour out. The first opportunity is during the removal of the perforated ring that surrounds the TAGSAM sample plate. If the particles are piled against one side of this ring, its removal may trigger a small mass wasting event which could produce a measurable angle of repose. This angle would be measured via photo-documentation of the slumped sample on the TAGSAM plate. Photographs from a side view (achieved directly or with mirrors in the glove box) can be used to measure the relative width and height of the pile, from which the angle of repose would be calculated. The second angle of repose measurement opportunity would be when the sample is poured out into sub-trays. Here, photo-documentation of the pour out would allow for the reconstruction of the angle at which granular flow initiates. If material flow initiates at the top-layer of the sample, then an internal angle of friction angle could be constrained. If material flow initiates at the sample/back-plate interface, then the sample-aluminum friction properties are measured, which is also valuable for understanding the interaction of man-made hardware with asteroidal materials.

Finally, the detailed shapes of the Bennu sample may also provide insight into the angle of friction through the measurement of particle roundness (Santamarina & Cho, 2004)

## Sample preparation:

No sample preparation is necessary as this is a passive measurement through photo-documentation of sample manipulation during preliminary examination.

## Sample impact:

No impact to the sample as this is a passive measurement through photo-documentation of sample manipulation during preliminary examination.

## Data quality:

The uncertainty in the angle of repose measurement will be a function of the pixel scale of the photographs, and the physical nature of the sample itself such as mass, volume, and particle size. The slope of the sample heap will not be perfectly planar, which will be the main contributing factor to uncertainty.

The result of the angle of repose measurement can furthermore be validated with a high-fidelity simulant (HFS). The simulant could later be used to build upon the results obtained with the Bennu sample by exploring different sample mass amounts and particle size frequency distributions. SPTAWG will facilitate the production of a HFS by providing Exolith Lab with the detailed physical characteristics of the Bennu sample.

## Data products:

Data products include text files describing angle measurements, photo sets used to extract these angles, and documentation of the camera specification and the geometry of their placement relative to the samples in the glovebox(es).

## Facility(ies):

NASA Johnson Space Center, Houston, TX, USA

- SRC disassembly glovebox for the side experiment
- TAGSAM disassembly glovebox for the ring removal measurement opportunity

## References:

Al-Hashemi H. M. and Al-Amoudi, O. S. (2018) A review on the angle of repose of granular materials. *Powder Technol.* 330, 397–417.

Ballouz, R.-L., et al. (2021) Modified granular impact force laws for the OSIRIS-REx touchdown on the surface of asteroid (101955) Bennu. Monthly Notices of the Royal Astronomical Society 507 (4), 5087-5105

Barnouin, O.S., et al. (2022) The formation of terraces on asteroid (101955) Bennu.

Journal of Geophysical Research: Planets 127 (4), e2021JE006927

Calle, C. I. and Buhler, C. R. (2020) Measurement of the angle of repose of Apollo 14 lunar sample 14163. The Impact of Lunar Dust on Human Exploration (LPI Contrib. No. 2141), Abstract #5030. Houston, TX.

Carrigy, M. (1970) Experiments on the angles of repose of granular materials. *Sedimentology* 14, 147–158.

Metcalf, J. (1966) Angle of repose and internal friction. *Int. J. Rock Mech. Min. Sci.* 3, 155–161.

Santamarina, J.C., Cho, G.C., 2004. Soil behaviour: The role of particle shape, in: Advances in Geotechnical Engineering: The SkemptonConference, Conference Proceedings. Thomas Telford Publishing, pp. 604–617. https://doi.org/10.1680/aigev1.32644.0035

Scheeres, D. J., et al. (2019) The dynamic geophysical environment of (101955) Bennu based on OSIRIS-REx measurements. *Nature Astronomy* 3, 352–361.

Walsh, K.J., Ballouz, R.-L., et al. (2022). Near-zero cohesion and loose packing of Bennu's near subsurface revealed by spacecraft contact. Science advances 8 (27), eabm6229.
255

# 5.51 Nanoindentation and Microindentation

## Overview:

Nanoindentation (NI) is a common technique for quantitative determination of mechanical properties of materials — including indentation modulus; indentation hardness; ductility; elastic work, inelastic work, and total work; creep curves; storage modulus; and loss modulus — at sub-micron length scales (tens to hundreds of nanometers). A diamond indenter tip is pressed into a sample surface until it has achieved the maximum desired depth (or force), at which point the load is decreased, leading to an elastic relaxation of the sample surface. The mechanical properties are then determined directly from the load versus displacement curve that was recorded during the test (see the figure). Individual grains and minerals may be probed with this technique using a well-polished geologic thin section. The test can be performed in a specific location, or can be done over an area in a grid pattern to determine how the properties change spatially. The properties can be spatially correlated with data from other tests, such as mineralogy in the sample.

Microindentation (MI) is similar but probes to depths between 1 and 15 microns. NI offers highly localized interrogation of the mechanical properties, whereas MI is more of a homogenized test. Functionally, they work the same way, described further below.

## How it works:

The test method for both NI and MI consists of pushing an indenter tip of known geometry and material properties, typically a diamond 3-sided pyramid Berkovic or a 4-sided pyramid Vickers tip, into the surface of the material. Though the Berkovic and Vickers probes are most common, the shape of the tip may be chosen based on expected sample strength. Softer samples (e.g., phyllosilicates) might be better analyzed with a spherical tip.

The mechanical properties of the indented material directly under and closely neighboring the probe are then extracted from the response of the material during the test. A typical test consists of a uniformly increasing loading phase, followed by a short holding period and finally unloading at a constant rate. The output from the test is a curve pictured below that measures the depth the probe traveled into the material ($h$) and the corresponding force at each depth ($P$). The plain strain modulus, also called indentation Modulus ($M$), is calculated from the unloading part of the test, when the material is relaxing. The indentation hardness ($H$), which is a measure of strength, is calculated by dividing the maximum force reached during the test by the area of the remaining footprint after the probe has been removed from the material.

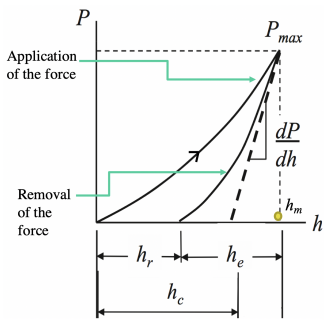

Nanoindentation may be used to characterize the strength properties of materials that contain nanoscale pores. This is possible if the size of the pores is smaller by at least a factor of ~5 than the width of the indentation footprint (which is typically about 7 times the depth of indentation). The indenter tip is pressed into the surface, causing some of the pores to collapse. It is then retracted and repressed into the surface repeatedly, each time triggering more collapse. The elastic response is calculated with each repeated indentation. The initial values will be representative of the strength properties of the porous material. As more pores collapse, the measured strength values will increase towards the value of the native, non-porous material. Because 0% porosity cannot be achieved, the true strength value of the material without pores can be extrapolated using models developed in the field of poromechanics (e.g. Bobko et al., 2011).

Indentation can also be used to determine other features of the material. The ductility can be calculated by dividing $M$ by $H$. The work performed by the machine can be determined by integrating the area under the loading and holding portion of the curve. The elastic work, responsible for the material relaxing, can also be quantified by integrating the area under the unloading curve. The inelastic work, the amount of work responsible for permanently deforming the material under the probe, is calculated as the difference between these two. If creep is of interest, this can also be determined by leaving the load on the sample surface for a long duration of time, then determining the compliance curve from the raw data using a well-established process in a solver such as MatLab. If dynamic analysis is of interest, nanoindentation can also be performed to get the storage modulus, related to the amount of elastic energy stored in the material, and the loss modulus, related to the amount of energy that goes into visco-elastic (or time-dependent) processes. Because these tests will be done across multiple scales, scaling relationships for these properties and features can be determined.

## Sample preparation:

The sample surface must be very well polished so that the RMS roughness is smaller by at least a factor of ~5 than the depth of indentation. Indentation depth is expected to be in the range of 50–350 nm, so the RMS roughness should be less than 10–50 nm. Longer-wavelength roughness (i.e.,, features that are wider than the indentation footprint) are less likely to affect the experiment, even if their height is greater than 1/5 the indentation depth. The sample should be at least 10 times as thick as the maximum depth of indentation. Thus, a standard 30-micron-thick geologic thin section is sufficiently thick to indent up to a depth of 3 microns, which is far deeper than would be practical or necessary for NI and is in the range for what is typical for MI. Given that the NI measurements are only sensitive to the properties of the material within the uppermost hundreds of nanometeres of the surface, they should be performed before any other measurements that might alter the surface (e.g., Raman spectroscopy). Because the MI indents go much deeper, to a few thousand nanometers, small imperfections due to these scanning techniques should be less influential.

## Sample impact:

MI and NI are destructive in that they leave indentation footprints on the sample surface that are approximately 7 times wider than the maximum indentation depth. The crystal structure of the material is disrupted to a depth of ~5 times the actual indentation depth. Typical indentation depths on the nanoscale are 50–250 nm and on the microscale are about 3–15 microns.

The size of the indentations left by NI are so small that this technique is commonly referred to as non-destructive.

The location of the indentations will be recorded spatially, including with magnified optical images, and provided as a metadata file that accompanies the sample. Because the machine is in a cooled room, the temperature is typically in the mid to low 70s (in Fahrenheit, or low 20s in Celsius). If exposure to humidity is a concern, the testing machines can be encased in a large glove box purged with nitrogen or helium gas.

## Data quality:

Before the test is performed, the machine is calibrated to the response of the indentation modulus ($M$) on a reference material. Typically, for minerals, clays, and polished porous cement, the standard deviation in $M$ is generally less than 7% and for $H$, less than 15%. For the three work quantities and creep curves discussed above, standard deviations of less than 15% are also expected.

The uncertainty can come from a few sources: (i) pores that lead to crack development; (ii) roughness that is too high; (iii) indenting that is too deep or too close to the edge of the sample; (iv) indenting on the boarder of two different phases, for example at the edge of a rock that is embedded in concrete; and (v) a tilted sample surface. A small tilt, with a slope of less than 3%, is acceptable. However, if the sample is tilted >3%, it can cause the probe to press at an angle, instead of perpendicular, to the sample.

If the sample is prepared flat and polished to an RMS discussed in sample preparation above, much of the standard deviations should be smaller than the aforementioned values. Because MI is on a much larger scale, the scatter is greatly reduced as it is not as negatively influenced by the uncertainty sources mentioned above. Normally, scatter to within less than 4% for $M$ and less than 9% for $H$ for work quantities and creep curves are seen on the microscale.

## Data products:

Derived values of strength and elastic properties and their associated uncertainties for each indentation site will be stored in simple ascii tables and/or MS Excel sheets. The location of each indentation site is in these files using a coordinate system defined by a few control indentations placed in an circular pattern along the perimeter of the sample surface. A series of optical raster images will also be provided to qualitatively show the locations of the indentations on the sample surface for quick reference.

## Facility(ies):

Arizona State University, Tempe, AZ, USA

In the Experimental Multiscale Mechanical Materials Laboratory (EM$^3$ Lab, PI Christian Hoover):

- NI: Ultra Nanoindentation Tester (UNHT$^3$) manufactured by Anton Paar for determination of nanoscale indentation modulus and hardness. The indenter is encased inside an acoustic sound isolation chamber and supported by a vibration dampening system consisting of a piezo table, which dampens vibrations similarly to how noise-canceling headphones work, and a large granite slab to attract any waves that move through the table away from the testing head. The chamber is also equipped with humidity controls ranging from less than 1% to over 65% relative humidity.
- MI: Micro Combi Tester (MCT$^3$) manufactured by Anton Paar for determination of microscale indentation modulus and hardness. The indenter is placed on a rigid frame, which is bolted into a granite slab to dampen vibrations. The entire assembly is isolated from the floor using air pistons, creating a floating table. The assembly is not isolated from the environment as small fluctuations in temperature and humidity are not as influential for indents that go several microns deep. If isolation is needed, the floating table is versatile enough to accommodate a sound- and climate-controlled chamber on top. The MCT$^3$ will also be used for grain compression experiments described in section 5.52 Compression test
- The laboratory personnel also have access an ECOMET/AUTOMET 250 Pro semi-automated polishing system for smoothing sample surfaces as needed for NI tests

## References:

Bobko, C. P., et al. (2011) The nanogranular origin of friction and cohesion in shale—A strength homogenization approach to interpretation of nanoindentation results. *International Journal for Numerical and Analytical Methods in Geomechanics* 35, 1854–1876.

# 5.52 Compression test

## Overview:

Compression testing involves squeezing a single mm to sub-mm sample particle between two rigid metallic pistons for quantitative determination of mechanical properties (elastic modulus, crushing strength, critical flaw length at failure, and fracture behavior). A microindentation machine is equipped with a "flat-punch" probe, which is a solid metallic cylinder that is pressed down onto the particle. The load (in millinewtons) is recorded as a function of displacement, and once the flat-punch probe has achieved the maximum desired displacement (or force), the load is decreased. During the test, either (i) the particle can sustain the prescribed load and remains undamaged; (ii) the particle cannot sustain the applied load and it crushes completely; or (iii) the particle can partially sustain the load, leading to partial crushing of the particle, where it remains partly intact. Depending on which of these occurs, the mechanical properties can be determined directly from the load versus displacement curve that was recorded during the test. The overall response of the sample can be understood in the context of mineralogy and structural defects detected in other tests.

## How it works:

A particle is placed on a mechanical metallic stage, which raises it up to an anvil. The particle is then compressed between these two surfaces. The mechanical properties of the particle are extracted from the response of the material during the test. A typical test consists of uniformly increasing the loading phase, followed by a short holding period, and finally unloading at a constant rate. Typically, one of the three outcomes highlighted above is desired. The compressive or crushing strength is then calculated using a simple function that uses the dimensions of the particle and the maximum force that caused the particle to crush, if applicable. The output from the test is a simple force versus displacement plot that measures the distance the anvils traveled ($d$) and the corresponding force at each distance ($P$).

The top anvil and the bottom substrate are both made of steel, whose material properties are well known. Therefore, the elastic response of the particle can be extracted using the elastic (linear) portion of the force displacement curve by applying the so-called "parallel model". An assumed value for Poisson's ratio is required to calculate the elastic modulus of the particles. In addition to measuring the elastic response, these tests may also be used to study the sample's fracture properties, or tendency to fragment. Using theoretical models, the fracture properties of each particle can be extracted from the experimental observations. Additionally, the probability of a particle fracturing can also be estimated.

The measurements are conducted at ambient atmospheric pressure and humidity. Because the machine is in a cooled room, the temperature is typically in the mid to low 70s (in Fahrenheit, or low 20s in Celsius), and the relative humidity is typically less than 20%. If exposure to humidity is a concern, the chamber containing the testing machine can be purged with nitrogen or helium gas.

## Sample preparation:

There are no specific sample polishing or grinding procedures needed. However, before a compression test is conducted, the size of the test particle must be measured. Mean diameter may be simply determined through a sieve test and/or with calipers. Shape and size may also be determined in greater detail by photogrammetry using photos taken with an optical microscope.

This test will work best on particles ranging from a few hundred microns to a few millimeters in diameter. Mechanical failure can occur because of flaws or imperfections in the surface, so compressive testing should be performed on particles before any other measurements that might alter or damage their surfaces. However, if prior destructive measurements must be performed first, then it would be best if their locations could be clearly marked to avoid putting the previously tested areas in contact with the substrate or the anvil.

## Sample impact:

The majority of compression tests are destructive in a sense that the test particles will be permanently deformed, cracked, and in some cases completely disaggregated into smaller fragments. However, all material should be recoverable. The maximum pressure experienced by a sample depends on the size and strength of the material and the applied load. Tests with simulants and meteorite materials could be performed ahead of sample analysis to estimate maximum pressure values, which can be used to predict whether phase changes would occur within the sample.

The samples will be exposed to standard atmosphere at room temperature during these tests and thus may become contaminated by moisture in the air unless the system is modified to include a containment box for gas purging.

## Data quality:

Before the test is performed, the machine is calibrated on the response of a spherical reference material. Deviation from the reference typically arises if a particle does not have a spherical shape. Though the output of the tests will be true responses, any mathematical models that calculate features or properties from that curve usually rely on the particle being as close to a spherical shape as possible. Thus, the particles should be judged for their shape before the testing takes place and only the ones closest to a sphere should be selected for testing. The error in the result due to non-sphericity can be estimated by repeating tests with samples of comparable composition and varying levels of sphericity. More sophisticated analyses of these experiments can also be performed using numerical techniques, such as finite element method, where the objective is to replicate the observed deformation and fracturing to more precisely determine the material properties.

The precise mineralogy and porosity will vary between test specimens. Increasing the number of test specimens will allow for the determination of the relationships between the mechanical response and the proportion of different mineral phases and pores.

## Data products:

The raw force versus displacement data as well as derived quantities (elastic modulus, crushing strength, critical flaw length at failure, fracture behavior, and their associated uncertainties) will be stored in simple ascii tables and/or MS Excel sheets. A series of optical raster images will also be provided to qualitatively show how the average diameter of the particle was calculated and the material appearance before and after the test.

## Facility(ies):

Arizona State University, Tempe, AZ, USA

At the Experimental Multiscale Mechanical Materials Laboratory (EM$^3$ Lab, PI Christian Hoover):

- Micro Combi Tester (MCT$^3$) manufactured by Anton Paar will be used for particle compression experiments. The MCT$^3$ is placed on a rigid frame and is bolted into a granite slab to dampen vibrations. The entire assembly is isolated from the floor using air pistons, creating a floating table. The assembly is not isolated from the environment as small fluctuations in temperature and humidity are not as influential for specimens this size. However, if isolation is needed, the floating table is versatile enough to accommodate a sound- and climate-controlled chamber on top. The MCT$^3$ will also be used for microidentation experiments using geologic thin sections, described in the section 5.51 Nanoindentation and Microindentation
- The laboratory personnel also have access to a scanning electron microscope (SEM) and optical microscopes with total objective zoom to 1000× magnification for characterizing the shape of the sample particles before and after compression experiments.

# 5.53 Seismic Velocities and Rock Ultrasonic Elastic Constants

## Overview:

To determine elastic and strength properties of the sample, the compression (p) and shear (s) wave velocities in rock may be measured using a pulse generator and p- and s-wave ultrasonic transducers. A suitably sized sample and frequency combination will provide the dilatational wave sample velocities for general application to rock bodies (as opposed to bar or rod velocities). The seismic velocities (in some cases coupled with rock density) may be used to calculate the elastic properties of rock samples, including Young's modulus, shear modulus, Poisson's ratio, Lamé's constant, and bulk modulus. The seismic velocities can be used to estimate unconfined compressive and shear strengths by comparison to relationships defined by other rock samples of similar mineralogy and microporosity. Also, the seismic velocity measurements may be used to search for anisotropy in rock samples; if anisotropy >2% exists, calculating the above elastic properties has greater uncertainty.

## How it works:

Determination of compression (p) and shear (s) wave velocities in rock samples using a pulse generator and ultrasonic transducers has been routinely done for several decades, particularly for engineering and exploration purposes; these "bench top" velocities have been widely validated through comparison to large-scale, in situ measured rock velocities. ASTM D-2845 (1995) describes typical procedures.

A rock sample is placed between two coaxially located transducers. For best results, the sample will have two parallel planar surfaces. Holding the transducers and sample in a jig aids in measurement efficiency. One transducer generates seismic waves to travel through the sample while the other transducer receives the waves. The interval between the transmission and received signals with an oscilloscope, together with the sample thickness, are used to determine sample velocity. Coupling of the transducers' generated sound waves to the rock surface is often aided by using a liquid couplant in general practice on terrestrial samples, but sufficient pressure will achieve measurable signals to avoid contaminating samples (and altering velocities in porous/permeable samples) whose context requires minimizing contamination. The pulse generator frequency is selected to provide optimum signals to pick first arrivals of the seismic waves on the oscilloscope. The sample size has to be considered: "…the travel distance of the pulse through the rock shall be at least 10 X the average grain size so that an accurate average propagation velocity may be determined" (ASTM D-2845) to obtain representative velocities.

## Sample preparation:

Aside from the cutting required to prepare the sample shape, seismic velocity measurement is nondestructive and, if performed in a clean environment, has no negative effect on samples measured. The equipment is portable and can be deployed and configured compatibly with clean room infrastructure such as a glove box.

## Sample impact:

Aside from the cutting required to prepare the sample shape, seismic velocity measurement is nondestructive and, if performed in a clean environment, has no negative effect on samples measured. The equipment is portable and can be deployed/configured compatible with clean room infrastructure such as a glove box.

## Data quality:

Measurement precision and reproducibility of p-wave velocities of several percent is achievable (ASTM D-2845), varying with the characteristics of individual rock lithologies. In comparison, s-wave velocity precision is poorer, as shear transducers produce some compressional wave energy, which obscures the first break of the shear wave.

The derived dynamic elastic constants are best compared to those of analogous natural and artificial materials for adjustment to static values. A substantial literature exists on elastic constant comparison. For example, Mashinskii (2003) noted that the dynamically derived Young's modulus is typically four to eight times larger than the static Young's modulus in natural rock samples. The derived dynamic elastic constants may also be compared to those of natural and artificial standards to estimate unconfined compression and shear strengths to <50% uncertainty (the accuracy of the estimate is influenced by the fidelity of analogs available for comparison).

## Data products:

Measurement data are tabulated in spreadsheets such as MS Excel. The derived p- and s-wave velocities are used along with sample density data to calculate the dynamic Young's modulus, shear modulus, Poisson's ratio, Lamé's constant, and bulk modulus. Coupled with sample mineralogy and microporosity, the velocity data allow estimation of unconfined uniaxial compression and shear strengths, as well as coefficients of restitution.

## Facility(ies):

University of Calgary*, Calgary, AB, Canada

- Pulse generator (Olympus Model 5077 PR with 35 MHz ultrasonic bandwidth)
- Oscilloscope (Tektronix DPO2014 with 100 MHz bandwidth)
- p & s wave ultrasonic contact transducers (13 mm diameter Olympus V103 and 153 are typical examples)
- Custom-built sample measuring jig to ensure coaxial alignment of transducers and controlled contact pressures

## References:

ASTM D-2845 (1995) Standard test method for laboratory determination of pulse velocities and ultrasonic elastic constants of rock. American Society for Testing Materials.

Mashinskii, E. (2003) Differences between static and dynamic elastic moduli of rocks: Physical causes. Geologiya I Geofizika 44, 953–959.

# 5.54 Capacitance Dilatometry

## Overview:

Capacitance dilatometry is a method for measuring the linear coefficient of thermal expansion in a solid material. The rate of expansion/contraction of a sample is measured along a single axis as a function of sample temperature between 5–300 K.

## How it works:

A parallelepiped or cylindrical sample is placed between a flat sample plate and a rounded probe that is fixed to a parallel plate capacitor. The capacitor is highly sensitive to changes in sample thickness that would cause the distance between the capacitor's plates to change (0.8 Å resolution). The temperature of the sample is varied and changes in the length of the sample along the axis of the capacitor probe and sample plate is recorded. The experiment is conducted in a liquid-helium cooled cryostat thermal vacuum chamber for thermal control. The linear coefficient of thermal expansion is simply determined from the rate of sample length change versus temperature change. This quantity is expected to be temperature dependent, i.e., the rate of sample expansion/contraction will vary depending on the instantaneous average sample temperature (e.g., Opeil et al., 2020).

## Sample preparation:

A sample should ideally be prepared to have two parallel, flat faces. Sample contraction/expansion is measured across the axis that is normal to these two parallel surfaces. The sample must be at least 2 mm in thickness along the axis of expansion/contraction and should be at least approximately 3 mm in width. It is expected that the sample can be cut to shape with a wire saw. The dilatometer is able to accommodate irregularities on the sample surface, as described in Schmiedeshoff et al. (2006), and as such additional polishing of the sample is likely unnecessary.

## Sample impact:

The cutting of the sample, described above, may be the most impactful step in the capacitance dilatometry measurement. A small force must be applied to the sample in order to hold it in place within the apparatus and to ensure continuous contact with the sample plate and capacitor probe. As such, there is a small risk that the sample could be unintentionally disaggregated, if it is extremely weak, as it is loaded into the apparatus. The sample will be exposed to the laboratory room air during experiment preparation. Measurements are performed within a thermal vacuum chamber at high vacuum, which could lead to sample outgassing. Finally, the sample will experience temperatures in the range of approximately 5–300 K, which could lead to potentially irreversible phase transitions (although they were not observed in CM chondrites, Opeil et al., 2020).

## Data quality:

The parallel plate capacitor that is used to determine changes in sample thickness has a resolution of 0.8 Å. Data are typically collected with a cooling/warming rate of 0.25 K/minute. Temperature is measured with a calibrated Cernox thermometer with a long-term calibration accuracy of ≤±0.180 K.

## Data products:

Data include raw capacitance and calibrated sample thickness values as a function of measured temperature. The coefficient of thermal expansion is provided alongside these values as a derived quantity. Data are provided in csv format.

## Facility(ies):

Boston College, Chestnut Hill, MA, USA

Capacitance dilatometer

- Instrument integrated into a Quantum Design Physical Property Measurement System (QD-PPMS)
- Also capable of applying a magnetic field to the sample, measuring expansion/contraction as a result.
- See Opeil et al. (2020) and Schmiedeshoff et al. (2006) for a detailed description of the instrument

# 5.55 Direct Shear Strength Measurement

## Overview:

A shear stress is applied to a cubic sample until it fails (subdivides) by shear fracturing. The ultimate shear strength of the sample is determined from the peak of the resulting shear stress versus shear displacement curve. The test is performed with the sample loaded into the cubic cavity of a two-sided shear box assembly. A load is applied to one side of the assembly in order to move it in a shearing motion relative to the other side, which is fixed. The sample is in the middle of the shear plane that results from the relative motion of the two sides of the assembly.

## How it works:

The direct shear strength measurement is performed using a shear box assembly/shear fixture. The assembly has two halves that may move independently, with a cubic cavity in the center that is split between the two halves. A cubic sample of comparable dimensions to the cavity is placed inside. When one fixture half is moved with respect to the other (using an electromechanical press), a shear force is created in one planar region within the sample. Increasing the force eventually leads to shear distortion and finally failure, typically on one plane across a known area. Applied force is digitally recorded throughout. The applied force may be converted to apparent shear stress with knowledge of the area of the plane of sample shear failure. The direct shear strength is then determined from the peak of the shear stress.

Sample size should be chosen so that it is representative of the bulk lithology — the general empirically established consideration is to use a sample whose dimensions exceed that of the average or largest grain by a factor of ten, although this can be complicated if a hierarchy of different "grain size" classes exist, e.g., mineral grains have one size range whilst comprising clasts of larger sizes in a breccia. Obvious flaws within a sample (e.g., veins or fractures) should also be avoided, as they may bias the result, unless the effects of the flaw are of interest to the investigator.

## Sample preparation:

The shear box assemblies currently in operation can accommodate test sample cubes of 2.5, 5 and 10 mm-size (other sizes could be manufactured at need). These cubes are typically cut from irregularly shaped clasts using a wire saw (to minimize sample loss to cutting). Samples are cut "dry"/without lubricant to avoid any unnecessary contamination; in the cutting process samples are held in aluminum, steel or Teflon grips. The cutting wires are steel embedded with natural diamonds, and typically kerfs are 0.14 – 0.15 mm width. Sample offsets are retained as well as dust generated by the cutting process. Cutting is typically done in a Class 100 laminar flow bench, but controlled atmospheres are possible. After cutting cubes are finished with fine grit sanding with dimensions and surface planarity monitored with a digital micrometer.

## Sample impact:

The cube sample fractures and slides on a rough plane at its center. The two remnants of the cube are intact and could be used for other analyses. A small amount of powder/fine fragments coats the sheared surfaces. The sample remnants and powder are not contaminated during the strength measurement for most purposes only having been exposed to the steel of the supporting fixtures. Samples are exposed to a somewhat larger range of potential contaminants during cube preparation (as noted above).

## Data quality:

The Test Resources 5 and 50 kN load cells are calibrated by the manufacturer before delivery and further monitored in house: the 5 kN load cell (at low loads) is compared to force meter measurements, while the 50 kN cell (at low loads) is compared to the output of the 5 kN using measurements of the same elastic sample. The digitally measured loading force is recorded to much higher precision (and reproducibility) than the natural strength variability found in rock samples (e.g., ASTM D5607 – 16).

## Data products:

Measurement data (e.g., force and displacement) are generated in csv files from the electromechanical press controller; these output data are then tabulated in spreadsheets such as csv or Excel.

## Facility(ies):

University of Calgary, Calgary, AB, Canada

The equipment at the U. of Calgary includes:

- Test Resources electromechanical press (Model 313Q) including 5 & 50 kN load cells and extensometer
- custom-built shear box/guillotine-style fixtures for 2.5-, 5-, and 10-mm cubes
- Well Diamond wire saw (model Well 3032)

## References:

ASTM D-5607 – 16 (2017) Standard test method for performing laboratory direct shear strength tests of rock specimens under constant normal force. American Society for Testing Materials.

## 5.57 Particle Size Frequency Distribution (PSFD) Measurement

### Overview:

The particle size frequency distribution (PSFD) is a measure of the number of particles versus size within the returned sample. The results may be presented in a raw form (e.g., table of individual particle sizes), binned and plotted form (e.g., cumulative PSFD), and an interpretive form (e.g., power-law fit, exponential-law, Weibull, etc. to the cumulative PSFD).

### How it works:

Particles are manually mapped in photos of the sample at various stages of the sample removal and processing. The longest axis of each identifiable particle is mapped as a line segment in the ArgGIS geographic information system software package. Typically, the largest particles, which are naturally less abundant, are counted first. Smaller particles are subsequently counted until a desired completeness limit (based on minimum particle size) is reached or until the remaining particles are too small to be adequately resolved (typically a three to five pixel sampling rule).

### Sample preparation:

No special sample preparation is necessary. However, the process of particle counting is greatly facilitated if the sample is spread into a thin layer so as to minimize overlap and/or burial of particles, which can affect the perceived size of a particles. Furthermore, images with high contrast (in hue and/or intensity) makes easier the process of distinguishing neighboring particles.

A scale bar within the field of view of the image is necessary to convert particle sizes in pixels to real units. A lens distortion map or equation is necessary to produce a global map of pixel scale.

### Sample impact:

The sample is not impacted by the actual PSFD measurement process. It may be affected by the sample preparation, where it is spread into a thin layer, during which time some particles may break due to manual manipulation.

### Data quality:

The accuracy of individual particle size measurements and the completeness limit of the PSFD will be a function of the pixel resolution of the camera and our ability to convert the pixel lengths to real lengths using an object with a known scale and knowledge of the camera lens distortion.

### Data products:

Particle line segments will be generated in ArcGIS (shapefile format), which are then exported as tabulated data in csv/ascii format (line end-point coordinates). Associated fields can be inserted inside the shapefile. Such fields can be populated with specific properties of each pebble/particle, such as color ratios, lithological class, etc. This could lead to the identification of SFD of different populations of particles.

### Facility(ies):

Johnson Space Center/University of Arizona

Quantitative Reflectance Imaging System (QRIS)

# 5.58 Lock-In Thermography

## Overview:

Lock-in thermography is non-contact method to directly measure the thermal diffusivity of a solid sample. This is gently heating the sample with a modulating laser and observing the phase lag of the resulting thermal waves with an infrared camera. The thermal conductivity of the sample may be extracted from the measured thermal diffusivity result if the sample density and specific heat are known.

## How it works:

A sample with a naturally flat or cut surface is heated periodically with a 638±5 nm laser with 7 μm beam diameter. The surface of the sample is heated by the periodic laser, creating a thermal wave that extends radially from the laser incidence point. This thermal wave on the surface of the sample is observed with an infrared camera. The thermal diffusivity of the sample is a direct function of the phase lag of the measured thermal wave:

$$\alpha = \frac{\pi f}{(d\theta/dr)^2}$$

where $\alpha$ is the thermal diffusivity, $f$ is the heating frequency, and $d\theta/dr$ is the first derivative of the phase lag as a function of distance from the heating point.

The heating frequency is selected specifically to avoid temperature wave reflection at the sample edge. As illustrated by the above equation, the thermal diffusivity result is independent of the intensity of the laser, the scattering properties of the sample surface, nor the sample emissivity. For more details on the measurement technique, see Ishizaki and Nagano, 2015; 2019 and Ishizaki et al. 2022.

Small-scale sample surface roughness may be ignored when it is less than the thermal diffusion length, $\sqrt{(\alpha/(\pi f))}$. For an example thermal diffusivity ($\alpha$) of 0.3 mm2/s and heating frequency is 4 Hz, the thermal diffusion length is about 150 μm. Larger-scale surface undulations can be ignored if they are less than the focal depth of the thermal camera (~15 μm).

## Sample preparation:

The sample must either have a naturally occurring surface that is approximately flat, or it must be cut to have a flat surface. No polishing or additional surface preparation steps are required. A 3d scan of the sample is needed so that the analyst can determine if the size and flatness of the sample surface is satisfactory.

The lock-in thermography system can accommodate samples as small as ~0.5 mm in all dimensions. The maximum sample thickness is ~10 mm and the maximum width is ~25 mm.

## Sample impact:

The sample is placed in a small vacuum chamber during the measurement and thus will be subjected to high vacuum conditions. The heating by the laser is modest (only a few kelvins) and thus the sample does not deviate far from room temperature. The sample temperature can intentionally be elevated to a desired value (≤350 K) in order to assess the temperature-dependence of the measured thermal diffusivity.

For measurements where the sample is heated from below and measured from above, it is placed on top of a sparse grid of Kevlar fibers that are stretched over the opening of a steel ring. For measurements where the sample is heated from the side and measured from above, it is simply resting on a stainless-steel stage.

## Data quality:

Ishizaki et al. (2019) summarize results of the system with calibration standards, where it is shown that measured thermal diffusivity values are within 6% of reference values.

The lock-in thermography technique was used to measure the thermal diffusivity of samples from Ryugu. The results were compared to another independent thermal diffusivity measurement method and found to be within 20% agreement (Nakamura et al., 2022).

## Data products:

Raw and calibrated data products are expected to include:

- Calibrated infrared images with timestamps
- Ascii files describing laser intensity and frequency setting as a function of time
- Ascii files containing 2d plot data of phase lag versus distance from heat source
- Ascii files containing polar plot data for phase lag and thermal diffusivity distribution
- Images converted to maps of phase lag, thermal diffusivity, thermal conductivity, and volumetric heat capacity on the sample.

## Facility(ies):

Nagoya University, Nagoya, Japan

Lock-in thermography system

- FLIR camera model SC5600
- InfRec ThermoHAWK H9000
- DCG Systems (FEI) ELITE
- FLIR camera model X6981

JAXA Institute of Space and Astronomical Science (ISAS), Sagamihara, Japan

Lock-in thermography system

- Infratec camera model Image IR® 8350 hp

## References:

Alasli, A., Fujita, R., and Nagano, H. Thermophysical Properties Mapping of Composites by Lock-in Thermography: Applications on Carbon Fiber Reinforced Plastics. International Journal of Thermophysics, 43, Article number: 176 (2022).

Ishizaki, T. and Nagano, H. (2015) Measurement of Three-Dimensional Anisotropic Thermal Diffusivities for Carbon Fiber-Reinforced Plastics Using Lock-In Thermography. International Journal of Thermophysics 36, 2577–2589.

Ishizaki, T. and Nagano, H. (2019) Measurement of 3D thermal diffusivity distribution with lock-in thermography and application for high thermal conductivity CFRPs. Infrared Physics and Technology 99, 248–256.

Ishizaki, T., Kawahara, T., Tomioka, K., Tanaka, S., Sakatani, N., Nakamura, T., and Nagano, H. (2022) Measurement of Thermal Diffusivity Distribution for Murray and Murchison Meteorites Using Lock-in Thermography. International Journal of Thermophysics 43, 97.

Nakamura, T., et al. (2022) Formation and evolution of carbonaceous asteroid Ryugu: Direct evidence from returned samples. Science, abn8671.

Elsila J. E., et al. . (2019) Developing an Efficient Coordinated Organic Analysis for Returned Samples. In 50th Lunar and Planetary Science Conference, pp. Abstract #1051. Lunar and Planetary Institute, Houston.

Eugster, O., et al. (2006). Irradiation records, cosmic-ray exposure ages, and transfer times of meteorites. In: *Meteorites and the Early Solar System II* (pp. 829-851). University of Arizona Press.

Friedrich, J. M., et al. (2018) Effect of polychromatic X-ray microtomography imaging on the amino acid content of the Murchison CM chondrite. *Meteoritics & Planetary Science 54*, 220-228.

Furukawa, Y., et al. (2019) Extraterrestrial ribose and other sugars in primitive meteorites. *Proceedings of the National Academy of Sciences* 116, 24440–24445.

Ghosh, A., et al. (2006) Asteroidal heating and thermal stratification of the asteroid belt. In: *Meteorites and the Early Solar System II* (pp. 555–566). University of Arizona Press.

Gillis-Davis, Jeffrey J., et al. (2017) Incremental laser space weathering of Allende reveals non-lunar like space weathering effects. *Icarus* 286, 1–14.

Gilmour, C. M., et al. (2019) Physical property measurement system and atomic force microscope thermal conductivity measurements of carbonaceous chondrites. In 50th Lunar and Planetary Science Conference, Abstract #2206. Lunar and Planetary Institute, Houston.

Glavin D. P., et al. (2006) Amino acid analyses of Antarctic CM2 meteorites using liquid chromatography-time of flight mass spectrometry. *Meteoritics & Planetary Science* 41, 889-902.

Glavin D. P., et al. (2010) The effects of parent body processes on amino acids in carbonaceous chondrites. *Meteoritics & Planetary Science* 45, 1948-1972.

Glavin, D.P., et al. (2018) The origin and evolution of organic matter in carbonaceous chondrites and links to their parent bodies. In: *Primitive meteorites and asteroids* (pp. 205–271). Elsevier.

Glavin, D.P., et al. (2019) The search for chiral asymmetry as a potential biosignature in our solar system. *Chemical Reviews*, 120, 4660−4689.

Glavin, D. P., et al. (2021) Extraterrestrial amino acids and L-enantiomeric excesses in the CM2 carbonaceous chondrites Aguas Zarcas and Murchison. *Meteoritics & Planetary Science* 56, 148–173.

Goodrich, C.A., et al. (2019) The first samples from Almahata Sitta showing contacts between ureilitic and chondritic lithologies: Implications for the structure and composition of asteroid 2008 TC 3. *Meteoritics & Planetary Science 54*(11), 2769–2813. Methods supplement.

Hage, F.S., et al., (2020) Single-atom vibrational spectroscopy in the scanning transmission electron microscope. *Science* 367, 1124-1127.

Hamilton, V. E., et al. (2019) Evidence for widespread hydrated minerals on asteroid (101955) Bennu. *Nature Astronomy* 3, 332–340.

Hamilton, V. E. et al. (2021) Evidence for limited compositional and particle size variation on asteroid (101955) Bennu from thermal infrared spectroscopy. *Astronomy & Astrophysics* 650, A120.

Hamilton, V. E. et al. (2022) GRO 95577 (CR1) as a mineralogical analogue for asteroid (101955) Bennu. *Icarus* 383, 115054.

Hanna, R. D. and Ketcham, R. A. (2017) X-ray computed tomography of planetary materials: A primer and review of recent studies. *Geochemistry* 77, 547–572.

Hartzell, C. M. (2019) Dynamics of 2D electrostatic dust levitation at asteroids. *Icarus* 333, 234–242.

Hartzell, C. E. et al. (2022) An Evaluation of Electrostatic Lofting and Subsequent Particle Motion on Bennu. *Planetary Science Journal. 3:85* doi:10.3847/PSJ/ac5629.

Hashiguchi, M. and Naraoka, H. (2018) High-mass resolution molecular imaging of organic compounds on the surface of Murchison meteorite. *Meteoritics & Planetary Science* 54, 452–468

Hayatsu, R. et al. (1980) Phenolic ethers in the organic polymer of the murchison meteorite. *Science*, 207, 1202-1204.

Hemmler, D., et al. (2018) Tandem HILIC-RP liquid chromatography for increased polarity coverage in food analysis. *Electrophoresis* 13, 1645–1653.
269